\documentclass[mnsc]{informs3}
\usepackage{booktabs} % For formal tables

\usepackage{mathtools}
\usepackage{subfig}
\usepackage{float}
\usepackage{endnotes}
\OneAndAHalfSpacedXI
\usepackage{url}            % simple URL typesetting

\usepackage{tcolorbox}
\usepackage{bbm}
\usepackage{dsfont}
\usepackage{amsmath,amsfonts,amssymb,commath,hyperref}
\usepackage{multirow}
\usepackage{natbib}
\usepackage{caption}
\usepackage[colorinlistoftodos,prependcaption,textsize=small]{todonotes}
\usepackage[shortlabels]{enumitem}
\usepackage{xspace}
\usepackage[multiple]{footmisc}

\usepackage{algorithm}
\usepackage[noend]{algpseudocode}
\newcommand{\EE}[1]{\mathbb{E}\left[#1\right]}

\newcommand{\PP}[1]{\mathbb{P}\left[#1\right]}
\newcommand{\parenthesis}[1]{\left(#1\right)}
\newcommand{\bracket}[1]{\left\{#1\right\}}
\newcommand{\squarebracket}[1]{\left[#1\right]}

\newcommand{\af}{\alpha^f}
\newcommand{\anf}{\alpha}

\newcommand{\qvec}{\mathbf{b}}
\newcommand{\ql}{b_l}
\newcommand{\qr}{b_r}
\newcommand{\mubar}{\mu^{\texttt{KS}}}
\newcommand{\munum}{\mu^{\texttt{KS}}}
\newcommand{\muemp}{\mu^{\texttt{EMP}}}

\newcommand{\gks}{g^{\texttt{KS}}}
\newcommand{\lonely}{$L$}
\newcommand{\popular}{$H$}
\newcommand{\flex}{flexible}
\newcommand{\nonflex}{regular}

\newcommand{\tree}{$\bar{G}$ }
\newcommand{\setl}{S_l}
\newcommand{\setr}{S_r}
\newcommand{\delete}{\Psi}
\newcommand{\degreef}[2]{z^{f}_{#2}(#1)}

\newcommand{\wonef}[1]{w_L^{f}(#1)}
\newcommand{\wonenf}[1]{w_L^{nf}(#1)}
\newcommand{\wonehatf}[1]{\hat{w}_L^{f}(#1)}
\newcommand{\wonehatnf}[1]{\hat{w}_L^{nf}(#1)}
\newcommand{\wtwof}[1]{w_H^{f}(#1)}
\newcommand{\wtwonf}[1]{w_H^{nf}(#1)}
\newcommand{\wtwohatf}[1]{\hat{w}_H^{f}(#1)}
\newcommand{\wtwohatnf}[1]{\hat{w}_H^{nf}(#1)}
\newcommand{\yonef}[1]{y_L^{f}(#1)}
\newcommand{\yonenf}[1]{y_L^{nf}(#1)}
\newcommand{\yonehatf}[1]{\hat{y}_L^{f}(#1)}
\newcommand{\yonehatnf}[1]{\hat{y}_L^{nf}(#1)}
\newcommand{\ytwof}[1]{y_H^{f}(#1)}
\newcommand{\ytwonf}[1]{y_H^{nf}(#1)}
\newcommand{\ytwohatf}[1]{\hat{y}_H^{f}(#1)}
\newcommand{\ytwohatnf}[1]{\hat{y}_H^{nf}(#1)}
\newcommand{\bin}{\text{Binom}}

\newcommand{\poisson}{\text{Poisson}}
\newcommand{\pf}{p^f}
\newcommand{\pnf}{p}
\newcommand{\vl}{v^l}
\newcommand{\vr}{v^r}
\newcommand{\budget}{B}

\newcommand{\lhs}{left-hand side }
\newcommand{\rhs}{right-hand side }

\newcommand{\xlb}{x^{lb}}
\newcommand{\xub}{x^{ub}}

\usepackage[capitalize]{cleveref}
\newtheorem{condition}{Condition}
\crefname{condition}{Condition}{Conditions}

\crefname{observation}{Observation}{Observations}
\usepackage[suppress]{color-edits}
\addauthor{sm}{olive}
\addauthor{df}{blue}
\addauthor{kz}{purple}

\usepackage{natbib}
 \bibpunct[, ]{(}{)}{,}{a}{}{,}%
 \def\bibfont{\small}%
 \def\newblock{\ }%

\TheoremsNumberedThrough     % Preferred (Theorem 1, Lemma 1, Theorem 2)
\ECRepeatTheorems

\EquationsNumberedThrough    % Default: (1), (2), ...
\MANUSCRIPTNO{}

\begin{document}
\RUNAUTHOR{Freund, Martin, and Zhao}
\RUNTITLE{Two-Sided Flexibility}  
\TITLE{Two-Sided Flexibility in Platforms}

\ARTICLEAUTHORS{%
\AUTHOR{Daniel Freund}
\AFF{Massachusetts Institute of Technology, Cambridge, MA\\ \EMAIL{dfreund@mit.edu}}%, \URL{}}
\AUTHOR{Sébastien Martin}
\AFF{Northwestern University, Evanston, IL\\ \EMAIL{sebastien.martin@kellogg.northwestern.edu}}
\AUTHOR{Jiayu (Kamessi) Zhao}
\AFF{Stanford University, Palo Alto, CA\\ \EMAIL{kamessi@stanford.edu}}
% Enter all authors
} % e

\ABSTRACT{Flexibility is a cornerstone of operations management, crucial to hedge stochasticity in product demands, service requirements, and resource allocation. In two-sided platforms, flexibility is also two-sided and can be viewed as the compatibility of agents on one side with agents on the other side. Platform actions often influence the flexibility on either the demand or the supply side. But how should flexibility be jointly allocated across different sides? Whereas the literature has traditionally focused on only one side at a time, our work initiates the study of two-sided flexibility in matching platforms. We propose  \dfreplace{a parsimonious}{an abstract} matching model in random graphs and identify the flexibility allocation that optimizes the expected size of a maximum matching. Our findings reveal that flexibility allocation is a first-order issue: for a given flexibility budget, the resulting matching size can vary greatly depending on how the budget is allocated. Moreover, even in the simple and symmetric settings we study, the quest for the optimal allocation is complicated. In particular, easy and costly mistakes can be made if the flexibility decisions on the demand and supply side are optimized independently (e.g., by two different teams in the company), rather than jointly. To guide the search for optimal flexibility allocation, we uncover two effects -- flexibility cannibalization and flexibility asymmetry -- that govern when the optimal design places the flexibility budget only on one side or equally on both sides. In doing so we identify the study of two-sided flexibility as a significant aspect of platform efficiency. 
}
\maketitle

\dfedit{\section{Introduction}

% =============================================================================
% PART 1: THE MATH PROBLEM
% =============================================================================

Motivated by applications in matching platforms, we introduce and study a model of flexibility in bipartite matching. Our model considers a bipartite graph with $n$ nodes on each side, where edges form randomly between nodes on opposite sides. We introduce two types of nodes: flexible and regular, such that edges incident to flexible nodes are more likely to realize. Specifically, with parameters $p_n^f > p_n$, the probability of an edge realizing between two regular nodes is $2p_n$, $p_n + p_n^f$ between one regular and one flexible node, and $2p_n^f$ between two flexible nodes (see \cref{fig:edge_prob}). Nodes on the left side of the bipartite graph realize as flexible with probability $\ql$ and nodes on the right side with probability $\qr$. 

\begin{figure}[ht]
    \centering\includegraphics[width=0.7\textwidth]{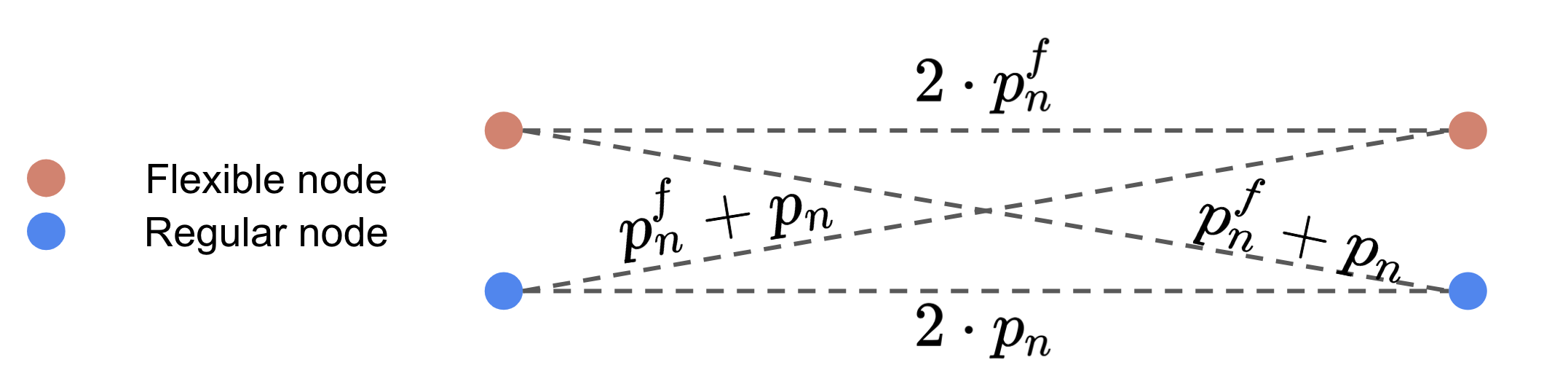}
    \caption{Illustration of edge probabilities between possible types of nodes. Flexible nodes (orange) have higher connection probabilities than regular nodes (blue).}
    \label{fig:edge_prob}
    % \vspace{-.25in}
\end{figure}

A decision maker controls $(\ql,\qr)$ subject to a \emph{flexibility budget} constraint $\ql + \qr = B$, where $B \in [0,1]$ corresponds to twice the total expected fraction of flexible nodes across both sides. The central question is: \emph{how should the budget be allocated between the two sides?} Should the decision maker concentrate all flexibility on one side (a \emph{one-sided} allocation with $\ql = B, \qr = 0$ or vice versa), spread it evenly (a \emph{balanced} allocation with $\ql = \qr = B/2$), or choose some intermediate design? Our goal is to understand how this higher-level allocation decision affects downstream matching outcomes {in the graph that results from edges realizing based on the above random process}.  Since platforms use various matching algorithms in practice, we study two metrics to evaluate the flexibility allocation: the expected size of a maximum matching $\mu$ (a complex combinatorial quantity that depends on the entire edge structure) and the {minimum} fraction of non-isolated nodes on one of the two market sides $\phi$ (which provides an algorithm-agnostic upper bound on matching performance and is more amenable to theoretical analysis). We study these quantities analytically in the traditional asymptotic random graph theory regime  where $n \to \infty$ and $p_n$ and $p_n^f$ scale as $\mathcal{O}(1/n)$; this yields sparse graphs in which the matching performance converges to well-defined limits. 

This is a deliberately simplified model. We study it not because  it captures the full complexity of real-world platforms, but because it isolates a fundamental question: \emph{when flexibility can exist on both sides of a bipartite system, how do the two sides interact?} Despite its apparent simplicity, the problem is mathematically rich. Characterizing maximum matchings in random graphs is a classical area with a long literature and many open questions; our work extends this line of research to graphs with a specific heterogeneous node type structure. As we will see, even this simple setting gives rise to nuanced behavior. Depending on the problem parameters $B$, $p_n$ and~$p_n^f$, two competing effects push towards either the one-sided or the balanced allocation being optimal: 

\begin{itemize}
    \item \textbf{Flexibility cannibalization.} 
    When flexible nodes exist on both sides, we expect them to form many edges between each other. Yet, intuitively, these edges are unlikely to significantly improve either metric: flexible nodes are likely to have high degree, at least compared to regular ones, so edges between them are unlikely to significantly decrease the number of isolated nodes. Similarly, we are unlikely to want to match them to each other rather than to hard-to-match regular nodes. Therefore, edges between flexible nodes are unlikely to increase the matching size and thus may be ``wasted.'' This effect favors instead concentrating flexibility on one side.

    \item \textbf{Flexibility asymmetry.} When all flexible nodes are on the left side, the regular nodes on the left side are ``disadvantaged'': they can only connect to regular nodes on the right side. Meanwhile, flexible nodes have a leg up by virtue of being flexible and regular nodes on the right side have potential flexible neighbors on the left side. This asymmetry can leave many regular nodes on the left side, the one with flexible nodes, isolated, which in turn limits the size of a maximum matching. To avoid this effect, one would want to spread flexibility across both sides. 
\end{itemize}

These two effects push in opposite directions, and exploring their interplay is the core contribution of this paper.
Crucially, these effects can only be managed because we allow for flexible nodes to occur on both sides, and thus interact. In contrast, if we only allowed for flexibility on one side, then there would not be any cannibalization, and asymmetry would be unavoidable. Only by allowing the decision-maker to allocate flexibility across sides, we identify a tradeoff between the two effects that needs to be managed. Indeed, this motivates our choice to frame our problem as an allocation question: by fixing the total budget~$B$ and optimizing over \emph{where} it is allocated,
we isolate effects that emerge purely from how flexibility is distributed.

Our goal, following the spirit of \citet{jordan1995principles}'s seminal work on process flexibility, is to uncover and characterize an important phenomenon in the design of flexibility structures. For $\phi$, we obtain a complete analytical characterization of the optimal allocation across all parameters $(p_n,p_n^f, B)$. For $\mu$, we provide analytical results for extreme parameters  where each effect is most pronounced, complemented by extensive numerical analysis across a wide range of parameters $(p_n,p_n^f, B)$. Together, these results paint a detailed picture of when and why different allocations dominate, and they suggest that the flexibility allocation is a first-order design decision in systems where flexibility can be influenced on both sides.

% =============================================================================
% PART 2: PLATFORM MOTIVATION
% =============================================================================

\subsection{Why Study This Problem?}

Our formulation is a deliberate abstraction of how flexibility operates in modern matching platforms. We now explain the connection. 

\noindent\textbf{Connecting the model to platforms.}
{Consider an interpretation of our model in the context of ride-hailing: nodes on one side represent drivers, and nodes on the other side represent riders.} An edge represents compatibility: whether a driver-rider pair can be matched at a given moment. For example, the driver and the rider should be close enough.

What does it mean for an agent to be \emph{flexible}? A flexible rider accepts longer wait times, {allowing for pickups from farther away or en-route drivers, which make her compatible with more drivers}. A flexible driver accepts rides he might otherwise decline—perhaps if they are further away or pay a lower rate. In both cases, flexibility increases the \emph{probability} of being compatible with agents on the other side, but does not guarantee compatibility, since factors like location vary randomly. This is precisely what our model captures: flexible nodes have higher edge probabilities, but edges still form stochastically. 

Platforms can encourage flexibility through incentives. \cref{fig:lever} shows two examples thereof from Lyft: ``Wait and Save'' offers riders a discount in exchange for longer wait times, while ``Ride Streak'' rewards drivers for completing consecutive rides. Both mechanisms increase the probability that the incentivized agent will match with agents on the other side. However, the platform can neither directly create edges nor set individual agents to be flexible. It can only influence the \emph{expected number} of agents who become flexible—through the design of incentives like discounts or bonuses. This is why our model optimizes over $(\ql, \qr)$, the probability that agents on each side are flexible, rather than over specific pairs of agents.

Of course, our model omits many features of real-world platforms: dynamic arrivals, strategic agent behavior, heterogeneous match values. We abstract from these to focus on the allocation question in its purest form. The hope is that the effects we identify—cannibalization and asymmetry—may persist in richer environments. Indeed, our numerics find them to be robust across several model extensions (\cref{sec:ext}), including spatial constraints, market imbalance, and heterogeneous preferences.

\begin{figure}[ht]
    \centering
    \subfloat[\centering ``Wait and Save'' Option]{{\includegraphics[width=0.25\textwidth]{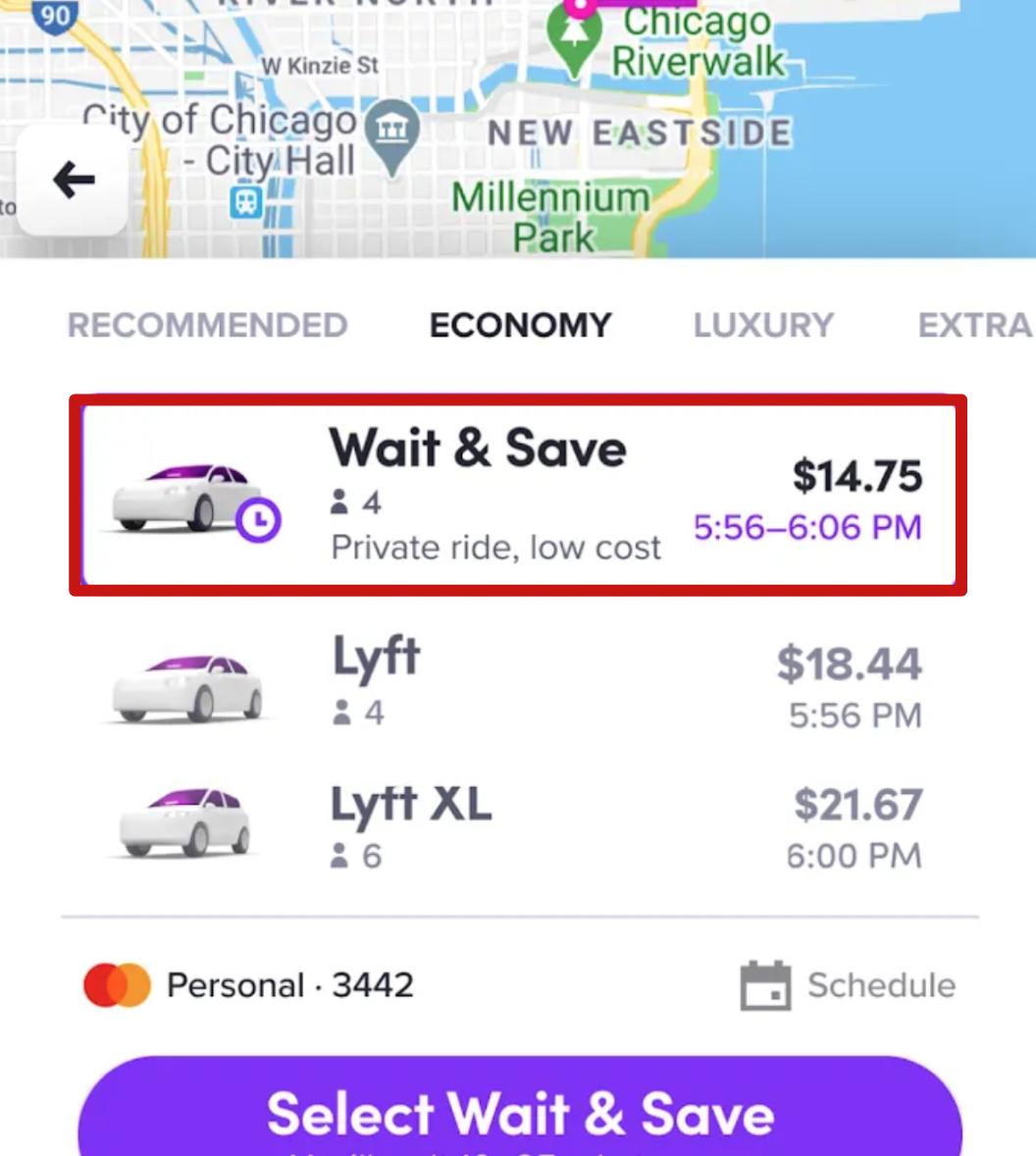} }}%
    \hspace{1.5cm}
    \subfloat[\centering ``Ride Streak'' Mode]{{\includegraphics[width=0.29\textwidth]{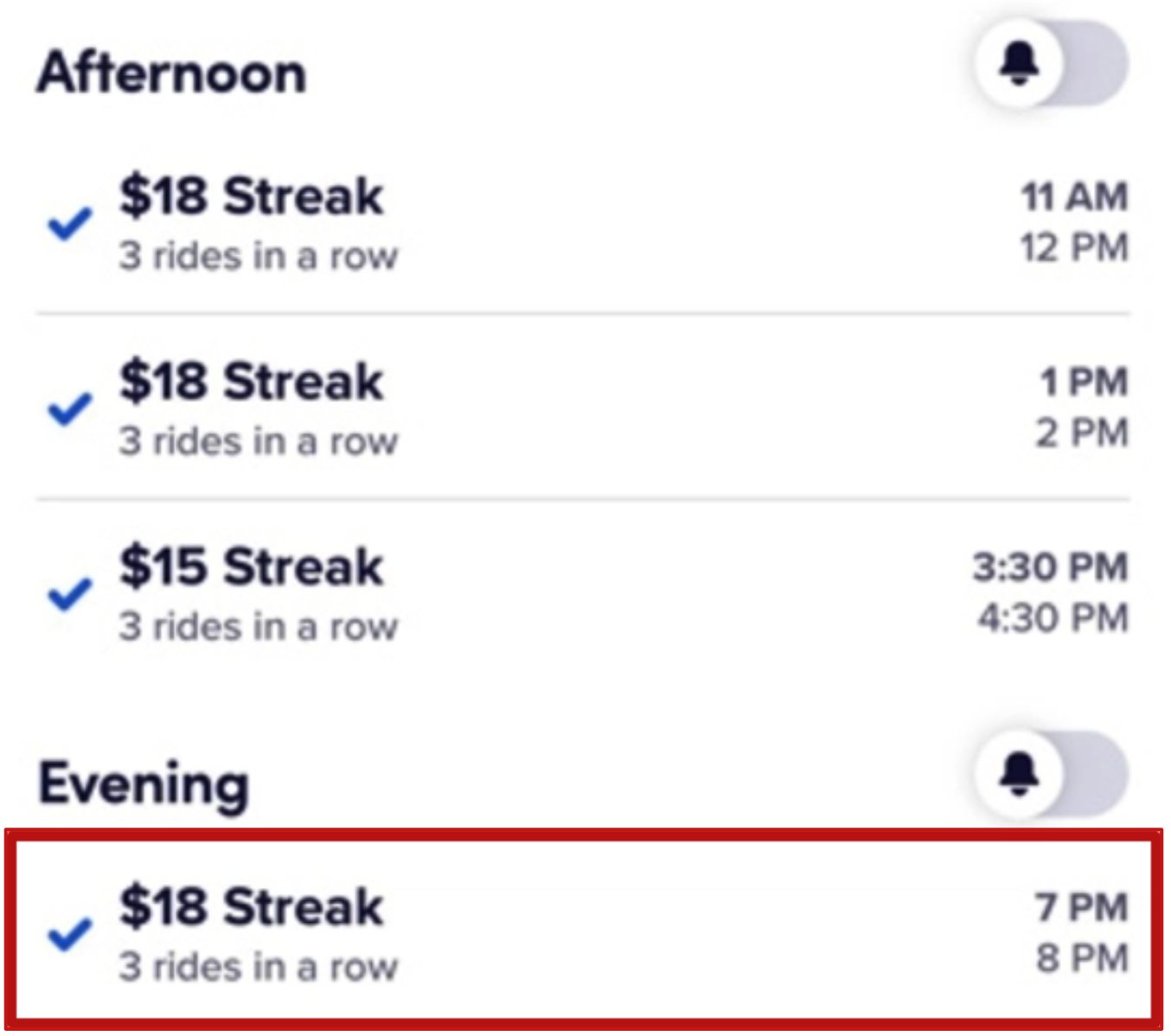} }}%
    \caption{Examples of flexibility incentives on the demand side (a) and supply side (b) of Lyft.}
    \label{fig:lever}
    \vspace{-.15in}
\end{figure}

\noindent\textbf{Two-sided flexibility.}
Crucially, flexibility levers exist on both sides of the market: platforms can incentivize riders, drivers, or both. This two-sided nature is central to our study: if flexibility can be encouraged on either side, how should a platform allocate its efforts? Throughout this paper, we use ride-hailing as our primary example. However, the \dfreplace{theory}{conceptual framework} applies more broadly. \cref{tb:platform} illustrates similar two-sided flexibility mechanisms in food delivery and freelancing {(see Appendix~\ref{app:examples} for detailed descriptions)}.

\begin{table}[ht]
\begin{center}
\begin{tabular}{|c|c|c|c|}\toprule
Industry & Platform(s) & Demand side lever & Supply side lever \\ \midrule
Ride-hailing & Lyft   & Wait and save   & Ride streak \\
Food delivery & Uber Eats &  No rush delivery & Surge incentives \\
Freelancing & Upwork &  Project catalog & Upwork academy \\
\bottomrule
\end{tabular}
\caption{Examples of platforms deploying flexibility incentives on both market sides.}
\label{tb:platform}
\vspace{-.15in}
\end{center}
\end{table}

\noindent\textbf{Differences from traditional studies of flexibility.}
In traditional operations, flexibility typically refers to supply-side resources handling multiple demand types—as in the ``long chain'' design of \citet{jordan1995principles}, where a central planner configures plant-product compatibilities \emph{deterministically}. \cref{fig:flexibility} illustrates the contrast: in manufacturing (a), the planner chooses specific edges; in platforms (b), {node} flexibility increases the \emph{probability of edges forming}. These settings differ in fundamental ways—stochastic versus deterministic compatibility, node-level versus edge-level control, two-sided versus one-sided flexibility—which we discuss in detail in \cref{sec:rel_work}.

\begin{figure}[ht]
    \centering
    \subfloat[\centering Process flexibility]{{\includegraphics[width=0.18\textwidth]{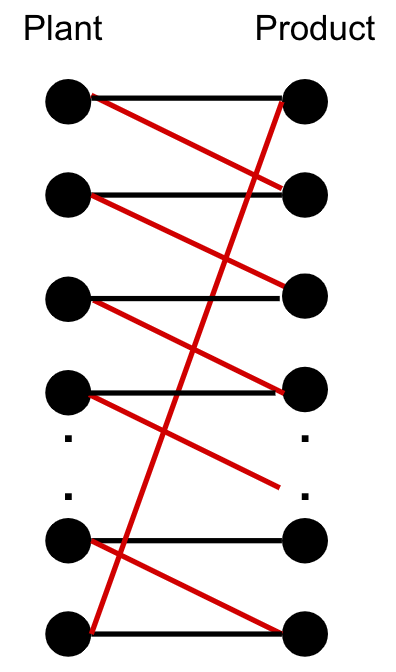} }}%
    \hspace{0.8cm}
    \subfloat[\centering Flexibility in platforms]{{\includegraphics[width=0.35\textwidth]{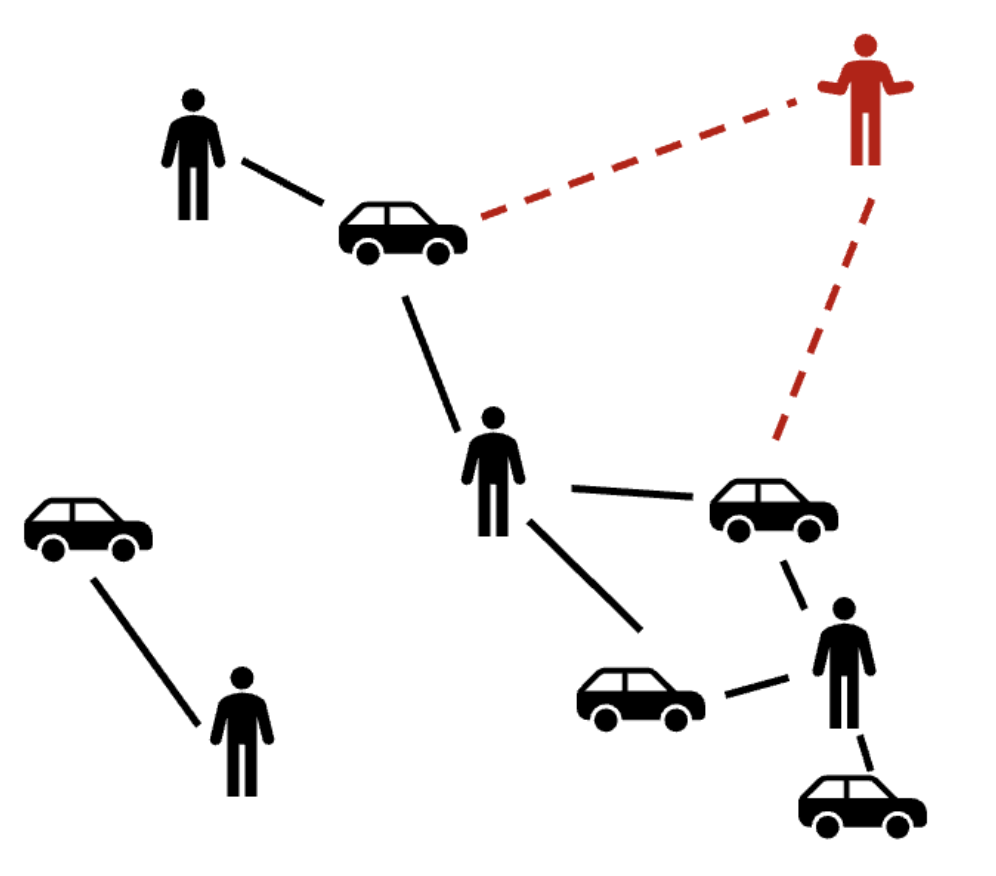} }}%
    \caption{Process flexibility in manufacturing (a) versus flexibility in platforms (b). In manufacturing, the red edges represent compatibilities configured deterministically. In platforms, flexibility increases the \emph{probability} of edge formation.}
    \label{fig:flexibility}

\vspace{-.35in}
\end{figure}

% =============================================================================
% PART 3: CONTRIBUTIONS
% =============================================================================

% \vspace{-.25in}

\subsection{Contributions}
% \dfcomment{this paragraph moves to the contribution subsection:
% }

Our work initiates the study of two-sided flexibility allocation in bipartite matching. We identify two competing effects—flexibility cannibalization and flexibility asymmetry—that govern the optimal allocation. Beyond characterizing \emph{when} each effect dominates, we strive throughout the paper to explain \emph{why}, connecting our theoretical results to realistic scenarios so that readers can build intuition for how these effects might manifest beyond our stylized model. Our contributions are threefold.

\vspace{.1cm}
\noindent \textbf{1. Structural insights: Identifying and characterizing the two effects.}
Our core conceptual contribution is to identify flexibility cannibalization and asymmetry as two forces governing the optimal flexibility allocation. We characterize the mechanisms behind each effect and provide intuition for when each dominates.

For the fraction of non-isolated nodes $\phi$, we obtain a \emph{complete analytical characterization} of the optimal allocation (\cref{thm:deg0_metric}): for any flexibility budget and any edge probabilities, $\phi$ is maximized by either the one-sided or the balanced allocation, and we exactly delineate the parameter regions where each dominates.
For the matching probability $\mu$, a complete characterization remains out of reach, but we establish analytical results in two extreme parameter regimes:
\begin{itemize}
    \item When half the nodes are flexible ($B=1$) and regular nodes connect to each other with probability zero, then the one-sided allocation strictly dominates (\cref{thm:compare_1}) no matter by how much flexibility increases the expected degree of a node. This is where cannibalization is most pronounced.
    \item When flexibility increases the expected degree of a node by a lot, regular nodes form a small but positive number of edges among each other,  and flexibility is scarce ($B < 1$), then the balanced allocation strictly dominates (\cref{thm:balanced_better}). This is where asymmetry is most damaging.
\end{itemize}
Beyond these regimes, we extend classical Karp-Sipser analysis to our random graph model, deriving nonlinear equations that closely relate to the asymptotic matching probability (\cref{sec:ks}). This theoretical foundation enables {a comprehensive computational study} across all parameters, {suggesting} that either the one-sided or balanced allocation is optimal everywhere—consistent with our analytical results for $\phi$.

\vspace{.1cm}
\noindent \textbf{2. Technical contributions: Comparing matching sizes across sparse bipartite random graphs.}
Our model gives rise to a class of sparse bipartite random graphs with heterogeneous node types. Characterizing maximum matching sizes in such graphs is a classical but difficult problem. {To compare the matching sizes across different flexibility allocations,} we develop three analytical approaches that apply in separate extreme parameter regimes: % \dfcomment{add a few words on ``''?}
\begin{itemize}
    \item A novel \emph{coupling construction} (\cref{sec:cannibalization}) that compares \emph{pairs} of graph realizations across allocations, allowing us to establish dominance without computing matching sizes explicitly. To our knowledge, this technique has not appeared in the random graph literature and may be of independent interest.
    \item {A \emph{2-stage matching approach} for which  {concentration bounds} ensure with high probability lower and upper bounds on the matching size (\cref{sec:asymmetry}).}
    \item A \emph{Karp-Sipser style analysis} (\cref{sec:ks}) extended to our heterogeneous random graphs, yielding nonlinear equations that characterize matching probabilities. We combine this with computer-aided proofs that solve approximately $6 \times 10^6$ equation systems with provable precision.
\end{itemize}

\vspace{.1cm}
\noindent \textbf{3. Practical implications: {The two sides must be optimized together.}} 
{The interaction of flexibility on the two market sides 
can easily cause inefficient market designs. 
In \cref{sec:experiment}, rather than assuming that the total flexibility level $B$ is a given parameter, we extend our model by letting the platform decide both the flexibility allocation $(\ql, \qr)$ and~$B$ (recall that $\ql + \qr = B$). In that part, we assume that the platform optimizes a profit function which incorporates both revenue from matches and a linear cost for~$B$.
In the resulting optimization landscape, suboptimal allocations can be locally optimal or saddle points. Moreover, these suboptimal allocations are often where coordinate-wise gradient descent methods converge to.
We would expect this to happen in practice when demand and supply teams experiment separately, a  common practice in industry. 
These interactions have first-order impact: our numerical results show that the wrong allocation can reduce the size of the maximum matching, i.e., revenue, by 8\% or more.
Worse, when incorporating the cost of flexibility, local optimization may converge to a solution that yields less than 10\% of the globally optimal profit.
When each side is optimized independently,  these interactions are missed entirely.}

In \cref{sec:ext}, we show numerically that our qualitative findings persist across several model extensions: spatial compatibility, market imbalance, and heterogeneous preferences.
They also persist when the matching outcome on the random graph is based on a greedy matching (see Appendix~\ref{sec:greedy}).  While we cannot claim that these effects appear in any specific real platform, their robustness across model variations suggests that they reveal fundamental insights about two-sided flexibility in bipartite matching.

\vspace{.1cm}
\noindent \textbf{Paper outline.}
In \cref{sec:model} we present our formal model. In
\cref{sec:cannibalization}, we study flexibility cannibalization and establish the dominance of the one-sided allocation in certain regimes.
\cref{sec:asymmetry} studies flexibility asymmetry and establishes the dominance of the balanced allocation in other regimes.
\cref{sec:ks} leverages a Karp-Sipser-style analysis to obtain numerical, and some analytical, results in other  parameter regimes.
\cref{sec:experiment} explores implications for platform experimentation and
\cref{sec:ext} demonstrates the robustness of our findings across model extensions. We conclude with open questions in
\cref{sec:conclusion}.

% =============================================================================
% PART 4: RELATED WORK
% =============================================================================

\subsection{Related Work}\label{sec:rel_work}

% \dfcomment{NEW VERSION OF RELATED WORK:}

Our work relates to several different streams of literature. The motivation for our study comes from both the literature on efficient flexibility structures in operations, and recent works on flexibility on platforms. Yet, on a technical level, our model is much more closely aligned with studies on matchings in random graphs. Lastly, some of our findings relate to recent work on platform experimentation and to classical work on manufacturing/marketing incentives. We discuss these connections in detail below.

\vspace{.1pt}
\noindent \textbf{Flexibility in operations.} Flexibility has a long history dating back to \citet{buzacott1986flexible} and \citet{fine1990optimal}. Much of this early literature studied the optimal \emph{amount} of flexible manufacturing capacity \citep{fine1990optimal, van1998investment, netessine2002flexible, chod2005resource}, optimizing over a single dimension on the supply side. The seminal work of \citet{jordan1995principles} introduced the ``long chain'' design: with $n$ plants and $n$ product types, adding just $2n$ carefully placed edges captures most of the benefits of full flexibility with $n^2$ edges. This finding has inspired extensive work on process flexibility in manufacturing \citep{iravani2005structural, akcsin2007characterizing, chou2011process, simchi2012understanding, chen2015optimal, desir2016sparse}. Similar approaches to flexibility designs have been used in call centers \citep{wallace2005staffing}, data centers \citep{tsitsiklis2017flexible}, and general parallel queueing networks \citep{bassamboo2012little}. What these works have in common, along with, e.g., literature on bipartite matching queueing systems \citep{afeche2022optimal,caldentey2025designing} and some papers related to stochastic matching \citep{feng2024designing, behnezhad2019stochastic}, is that they explicitly construct a bipartite graph on demand and supply, i.e., the graph topology.
{Our work shares the focus on optimal flexibility \emph{designs}, but differs in three key ways: (i) edges in our model arise stochastically rather than being chosen \emph{deterministically}, (ii) we optimize over the \emph{fraction} of flexible nodes on each side rather than over specific edges, and (iii) our model distinguishes between flexibility occurring on one side or another, and we focus on the structural properties that arise from their interplay.}\endnote{Because process flexibility optimizes over edges, it is not concerned with the side on which flexibility is created, i.e., the setting is agnostic as to whether a customer type becomes willing to accept a different product or a plant is equipped to produce a different product.
In contrast in our setting the interplay between flexibility on \emph{both} sides plays a role.}

\vspace{.1pt}
\noindent \textbf{Flexibility on platforms.}
There is a growing literature on flexibility mechanisms in online platforms such as ridehailing or e-commerce. Oftentimes, these papers do not explicitly frame their findings as relating to flexibility, and, as far as we know, all of these papers only study flexibility on one side at a time.
Supply-side studies in ridehailing examine driver repositioning and incentives \citep{ong2021driver} and priority modes \citep{krishnan2022solving}, both of which incentivize drivers to be more willing to serve any given ride. Demand-side studies analyze waiting mechanisms \citep{freund2021pricing} or subscriptions \citep{bergeroptimal} in ridehailing and opaque selling \citep{elmachtoub2019value,freund2025power} or flexible delivery windows \citep{zhou2021supply} in e-commerce --- all of which provide demand with options that make them more flexible with respect to the exact supply through which their request is fulfilled. However, these papers, and many like it, focus on adequately modeling and studying a particular mechanisms that operates on a single market side. In contrast, we abstract away from the details of particular mechanisms, and instead investigate how levers on two sides interact; this abstraction makes our contribution less practical, but it enables us to identify fundamental trade-offs   that one-sided analyses fail to capture.

\vspace{.1pt}
\noindent \textbf{Maximum matchings in random graphs.} The literature on maximum matchings in random graphs is predominantly occupied with two questions: in a given (random) graph, does there exist a perfect matching (that involves all nodes in the graph) with constant/high probability and, if not, what is the (expected) size of a maximum matching. Most famously,
\cite{erdHos1966existence} show that in a Erd{\H{o}}s-R{'e}nyi random graph on $n$ nodes, with average degree $d \geq \log n +\omega(1)$, a perfect matching exists with high probability; \cite{erdHos1963random} prove an analogous result for bipartite random graphs, which \cite{balachandran2019normalized} generalize to appropriately defined asymmetric bipartite random graphs.
For a slightly different class of random graphs \cite{bollobas1983matchings} prove a strictly stronger property (existence of Hamiltonian cycles) under a similar type of threshold. Many other types of random graphs have also been studied through the lens of \emph{does there exist a perfect matching} \citep{frieze1986maximum,gao2022perfect,karonski2020perfect}.

Our results are more closely connected to papers that tackle the second question. \citet{karp1981maximum}, and later \cite{aronson1998maximum}, consider Erd\H{o}s-R'enyi random graphs in 
a sparser regime in which the expected number of edges is linear in $n$. They develop a simple algorithm, nowadays referred to as the Karp-Sipser (KS) algorithm, that allows them to compute the expected matching size.
Later authors 
have extended the KS-style analysis to other models of random graphs, e.g., \cite{bohman2011karp}  for graphs in
which no vertices have degree smaller than~2. Recently,  these  results have been refined through non-algorithmic approaches such as the cavity method \citep{zdeborova2006number} and results have been obtained for random graphs with a fixed degree distribution  \citep{bordenave2010rank,bordenave2013matchings,salez2013weighted} (under some assumptions on the
distributions). Similarly, \citep{balister2015controllability} study the so-called ``configuration model'' in bipartite graphs. These models seem superficially similar to ours but neither strictly generalizes the other (e.g., our degree distribution is more restricted but our setting permits for two positive-degree nodes to have probability zero to have an edge between them). Moreover, our goal also differs from this stream of literature: rather than characterizing matching sizes, we aim to \emph{compare} them across allocations. Of course, characterizing/bounding matching sizes under different allocations is one way to conduct such comparisons, and we follow that approach in Sections \ref{sec:asymmetry} and Section~\ref{sec:ks}.\endnote{Our analysis in \Cref{sec:ks} relies on computer-aided proofs which also draws similarities to famous results in combinatorics such asthe four-color theorem \citep{appel1977solution, robertson1996new}. However, within the literature on maximum matchings in random graphs, we know of no other papers that use these tools towards provable comparisons of the limiting behavior. In that regard, \citet{gamarnik2006maximum} may be closest to our approach, though they (i) compute a single explicit solution to a nonlinear equation, whereas our grid search requires solving approximately $6 \times 10^6$ systems of nonlinear equations to provable precision, and (ii) they are also not concerned with comparisons of the type we focus on.} However, our analysis in \Cref{sec:cannibalization} exemplifies how the types of comparisons we are interested in can be done without explicitly computing expected matching sizes. Since we first published a draft of our paper, \cite{ameen2026uniformity} extended our line of work by studying service range allocation in a geometric random graph model (similar to ours in Section~\ref{sec:spatial}) where the budget on service ranges resembles our flexibility budget.

\vspace{.1pt}
%\smcomment{I really like it, great job -- let me know if you need help with the additional references (see your comment), but I feel conflicted adding too many about interference references, given how tenuous the link is to us (still worth writing this part, though!)}
\noindent \textbf{Platform experimentation and in-organization incentives.} Our discussion in Section \ref{sec:experiment}
 explores how different teams may independently experiment and thus fail to optimize the deployment of  different flexibility levers. 
Intuitively, this setup is reminiscent of both the classical literature on manufacturing/marketing incentives \citep{shapiro1977can,porteus1991manufacturing}, in which different teams make seemingly ``optimal'' decisions that jointly yield inefficiencies, and the recent literature on biases and interference in experimentation leading to poor operational decision-making \citep{johari2022experimental,farias2022markovian,holtz2025reducing}. However, the mechanism in our setup differs from both. In the former setting, inefficiencies arise from misaligned incentives across departments. To overcome these, solutions in the literature aim on align incentives by internally integrating functional teams \citep{weir2000empirical}, increasing the interface between manufacturing and marketing management \citep{hausman2002should}, or achieving strategic alignment between external positioning and internal arrangement \citep{henderson1999strategic}. In contrast, our setting already assumes that incentives are aligned, i.e., both teams optimize over the same objective function. In the latter setting, interference leads to biased estimates of the treatment effect of an intervention; to avoid this outcome, the literature aims to improve experimental design \citep{bojinov2023design} and develop better estimators \citep{shirani2024causal}. These solutions do not apply in our setting, in which inefficiencies occur even if we assume that the effect of each intervention can be exactly measured experimentally without any interference. Instead, the inefficiency arises from lack of visibility: absent joint experimentation, the teams fail to observe the interaction effects. This suggests that solutions need to focus on coordinating joint experimentation in addition to aligning incentives and avoiding interference.
}

\dfedit{\section{Preliminaries }\label{sec:model}

% \dfcomment{Removed the word ``parsimonious'' to better align with Rene's suggested positioning}

We study two-sided flexibility in platforms through a deliberately simplified matching model in a random bipartite graph $G$, wherein nodes on both sides are either \textit{\flex} or \textit{\nonflex}. Here, we formally introduce our model and objectives, which we then analyze throughout Sections \ref{sec:cannibalization}--\ref{sec:experiment}.} Section \ref{sec:ext} and its appendices will study closely related variants of our main model and show how our findings apply more broadly.

% \noindent\textbf{Random Graph {Generation}. }
\subsection{Random Graph Generation}
We begin by describing the random bipartite graph $G$ through which we model flexibility. Denote the set of nodes on the \lhs and \rhs of $G$ by $V_l$ and~$V_r$, respectively, and let $V = V_l \cup V_r$. 
\dfedit{With our platform interpretation in mind, we will often view} each node on one side as representing a demand agent (e.g., a rider) and each node on the other as representing a supply agent (e.g., a driver). We assume that there are $n\in \mathbb{N}^+$ nodes on each side \dfedit{(in \cref{sec:imbalanced}, we also consider imbalanced markets)}. and index them such that %index nodes from $1, ..., n$ on each side, so that
$V_l = \bracket{\vl_1, ..., \vl_n}$ and $V_r = \bracket{\vr_1, ..., \vr_n}$; $[n]$ denotes the set $\bracket{1,...,n}$.  Whether a node is \textit{\flex} or \textit{\nonflex} is determined by the decision variable $\qvec = \parenthesis{\ql,\qr} \in [0,1]^2,$ where $\ql$ and $\qr$ respectively specify the probability that a node on the \lhs and \rhs opts into being \textit{flexible}. Formally, for all $i\in[n], k \in \bracket{l,r}$ we independently sample a Bernoulli random variable $F^k_i \sim \text{Bernoulli}(b_k)$ and node $v_i^k\in V_k$ is \textit{\flex} if $F^k_i = 1$; otherwise, it is \textit{\nonflex}. 

We model compatibility, i.e., the presence of an  edge $(\vl_i,\vr_j)$, as an independent Bernoulli random variable~$R_{ij}$, such that the edge realizes if and only if $R_{ij} = 1$. In line with the examples in the introduction, we want edges with flexible nodes to be more likely to exist. Therefore, the probability that $R_{ij} = 1$  must be increasing in $F_i^l$ (the left node is flexible) and $F_j^r$ (the right node is flexible). We focus on the \textit{sparse} random graph regime, where the expected degree of each node remains constant as the size of the graph $n$ scales large.\endnote{\citet[Theorem 2]{erdHos1963random} proved that a random bipartite graph with $n$ nodes and i.i.d. edge probability $c \cdot \log(n)/n$ almost surely (as $n \to \infty$) possesses a perfect matching if $c>1$.}% Thus, subsequent studies \citep{karp1981maximum,balister2015controllability} often focus on the case where the edge probability is in $\mathcal{O}(1/n)$ and each node's expected degree is~$\mathcal{O}(1)$.} 
Specifically, we take non-negative parameters $\af>\anf$ as given 
and define $\pf_n = \af/n$ and $\pnf_n = \anf/n$. Then, the probability of an edge forming between two regular nodes is $2\pnf_n$, the probability of an edge between two flexible nodes is $2 \pf_n$, and the probability of an edge between a flexible and a regular node is $\pnf_n + \pf_n$. Formally, this can be written as

$$\PP{R_{ij} = 1|F^l_i,F^r_j} =  2\pnf_n +(F^l_i + F^r_j) \cdot (\pf_n - \pnf_n), \forall i,j \in [n].$$ 
Intuitively, $\alpha$ and $\alpha^f$ control how likely regular and flexible nodes are to be compatible with nodes on the other side, which is why we have $\alpha^f>\alpha$. We illustrate the resulting edge probabilities in \cref{fig:edge_prob}. This setting is symmetric, as flexibility on either side contributes equally to form an edge, with such contributions being additive and independent of $i$ and~$j.$ This additive edge probability is a careful modeling choice: indeed, note that the expected total number of edges in the graph is given by $n^2(2\pnf_n + (\ql+\qr)(\pf_n-\pnf_n))$. This is a function of $B=b_l+b_r$, which can be interpreted as the total ``flexibility level'' in the platform. Therefore, the expected number of edges in the graph is invariant to~$(\ql,\qr)$ for fixed $B$ -- it only depends on the total flexibility, not how we allocate it \dfedit{(our numerical extensions in Sections \ref{sec:spatial} and \ref{sec:imbalanced} relax this invariance assumption)}. As discussed later, this feature allows us to investigate effects that are driven by the \textit{distribution} of edges within a graph rather than the \textit{number} of edges.

\dfedit{
\subsection{Performance measures}
In the resulting $n \times n$ random bipartite graph $G$, we investigate two different measures that closely relate to the number of nodes a platform may be able to match (as a reminder, a matching is a set of edges in $G$ such that each node can be in at most one edge): 
\begin{itemize}
    \item 
\textit{{the} maximum matching size}: a maximum matching contains the largest possible number of edges across all matchings. We denote by the random variable $\mathcal{M}_n(\ql,\qr)$ the size of a \textit{maximum matching} and define the \textit{matching probability} $\mu_n(\ql,\qr)$ as the expected fraction of nodes that are part of a maximum matching, i.e., $\mu_n(\ql,\qr) = \EE{\mathcal{M}_n(\ql,\qr)/n}$. 
\item {\textit{{the minimum of both sides' }fraction of non-isolated nodes }}: a node is called isolated if it has degree 0 and consequently cannot be part of any matching. Thus, the fraction of non-isolated nodes in either~$V_l$ or~$V_r$ poses another upper bound on the resulting matching size that is agnostic to the matching process. We denote this quantity, the expected minimum (across both sides) fraction of non-isolated nodes in~$G$, as ${\phi}_n(\ql,\qr)$, which gives us that $\mu_n(\ql,\qr)\leq {\phi}_n(\ql,\qr)$.
%\dfcomment{why do we need both $\phi_n$ and ${\phi}_n$? I know this relates to what we talked about last time, where you were suggesting a different objective, and I am cool with that, but can't we just look at ``non-isolated nodes'' instead of doing this roundabout way that requires twice the notation?} \kzcomment{Yes; totally fine by me to just keep $\bar{\phi}$ to track the ``non-isolated nodes''; revised accordingly}
\end{itemize}
\dfedit{Our focus is the upstream flexibility decision; we aim for results that are robust to the downstream matching process. We study $\phi$ alongside $\mu$ for three reasons. First, $\phi$ is a relevant bound in practice because many platforms do not achieve the maximum matching due to decentralized decisions, limited  information available online, or suboptimal matching algorithms. Second, as we will see, many paramter regimes yield similar results for both metrics, demonstrating robustness to the matching technique (in Appendix~\ref{sec:greedy}, we further explore two greedy heuristics). Third, $\phi$ admits a complete analytical characterization in closed form, which complements our theoretical and analytical results for the combinatorially more complex metric $\mu$.}

Our theoretical and numerical results explore the asymptotic behavior of $\mu_n(\ql,\qr)$ and ${\phi}_n(\ql,\qr)$ with respect to $(\ql,\qr)$ as $n~\to~\infty,$ a conventional scale of interest in the study of random graphs. Accordingly, we denote ${\phi}(\ql,\qr)=\lim_{{n \to \infty}} {\phi}_n(\ql,\qr)$ and $\mu(\ql,\qr) := \lim\sup_{{n \to \infty}} \mu_n(\ql,\qr)$; this equals $\lim_{{n \to \infty}} \mu_n(\ql,\qr)$ when the latter exists.
% \smdelete{Maximizing $\mu(\ql,\qr)$ is a natural objective for a platform that makes an upstream flexibility decision with the goal of having as many matches as possible; indeed, if downstream matches are centrally decided with full information, the platform can exactly achieve $\mu(\ql,\qr)$.
% However, when matches form decentrally or matching decisions are made online, then a maximum matching may not be attainable downstream.  Nonetheless, $\mu(\ql,\qr)$ remains a natural proxy objective that is agnostic of the exact matching process. We view  maximizing $\phi(\ql,\qr)$ as another appropriate proxy objective for a matching platform that allocates chooses a flexibility design upstream with the goal of making as many matches as possible while remaining agnostic of the downstream matching process. Lastly, in Appendix \ref{sec:greedy}, we also explore numerically the impact of the flexibility allocation on the downstream matching size that arises under two greedy matching heuristics.}
} 

% \dfcomment{notice: I added the last sentence above.}

\dfedit{\subsection{Performance of Flexibility Allocations}} Our work examines whether and when platforms should invest in flexibility across both market sides. As we are particularly interested in the flexibility allocation problem (i.e., what is the best way to allocate a given amount of flexibility), most of the paper assumes a fixed flexibility budget $B=\ql+\qr \geq 0$. %\footnote{{We could study all allocations such that $\ql+\qr = B$ instead of limiting ourselves to the balanced and one-sided allocations. However, (i) these are the two most natural choices given the symmetry of the problem,  (ii) these allocations have nice symmetry properties that simplify our already arduous analysis, and  (iii) our numerical results suggest, as displayed in Figure \ref{fig:diagonal}, that the optimal allocation is always either the one-sided or the balanced allocation.}} 
At first sight, fixing the budget $B$ may not seem realistic. However, the problem of deciding both $\ql$ and $\qr$ can be decomposed into first choosing $B$ and then solving the allocation problem. Most of the paper focuses on the latter, as our goal is to study the interplay between the two types of flexibility. Then, Section \ref{sec:experiment} will study the general problem where the platform can choose $\ql$ and $\qr$ to maximize profit. 

Our main results compare two different allocations $(\ql, \qr)$ of the budget: the \textit{one-sided} allocation does not make use of flexibility across both market sides and instead invests the entire budget on one side, whereas the \textit{balanced} allocation allocates the flexibility budget equally on both sides.

\begin{definition}\label{def:flex_profile}
    For a given budget $\budget \in (0,1],$ the flexibility allocation $\qvec = (\budget, 0)$ or $\qvec = (0, \budget)$ is called the \textit{one-sided allocation}, whereas $\qvec = (\budget/2, \budget/2)$ is called the \textit{balanced allocation}.
\end{definition}

\begin{figure}[ht]
    \centering
    \includegraphics[width=0.9\textwidth]{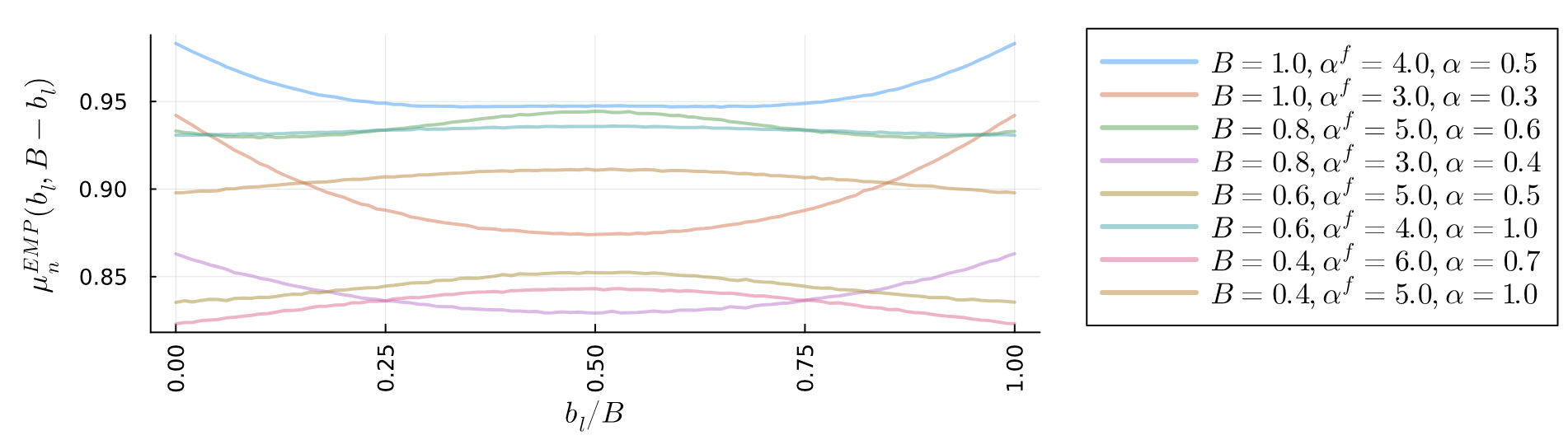}
    \caption{\dfedit{Illustration of $\muemp(\ql,\budget-\ql)$ with respect to $\ql/\budget$ for varying values of $\budget,\af$ and $\anf$. $\muemp$ is an empirical estimate of $\mu$, averaging over 10,000 realizations of a random graph of size $n=100$.}}
    \label{fig:diagonal}
    \vspace{-.25in}
\end{figure}

\dfedit{
Our first result shows that it is always one of these two allocations that maximizes the metric $\phi$. It is proven in Appendix \ref{app:deg0} and illustrated in Figure \ref{fig:deg0}. 
\begin{theorem}\label{thm:deg0_metric}
Either the one-sided allocation or the balanced allocation is optimal for $\phi$:
\begin{equation*}
    \phi(\ql,\qr) \le \max\left(\phi(\budget,0),\phi(\budget/2,\budget/2)\right)  \qquad \forall \af \geq \anf,\, \budget \in [0,1],\, \ql, \qr \ge 0 \text{ s.t. } \ql+\qr = \budget\
\end{equation*}
Furthermore, for any $B \in [0,1]$ and $\af \geq \anf$, we characterize when the one-sided allocation is optimal:
$$
\phi(\budget,0) \ge \phi(\budget/2,\budget/2) \iff e^{-\frac{\budget}{2}(\af-\anf)}\!\left(1-\tfrac{\budget}{2}+\tfrac{\budget}{2}\,e^{-(\af-\anf)}\right) - (1-\budget)-\budget\,e^{-(\af-\anf)}\ge 0;
$$%\dfcomment{Kamessi to check that Sebastien's ge 0 is not supposed to be le 0...}
In particular, when $B<2/3$, the balanced allocation always maximizes ${\phi}(\ql,\qr)$ \dfedit{and when $B=1$ the one-sided allocation maximizes ${\phi}(\ql,\qr)$}.
\end{theorem}
}
% \begin{theorem}\label{thm:deg0_metric}
% For any $B \in [0,1]$ and $\af \geq \anf$,  the one-sided allocation maximizes ${\phi}(\ql,\qr)$ if and only if
% $$
% e^{-\frac{\budget}{2}(\af-\anf)}\!\left(1-\tfrac{\budget}{2}+\tfrac{\budget}{2}\,e^{-(\af-\anf)}\right) - (1-\budget)-\budget\,e^{-(\af-\anf)}\smedit{\ge 0};
% $$
% otherwise, the balanced allocation maximizes ${\phi}(\ql,\qr)$. In particular, when $B<2/3$, the balanced allocation always maximizes ${\phi}(\ql,\qr)$.
% \end{theorem}
\dfedit{
\begin{figure}[ht]
    \centering
    \includegraphics[width=0.6\textwidth]{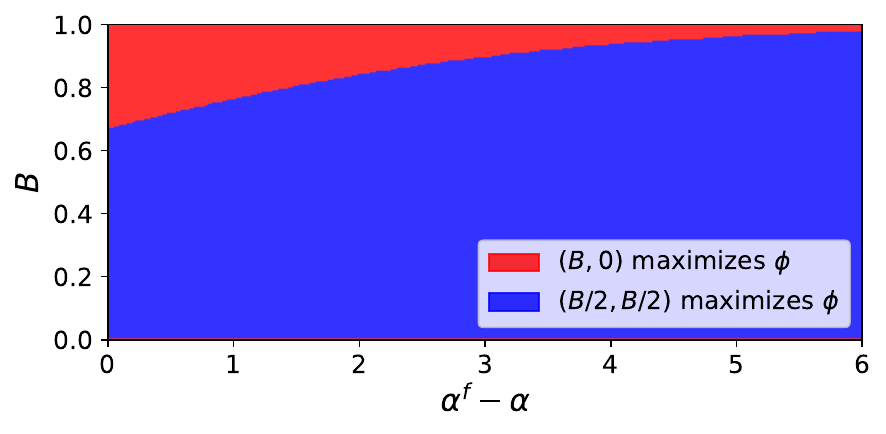}
    \caption{Illustration of the regions where the one-sided (resp. balanced) allocation maximizes $\phi(\ql,\qr)$.}
    \label{fig:deg0}
    \vspace{-.15in}
\end{figure}

Numerically, we similarly identify that the maximizer for the maximum matching $\mu$ is always found at either the one-sided or the balanced allocation --- \Cref{fig:diagonal} illustrates this for eight different parameter combinations  of $B,\anf,$ and $\af$ that yield a wide range of different behaviors in terms of how the flexibility allocation affects the matching size. 
% Since the expected number of edges in our model is invariant to different allocations of a fixed budget $B$, the differences in the objectives, must be driven by the \textit{distribution} of edges within a graph rather than the \textit{number} of edges. We will uncover that this holds true because some allocations of flexibility yield many realized edges that do not reduce the number of isolated nodes or produce additional matches.

We know that, for a fixed budget $\budget$, all flexibility allocations have the same expected number of edges. Therefore, what makes an allocation optimal must be its influence on the \textit{distribution} of edges within a graph rather than the \textit{number} of edges, and how this distribution influences the maximum number of matches (for~$\mu$) or the number of isolated nodes (for $\phi$).
%Specifically, we uncover two effects that are driving forces for when each flexibility allocation dominates. 
In the next two sections, we uncover two effects -- \emph{flexibility cannibalization} and \emph{flexibility asymmetry} that help explain \emph{why} either the one-sided or the balanced allocation may be superior; understanding the mechanisms behind these effects will help us see where they might apply beyond our simplified model.}

\begin{figure}
    \centering
    \subfloat[\centering $\budget = 0.6$]{{\includegraphics[width=0.48\textwidth]{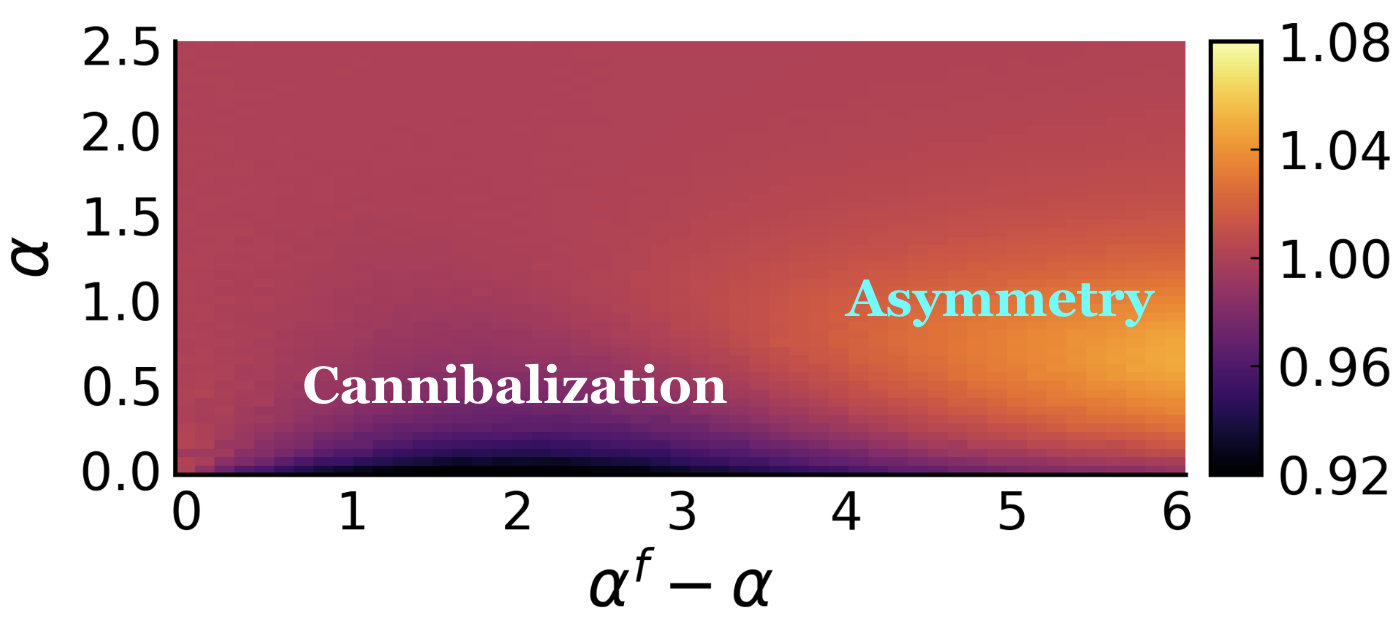} }}%
    \hspace{0cm}
    \subfloat[\centering $\budget = 1$]{{\includegraphics[width=0.47\textwidth]{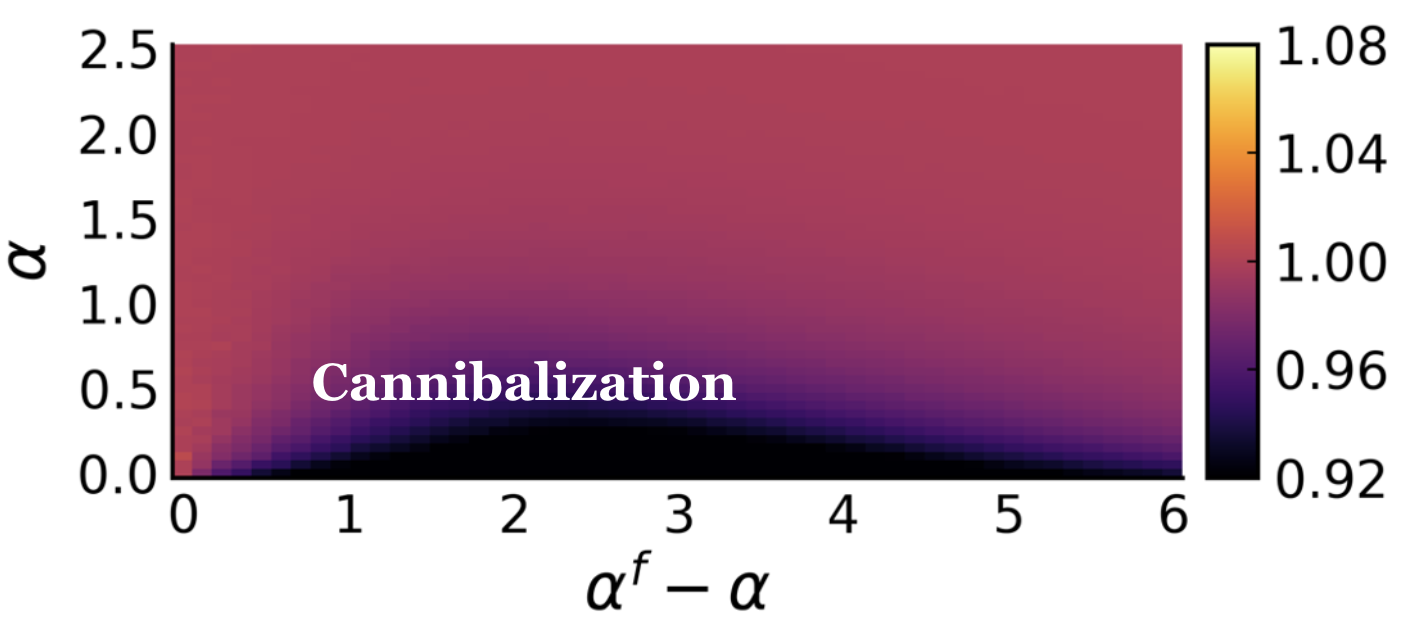} }}%
    \caption{The heatmaps present values of $\frac{\muemp(\budget/2,\budget/2)}{\muemp(\budget,0)}$ for varying $\anf$ and $\af - \anf$ and highlight the parameter regimes where the ratio is highest or lowest due to the dominance of the flexibility cannibalization or the asymmetry effect. \dfedit{$\muemp$ is an empirical estimate of $\mu$, averaging over 10,000 realizations of a random graph of size $n=100$.}% and identify regions (I)-(III) where the two allocations have comparable performances.
    }
    \label{fig:ratio}
    \vspace{-0.1in}
\end{figure}

\section{Flexibility Cannibalization}\label{sec:cannibalization}
\dfedit{Under any flexibility allocation that is not one-sided, there may be flexible nodes on both sides, and these nodes may share edges. We would expect such edges to not be as useful as edges involving at least one regular node. This is because flexible nodes typically have a larger degree than regular nodes, and thus edges to regular nodes are more likely to increase the number of nodes that can be matched or reduce the number of isolated nodes. Therefore, as the expected total number of edges is fixed (given $B$), edges between flexible nodes cannibalize the existing benefit of these nodes being flexible (they could have made an edge with regular nodes instead);  we refer to this effect as \emph{flexibility cannibalization}. Naturally, the one-sided allocation minimizes such cannibalization by avoiding all edges between flexible nodes.}

\begin{figure}[h]
    \centering
    \includegraphics[width=0.8\textwidth]{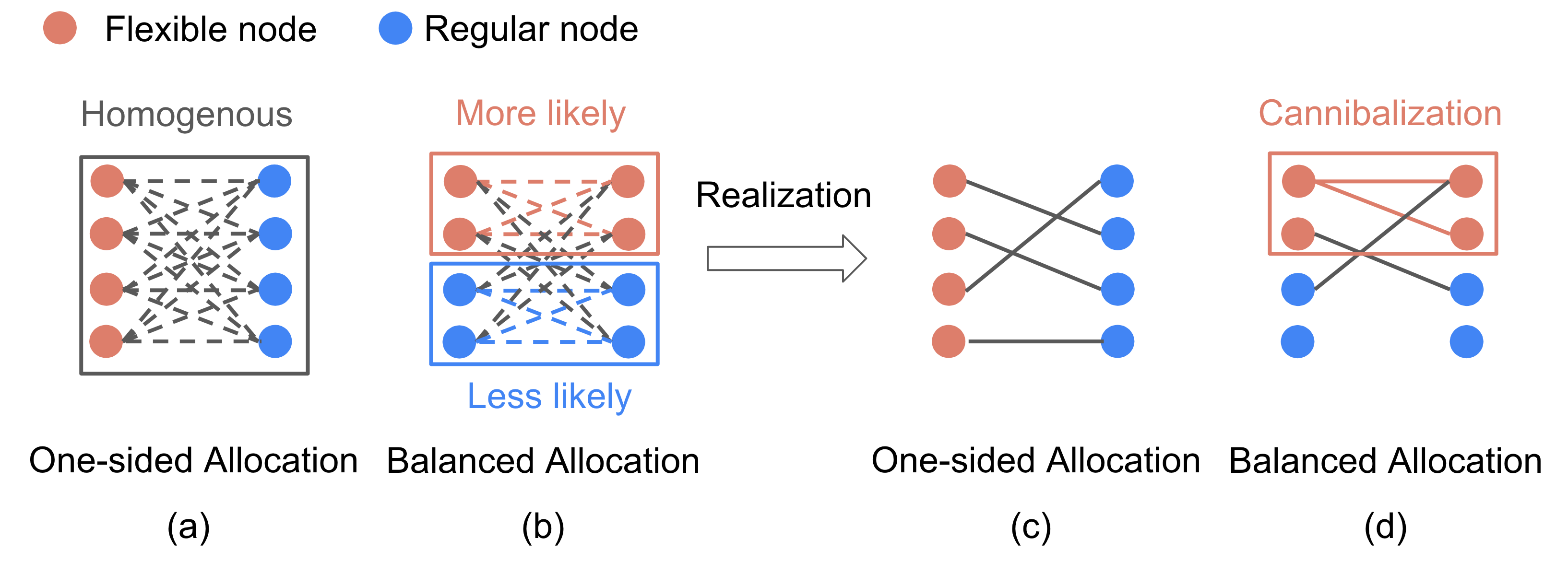}
    \caption{Illustration of flexibility cannibalization. Plots (a)-(b) compare edge probabilities and plots (c)-(d) illustrate possible realizations of one-sided and balanced allocations. In the balanced allocation, edges are more likely to realize in the upper subgraph of flexible nodes than in the lower subgraph. Consequently, despite plots (c) and (d) containing the same number of realized edges, (d) leads to fewer matches due to the cannibalization of edges in the upper subgraph.}
    \label{fig:intuition1}
    \vspace{-0.15in}
\end{figure}

The left side of \cref{fig:intuition1} illustrates how flexible nodes tend to have a higher average degree in the balanced allocation than in the one-sided allocation. In the balanced allocation, it shows that edges concentrate in the subgraph of flexible nodes. The right side of the figure shows that this may create edges that are incident to the same flexible nodes, which \textit{cannibalize} each other while leaving many regular nodes unmatched.
% \vspace{-.1in}

This intuition is confirmed with a simple computation: in expectation, the average degree of flexible nodes in allocation $(\ql, B-\ql)$ is $(\af + \anf) + (\af - \anf) \cdot 2 \ql (B-\ql)/B$. The first term,  $(\af + \anf)$, is independent of~$\ql$, whereas the term $(\af - \anf) \cdot 2 \ql (B-\ql)/B$  captures the contribution of edges between flexible nodes. This contribution is maximized when $\ql=B/2$, i.e., in the balanced allocation, which, therefore, is particularly likely to have edges incident to flexible nodes cannibalizing each other. Of course, this is just a description of the flexibility cannibalization effect, not proof that the balanced allocation is suboptimal \dfedit{for either objective}. Before diving into the mathematical formalization to understand when this is the case, we now provide an example of how flexibility cannibalization may arise in practice.

\dfedit{This effect, while derived from our simplified model, may play out in real platforms. To illustrate, consider} the ride-hailing platform from the introduction and \cref{fig:lever}. It employs flexibility incentives on both market sides: ``Wait and Save'' on the demand side and ``Ride Streak'' on the supply side -- this can be interpreted as the balanced allocation in our model. Flexibility cannibalization occurs when flexible drivers and flexible riders end up clustered in the same area, which often happens naturally due to the stochastic nature of supply and demand. The platform then has no choice but to match flexible drivers with flexible riders, as there are no regular alternatives around. The platform also has difficulty finding feasible matches in other areas without flexible drivers and riders. In our model, this is analogous to edge realizations such that flexible nodes form edges with each other rather than with regular nodes in the balanced allocation.
In that situation, the platform has to pay twice the cost of flexibility for each flexible-flexible match (Wait and Save discount and Ride Streak bonus), and this flexibility does not help with the harder-to-match regular nodes. In contrast, by incentivizing only one side of the market, the platform avoids ever ending up paying twice for flexibility, and all the flexibility is guaranteed to be used to help match regular nodes.

\dfedit{Returning to our theoretical results, consider the case $B = 1$ and $\anf = 0$, where half the nodes are flexible and regular nodes cannot match with regular nodes. Theorem \ref{thm:deg0_metric} (recall Figure \ref{fig:deg0}) already implies that the metric $\phi(\ql,\qr)$ is maximized by the one-sided allocation in this regime. In \Cref{thm:compare_1} we extend our understanding of this regime by showing that the one-sided allocation also outperforms the balanced one under  objective $\mu$. The proof constructs a coupling that specifically exploits flexibility cannibalization.}% to prove that the one-sided allocation dominates the balanced~one.}

\begin{theorem}\label{thm:compare_1}
    If (i) $B = 1$ (``half of the nodes are flexible"), and (ii) $\anf = 0$ (``no edges between regular nodes"), then \dfedit{$\mu(B,0) - \mu(B/2, B/2) \geq \parenthesis{\af}^3/2^5 \cdot e^{-7\af}$.} %$\mu(B,0) \geq \mu(B/2, B/2)$.
\end{theorem}

\begin{figure}[ht]
    \centering
    \includegraphics[width=0.6\textwidth]{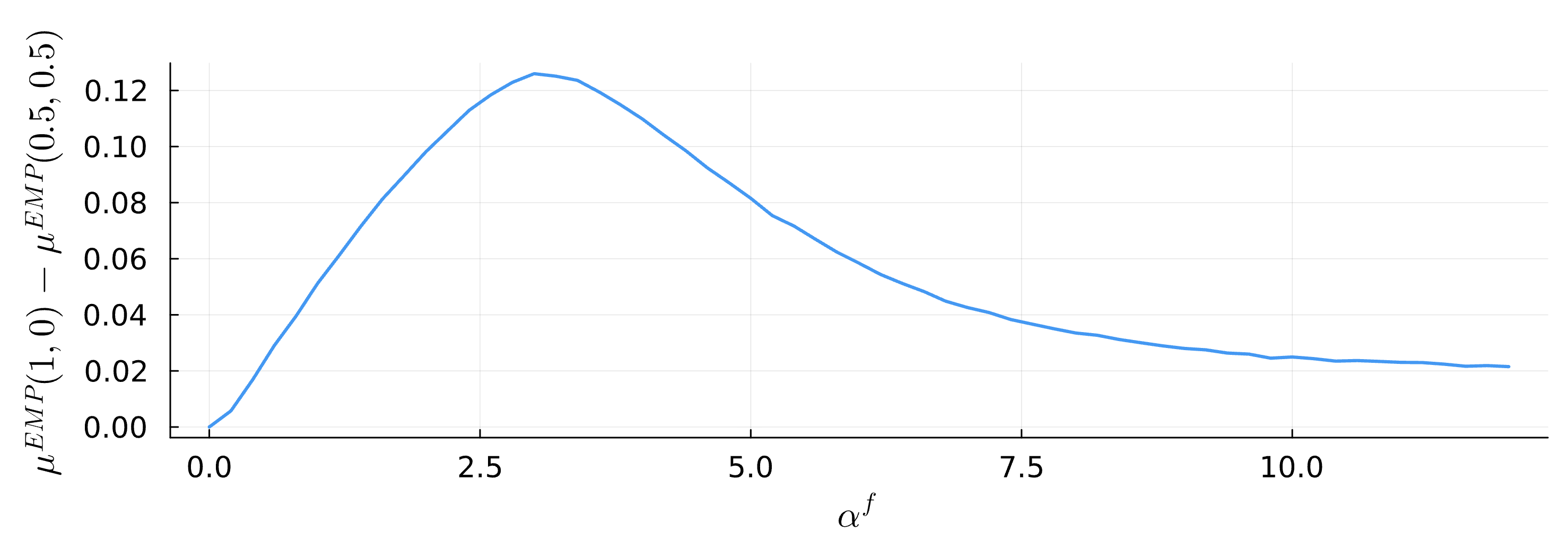}
    \caption{The difference between one-sided and balanced allocations for varying $\af$ when $B = 1$ and $\anf = 0$.}
    \vspace{-0.2cm}
    \label{fig:alpha0b1}
    \vspace{-.1in}
\end{figure}

\begin{figure}[ht]
    \centering
    \includegraphics[width=0.8\textwidth]{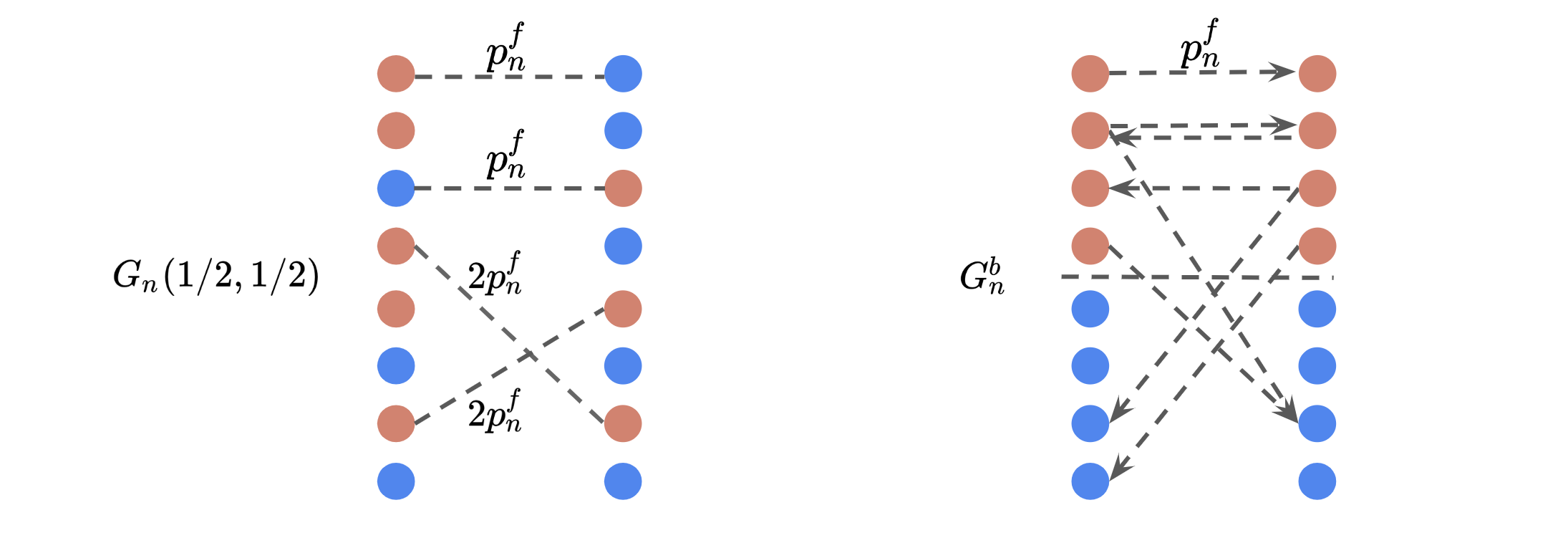}
    \vspace{-0.15in}
    \caption{Illustration of how $G^b_n$ replaces the graph arising from the balanced allocation: In $G^b_n$ the top $n/2$ nodes (in red) on each side are ``flexible" and generate a directed edge towards any node on the opposite side with probability $\pf_n.$}
    \label{fig:coupling0}
    \vspace{-0.1in}
\end{figure}

\begin{figure}[ht]
    \centering
    \includegraphics[width=0.8\textwidth]{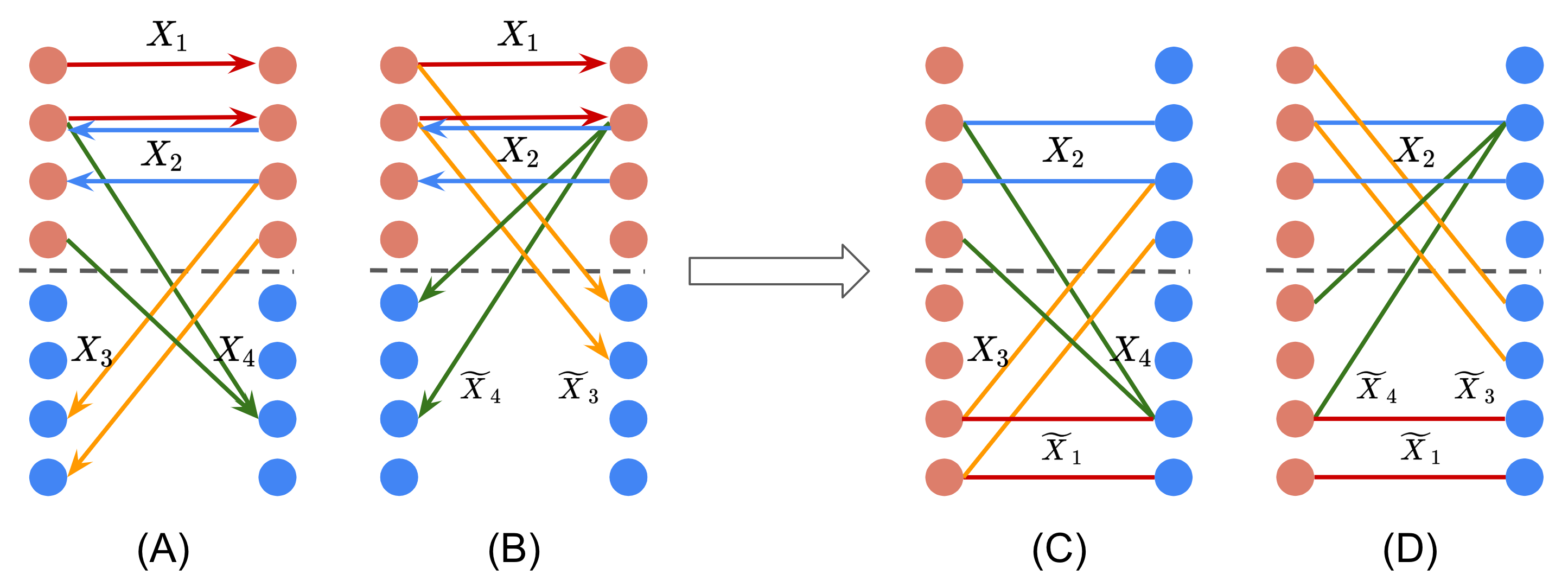}
    \vspace{-0.15in}
    \caption{Illustration of the edges' coupling in graph (A) - (D)}
    \label{fig:coupling1}
    \vspace{-0.1in}
\end{figure}

\dfedit{Even though cannibalization exists in regimes with $B < 1$ and $\anf>0$, Section \ref{sec:asymmetry} will show that another effect can counteract it.  Before diving into that second effect, we use the rest of this section to sketch out the proof  \cref{thm:compare_1}, wherein we develop a novel coupling technique that highlights the role of flexibility cannibalization.  To the best of our knowledge, our proof technique---coupling \emph{pairs} of graph realizations---has not been previously applied in random graph theory, and we think our result would be difficult to establish without it. Though we view this proof as our most elegant technical contribution, we defer the details to \Cref{sec:coupling_proof}, along with a more detailed proof sketch, and only include a brief discussion here.

\textbf{Brief proof sketch of \cref{thm:compare_1}.} Our proof first replaces the random graph arising from the balanced allocation, denoted $G_n(1/2,1/2)$, by a random graph $G_n^b$ with the same asymptotic matching size but a slightly different generating process (see \cref{fig:coupling0}). In $G_n^b$,  there are exactly $n/2$ flexible nodes on each side and edges are directed. The edge generating process can be thought of as occurring at each flexible node: a flexible node independently creates a directed edge with probability $\pf_n$ to each node on the other side of the graph (this contrasts our actual model in which undirected edges realize independently with probability 0, $\pf$, or $2\pf$); these changes affect $o(n)$  nodes and edges, and thus do not asymptotically affect~$\mu$. 

Our goal is then to show that the expected maximum matching size in $G_n^b$ is smaller than what it would be in the graph $G_n^o$ that realizes under a one-sided allocation. To show this, we couple two realizations of $G_n^b$ as displayed in \cref{fig:coupling1} (A) and (B). Intuitively, the difference between (A) and (B) is that the edges to the regular nodes in the bottom half are ``flipped'' across the vertical axis. This provides us with two realizations of $G_n^b$ that occur with equal probability. We then couple these two realizations of $G_n^b$ with two realizations of $G_n^o$ --- see \cref{fig:coupling1} (C) and (D). Intuitively, (C) is created by moving the edges that go from left to right in the top half of (A) to the bottom half; (D) is similarly created from (B). Our proof verifies that this is a valid coupling, i.e.,  (A) and (B) realize with the same probability in $G^b_
n$ and this is the same probability with which (C) and (D) realize in $G^o_n$.
% \smcomment{I don't understand what you mean by weights and weighted sums: isn't it just that, asymptotically, both realizations of (A) and (B) can be coupled to realizations of balanced and (C) and (D) to one-sided (with the slightly different edge generation)?} with identical weights, $\mu(1/2,1/2)$ can be expressed as a weighted sum of matching sizes over realizations (A) and (B) and $\mu(1,0)$ over realizations (C) and (D).

Given these coupled realizations, the result requires us to show that the sum of the maximum matching sizes in (C) and (D) is larger than their sum in (A) and (B). To do so, we (i) identify maximum matchings in (A) and (B), (ii) show that all of the edges within them can be injectively mapped into valid matchings within (C) and (D), and (iii) show that some additional edges in (C) and (D) can {still} be added to the matchings. For (ii), we need to make a careful combinatorial argument: {the matches in (A) and (B) that are formed by edges between the flexible nodes, e.g., the red and blue edges in \cref{fig:coupling1}, can simply be copied to either (C) or (D).} For the remaining edges\smedit{,} it is not guaranteed that just copying them over yields valid matchings. Instead, we construct a dynamic mapping based on an auxiliary graph to ensure a valid matching for edges mapped into both (C) and (D). Intuitively, as we construct the mapping, we use the fact that if edges form valid matches, despite cannibalization, in the denser top subgraph of (A) and (B), then they can also form valid matches in (C) and (D), where the edges are distributed across more distinct endpoints.

The above reasoning only argues that the asymptotic matching probability under the one-sided allocation is greater-equal to that under the balanced allocation. The final step in the proof to obtain the bound in \Cref{thm:compare_1} requires us to identify that each of the nodes on the left side of (C) has a constant probability, of at least $\parenthesis{\af}^3/2^5 \cdot e^{-7\af}$, to have another edge that can be added to the matching.}

\section{Flexibility Asymmetry}\label{sec:asymmetry}
\dfedit{Flexibility cannibalization favors the one-sided allocation, yet we have seen in \cref{fig:diagonal,fig:ratio} that the balanced allocation can be optimal. There must be an opposing effect. We call it \emph{flexibility asymmetry}, and this section provides intuition---backed by theory---for when and why the balanced allocation dominates. (We note that the intuition for asymmetry is somewhat more subtle than for cannibalization.)}

\begin{figure}[h]
    \centering \includegraphics[width=0.8\textwidth]{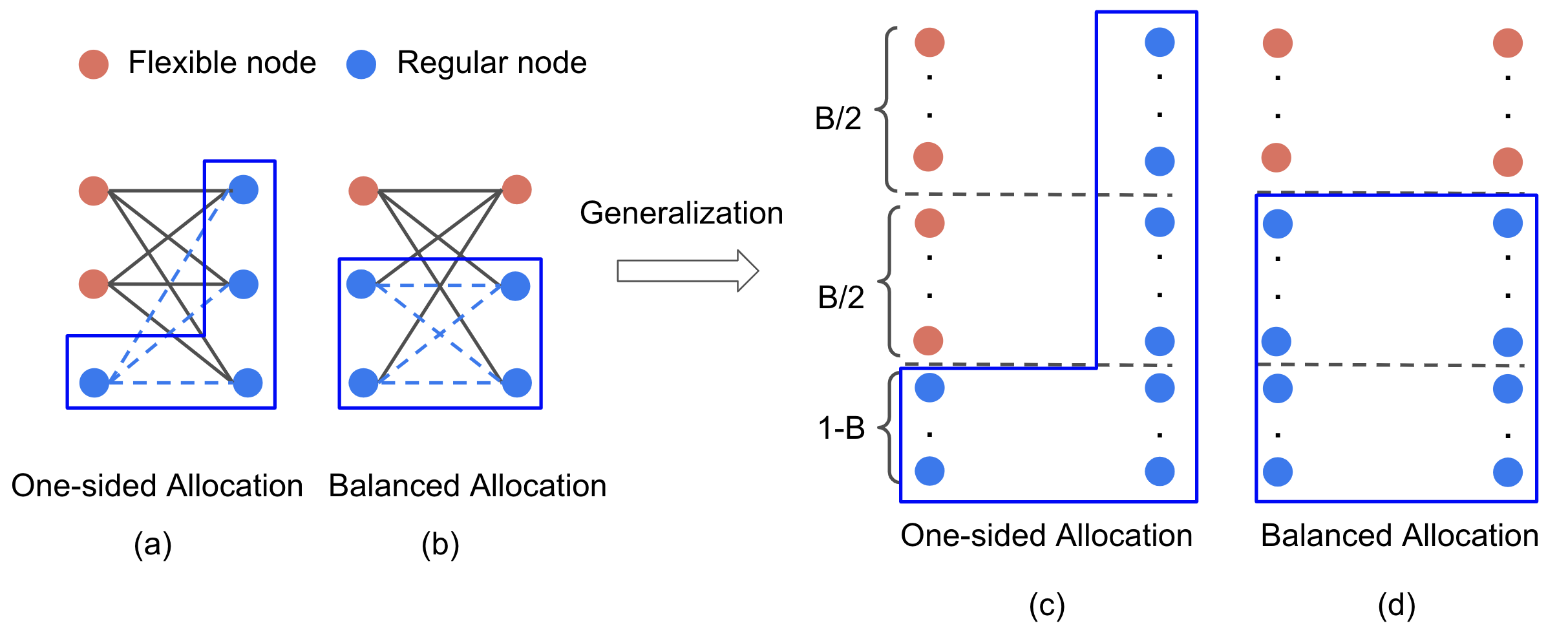}
    \vspace{-.1in}
    \caption{To gain intuition for the flexibility asymmetry effect, plots (a) and (b) assume that each flexible node is connected to all nodes on the other side of the graph, which resembles the case of $\af$ being large. A perfect matching in (a) requires the realization of one of the three dashed edges, while (b) requires one of four dashed edges to realize. This intuition holds at a larger scale: the expected number of edges in the subgraph of regular nodes is $(1-B/2)^2/(1-B)>1$ times greater for the balanced allocation in (d) than for the one-sided allocation in (c) \dfedit{when $\anf > 0$. Since the size of a maximum matching in that subgraph is close to the number of edges when $\anf$ is small, the balanced allocation is more conducive to matching the regular nodes among themselves when $\anf$ is a small  positive number.}}
    \label{fig:intuition2}
    \vspace{-0.2in}
\end{figure}

\dfedit{Whereas flexibility cannibalization is strongest when $\anf=0$, $B=1$, and $\af$ is of moderate size, flexibility asymmetry} requires $B<1$ and is strongest when $\af$ is large whereas $\anf$ is positive but small. By \Cref{thm:deg0_metric} $\phi$ is maximized by the balanced allocation in such a regime (see \Cref{fig:deg0}). To illustrate why this holds true, it is useful to consider the following extreme: imagine $\af$ being so large that every flexible node has degree $n$, i.e., flexible nodes are adjacent to all nodes on the other side (strictly speaking our model does not allow for $\af$ being that large but this will help with intuition). In a one-sided allocation, with $B<1$, there are $(1-B)n$ regular nodes on the side of the flexible nodes (in expectation), some of which are likely isolated given that their expected degree is $2\anf$. In contrast, in this extreme example where flexible nodes have degree $n$, having just one flexible node on each side already suffices to have no isolated nodes on either side (see Figure \ref{fig:intuition2} (b)). Simply put, this occurs because the one-sided allocation yields an asymmetry in potential neighbors for regular nodes on the two sides: the $n$ regular nodes on the side without flexible nodes have $Bn$ (potential) flexible nodes as neighbors whereas the $(1-B)n$ regular nodes on the side with flexible nodes have no flexible nodes as potential neighbors. Notice also that this requires $B<1$: when $B=1$, the side with flexible nodes has no regular nodes.

\dfedit{For the maximum matching size $\mu$, the asymmetry effect plays out in a similar but more complicated manner. We again obtain intuition by considering an example where flexibility is so strong that flexible nodes can be matched with any node on the other side (see Figure~\ref{fig:intuition2} (a) and (b)). In that case, a maximum matching can be found through a simple two-stage approach: First, find a maximum match in the subgraph of regular nodes  (marked in blue in the figure), then add the flexible nodes to make as many additional matches as possible.
Interestingly, when $\anf$ is small enough, regardless of the allocation of flexibility, there are almost always sufficiently many regular nodes unmatched after the first stage for the second stage to add exactly~$Bn$ matches by matching all flexible nodes to available regular nodes. Therefore, when $\anf$ is small enough, the allocation that maximizes the probability of a match is the one that is best in the first stage, i.e., the allocation that can maximize the probability of a match in the regular node subgraph. \dfedit{Figure~\ref{fig:intuition2} and its caption illustrate that the one-sided regular node subgraph is \emph{asymmetric} and explain why asymmetric graphs lead to fewer matches when $\anf$ is small but positive. %In summary, when flexibility is strong but limited, and when it is not too easy to match regular nodes with regular nodes, the best allocation is the one that enables the maximum number of regular-regular matches, making the balanced allocation dominate the one-sided allocation. 
The following theorem formalizes these intuitions by characterizing an upper bound on $\anf$ and a lower bound on $\af$ under which balanced flexibility dominates one-sided flexibility for $\mu$.
\begin{theorem}\label{thm:balanced_better}
Define \begin{align}
    \anf_\star(B) 
&:= \min\bracket{\tfrac{B^2}{8(1-B/2)^3}, \;\tfrac{1}{2(1-B/2)} \ln\parenthesis{\tfrac{2-B}{B}}}, \quad\text{and} \label{eq:alpha_ub}\\
\af_\star(B,\anf) &:= \tfrac{\ln\parenthesis{B}-\ln\parenthesis{2 \anf \squarebracket{\parenthesis{B/2}^2 - 2 \anf \parenthesis{1-B/2}^3}}}{\parenthesis{1-B/2} e^{-2 \anf \parenthesis{1-B/2}}-B/2}\quad\text{which is well-defined if}\; 0 < \anf < \anf_\star(B).\label{eq:alphaf_lb}
\end{align} Then,
    $\forall B \in (0,1):$ $\mu(\budget/2,\budget/2)  > \mu(\budget,0)$ whenever $0 < \anf < \anf_\star(B)$ and $\af > \af_\star(B, \anf)$.
\end{theorem}
\dfedit{\cref{thm:balanced_better} confirms our intuition on when the asymmetry effect is particularly strong: (i) $B$ should be strictly smaller than 1 (otherwise, for large $\af$, all allocations are equivalent because flexible nodes are sufficient to match everyone) and strictly larger than $0$ (otherwise, all allocations are equivalent because there is no flexibility); (ii) regular nodes should be hard -- but not impossible -- to match so that regular-regular matches matter ($0 < \anf < \anf_\star(B)$); (iii) flexible nodes must be especially easy to match ($\af \geq \af_\star(B, \anf)$).}
}
}
\dfedit{This effect, while derived from our simplified model, may play out in real platforms. We illustrate with an example to show how the intuition might generalize.}  For instance, the freelancing platform Upwork offers a digital learning program called ``Upwork Academy'' to train skilled freelancers to handle a wide range of tasks. This type of flexibility is valuable in allowing these freelancers to be matched with demanding customers who otherwise cannot be served. On the demand side, similarly, the platform uses a feature called ``Project Catalog'' to incentivize users to choose from a standardized pool of tasks that most freelancers can fulfill. Essentially, employing flexibility on both market sides allows the platform to match flexible agents on each side with ``difficult'' agents on the opposite side. With flexible agents on only one side of the platform, it can be difficult to match regular agents on that side of the platform. In particular, when some freelancers are well-trained but all customers are quite demanding, it becomes difficult for freelancers with a limited skill set to find a job. Similarly, when some customers have standard requests but no freelancers receive specialized training, it becomes difficult for the platform to serve the demanding customers.

\noindent\textbf{Proof sketch of \cref{thm:balanced_better}.}
{Our proof (Appendix \ref{app:balanced_better}) leverages the two-stage matching procedure to analyze $\mu(B/2, B/2)$: we first study the size of a maximum matching among the regular nodes (stage 1), and then quantify the additional matches that can be added using the flexible nodes (stage 2). This allows us to derive a lower bound on the number of matched nodes under balanced flexibility. Similarly, we derive an upper bound on the maximum matching size under one-sided flexibility by bounding the number of isolated nodes in the graph. We then prove the theorem by verifying that the upper bound is dominated by the lower bound in the specified parameter regime.}

\emph{Lower bound on the number of matched nodes for balanced flexibility.} For balanced flexibility, we lower bound the size of a maximum matching among the regular nodes, and then show that almost all flexible nodes can be matched afterward. As illustrated in \cref{fig:intuition2} (d), in stage 1, the balanced allocation faces an equal number of regular nodes on both sides. We prove (Appendix \ref{app:n1_lb}) the following bound on the maximum matching size in a $(1-B/2)n \times (1-B/2)n$ graph of regular nodes, denoted by random variable~$m_1$ (while the number of regular nodes on each side of the graph is not deterministically $(1-B/2)n$, it concentrates around $(1-B/2)n \times (1-B/2)n$ as $n$ scales large and we assume this deterministic number for the purpose of this proof sketch):

\begin{lemma}\label{lem:n1_lb}
    $\EE{m_1} \geq 2 \cdot \parenthesis{1-B/2} n \squarebracket{1-(1-B/2) \anf - e^{-2 \anf (1-B/2)}}$ as $n \to \infty.$
\end{lemma}
Intuitively, for small \(\anf\) the expected maximum matching size should be close to the expected number of edges because very few nodes have a degree more than $1.$ Our proof explicitly characterizes this, and lower bounds~$m_1$ by subtracting the number of ``redundant edges" (those incident to nodes with degree $> 1$) from the total number of edges. This allows us to derive the lower bound for $\EE{m_1}$ in \cref{lem:n1_lb}.

\emph{Upper bound on the number of matched nodes for one-sided flexibility.} To upper-bound the matching size under one-sided flexibility, we simply quantify the expected number of isolated regular nodes on the side on which there is flexibility. As illustrated in \cref{fig:intuition2} (c), since there are about $(1-B)n$ regular nodes on this side, about $(1-B)ne^{-2\anf}$ of these are isolated. It follows that at most $(1-B)(1-e^{-2\anf}) n$ of these regular nodes can be matched. 

\emph{Combining the bounds.}
For the purposes of this proof sketch, we assume that all flexible nodes can be matched to regular nodes in stage 2 of the algorithm. Then, the number of matches under the one-sided allocation is at most $Bn+(1-B)(1-e^{-2\anf}) n$ whereas the number of matches under the balanced allocation is at least $Bn+\EE{m_1}$. By verifying that $$Bn+(1-B)(1-e^{-2\anf}) n < Bn+ 2 \cdot \parenthesis{1-B/2} n \squarebracket{1-(1-B/2) \anf - e^{-2 \anf (1-B/2)}}$$ \dfedit{for any $B \in (0,1), 0 < \anf < \anf_\star(B)$ and $\af > \af_\star(B, \anf)$, we confirm that the balanced allocation creates more matches than the one-sided allocation in the stated regime.} In Appendix \ref{app:balanced_better} we show that the gap in the above inequality is sufficiently large to account for the fact that, in the balanced allocation, some flexible nodes may not be matched in stage $2$. We highlight that $\anf > 0$ is necessary for this comparison as otherwise no regular nodes could be matched in stage 1 (under either allocation) and the two sides of the inequality would be equal.

\section{Identifying the Right Allocation Across All Parameters}\label{sec:ks}

\dfedit{
In this section, we use the Karp-Sipser (KS) algorithm to analyze $\mu(\ql,\qr)$ for broader ranges of the parameters $\ql,\qr,\anf$, and $\af$. The KS analysis is based on a less intuitive but more classical tool in the study of sparse random graphs; as it involves a distinct machinery with additional technical challenges, we defer the formal description of the algorithm and the associated derivations to Appendix~\ref{sec:proof_global_add} to keep the exposition in the main body focused on the resulting insights. The KS-style analysis yields a quantity $\mubar(\ql,\qr)$, which is \dfedit{defined as the solution to} a set of nonlinear equations (see Equation \eqref{eq:w} and \cref{thm:global_matching} in the appendix).
\dfedit{Importantly, $\mubar$ is not a closed-form expression, but solving these equations is much more computationally efficient than simulating maximum matchings in large graphs.} Such equations are common in KS-based analyses \citep{karp1981maximum}. For a limited parameter regime—most notably the \emph{subcritical regime} where $\af+\anf<e$—we can show for the one-sided and the balanced allocations that $\mubar(\ql,\qr)$ equals $\mu(\ql,\qr)$  (see \cref{thm:equivalence}; it is common for KS-style analyses to require fundamentally different approaches outside the subcritical regime \citep{Bollobas1995, mastin2013greedy}). On the theoretical side, this equivalence allows us to use $\mubar$ to establish the dominance of the one-sided allocation over additional ranges of $\af$ and $\anf$ (see \cref{thm:compare_q} and~\cref{fig:compare}).
}

%that can be solved to provable numerical precision
% \footnote{{The geometric properties stated in \cref{thm:convex_concave} follows from a similar computer-aided proof. For concavity, we construct local upper bounds to show that the second-order derivative (SOD) of $\mubar(1/2,1/2)$ is negative in the direction $(0,1)$; for convexity, we use local lower bounds to show that the SOD is positive in the direction $(1,-1)$.}}

\begin{figure}[h!]%
    \centering
    \includegraphics[width=1\textwidth]{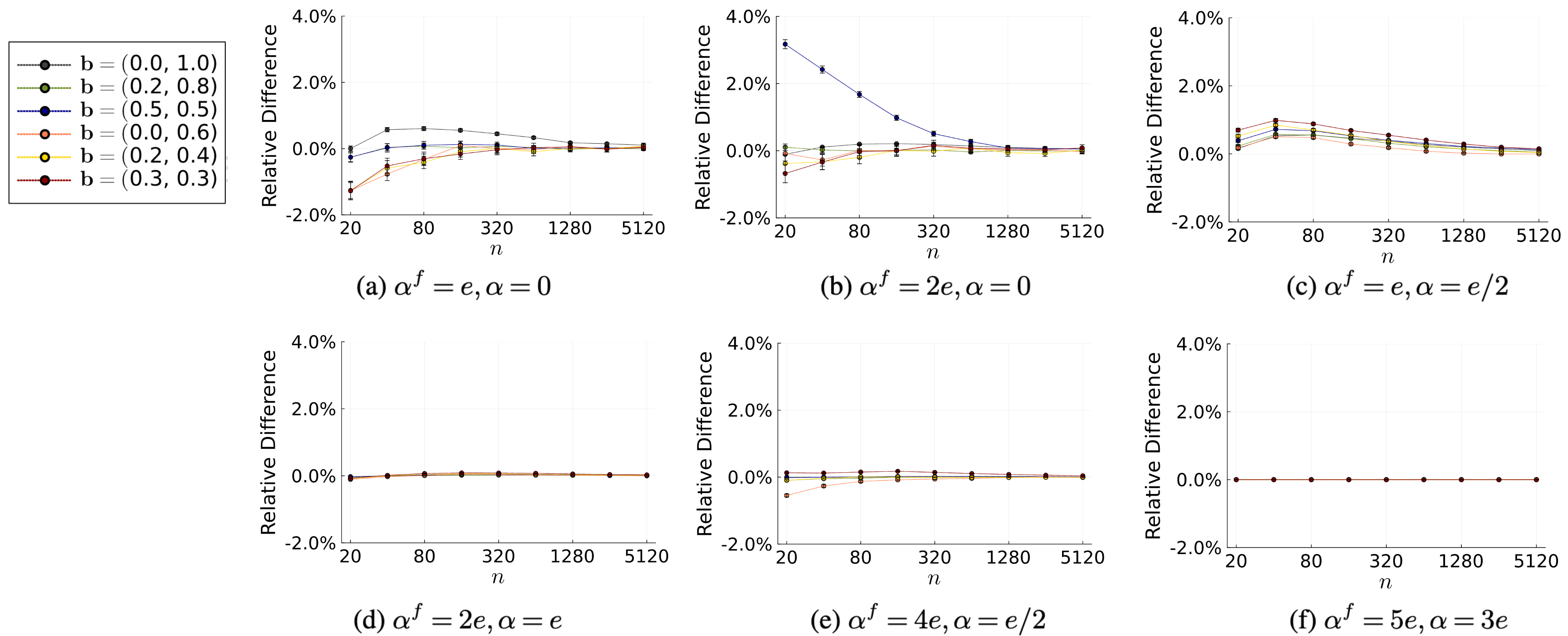}
    \caption{The plots present $(\munum(\ql,\qr) - \muemp_n(\ql,\qr))/ \muemp_n(\ql,\qr)$ across varying $(\ql,\qr)$ as $n$ scales large.}
    \label{fig:convergence}
    \vspace{-0.1in}
\end{figure}

%Despite the analytical challenges in extending \cref{thm:equivalence} and \ref{thm:compare_q} to other parameter regimes, 

\dfedit{
Our numerical results (see \cref{fig:convergence}) suggest that \(\muemp_n(\ql,\qr)\)\dfedit{---the empirical estimate of $\mu$ from simulations, as defined in the caption of \cref{fig:diagonal}---}approaches~\(\munum(\ql,\qr)\) as \(n \to \infty\) for a much wider set of parameters of $\af, \anf, \ql$ and $\qr$.} We, therefore, use \(\munum(\ql,\qr)\) as a surrogate function to evaluate different flexibility allocations, not just the balanced and one-sided one, across a wider range of parameters. {This is not doable with $\muemp$ due to the heavy computations needed to evaluate it with high precision.} Specifically, we conduct a grid search over $B,\af,\anf,\ql,\qr$ with the set of parameters denoted $S$ (details in Appendix \ref{sec:num_construction}); we trust this to give a better estimate of the true asymptotic matching probability while also being computationally more efficient. We highlight the following observations:

\begin{figure}%
    \centering
    \includegraphics[width=0.9\textwidth]{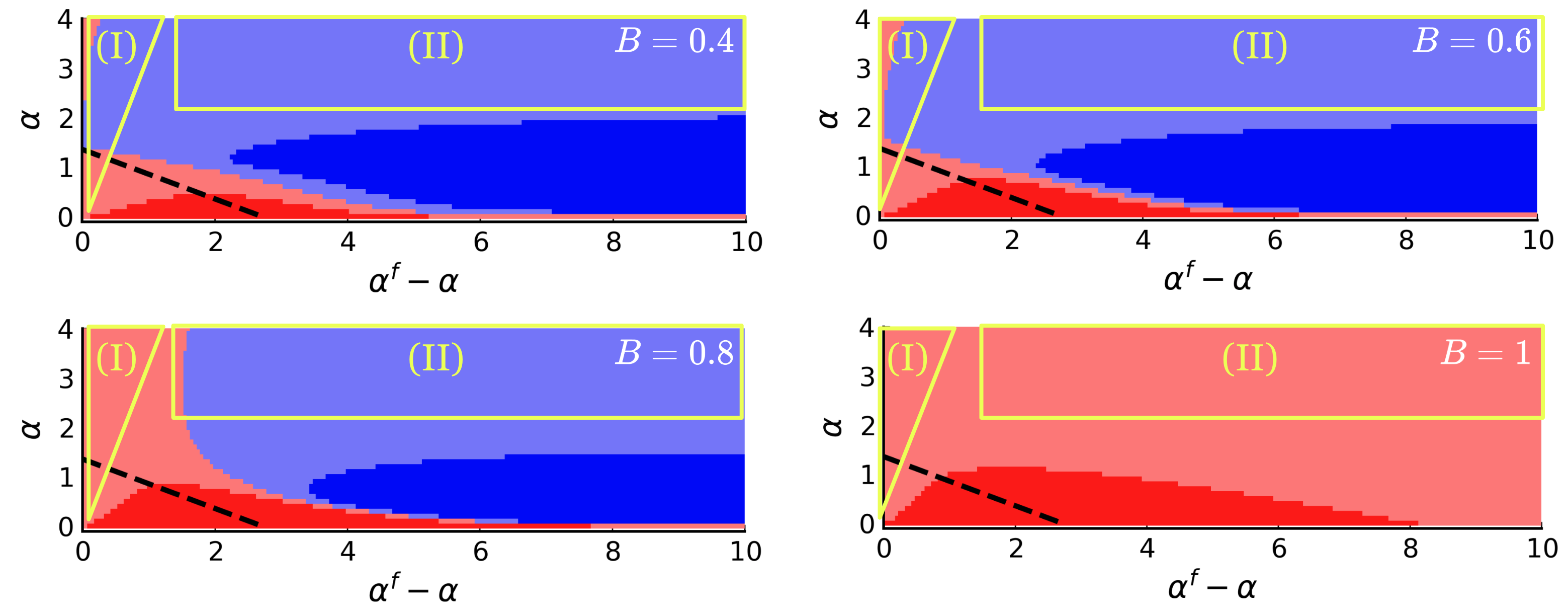}
    \vspace{-0.1in}
    \caption{The plots present the values of $\frac{\munum(\budget/2,\budget/2)}{\munum(\budget,0)}$ across varying $\anf,$ $\af-\anf$ and $B$: the ratio is smaller than $1$ in the red region (light red if between $0.99$ and $1$) and greater than $1$ in blue region (light blue if between $1$ and $1.01$). The dashed line highlights the boundary of the subcritical regime, and regions (I)-(II) indicate where the two allocations have comparable performances.}
    \label{fig:ratio2}
    \vspace{-0.1in}
\end{figure}

\noindent\textbf{Either the one-sided or the balanced allocation is optimal. } \dfedit{For the metric $\phi$, \cref{thm:deg0_metric} established that the optimal allocation is always either one-sided or balanced. For $\mu$, we observe the same pattern:} in line with the findings in \cref{fig:diagonal}, our more extensive numerical results show that, across our grid search, \(\munum(\ql,\qr)\) is always maximized by one of \dfedit{these two allocations. This justifies our focus on comparing them throughout the paper.}

\noindent\textbf{In the subcritical regime, the one-sided allocation is better. } \dfedit{In \cref{thm:compare_q} (Appendix~\ref{app:compare_q2}), we leverage the Karp-Sipser methodology in the regime where $\mu = \mubar$ to prove this result for most of the subcritical regime and $B = 1$.}
% \smdelete{In \cref{thm:compare_q}, we proved this result for most of the subcritical regime and $B = 1$.}
{When $B<1$, our computer-aided proof breaks because we cannot prove that the nonlinear equations in \eqref{eq:w} have a unique set of solutions.} However, we still find numerically that the one-sided allocation is better within the subcritical regime for all tested values of $B$ {(see \cref{fig:ratio2})}. This also matches our theoretical findings in that $\af$ cannot be too large in the subcritical regime, which naturally limits the effect of flexibility asymmetry (see \dfedit{lower bound on  $\af$ in} Theorem~\ref{thm:balanced_better}). %\dfcomment{just delete the rest of this paragraph?}\kzcomment{Kamessi will fix}In particular, even though Proposition~\ref{lem:degree_0} shows that the balanced allocation may be minimizing and the one-sided allocation may be maximizing the fraction of isolated nodes in this regime, this is outweighed by the effect of flexibility cannibalization.

\noindent\textbf{When $B = 1$ or $\anf = 0$ the one-sided allocation is better. } We find that $B = 1$ and $\anf = 0$ are the special cases where the one-sided allocation always dominates, regardless of the value of~$\af$ (for $\anf=0$, though hard to see in the plots, there is always a thin red line just above the x-axis). Comparing this with our two-stage matching procedure in \cref{sec:asymmetry}, we find that these are exactly the cases where the flexibility asymmetry effect dissipates. 
This also explains why the coupling technique presented in \cref{sec:cannibalization} is specific to $B = 1$ and~$\anf = 0$: for large $\af$, \cref{fig:ratio2} shows that the dominance of the one-sided allocation breaks down very close to the regime where $B=1$ and $\anf=0$.

\noindent\textbf{Characterizing when the flexibility allocation matters. } 
Finally, our numerical results in \cref{fig:ratio2} also allow us to characterize the regions in which the flexibility allocation is of second-order importance. 
In region (I), flexibility does not notably increase the edge probability as $\af/\anf\approx1$; thus, neither the budget nor the allocation of flexibility has a sizable effect. \dfedit{In region (II), as $\anf$ and $\af$ grow larger,} almost all nodes are matched irrespective of the flexibility allocation. Thus, in (I) and (II), any allocation of a fixed flexibility budget results in a similar matching size. \dfedit{In contrast, in regions with more prominent flexibility cannibalization (moderate values of $\anf$ and $\af$) or flexibility asymmetry (small $\anf$, large $\af$, and $B < 1$), the surrogate function identifies a larger gap between the one-sided and the balanced allocation.} This mirrors our observation in \cref{fig:ratio} that either of the two allocations can dominate the other by at least $8\%$ in these~regions.

\section{Managerial Implications for Platform Experimentation}\label{sec:experiment}
% \dfcomment{I am wondering if we can shorten this by making it, from the onset, about $g^{KS}$ rather than starting by talking about $g$... I'll make that change!} \kzcomment{just talk about $g^{KS}$}
Flexibility cannibalization and asymmetry can have important managerial implications. \dfedit{In the preceding sections, we fixed the flexibility budget $B$ and studied how to allocate it. Here, we make the model slightly more realistic by letting the platform decide \emph{both} the total amount of flexibility \emph{and} its allocation across the two sides. To highlight the resulting implications, we extend  our} surrogate function $\mubar(\ql,\qr)$  to study the geometry of a platform profit function that incorporates both the benefit and the cost of flexibility.
Formally, we assume that the flexibility decision $\parenthesis{\ql,\qr}$ incurs a linear cost of $c \cdot \parenthesis{\ql+\qr}$ for some constant~$c > 0$. \dfedit{That is, we aim to maximize the objective function
\begin{align}\label{eq:opt}
    \gks(\ql,\qr):= \mubar(\ql,\qr) - c \cdot \parenthesis{\ql+\qr}
\end{align}}
% We use the notation $g(\ql,\qr) := \mu(\ql,\qr) - c \cdot \parenthesis{\ql+\qr}$ to refer to this objective function and $\gks(\ql,\qr)$ to denote $\mubar(\ql,\qr) - c \cdot \parenthesis{\ql+\qr}$.
\dfedit{over $\qvec \in [0,1]^2$.} Such a linear cost model reflects a setting wherein the marginal cost of more agents becoming flexible is approximately constant. \dfedit{This simplicity allows us to focus on first-order effects in the design of bipartite flexibility structures; at the end of this section we also discuss the extent to which our high-level observations may translate to other cost structures}.
% \footnote{For example, if each node $v_i^k$ draws a value from $U(c,c+\epsilon)$ and becomes flexible if and only if an incentive has at least that value, then the platform can set the incentive value as $c+b_k/\epsilon$ to ensure that the probability of a node on side $k$ being flexible is $b_k$.} SEB: I'M removing this because it complicates things more than it helps (this supply model is weird, no?) I think the main argument is that the more natural convex costs create another curvature effect that adds up to ours, making the profit function more "concave". But cannibalization may still win, and a linear cost allows us to focus on what matters.

\begin{figure}%
    \centering
    \subfloat[\centering $\af = e/2, \anf = 0$ and $c = 0.4$]{{\includegraphics[width=0.4\textwidth]{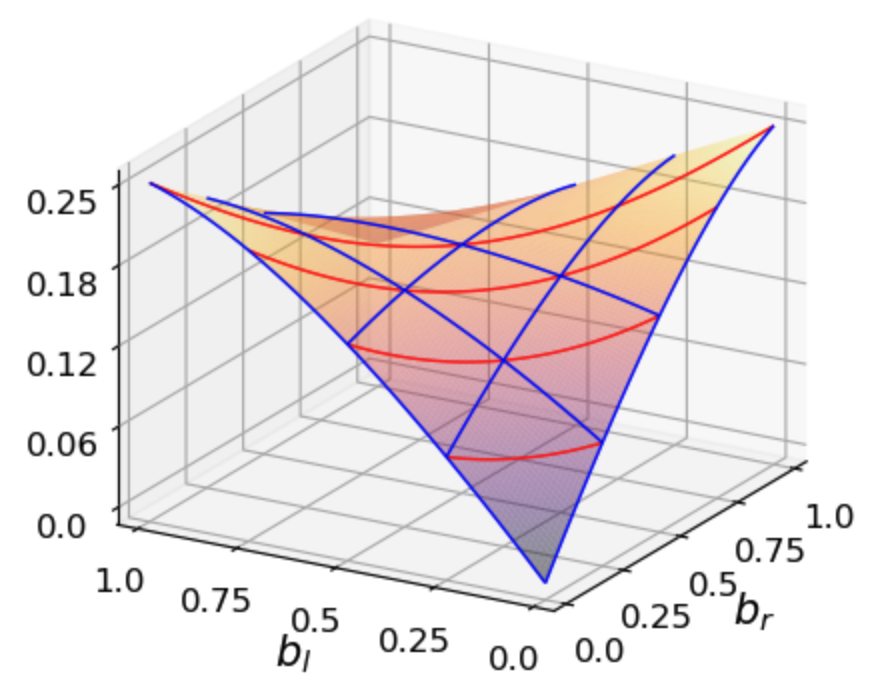} }}%3d3_emp.png
    \qquad
    \subfloat[\centering $\af = 3 e, \anf = e/8$ and $c = 0.3$]{{\includegraphics[width=0.4\textwidth]{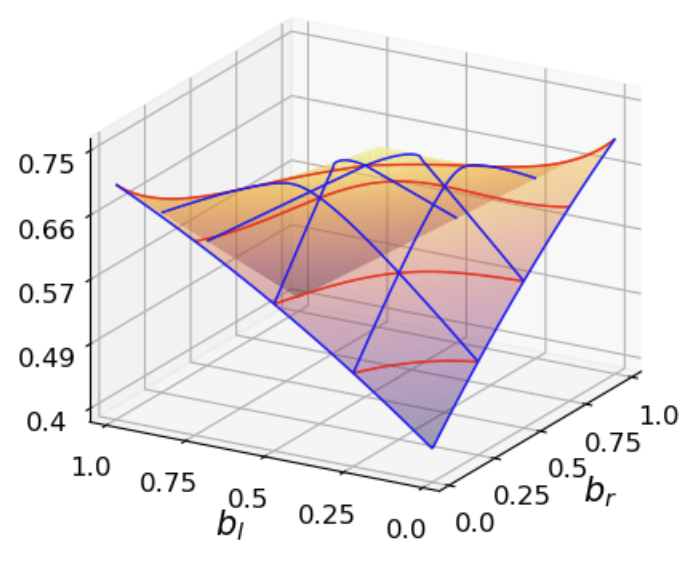} }}%
    \caption{The plots present the value of $\gks(\ql,\qr)$ for varying $\af, \anf$ and $c.$ We highlight the values of the function in the directions $(1,0)$ and $(0,1)$ in blue, and in the direction $(-1,1)$ in red.
    }
    \label{fig:convex_concave}
    \vspace{-.2in}

\end{figure}

In \cref{fig:convex_concave} we plot $\gks(\ql,\qr)$  for two different sets of parameters of $\af,\anf$ and $c$. The plots reveal distinct convexity and concavity properties in the directions of $(1,0)$, $(0,1)$, and $(1,-1)$. In the directions of $(1,0)$ and $(0,1)$, we consistently observe a concave function $\gks$ regardless of the starting point, i.e., the value of flexibility in our matching model exhibits decreasing returns. \dfedit{Decreasing returns are common in studies of flexibility, as exemplified by the insight that \emph{a little flexibility} is almost as valuable as full flexibility \citep{bassamboo2012little, elmachtoub2019value}.}
However, we also observe interesting geometric effects in the direction of \( (1,-1) \); this direction captures the tradeoff between investing a fixed budget of flexibility on one side or the other. Depending on the values of $\af$ and $\anf$, \cref{fig:diagonal} and \cref{fig:convex_concave} show that $\mubar(\ql, \qr)$, respectively $\gks(\ql, \qr)$, can be convex, concave or neither in the direction $(1,-1)$. Both the concavity in directions $(0,1)$ and $(1,0)$ and the potential convexity in direction $(1,-1)$ are supported by \dfedit{limited theoretical findings in 
the subcritical regime (see \cref{thm:convex_concave} in Appendix \ref{sec:ne})}.

In the remainder of this section, we first discuss two serious practical ramifications of such geometric properties and then argue how an understanding of flexibility cannibalization and asymmetry can help avoid a potential pitfall of current platform experimentation designs.
Many platforms today operate with separate teams dedicated to controlling flexibility incentives on the demand and supply sides. Through frequent experimentation, including continuous local improvement of algorithmic parameters, these teams aim to optimize the flexibility investment on each respective side of the market. For illustrative purposes, suppose the supply and demand teams iteratively vary the flexibility investment on their own side ($\ql$ and $\qr$) by $\gamma$ whenever doing so improves the objective. The teams would eventually settle at a point where neither has an incentive to further change its flexibility investment. In \cref{fig:experiment1}, we build upon $\gks(\ql,\qr)$ to illustrate the outcomes of such experiments, which yields the following two observations:

\begin{figure}[h!]%
    \centering
    \includegraphics[width=0.9\textwidth]{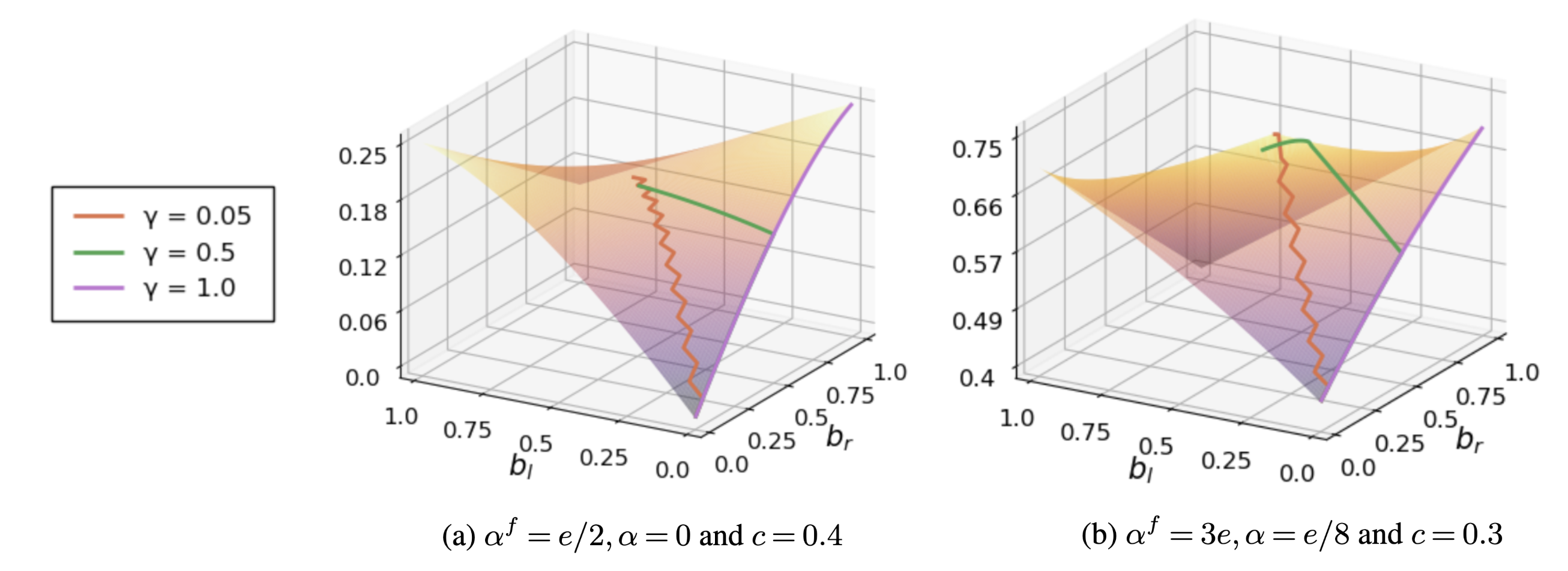} %experiment_combined.png
    \caption{The colored lines illustrate the trajectory of experimentation on the surface of $\gks(\ql,\qr)$ for different choices of~$\gamma$. The experiment terminates around a suboptimal balanced allocation in (a) when $\gamma$ is too small, and around a suboptimal one-sided allocation in (b) when $\gamma$ is too large.}
    \label{fig:experiment1}
    \vspace{-.2in}
\end{figure}

\noindent\textbf{Suboptimality due to the lack of joint experimentation.} 
Teams may settle at a suboptimal flexibility design due to not experimenting jointly. In other words, they may find themselves at a point where neither team can unilaterally improve the joint objective even though a joint experiment in the direction of either $(1,-1)$ or $(-1,1)$ would yield a strict improvement. \cref{fig:experiment1} (a) illustrates that this can arise, when $\gamma$ is small, %with the platform terminating its experiments 
near a suboptimal balanced allocation. At this suboptimal allocation, each team faces a concave objective and optimally manages its own levers (i.e., $\gks(\ql,\qr)$ is locally optimal in the directions $(1,0)$ and $(0,1)$), but the overall flexibility decision remains suboptimal for the platform. In Appendix \ref{sec:ne}, we formalize this observation by modeling the flexibility optimization along two axes as a game between two verticals within an organization  (see \cref{sec:rel_work}); even though the verticals have the same payoff functions, we show that the suboptimal allocation can emerge as a local Nash Equilibrium (see Appendix \ref{sec:ne}). Though the suboptimal allocation in \cref{fig:experiment1} (a) can be avoided by setting $\gamma$ sufficiently large,  \cref{fig:experiment1} (b) illustrates that this does not generally address the suboptimality. Indeed, such large $\gamma$ may cause an outcome wherein one team operates with more flexibility than is jointly optimal, leading the other team to not invest in flexibility at all. Thus, the platform becomes trapped in a one-sided flexibility allocation, even though adopting flexibility on both market sides would be more profitable.

\dfedit{Moreover, the cost of suboptimality can be large. Depending on the underlying cost structure, even a few percentage points of difference in matching size can determine whether the platform operates profitably or at a loss. Indeed, the locally optimal balanced allocation in \cref{fig:experiment1} (a) achieves just 73\% of the profit of the globally optimal one-sided allocation, and for other parameters the gap can be even larger, especially when the platform's margins are small. For instance, with $\af = 2e, \anf = 0$ and $c = 0.99,$ the profit of a locally optimal balanced allocation is less than 10\% of that of the globally optimal allocation. Similarly, among instances where the balanced allocation is optimal (e.g., \cref{fig:experiment1} (b)), the profit of a locally optimal one-sided allocation can be just 6\% of that of the global optimum ($\af = 50 e, \anf = e/100, c = 0.99$).} 

\noindent\textbf{Suboptimality due to the existence of saddle points.} Surprisingly, joint experimentation is not enough to avoid suboptimal flexibility designs. 
When $\anf$ and $\af$ are small, the concavity of $\gks(\ql,\qr)$ in the direction $(1,0)$ and convexity in the direction $(1,-1)$ give rise to a saddle point on the surface of $\gks(\ql,\qr)$, as illustrated in \cref{fig:experiment1} (a). 
Near the saddle point, the gradient of $\gks(\ql,\qr)$ is close to $0$ in all directions. 
As a result, when~$\gamma$ is too small, even with joint experimentation, the platform may fail to capture value in the $(1,-1)$ direction.
%The profit at the saddle point can be vastly smaller than that at the globally optimal one-sided allocation: in \cref{fig:experiment1} (a), the locally optimal balanced allocation achieves just 73\% of the profit of the globally optimal one-sided allocation, and for other parameters the gap can be even larger, especially when the platform's margins are small. For instance, with  $\af = 2e, \anf = 0$ and $c = 0.99,$ the profit of a locally optimal balanced allocation is less than 10\% of that of the globally optimal allocation.

As any local experimentation scheme, joint or not, may face obstacles in exploring the complicated geometry of $\gks$, an understanding of the cannibalization and asymmetry effects can also help platforms design non-local experiments. 
Consider a digital matching market with high $\af$ that operates with just one lever of flexibility at $(B,0)$; introducing a second lever on the opposite side of the market, one would usually invest in a little flexibility by experimenting with $(B,\epsilon)$. In contrast, our study shows that $(B,0)$ may be a local optimum of $\gks$ (see \cref{fig:diagonal}) whereas $(B/2,B/2)$ would yield a much greater profit. Similarly, a ride-hailing platform with small $(\anf,\af)$ may be locally stuck at a balanced allocation despite a one-sided one being much more profitable. To avoid both of these outcomes, our findings suggest that platforms may want to supplement local experimentation with experiments on qualitatively different flexibility designs, i.e., moving from one-sided to balanced or vice versa. Given the high cost of experimentation, it may make sense in practice to first leverage non-local simulation before then attempting experiments in significantly altered parameter regimes; nonetheless, our results emphasize the need to explore non-locally, which stands in contrast to common industry practices.

\begin{figure}[ht]
    \centering
    \includegraphics[width=0.8\textwidth]{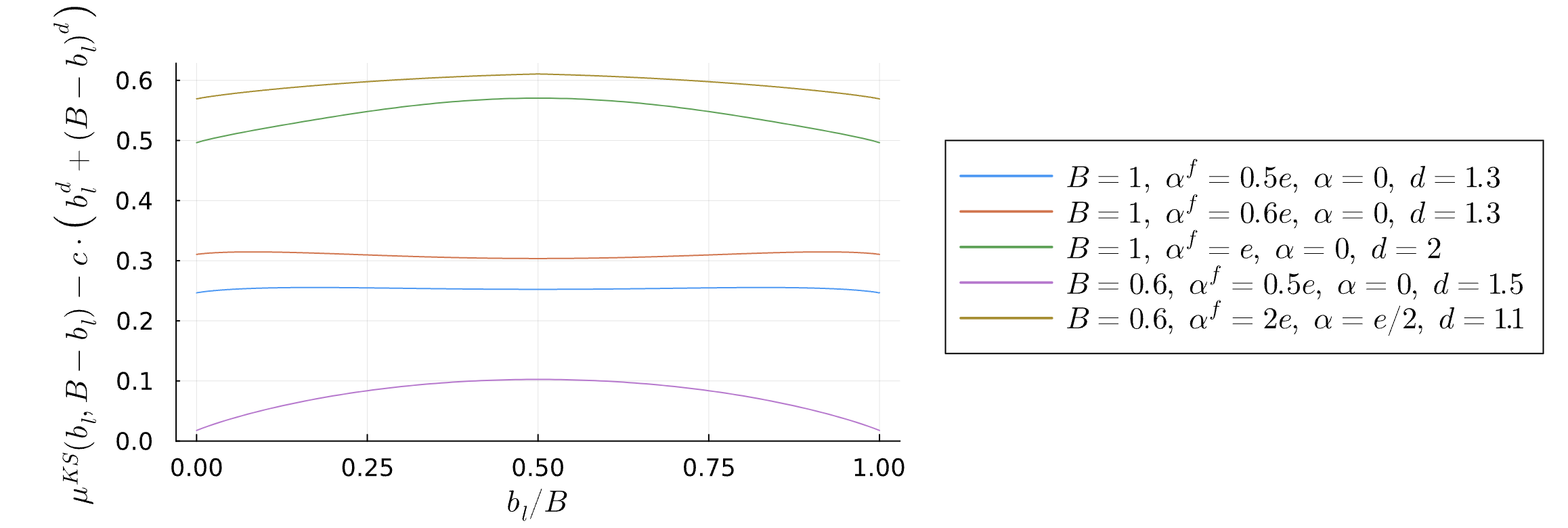}
    \caption{\dfedit{Illustration of the platform's profit with respect to $b_l/B$ for varying values of $B, \af, \anf$ and $d,$ where $d$ controls the convexity of the cost with $(\ql,B-\ql)$ incurring a cost of $c \cdot (\ql^d + (B-\ql)^d).$}}
    \label{fig:convex_cost}
    \vspace{-.2in}
\end{figure}

\dfedit{\noindent\textbf{Beyond a linear cost for flexibility.}
The analysis above assumes that the cost of flexibility is linear in the fraction of flexible agents.
In practice, however, making agents flexible likely becomes harder as more of them are flexible: some of them may respond to small incentives, while convincing more agents require progressively larger ones.
This suggests, as a natural extension, that one may want to model the costs as convex.
To get a sense of how convexity affects our findings, \cref{fig:convex_cost} shows the platform's profit for a few representative examples of convex cost structures.
Two patterns emerge in these examples.
First, convex costs can strengthen the case for a balanced allocation: in some parameter regimes where the one-sided allocation dominates under linear costs (e.g., $\anf = 0$ and $B = 1$), the balanced allocation becomes optimal under convex costs.
Second, the global optimum can arise at allocations other than the one-sided or balanced ones; when it does, the objective function appears quite flat along the $(1,-1)$ direction, suggesting that the precise allocation matters less in these cases.

More broadly, the central message of our analysis is not that saddle points or local optima always exist, but that flexibility on two sides interacts in ways that platforms should study carefully.
Under linear costs, cannibalization and asymmetry create a landscape with saddle points and local optima that trap independent experimentation.
Under more complex cost structures, these geometric features may change or new ones may emerge, but the fundamental insight persists: by optimizing each side independently, platforms risk missing interaction effects that only a joint analysis can reveal.}

\section{Robustness in Alternative Graph Models}\label{sec:ext}
\dfedit{Having identified flexibility cannibalization and asymmetry in our baseline model, we now verify that these effects persist in alternative settings. Though we also identify additional effects that arise in specific models, our two main effects remain dominant forces in the \dfedit{previously identified} regimes}.% we previously identified.}

\dfedit{In this section we show that the intuitions behind flexibility cannibalization and asymmetry effects extend to alternative models of matching in random graphs. The goal of these models is to capture additional market features such as (i) dependencies among the edges that realize (either due to ``local'' constraints where each node can only be matched to a predetermined set of potential neighbors or due to spatial constraints wherein the realization of edges is governed by the distances between supply and demand nodes), (ii) edges having heterogeneous value, and (iii) markets being imbalanced with one side having more participants than the other. 
In line with our previous analysis of the maximum matching metric $\mu$, this section focuses on the optimal matching under each graph realization; we refer the reader to Appendix B for outcomes under suboptimal matching algorithms.
%This section focuses on the maximum matching metric $\mu$; we refer the reader to Appendix~\ref{sec:greedy} for outcomes under different matching metrics. % which is common in many platforms that solve matching problems in the physical world (e.g., when edges are proximity-based).}\dfcomment{needs to be workshopped a bit: in the imbalanced model, and to a lesser extent in the choice model, the goal is a bit different...}
%We first consider a . % Finally, we use the spatial model to investigate imbalanced markets that allow an uneven number of supply and demand nodes. 
Across all of the models we study, we find that, even as different models create new effects, flexibility cannibalization and asymmetry continue to play  a crucial role for the optimal flexibility allocation and they remain dominant in the same regimes we previously identified. Throughout this section we abuse notation in that we denote by $\muemp(\ql,\qr)$ the empirical estimate of the respective objectives, which is always based on $1000$ samples on graphs with $n=100$. }

\subsection{Local Model}\label{sec:local}

\begin{figure}[h!]%
    \centering
    \includegraphics[width=0.85\textwidth]{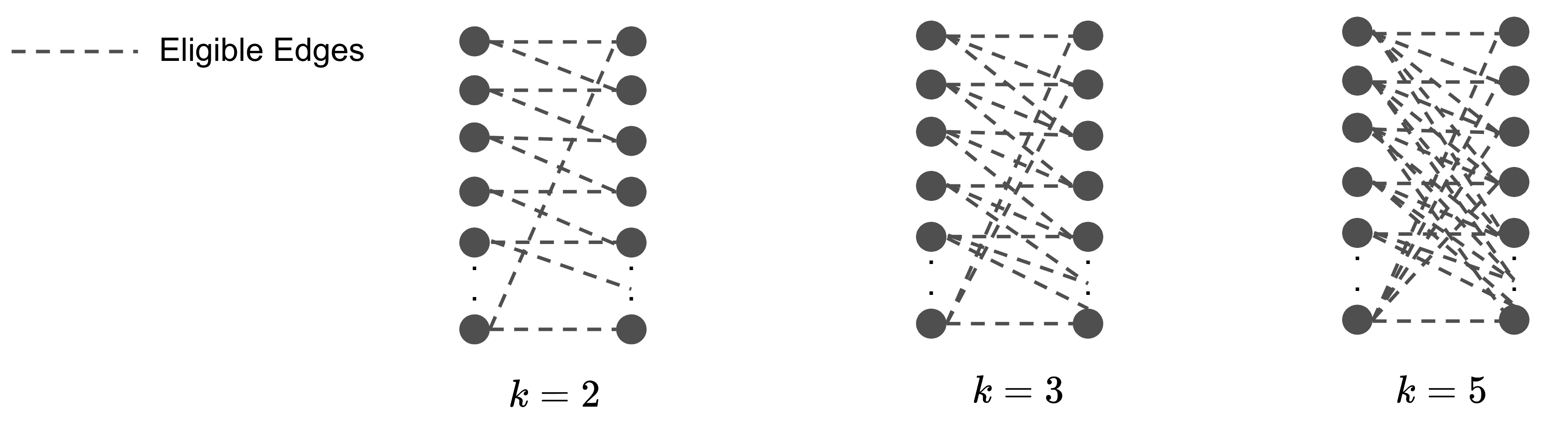} 
    \caption{Illustration of local models with different values of $k.$}
    \label{fig:local_model}
    \vspace{-0.15in}
\end{figure}

\dfedit{We first define and analyze the \emph{local model}, which is the simplest model to capture dependencies among the edges}. As illustrated in \cref{fig:local_model}, we assume that for any $i \in [n]$, $\vl_i \in V_l$ is only eligible to connect to $\vr_i, \vr_{(i+1)\text{ mod }n}, ... \vr_{(i+k-1)\text{ mod }n}$ in $V_r$, where $k$ is a constant that specifies the number of eligible neighbors of a node. 
In particular, in line with our previous model in Section \ref{sec:model}, we assume that
the probabilities associated with flexible and regular nodes now scale with $k$ (rather than $n$), giving $\pf = \af/k$ and $\pnf = \anf/k$ for constant $\af$ and $\anf$. We require that $0 \leq \pnf < \pf \leq 1/2$ for the edge probabilities to remain in $[0,1]$ and obtain the following additive model of conditional edge generation:
\begin{align*}
\PP{R_{ij} = 1\mid F^l_i, F^r_j} = 
\begin{cases} 
2\pnf +(F^l_i + F^r_j) \cdot (\pf - \pnf) & \text{if } ((j - i)\text{ mod } n) \leq k-1 \\
0 & \text{otherwise}
\end{cases}
\end{align*}
With a slight abuse of notation, we use $\mathcal{M}_n(\ql,\qr)$ to denote the size of a maximum matching in the local model, as we did in previous sections, and $\mu(\ql,\qr)$ to denote $\lim_{n \to \infty} \EE{\frac{\mathcal{M}_n(\ql,\qr)}{n}}$ (the proof of \cref{thm:compare_local_model} also shows that this limit exists). 

%We have assumed that every node is eligible to form edges with any node on the opposite side. However, many platforms address physical matching problems with locality constraints,  and a local model where nodes can only be matched with their neighbors is a more appropriate choice to model such platforms.

% Notice that these conditional probabilities follow the same additive properties as in our main model. 

For small $k$, we find that flexibility asymmetry cannot be a significant effect in the local model. In particular, as in Section \ref{sec:asymmetry}, the one-sided allocation yields expected degrees among regular nodes that are equal to 
$B (\af-\anf) + 2 \anf$ and $2 \anf$ on the two sides respectively. The gap between these is small when $\af-\anf$ is small; however,  under small $k$ in the local model, $\af$ must also be small to ensure $\af/k = \pf \leq 1/2$. Thus, intuitively, we expect flexibility cannibalization to dominate in that regime. Indeed, when $k = 2$, the next result shows the dominance of the one-sided allocation across the entire feasible parameter space:
\begin{theorem}\label{thm:compare_local_model}
    When $k = 2,$ $\mu(\budget,0) > \mu(\budget/2,\budget/2)$ for any $\budget \in (0,1]$ and $\pnf, \pf$ with $0 \leq \pnf < \pf \leq 1/2$.
\end{theorem}

\vspace{-.05in}

%When $k$ is small in the local model, $\af$ must also be small to ensure $\af/k = \pf \leq 1/2$, and thus the value of flexibility is limited. Intuitively, in the one-sided allocation the regular nodes on the two sides respectively have %$\ql \af + (2-\ql) \anf$ and $(B-\ql) \af + (2-B+\ql) \anf$ 
% $B (\af-\anf) + 2 \anf$ and $2 \anf$ neighbors in expectation, and the asymmetry between the two sides increases as $\af - \anf$ increases. Thus, when $\af$ is small, the flexibility asymmetry effect should also be weak. 

\begin{figure}[h]
    \centering
    \subfloat[\centering $k = 2$]{{\includegraphics[width=0.335\textwidth]{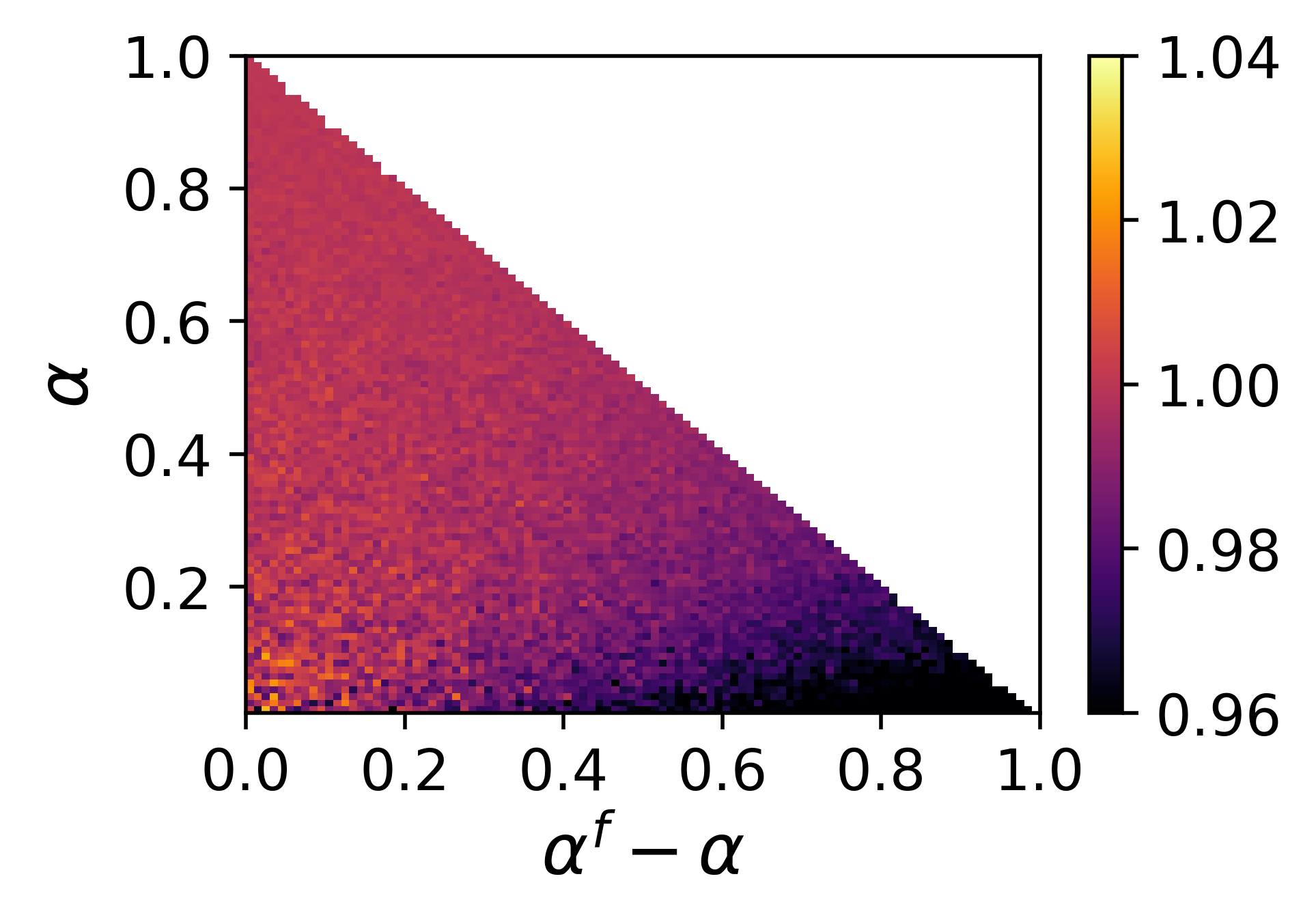} }}%
    \hspace{0cm}
    \subfloat[\centering $k = 5$]{{\includegraphics[width=0.55\textwidth]{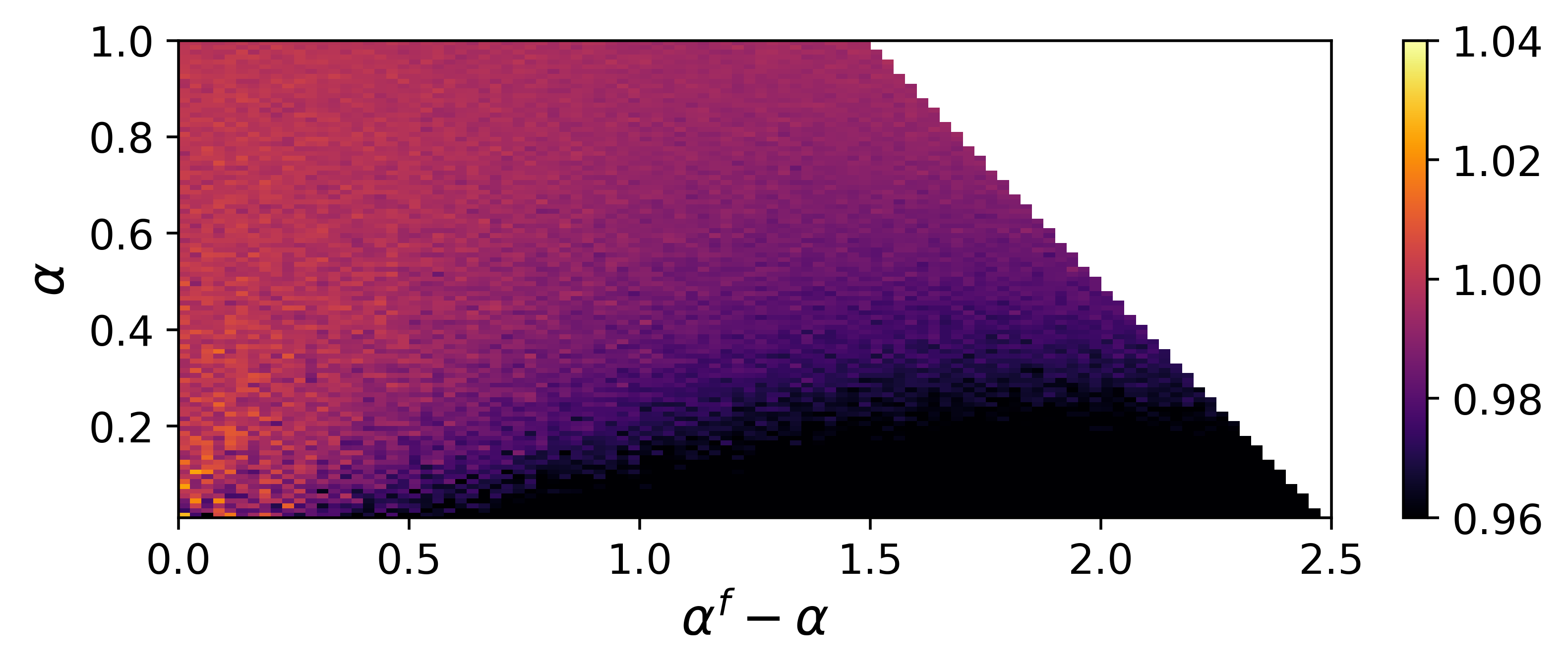} }}%
    \hspace{0cm}
    \subfloat[\centering $k = 10$]{{\includegraphics[width=0.9\textwidth]{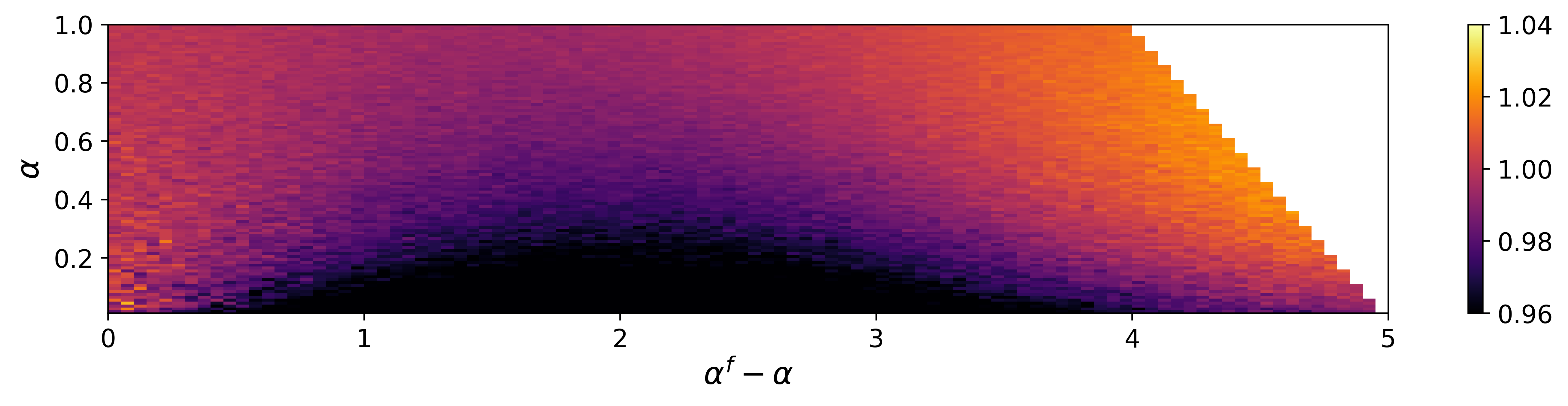} }}%
    \caption{The heatmaps present values of $\frac{\muemp(\budget/2,\budget/2)}{\muemp(\budget,0)}$ in the local model when $B = 0.6$ for varying $k$, $\anf$ and $\af - \anf$. The parameter regimes that violate $0 \leq \pnf < \pf \leq 1/2$ are left blank.}
    \label{fig:ratio_local}
    \vspace{-0.2in}
\end{figure}

To prove \cref{thm:compare_local_model}, we derive a closed-form solution for $\mu(\ql,\qr)$ through the combinatorial analysis of each possible local structure (see Appendix \ref{app:compare_local_model}). However, such an analysis becomes intractable for larger~$k$. Our numerical results in \cref{fig:ratio_local} indicate that as we increase $k$, the resulting plots begin to closely resemble \cref{fig:ratio} with a gradual emergence of the flexibility asymmetry effect. In the parameter regime where cannibalization previously dominated, which captures the entire parameter regime for small $k$, we observe that the one-sided allocation continues to dominate. 
In contrast, when $k$ becomes large enough for $\af-\anf$ to create significant asymmetry, we again observe the balanced allocation performing better.

\vspace{-.05in}
\subsection{Spatial Matching}\label{sec:spatial}

\dfedit{In this subsection, we focus on a model with spatial dependencies to better capture the matching problems faced by ride-hailing and food delivery platforms.} In such platforms, the compatibility between pairs of agents is primarily governed by the distances between them. Thus, we start by considering a two-dimensional cell $[0, 1]^2$ with  uniformly distributed drivers and riders. The $n$ drivers are at locations denoted by vectors $\mathbf{d}_1, \mathbf{d}_2, \ldots, \mathbf{d}_n$, and the $n$ riders at $\mathbf{r}_1, \mathbf{r}_2, \ldots, \mathbf{r}_n$. For a given flexibility allocation $\qvec = (\ql, \qr)$, driver $i$ is flexible if random variable $F^l_i \sim \text{Bernoulli}(\ql)$ takes the value of $1$, otherwise the driver is regular. Similarly, each rider $j$ is associated with $F^r_j \sim \text{Bernoulli}(\qr)$, and the rider is flexible if and only if $F^r_j = 1$. We take constants \(\af\) and \(\anf\) such that \(0 \leq \anf < \af\) and define \(\pf_n = \af/\sqrt{n}, \pnf_n = \anf/\sqrt{n}\), respectively. We assume that an edge exists between a driver $i$ and a rider $j$ if their distance is within a threshold decided by their respective flexibility types:
\begin{align*}
\PP{R_{ij} = 1\mid F^l_i, F^r_j} = 
\begin{cases} 
1 & \text{if } \norm{\mathbf{d}_i - \mathbf{r}_j}_2 \leq 2\pnf_n +(F^l_i + F^r_j) \cdot (\pf_n - \pnf_n) \\
0 & \text{otherwise}
\end{cases}
\end{align*}
In other words, $\mathbf{r}_j$ has an edge with $\mathbf{d}_i$ if their distance is within $2\pnf_n +(F^l_i + F^r_j) \cdot (\pf_n - \pnf_n).$ The asymptotic set-up $\pf_n,\pnf_n \in \Theta(1/\sqrt{n})$ ensures that the expected number of edges in the spatial graph is the same  $\Theta(n)$ that we considered in our previous asymptotic regime. 

\begin{figure}%
    \centering
    \subfloat[\centering $\budget = 0.6$]{{\includegraphics[width=0.35\textwidth]{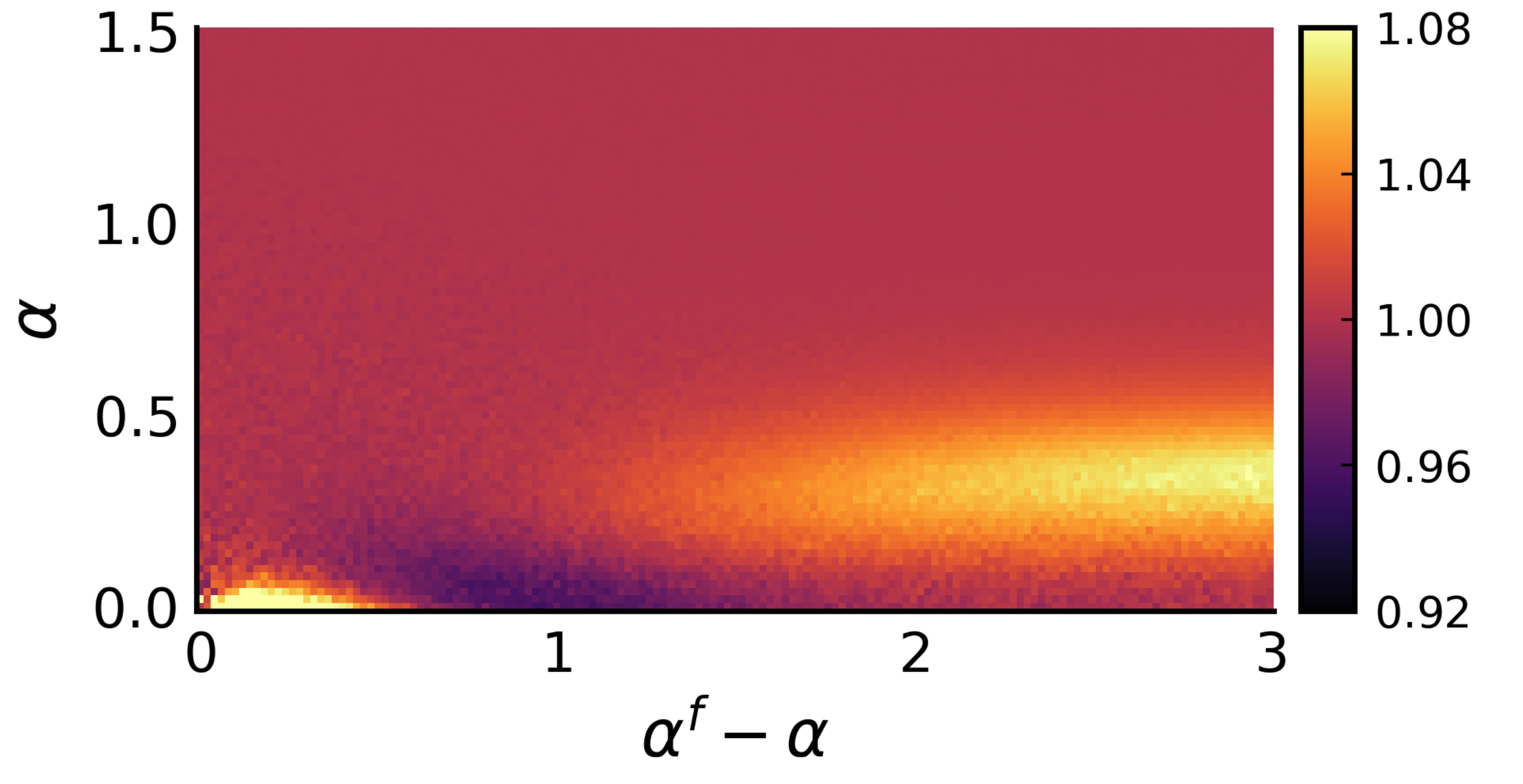} }}%
    \hspace{0cm}
    \subfloat[\centering $\budget = 1$]{{\includegraphics[width=0.35\textwidth]{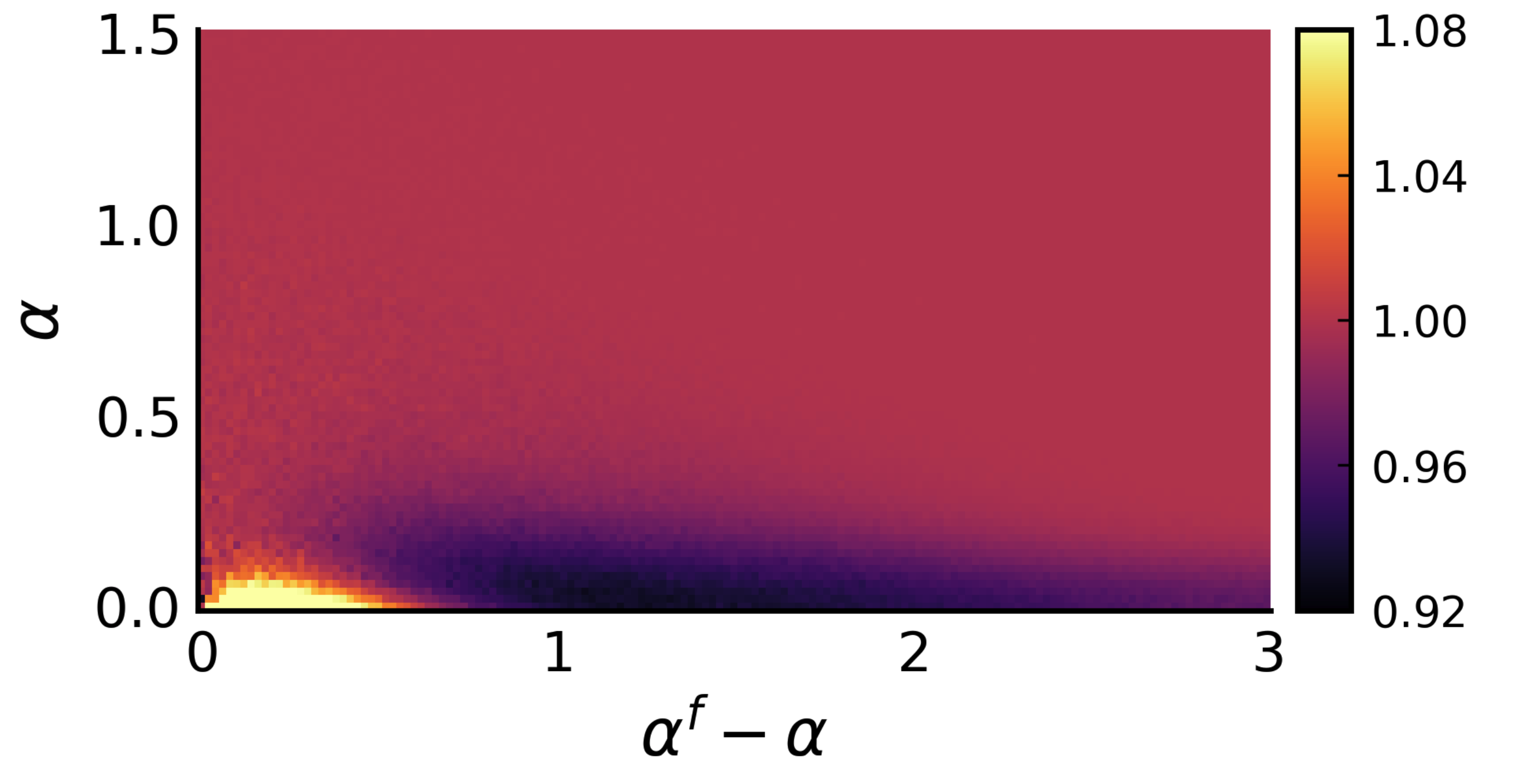} }}%
    \caption{The heatmaps present values of $\frac{\muemp(\budget/2,\budget/2)}{\muemp(\budget,0)}$ in the spatial matching model across varying $\anf$ and $\af - \anf$.}
    \label{fig:ratio3}
    \vspace{-0.2in}
\end{figure}

The spatial model relaxes two assumptions common to the previous models we have examined: (1) the conditional independence assumption on edge realization $R_{ij}$ with respect to indices $i$ and $j$, and (2) the equivalence of different flexibility allocations in expected edge counts. In particular, in the one-sided allocation the expected number of riders that connects to a random driver is $$\parenthesis{B \parenthesis{ \pf_n + \pnf_n}^2 + \parenthesis{1-B} \parenthesis{2 \pnf_n}^2} \cdot \pi \cdot n$$
\dfedit{if we assume for simplicity that the driver is at least $2\pnf_n$ away from the boundary of the $[0,1]^2$ cell (an event that occurs with probability $1$ as $n \to \infty$).}
This is smaller than the expected number of riders that connects to a random driver under the balanced allocation, which equals $$\parenthesis{\parenthesis{B/2}^2 \parenthesis{2 \pf_n}^2 + 2 \cdot B/2 (1-B/2) \parenthesis{ \pf_n + \pnf_n}^2 + \parenthesis{1-B/2}^2 \parenthesis{2 \pnf_n}^2} \cdot \pi \cdot n.$$ As such, we expect the balanced allocation to have an advantage over the one-sided allocation in the spatial setting. Indeed, in \cref{fig:ratio3} we find that in a parameter regime with small $\af$ and $\anf,$ the balanced allocation outperforms the one-sided allocation. This follows because the maximum matching size is close to the number of edges in this very sparse regime, and the latter is higher in expectation in the balanced allocation. In other parts of the heatmap, we find consistency with the results in our main model: the one-sided allocation can outperform the balanced allocation by over 8\% when $B = 1$ or when $\af$ is moderate; moreover, it can be worse by more than 8\% when $\af$ is very large, $\anf$ is a small positive number, and $B < 1$. 

We highlight that $\af$ and $\anf$ capture the density of the spatial market: 
multiplying $\af$ and $\anf$ by a factor of~$\eta > 1$ in a cell with side length $1$ is equivalent to maintaining the number of uniformly distributed agents in the two-dimensional cell  $[0,1/\eta]^2$ but with the same compatibility as before, i.e., with $\af$ and $\anf$ kept constant. As a result, we can interpret 
%Intuitively, in this spatial setting, $\af$ and $\anf$ capture the density of the market: multiplying $\af$ and $\anf$ by a factor of $\eta > 1$ in a grid with side length $1$ is equivalent to holding $\af$ and $\anf$ constant while maintaining the number of agents but shrinking the side length via division through $\eta$. 
the setting with very small $\af$ and $\anf$ as a spatial market in which (i) the density is very low, but (ii) drivers and riders nonetheless ``expect'' to be matched with agents that are very close. In contrast, as the market density increases, agents form more edges, which leads, at first, to flexibility cannibalization. With a further increase in density, we observe that when $B<1$ flexibility asymmetry becomes a dominant effect and the balanced allocation yields a much larger matching size than the one-sided allocation. Therefore, a natural interpretation of our results is that the optimal flexibility allocation depends on the market density, the flexible/regular agents' acceptable dispatch radius, and the flexibility penetration $B$.

\subsection{Imbalanced Market}\label{sec:imbalanced}
\begin{figure}[t]%
    \centering
    % -------- Row 1--------
    \subfloat[\centering $\frac{\muemp(0.3, 0.3)}{\muemp(0.6, 0)}$]{{\includegraphics[width=0.35\textwidth]{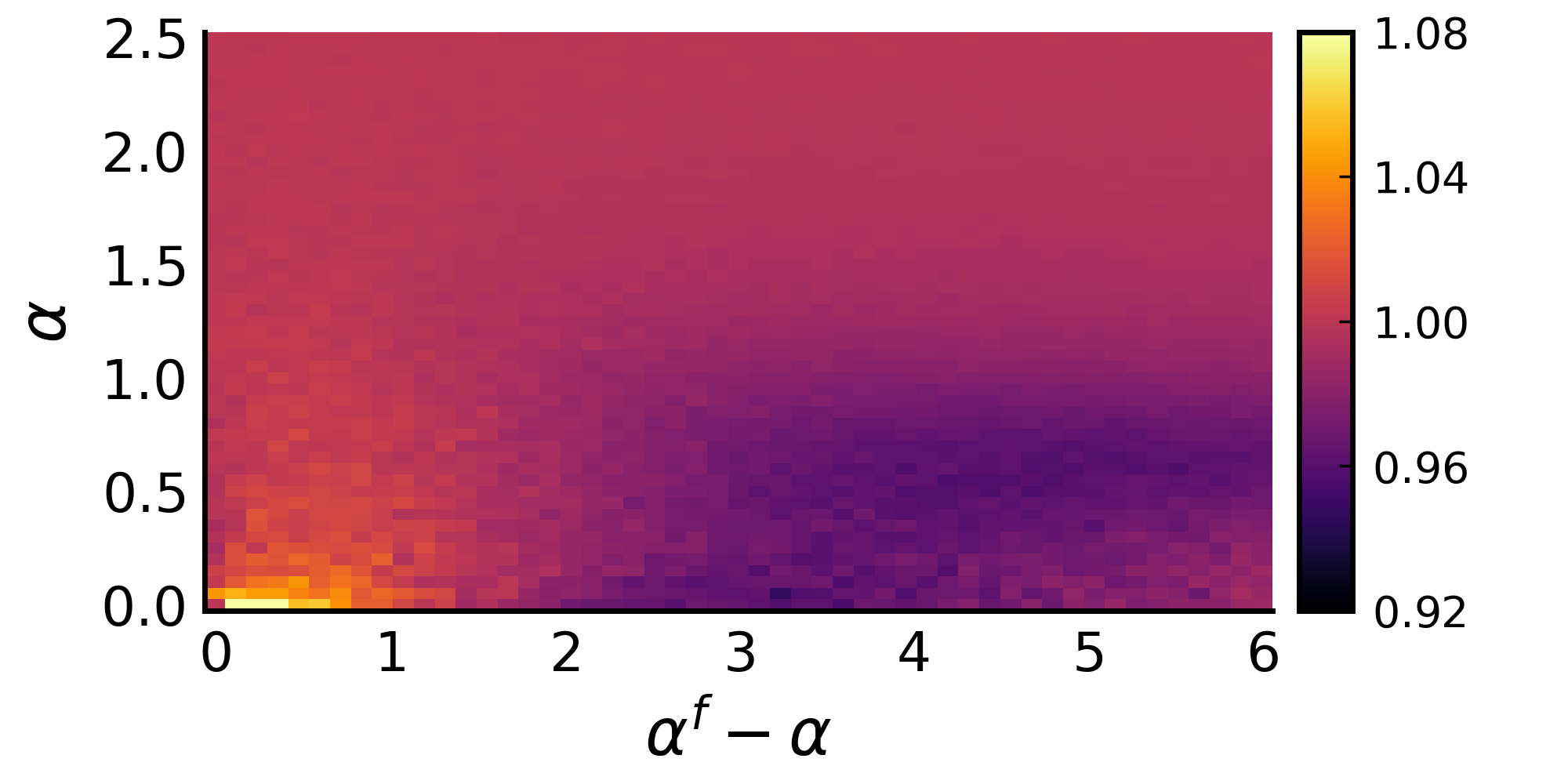} }}%
    \hspace{0cm}
    \subfloat[\centering $\frac{\muemp(0.5, 0.5)}{\muemp(1, 0)}$]{{\includegraphics[width=0.35\textwidth]{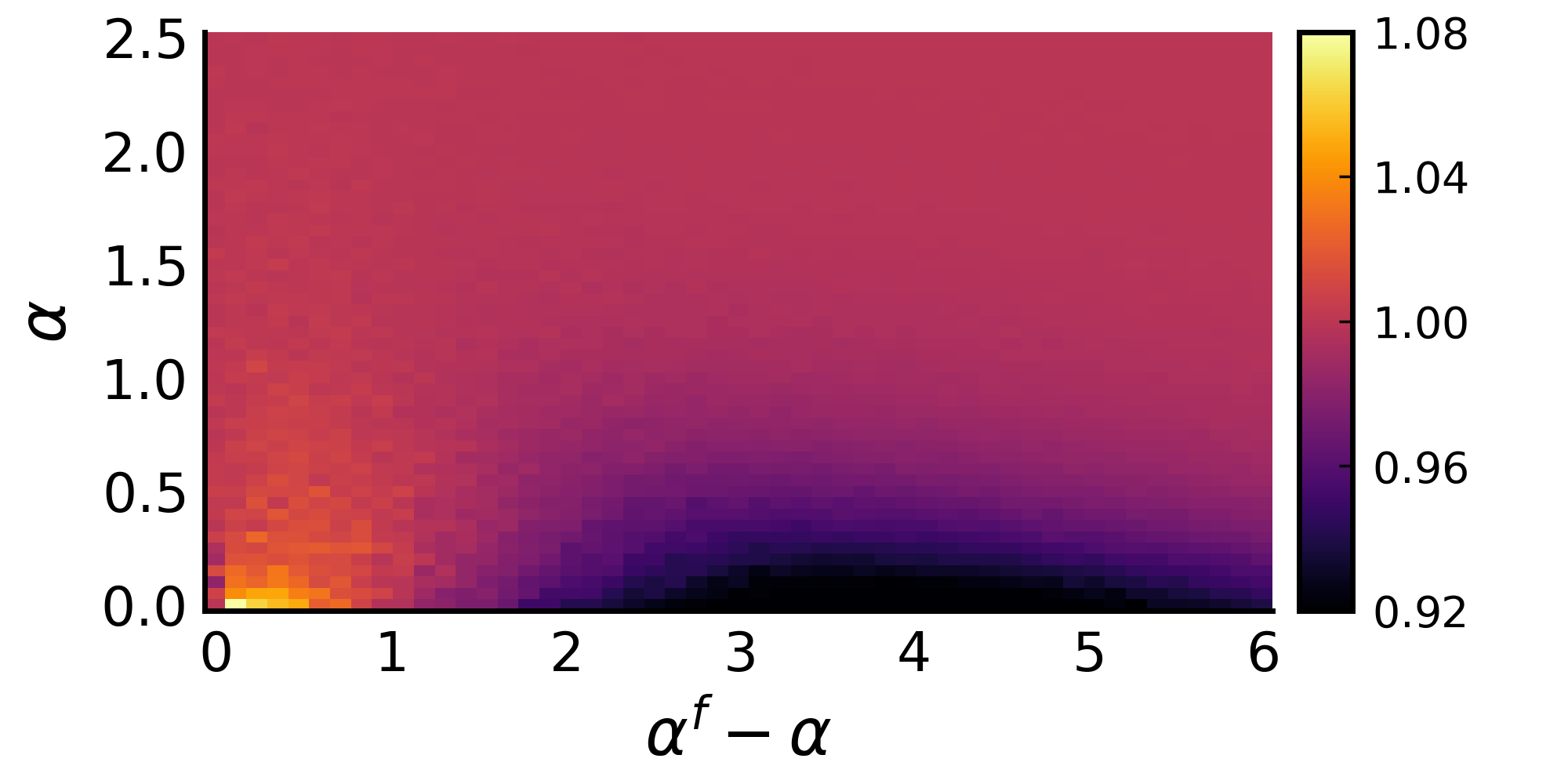} }}%
    \vspace{0em}
    % -------- Row 2--------
    \subfloat[\centering $\frac{\muemp(0.3, 0.3)}{\muemp(0, 0.6)}$]{{\includegraphics[width=0.35\textwidth]{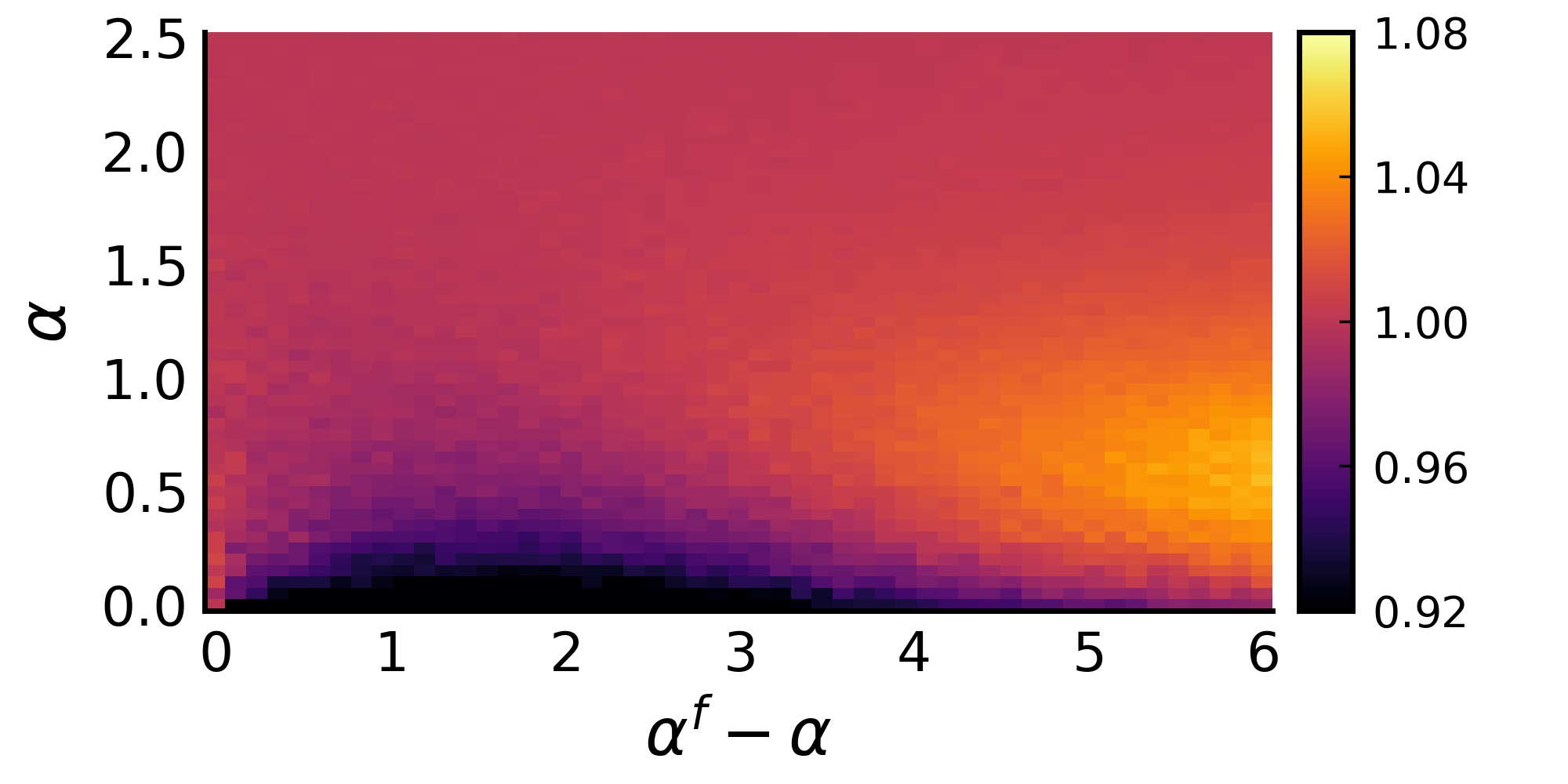} }}%
    \hspace{0cm}
    \subfloat[\centering $\frac{\muemp(0.5, 0.5)}{\muemp(0, 1)}$]{{\includegraphics[width=0.35\textwidth]{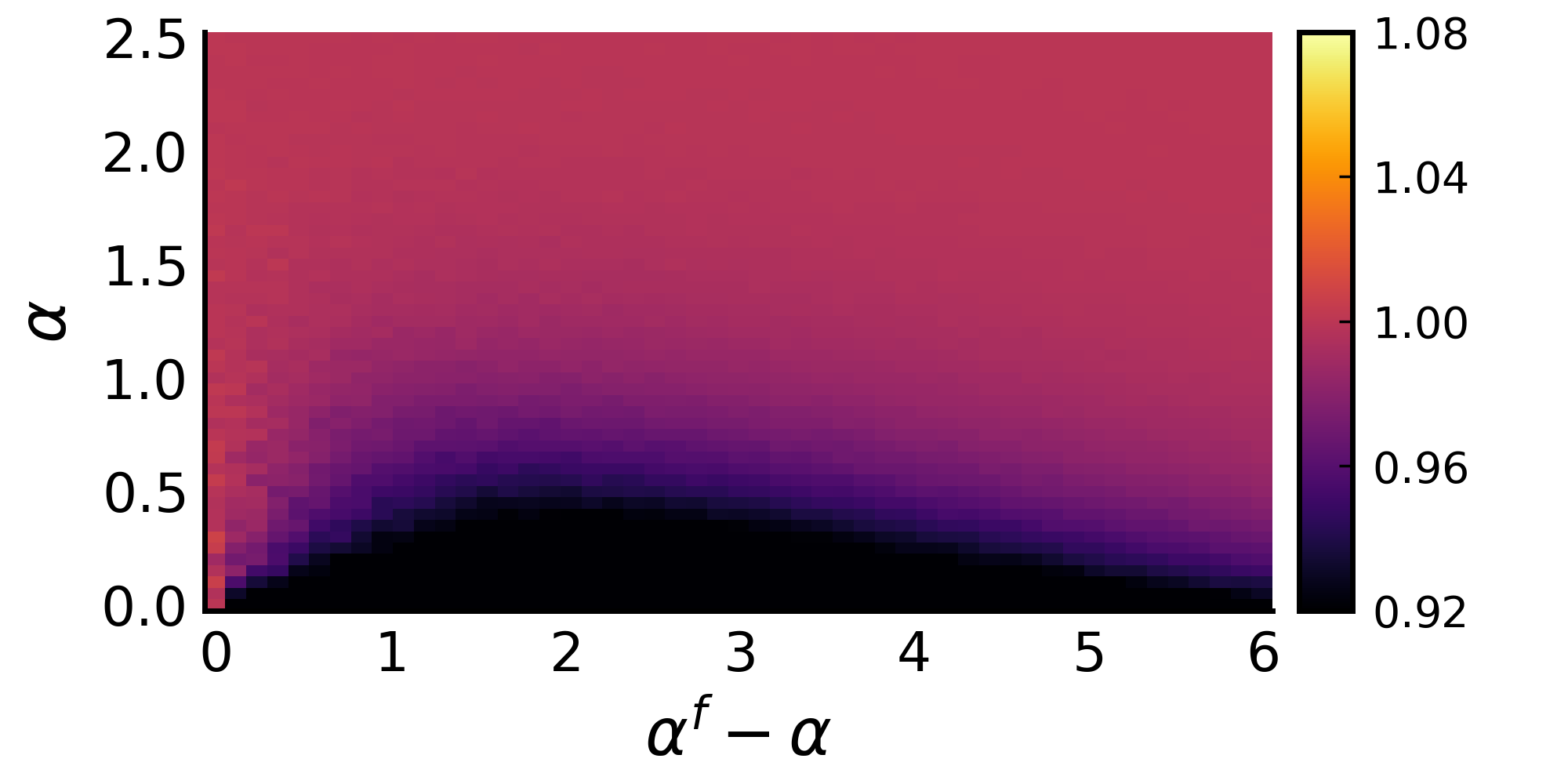} }}%
    \vspace{0em}
    % -------- Row 3--------
    \subfloat[\centering $\frac{\muemp(0.3, 0.3)}{\max\bracket{\muemp(0.6, 0),\muemp(0, 0.6)}}$]{{\includegraphics[width=0.35\textwidth]{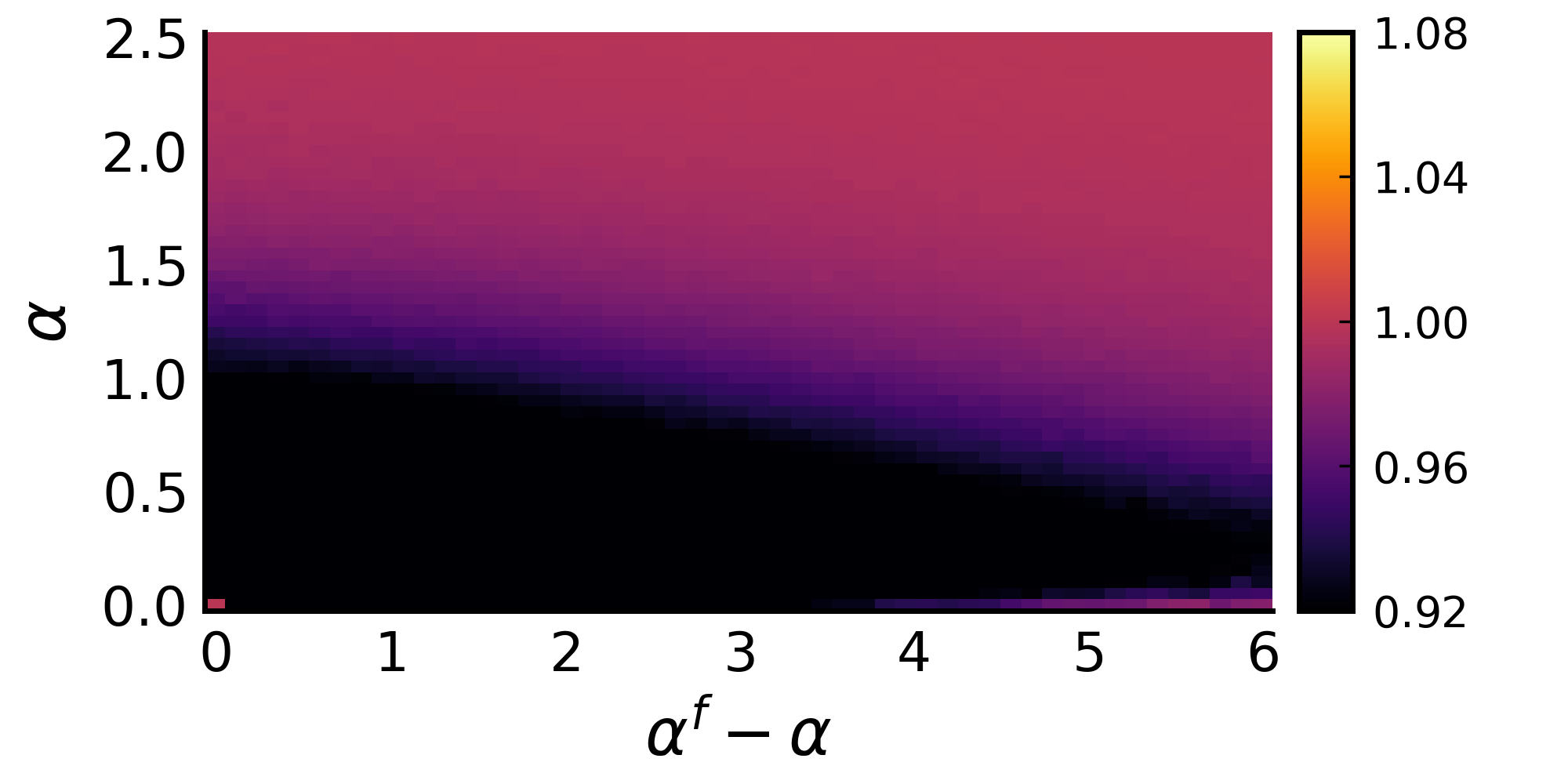} }}%
    \hspace{0cm}
    \subfloat[\centering $\frac{\muemp(0.5, 0.5)}{\max\bracket{\muemp(1, 0),\muemp(0, 1)}}$]{{\includegraphics[width=0.35\textwidth]{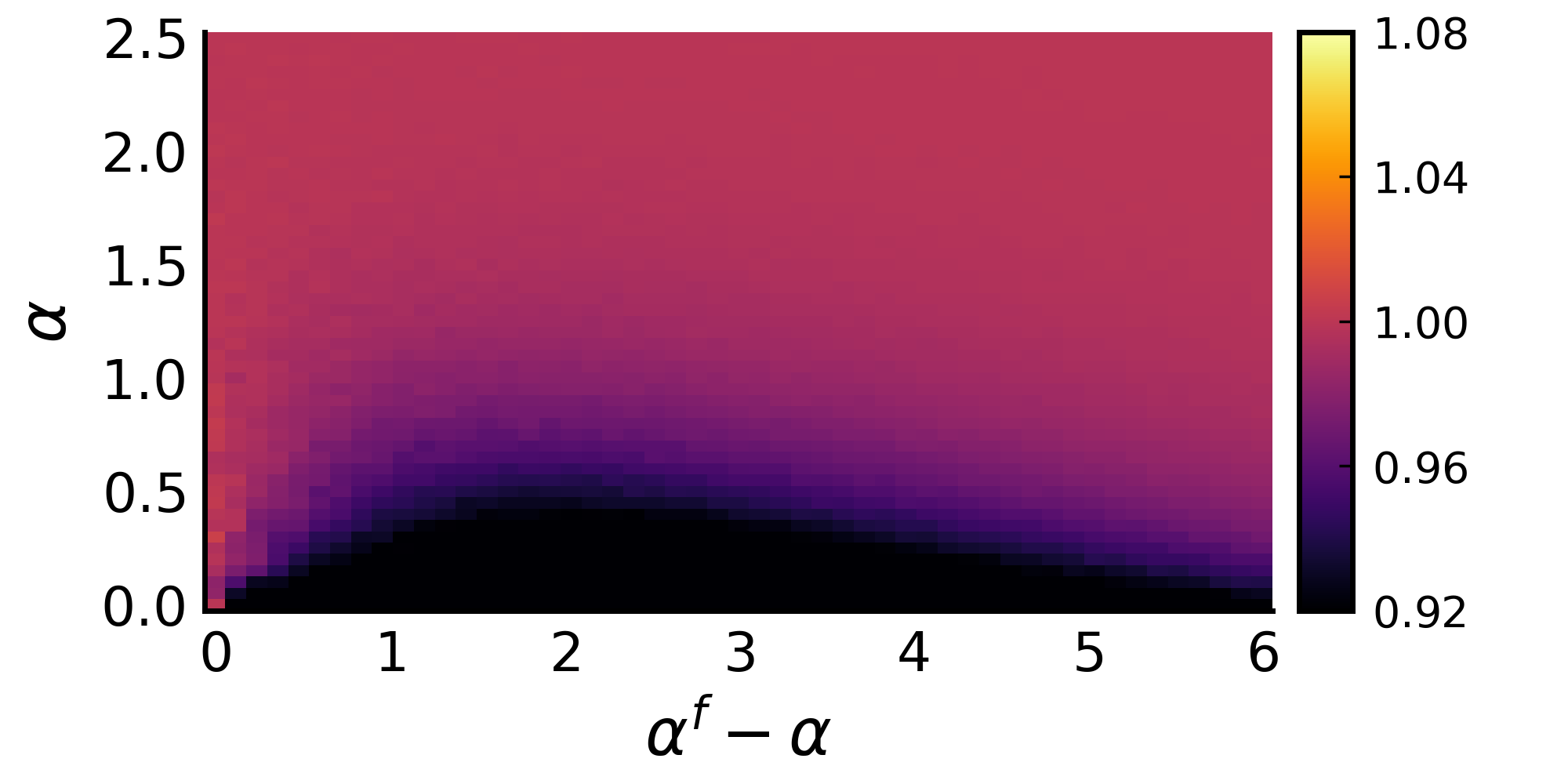} }}%
    % \vspace{0em}
    \caption{Comparing allocations $(B,0), (B/2, B/2)$ and $(0,B)$ in the imbalanced model when $\lambda = 0.8$.}
    \label{fig:ratio4}
    \vspace{-0.2in}
\end{figure}

In this subsection we numerically investigate whether the structural insights we identified in fully symmetric settings are robust to imbalances in matching markets. \dfedit{Specifically, we extend our model from \cref{sec:model} by allowing $\lambda \cdot n$ nodes in $V_r$, where $\lambda \in (0,1]$ (we ignore for symmetry the setting where $V_l$ is the short side). We assume that, for a given flexibility allocation $(\ql,\qr)$, each node in $V_l$ is flexible with probability $\ql \cdot \lambda$ and each node in $V_r$ is flexible with probability $\qr.$ This set-up ensures that the cost of incentivizing an equal number of nodes on the left- or right-hand side remains the same.} \dfedit{However, in this imbalanced setting, the one-sided allocations $(B,0)$ and $(0,B)$ are no longer equivalent. Comparing the allocations $(B,0)$, $(B/2,B/2)$, and $(0,B)$ reveals two opposing forces. First, similar to Section \ref{sec:spatial}, the expected number of edges is not fixed by $B$. Rather, it increases with the amount of flexibility placed on the right side, i.e., under $(0,B)$ we expect to observe more  edges than under $(B/2,B/2)$ and we expect even fewer under $(B,0)$.  Similar to Section \ref{sec:spatial}, in \cref{fig:ratio4} (a)-(d) we observe for the setting with very small $\anf,\af$, where the number of matches is close to the number of edges, that the resulting matching is larger the more flexibility is placed on the right. Second, due to the imbalance, even in a complete graph, there are always $(1-\lambda) n$ nodes in $V_l$ that cannot be matched; thus, we can better afford having hard-to-match or isolated nodes in $V_l$ than in $V_r$. Yet, 
under $(0,B)$ the hard-to-match regular nodes (i.e., those with an expected degree of $2 \anf$ because all of their neighbors are also regular nodes) reside in $V_r$. This becomes the dominating effect when $\anf$ is small, so the regular nodes are actually hard to match, and $\af \gg \anf$, so the matching size is not proportional to the number of edges, and the  benefit of additional edges under $(0,B)$ becomes smaller. In that regime we find that $(0,B)$ becomes the worst-performing allocation, followed by $(B/2,B/2)$ and then $(B,0)$. As a result, we find in \Cref{fig:ratio4} (e) and (f) that the better of the two one-sided allocations outperforms the balanced allocation across the board. These results suggest that, by carefully selecting the market side to incentivize flexibility based on the values of $\af$ and $\anf$, an imbalanced market benefits from asymmetric flexibility. In Appendix \ref{sec:simu_additional} we extend the  spatial model from \cref{sec:spatial} to an imbalanced setting and find that the better of the one-sided allocations dominates except for when $\anf,\af$ are extremely small, in which case the reasoning from Section \ref{sec:spatial} explains why the balanced allocation dominates.}
%\dfcomment{sooo... question: how does $(B,0)$ compare to the balanced allocation? Roughly speaking, I am wondering if the takeaway here should be that ``when one side is larger, then we should only incentivize the other side''?} \kzcomment{$(B,0)$ can still outperform the balanced allocation (in the same region where flexibility cannibalization previously dominates) but the extent is much weaker because it provably has fewer edges (in expectation) than the balanced allocation. Shall I include a plot here/how we just explain it in text? I think the overall message is that such an imbalanced market favors $(0,B)$ and hurts $(B,0)$}
%the resulting advantage of $(0,B)$ over $(B/2,B/2)$ in the expected edge counts weakens the flexibility asymmetry effect we previously observed for $B< 1$. 
%In contrast, the regions with the strongest advantage for the one-sided allocation are consistent % strongest in the same regions that we observed the regions where flexibility cannibalization dominates align 
%with all of our previous findings.

\vspace{-.05in}

\subsection{Heterogeneous Preferences}\label{sec:heterogeneous}

\begin{figure}[H]
    \centering
    \subfloat[\centering $B = 0.6$]{{\includegraphics[width=0.4\textwidth]{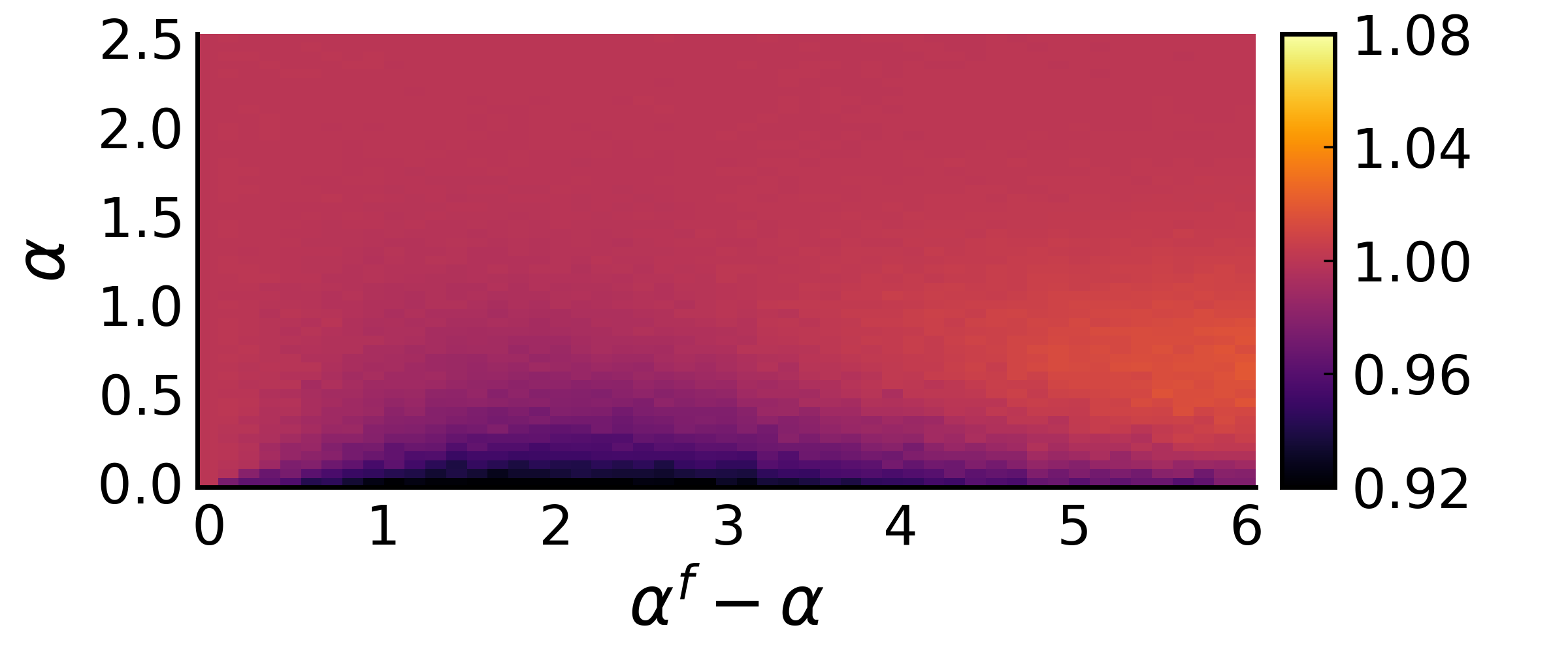} }}%
    \hspace{0cm}
    \subfloat[\centering $B = 1$]{{\includegraphics[width=0.4\textwidth]{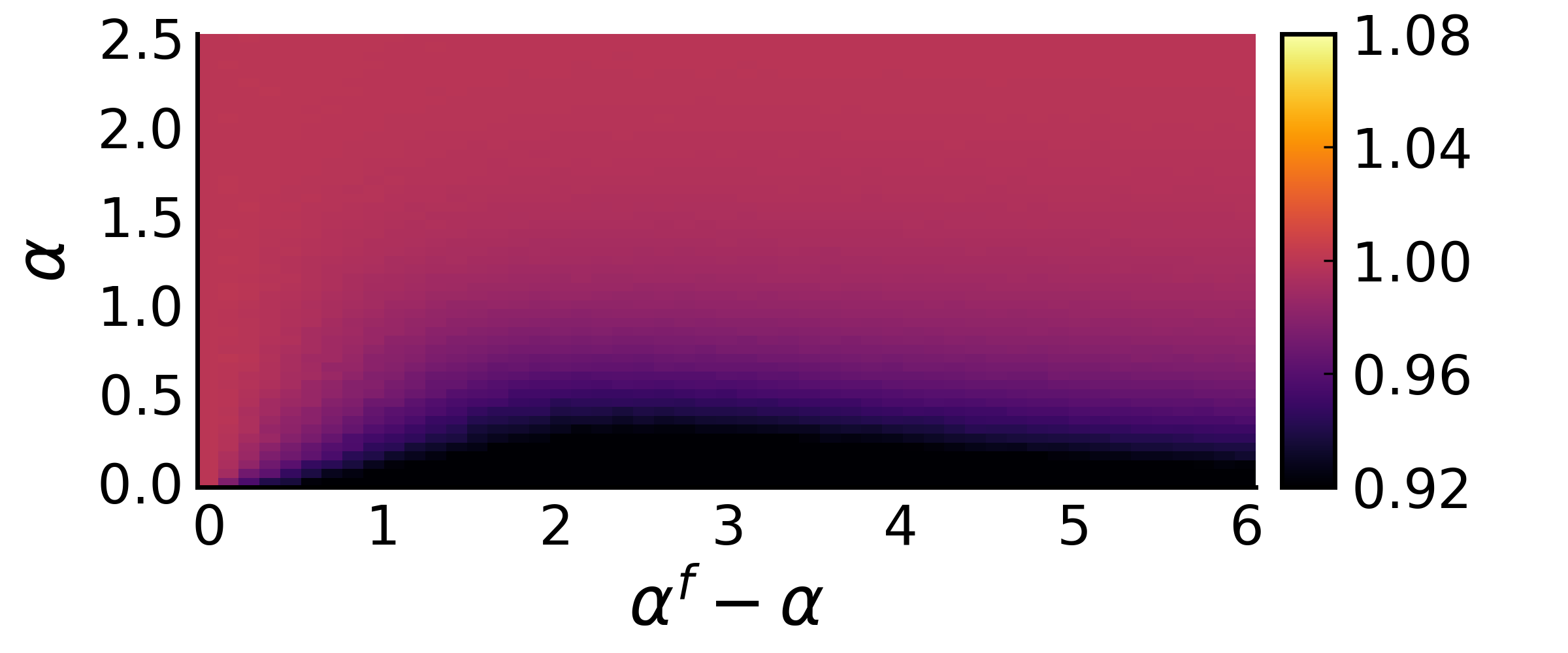} }}%
    \caption{The heatmaps present values of $\frac{\muemp(\budget/2,\budget/2)}{\muemp(\budget,0)}$ in the heterogeneous preference model across varying $\anf$ and $\af -\anf$.}
    \label{fig:ratio5}
    \vspace{-0.2in}
\end{figure}

\dfedit{
We next extend the model from Section \ref{sec:model} by allowing agents to exhibit heterogeneous preferences over potential matches. This modeling extension reflects that even when a rider–driver pair is feasible, agents may still prefer some matches to others (e.g., preferring a nearby regular driver over a more distant flexible one). To capture heterogeneous match quality, we introduce a random utility model: for every pair $(i,j)$, we draw a utility $w_{ij} \sim U(0,1)$ independently across pairs. Eligibility of an edge is still governed by the same flexibility-based thresholds as in the baseline model—namely, an edge is eligible if $w_{ij} > 1-2\anf/n$, $w_{ij} > 1-(\af+\anf)/n$, or $w_{ij} > 1-2\af/n$, %$w_{ij} > 1-2\pnf_n$, $w_{ij} > 1-(\pf_n+\pnf_n)$, or $w_{ij} > 1-2\pf_n$, 
depending on the flexibility types of $i$ and $j$. Among all eligible edges, the platform now seeks a maximum weight matching, where edge weights scale with utility (in the numerics below, the weight is taken as {$w_{ij} - 0.8$}). This extension enriches the framework by jointly modeling type-based feasibility and heterogeneous match desirability.%\dfcomment{now that i am looking at it, i just want to make sure that our rescaling did not affect the alpha-values? if it did not, i.e., if Kamessi is sure that what is written here is correct on a technical level/aligned with what is implemented, feel free to delete this comment.}

Across the full range of $(\anf,\af,\ql,\qr)$, our qualitative insights remain robust: we numerically find that either the one-sided or the balanced allocation seem to perform best (though the weights increase variance, we find across 300 tested values of $(\anf,\af)$ that other allocations never improve by more than $0.7\%$), and the parameter regions in which each allocation is superior closely parallel those in the base model (see \cref{fig:ratio5}). Quantitatively, however, the region in which the balanced allocation dominates becomes smaller after one accounts for agents' heterogeneous preferences. The reason for this is that additional flexible–flexible edges created under balanced flexibility result in low–$w_{ij}$ matches, which reduces, but does not eliminate, their contribution to the weighted objective. This weakens the relative benefit of the balanced allocation compared to the unweighted maximum matching setting.
}

\section{Conclusion}\label{sec:conclusion}

\dfedit{
In summary, our work provides a conceptual starting point for the study of two-sided flexibility in bipartite matching structures. Our model is not meant to represent any specific platform accurately but to gather structural insights likely to generalize to other settings. In particular, we identify flexibility cannibalization and asymmetry, which are respective drawbacks of the balanced and one-sided allocations, showcasing that flexibility on both sides can interact in complex ways. We characterize the typical parameter regimes where each effect dominates and numerically identify that an optimal flexibility allocation can affect the matching size by around 8\% --- in Section \ref{sec:experiment} we highlight that, depending on a platform's cost structure, the impact on profit can be even more substantial. 
Our main practical recommendation for platforms is that their various products affecting flexibility on both sides interact and should therefore not be considered independently. The teams in charge of these products should communicate and conduct experiments and simulations that jointly vary different levers to avoid suboptimal outcomes that can easily arise otherwise.
\\
On the theoretical side, comparing the expected maximum matching sizes in graphs with different flexibility allocations is a challenge, and we leverage a coupling construction, employ concentration bounds, and generalize KS algorithm-based analyses with computer-aided proofs. Nonetheless, our work gives rise to ample directions for future research:
\begin{itemize}
\item Though our numerical results, both with regards to $\mubar$ and $\muemp$, strongly suggest that $\mu$ is always maximized under either the balanced or the one-sided allocation, we do not have a proof of that being in the case in any parameter regime. Moreover, as we do not see a path to apply any of our techniques towards such a result, we view this as a natural open question posed by our work. 
\item Our analytical comparisons between the one-sided and the balanced allocation are also restricted to edge cases of the parameter space. Though these are most intuitive to illustrate flexibility cannibalization and asymmetry, it begs the question if $\mu$ admits an analytical delineation between the one-sided and the balanced allocation being optimal in the way that $\phi$ does (see \Cref{fig:deg0}).
\item Our work provides a full analytical treatment of the metric $\phi$, and a comprehensive numerical study along with analytical results of limited special cases, of $\mu$, but there are many other metrics by which one could evaluate flexibility allocations in bipartite matching settings. As an example, Section \ref{sec:heterogeneous} involves a utility model and a weighted matching whereas  Appendix~\ref{sec:greedy} provides numerical results for two natural greedy heuristics. These sections identify both similarities and differences to what we find for $\mu$ and $\phi$; future work could explore flexibility allocations under these and further additional~metrics.
\item Next, our model intentionally focuses on a particular type of edge probability distribution that keeps the expected number of edges invariant for a given budget~$B$. The advantage of this modeling choice is that it guarantees that all differences in our metrics are driven solely by the \emph{distribution of realizing edges}, not by the \emph{number of realizing edges}. Yet, in practice we would expect that the flexibility allocation affects the number of realizing edges and analyses that capture this (e.g., based on random geometric graphs or as described in \cref{sec:heterogeneous}) would be of both theoretical and practical interest.
\end{itemize}
Finally, though we focus on a matching model, two-sided flexibility may also appear in queueing and manufacturing settings. Investigating these could be fruitful and lead to a general theory of two-sided flexibility.
}

\vspace{-.25in}

\theendnotes  

\bibliographystyle{plainnat}

  % <- This includes the precompiled bibliography

\newpage
\setcounter{page}{1}
% Appendix

\begin{APPENDICES}{}

{

\section{Additional Examples of Two-sided Flexibility in Platforms}\label{app:examples}

{In this appendix, we provide additional examples of two-sided flexibility in platforms. As illustrated in \cref{fig:lever_ubereats}~(a), the ``No Rush Delivery" option on Uber Eats works similarly to the ``Wait and Save" feature on Lyft. This flexibility incentive allows eaters to receive a discount if they are willing to accommodate a delay in their food delivery. By opting into this feature, eaters allow Uber Eats more flexibility in matching them with delivery drivers, as the platform is given a broader pool of eligible drivers with which to match orders. On the supply side, Uber Eats also offers surge incentives, such as small bonuses (\cref{fig:lever_ubereats} (b)), to encourage drivers to remain committed to the platform and fulfill deliveries during busy periods. Similar to Lyft’s ``Ride Streak" bonuses, surge incentives make drivers compatible with a wider of assigned deliveries as they incentivize the acceptance of jobs that would otherwise be declined. % likelihood of accepting assigned deliveries because, with the bonus, drivers may accept assignments that they would otherwise decline.

\begin{figure}[H]
    \centering
    \subfloat[\centering No Rush Delivery]{{\includegraphics[width=0.23\textwidth]{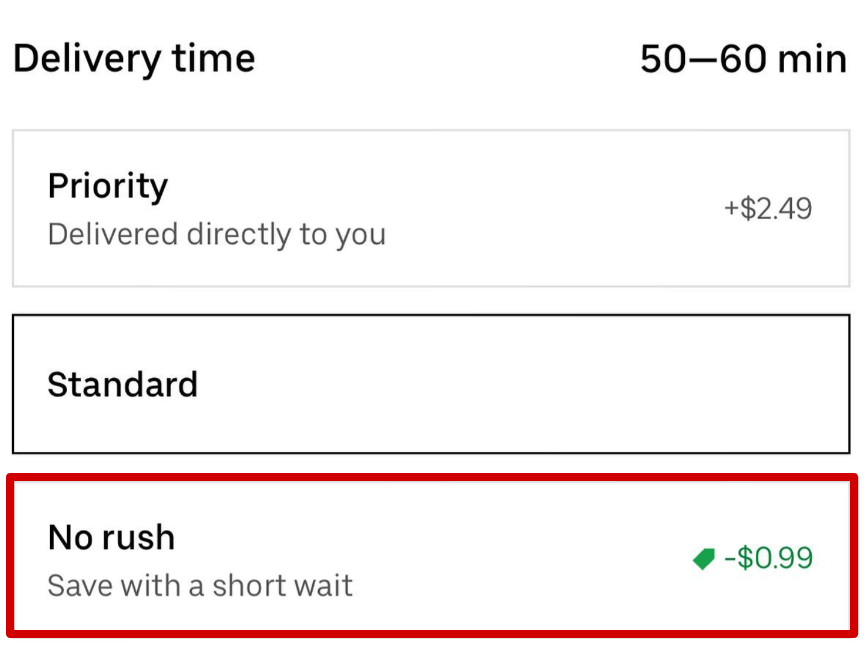} }}%
    \hspace{1.5cm}
    \subfloat[\centering Surge incentives]{{\includegraphics[width=0.24\textwidth]{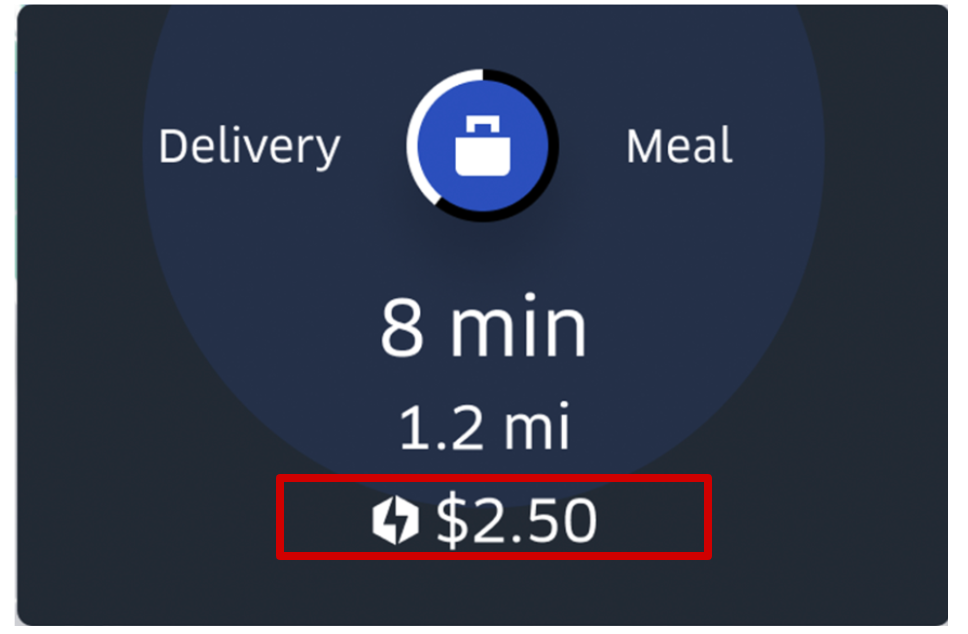} }}%
    \caption{Examples of flexibility incentives on demand and supply sides of Uber Eats, a food-delivery platform.}
    \label{fig:lever_ubereats}
    \vspace{-0.1in}
\end{figure}

\begin{figure}[H]
    \centering
    \subfloat[\centering ``Project Catalog" Design]{{\includegraphics[width=0.27\textwidth]{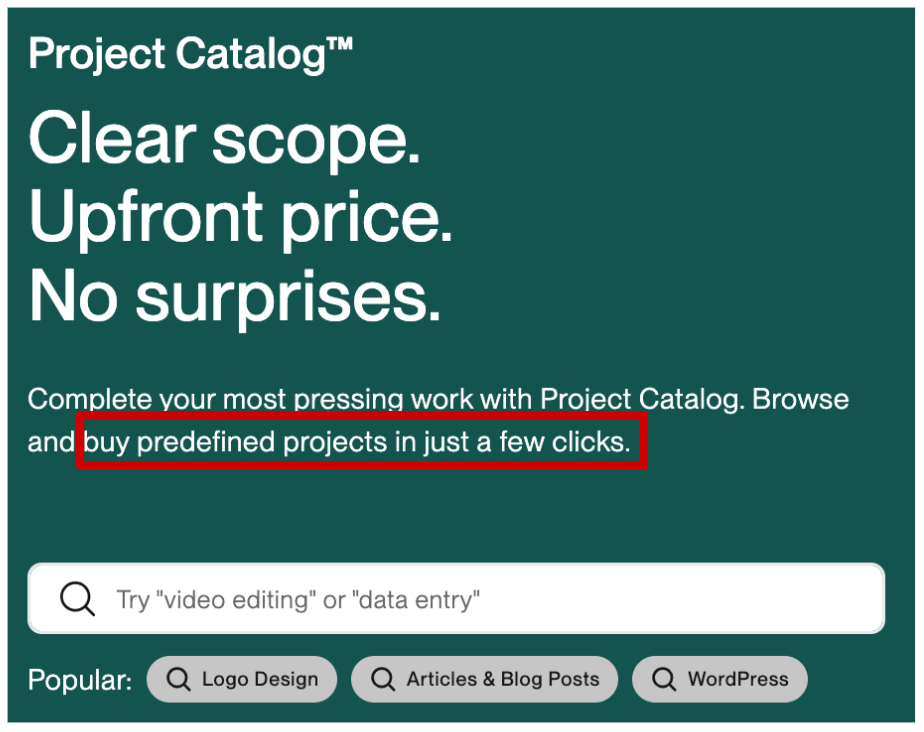} }}%
    \hspace{1.5cm}
    \subfloat[\centering ``Upwork Academy" Program]{{\includegraphics[width=0.29\textwidth]{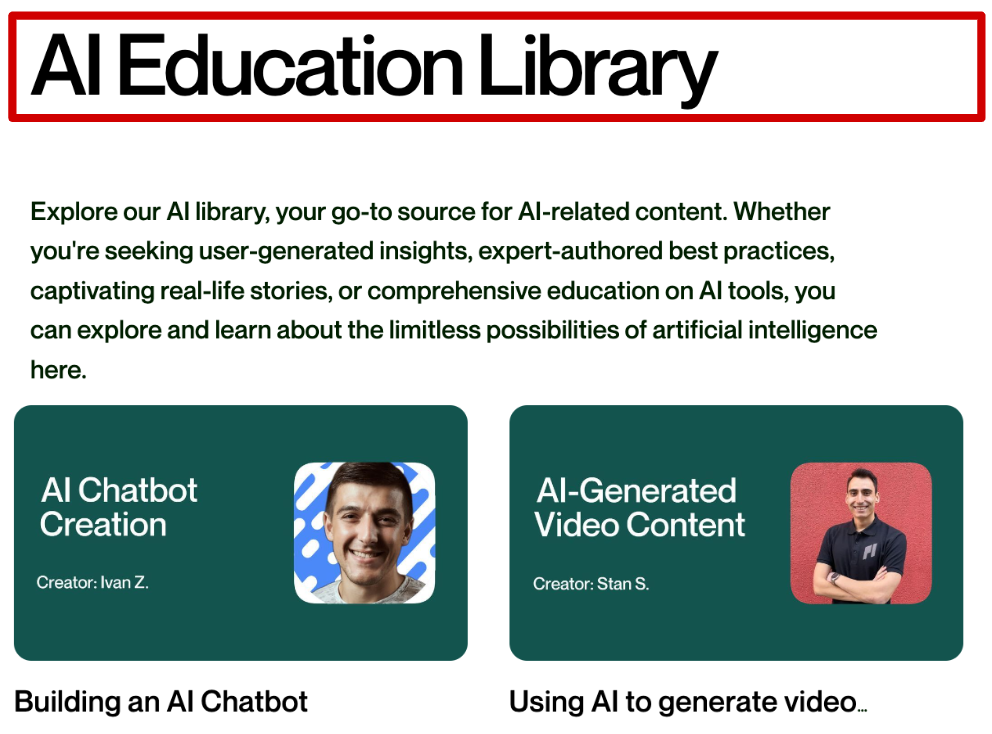} }}%
    \caption{Examples of flexibility incentives on demand and supply sides of Upwork, an online freelancing platform.}
    \label{fig:lever_upwork}
    \vspace{-0.1in}
\end{figure}

Examples of two-sided flexibility can also be found in online freelancing platforms. The platform Upwork has designed a ``Project Catalog" that allows freelancers to post predefined project offerings (\cref{fig:lever_upwork} (a)). This contrasts with the platform's traditional business model where customers post specialized requests and freelancers bid on them. With Project Catalog, less picky customers (i.e., those that do not have specialized requests in mind) can opt for standardized services, such as video editing or data entry, that can be fulfilled by a broad range of freelancers. By limiting the scope of customization, the platform makes it easier for these customers to match with freelancers. On the supply side, Upwork promotes flexibility through its Upwork Academy, which offers training courses in specialized skills, such as AI chatbot creation and video content generation (\cref{fig:lever_upwork} (b)), to enable freelancers to expand their skill sets. These courses then allow the freelancers to take on more types of jobs and increases their compatibility with customer requests. Through these examples we find that platforms commonly enhance the overall matching efficiency by steering agents on both market sides to be more flexible.}

\dfedit{
\section{Flexibility Designs under Greedy Matching Heuristics}\label{sec:greedy}
\dfedit{
In this appendix, we numerically explore the effect of flexibility allocations on the matching sizes that arise under two natural greedy heuristics. This complements the study of the maximum matching metric $\mu(\ql,\qr)$ and the fraction of non-isolated nodes $\phi(\ql,\qr)$ introduced in the main body, which both provided upper bounds on the achievable matching size under a given flexibility allocation. In contrast,  both greedy heuristics are easily implementable even in an environment in which nodes on one side arrive online over time; as such, the metrics in this section yield lower bounds on downstream matching sizes as a function of the upstream flexibility allocations. Across all of our numerical results we find that our insights from the main body continue to hold under the greedy heuristics. Specifically, (i) it seems that the greedy matching sizes seem to still be maximized under either the one-sided or the balanced allocation and (ii) the regions where each one dominates closely resemble those we identified in the main body.

\emph{Greedy heuristics.} We consider matching sizes that arise under two greedy heuristics. The underlying graphs are created as described in Section \ref{sec:model}. The greedy heuristics process nodes from $V_l$ in order and assign them to nodes from $V_r$ as follows:
\begin{figure}[t]%
    \centering
    % -------- Row 1: mu --------
    \subfloat[\centering $\frac{\muemp(0.3,0.3)}{\muemp(0.6,0)}$]{
        {\includegraphics[width=0.45\textwidth]{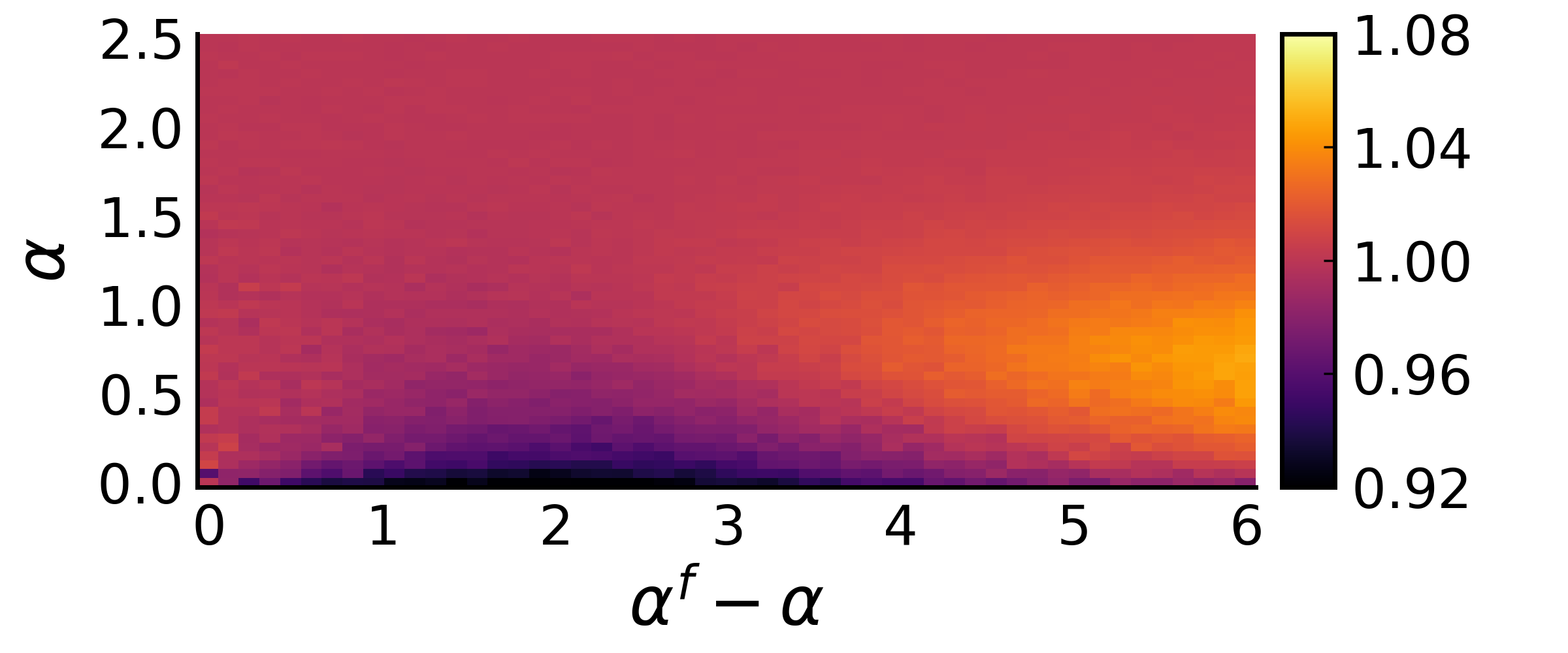}}
    }
    \subfloat[\centering $\frac{\muemp(0.5,0.5)}{\muemp(1,0)}$]{
        {\includegraphics[width=0.45\textwidth]{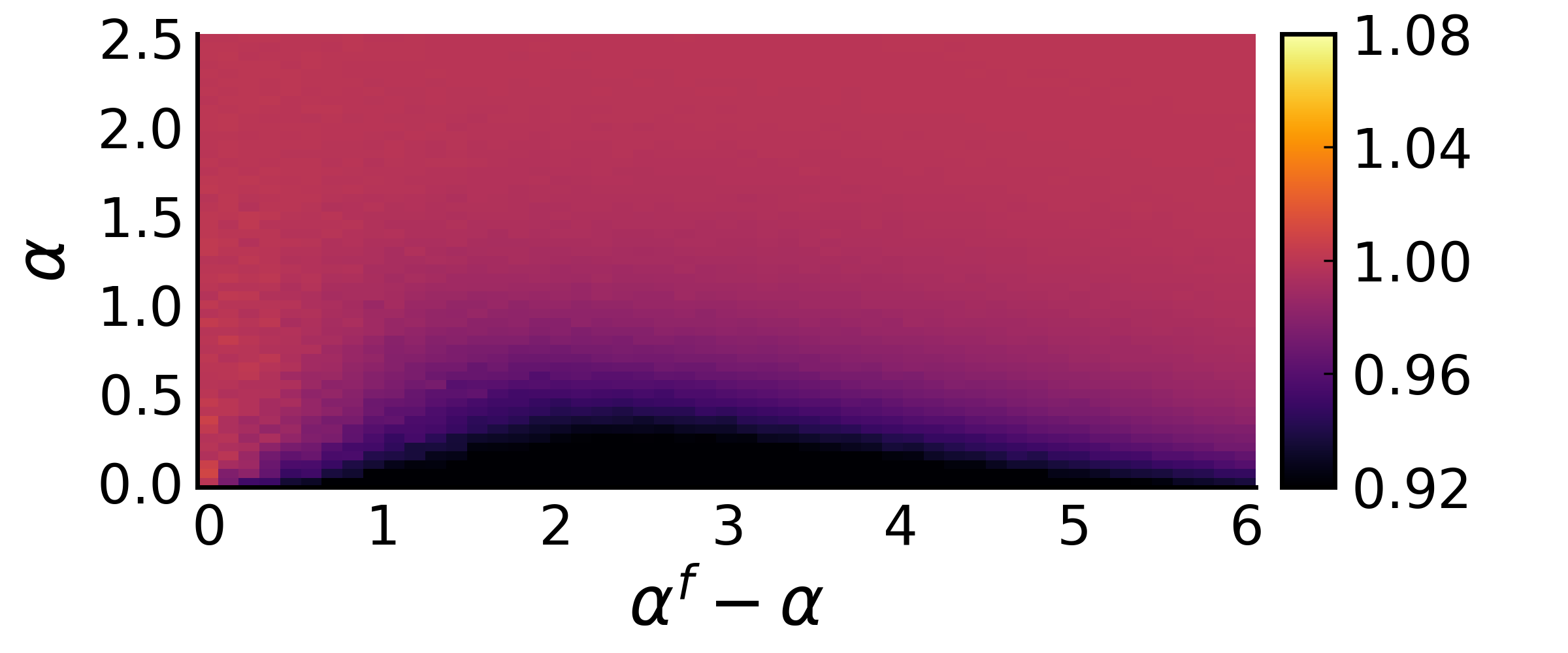}}
    }
    \vspace{0em}
    % -------- Row 2: phi --------
    \subfloat[\centering $\frac{\phi^{\texttt{EMP}}(0.3,0.3)}{\phi^{\texttt{EMP}}(0.6,0)}$]{
        {\includegraphics[width=0.45\textwidth]{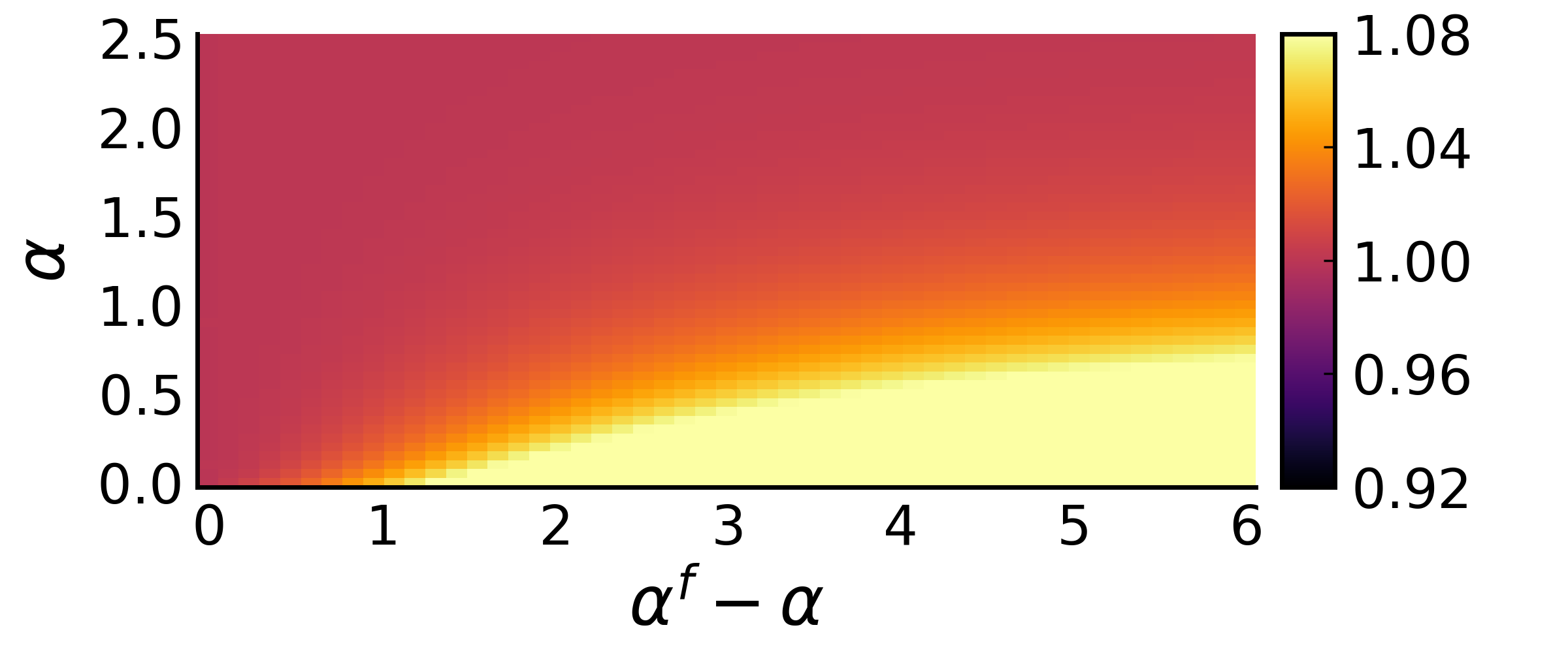}}
    }
    \subfloat[\centering $\frac{\phi^{\texttt{EMP}}(0.5,0.5)}{\phi^{\texttt{EMP}}(1,0)}$]{
        {\includegraphics[width=0.45\textwidth]{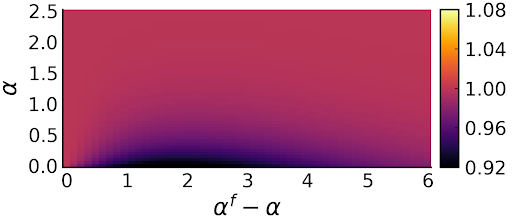}}
    }
    \vspace{0em}
    % -------- Row 3: psi --------
    \subfloat[\centering $\frac{\psi^{\texttt{EMP}}(0.3,0.3)}{\psi^{\texttt{EMP}}(0.6,0)}$]{
        {\includegraphics[width=0.45\textwidth]{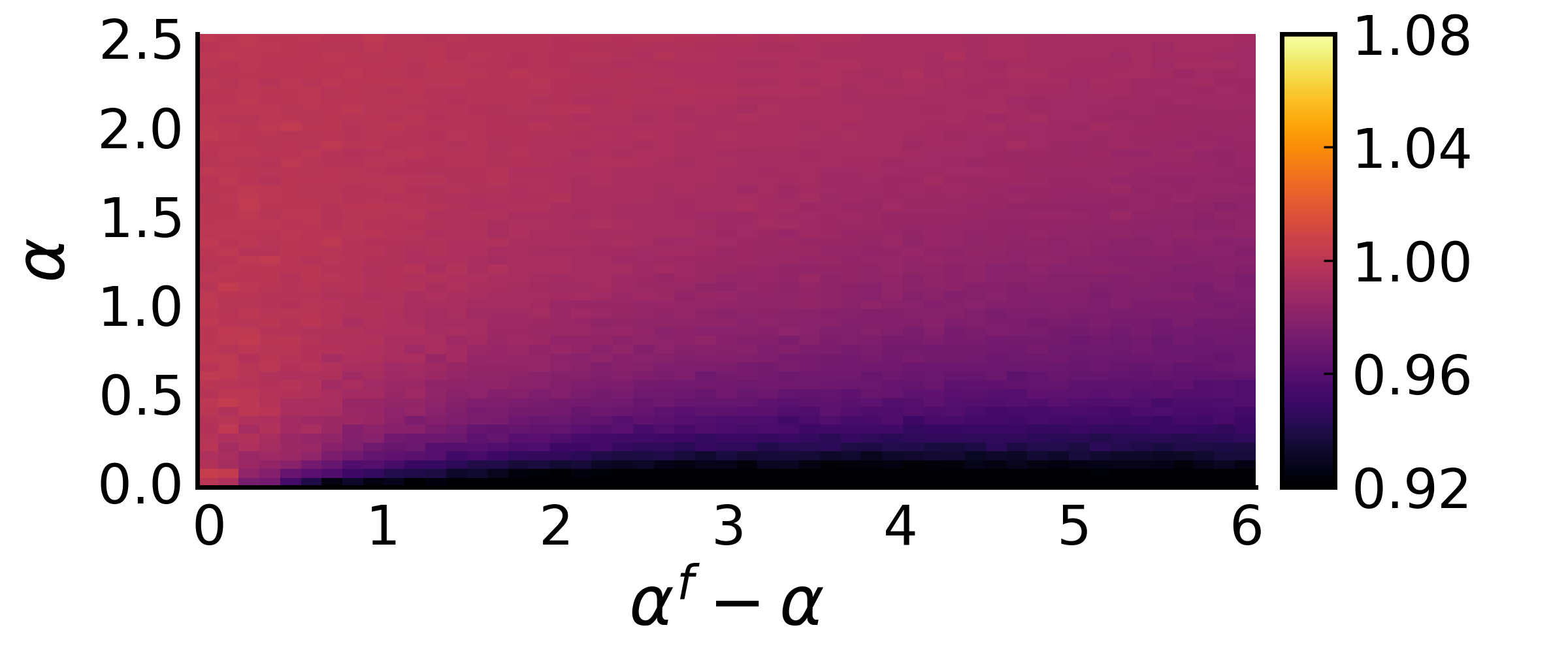}}
    }
    \subfloat[\centering $\frac{\psi^{\texttt{EMP}}(0.5,0.5)}{\psi^{\texttt{EMP}}(1,0)}$]{
        {\includegraphics[width=0.45\textwidth]{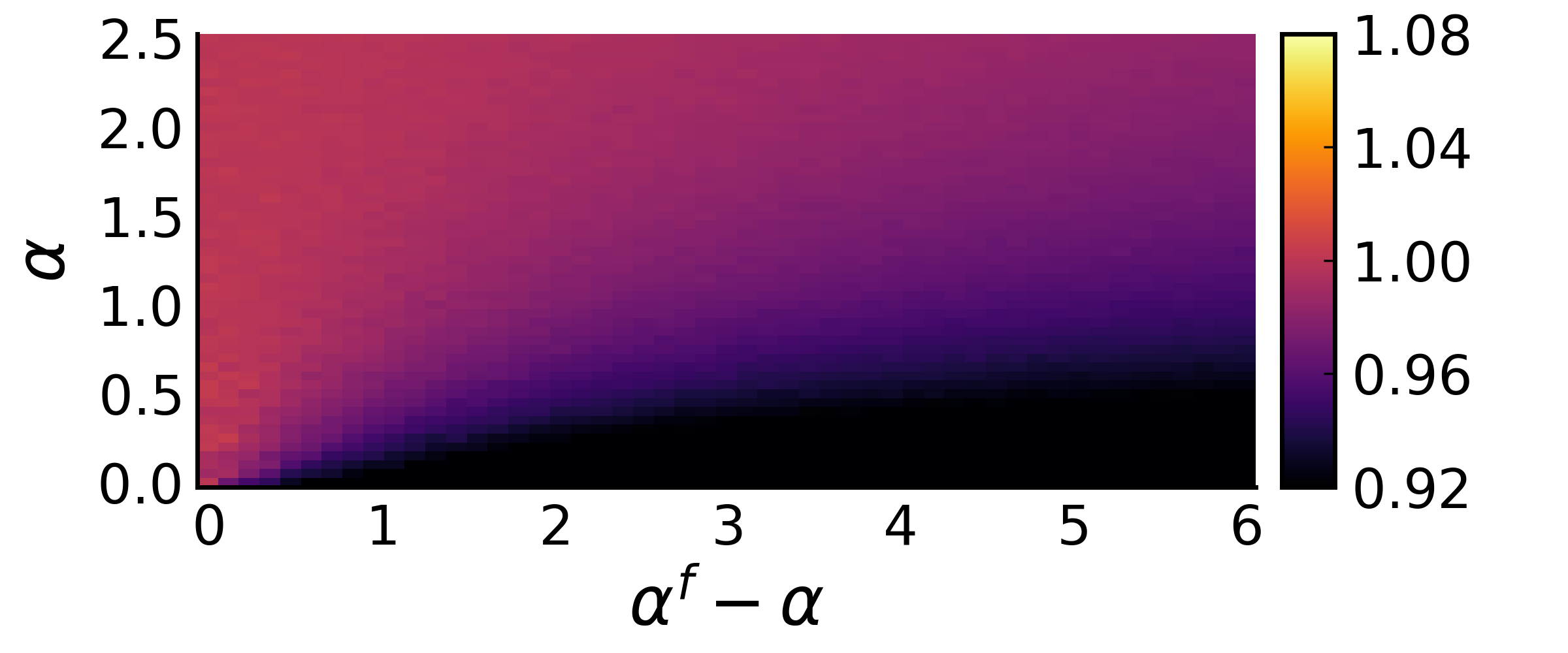}}
    }
    \vspace{0em}
    % -------- Row 4: psibar --------
    \subfloat[\centering $\frac{\bar{\psi}^{\texttt{EMP}}(0.3,0.3)}{\bar{\psi}^{\texttt{EMP}}(0.6,0)}$]{
        {\includegraphics[width=0.45\textwidth]{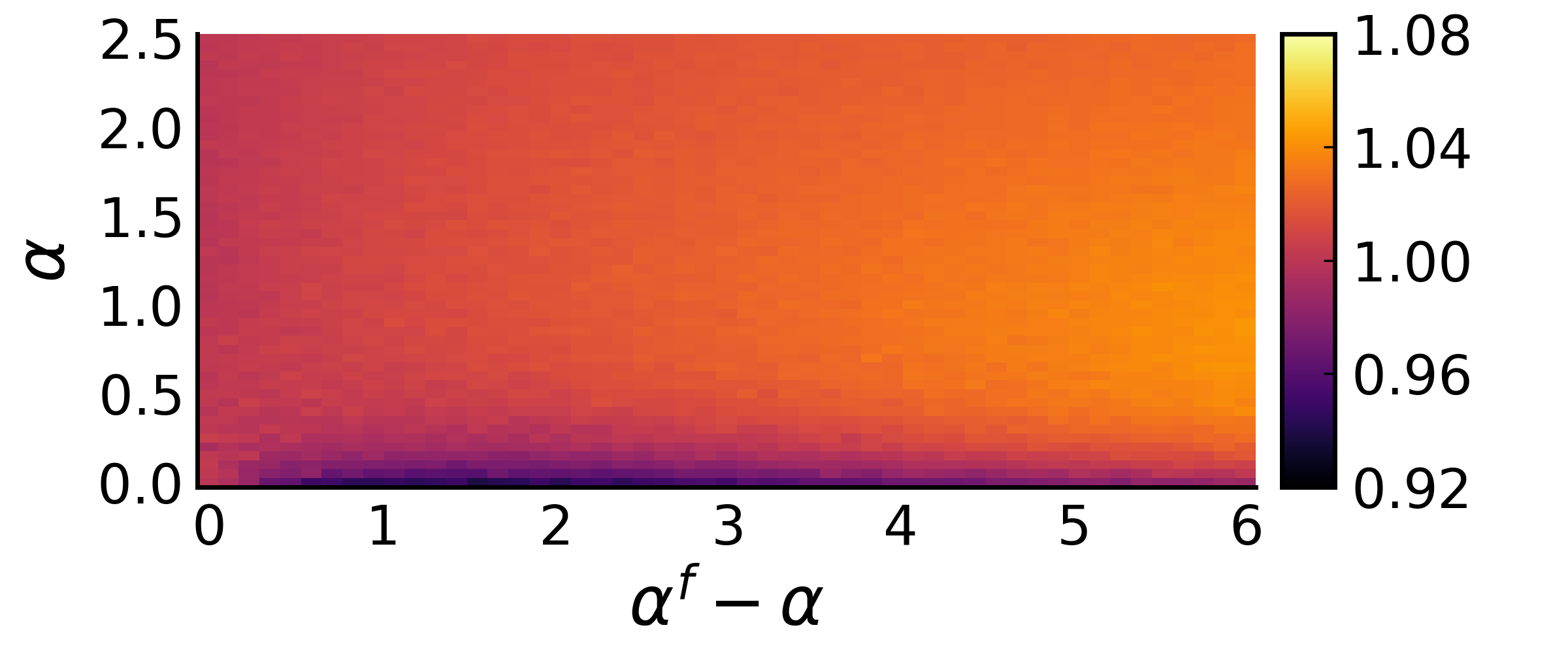}}
    }
    \subfloat[\centering $\frac{\bar{\psi}^{\texttt{EMP}}(0.5,0.5)}{\bar{\psi}^{\texttt{EMP}}(1,0)}$]{
        {\includegraphics[width=0.45\textwidth]{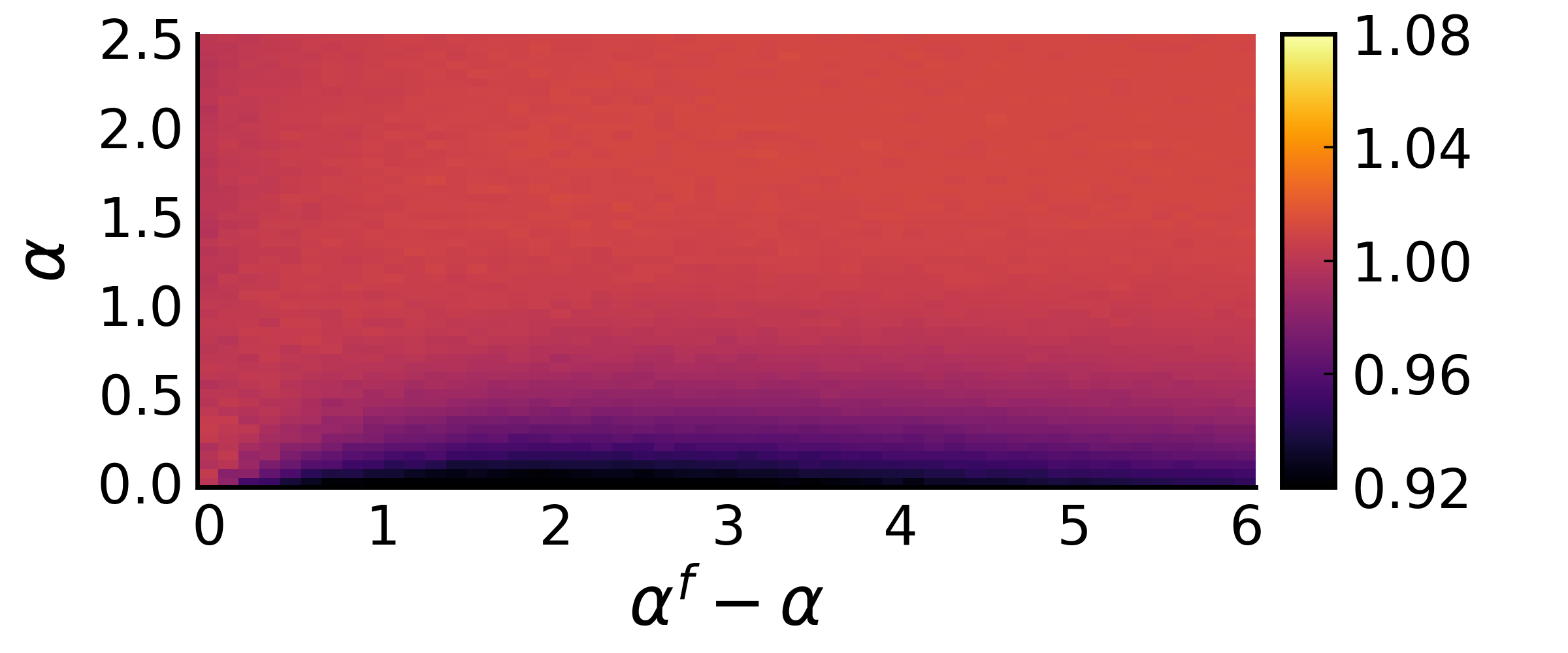}}
    }
    \caption{The heatmaps present the relative value of the balanced allocation over that of the one-sided allocation across varying $\anf$ and $\af-\anf$ under different matching objectives.}
    \label{fig:greedy_comparison}
\end{figure}
\begin{itemize}
    \item \emph{Naive greedy}, the outcome for which we denote by $\psi(\ql,\qr)$, processes nodes from $V_l$ by picking an unmatched neighbor from $V_r$ uniformly at random, adding that edge to the matching, and marking both nodes as matched;
    \item \emph{Prioritizing greedy}, the outcome for which we denote by $\bar{\psi}(\ql,\qr)$,  modifies the neighbor selection step by prioritizing regular neighbors whenever possible: when a node from $V_l$ has both regular and flexible neighbors, it first attempts to match to an unmatched regular neighbor (chosen uniformly at random), and only matches to a flexible neighbor if no regular neighbor is available. %This priority rule is applied symmetrically on both sides.
\end{itemize}
Both greedy heuristics leave nodes unmatched if no unmatched neighbor exists. We highlight that both heuristics can be viewed as online algorithms, where nodes from $V_r$ are known from the onset whereas nodes from $V_l$ arrive online over time. Further, the naive greedy heuristic can be implemented even if the platform only knows the edges that exist but does not know which nodes in $V_r$ are flexible.
\\
\emph{Findings.} Our numerical results suggest that, like $\mu$ and $\phi$, $\psi$ and $\bar{\psi}$ are always maximized by either the one-sided or the balanced allocation. As such, in \cref{fig:greedy_comparison}, we focus on the comparison between these two allocations for all four metrics. We consider $B\in\{0.6,1\}$. For both $\psi$ and $\bar{\psi}$ we find that {the one-sided allocation can significantly outperform the balanced allocation, and for $\psi$ the dominance region is even larger than that under $\mu$.} This is because the greedy matching, in a sense, amplifies the cannibalization effect by picking edges between flexible nodes even when regular nodes would have been available. Intuitively, we expect this to occur frequently for large $\af-\anf$ (when $B=1$, conditioned on a flexible node having a neighbor, the probability of that neighboring node being flexible is $2\af/(3\af+\anf)$). As a result, under the naive greedy algorithm, the one-sided allocation dominates throughout. In contrast, the prioritizing greedy heuristic alleviates this inefficiency by selecting regular nodes whenever possible; as a result (i) the region where the one-sided allocation performs better becomes much smaller, and (ii) like for $\mu$, larger values of $\anf$ allows for the balanced allocation to become optimal (though, in contrast to the asymmetry effect under $\mu$, this seems less dependent on $B$ and~$\af-\anf$).
%Across all metrics, we find that the two central forces identified in the main body—\emph{flexibility cannibalization} and \emph{flexibility asymmetry}—remain the fundamental drivers of the optimal allocation structure. In particular, we find that the metrics $\phi$ and $\psi$ respectively amplify these two forces. 
}

}

\section{Implications of the Structural Properties}\label{sec:ne}
{In this appendix, we demonstrate that in various models considered in the main body of our paper, the platform's objective can be concave in the directions $(1,0)$ and $(0,1)$ yet convex in the direction $(-1,1).$ We then interpret these results within the framework of a game between two players who set the flexibility on each side. This framework reflects two separate verticals or teams within an organization, each independently controlling one lever of flexibility. Our characterizations in this appendix illustrate that a balanced allocation may emerge as a suboptimal (local) Nash Equilibrium (NE) in such a context.

We start by establishing the geometric properties of the surrogate function $\mubar(1/2,1/2)$ for our main model in \cref{sec:model}:

\begin{theorem}\label{thm:convex_concave}
    When \(10^{-4} < \anf < 0.64 \af - 0.03\) and \( 0.62 \af + \anf < 1.68 \), $\mubar(1/2,1/2)$ is (i) strictly concave in the directions $(0,1)$ and $(1,0)$, and (ii) strictly convex in the direction $(1,-1)$.
\end{theorem}

The above result is proved in Appendix \ref{app:convex_concave} using KS-style analyses, and as illustrated in \cref{fig:verify_convex_concave}, the conditions cover most of the subcritical regime. Beyond the subcritical regime, the geometries become more nuanced, and \cref{fig:diagonal} shows that $\muemp(1/2,1/2)$ is not necessarily convex in the direction $(1,-1)$. 

Furthermore, we prove that for the local model (with $k=2$) considered in \cref{sec:local}, a simpler setting where the one-sided allocation always dominates, similar geometric properties hold across the entire parametric space of $\af$ and $\anf$ (see proof in Appendix \ref{app:curvature_local_model}):

\begin{theorem}\label{thm:curvature_local_model} When $k = 2$ in the local model, $\mu(\ql,\qr)$ is strictly concave in the direction $(0,1)$ when $\ql\in\{0,\frac{1}{2}\}$ and in the direction $(1,0)$ when $\qr\in\{0,\frac{1}{2}\}$. Moreover, $\mu(\ql,\qr)$ is strictly convex in the direction $(1,-1)$ when~$\ql + \qr = 1$.
\end{theorem}

We next formalize the interpretations of these geometric properties in the context of a game.}

\subsection{Nash Equilibrium and Saddle Point}\label{app:ne}

We start by providing, in \cref{def:game}~-~\ref{def:ne}, the textbook definition of NE and local NE for a general payoff function $g(\ql,\qr)$; we also define saddle points in \cref{def:sp}. We then verify these conditions for the platform's objective function defined in our paper. Notably, the suboptimal outcomes occur in our settings in spite of both teams sharing the same objective, i.e., there is no misalignment of incentives.

\begin{definition}[Game $\Gamma$]\label{def:game}
A game $\Gamma$ is defined by:
\begin{enumerate}
    \item The set of players $\{1, 2\}$;
    \item For each player $i \in \{1, 2\}$, $\mathcal{B}_i =[0,1]$ is the set of strategies available to player $i$;
    \item $g: \mathcal{B}_1
\times \mathcal{B}_2 \rightarrow \mathbb{R}$ is the payoff function, with $g(\ql,\qr)$ representing the payoff to each player for the strategy profile $\qvec = (\ql,\qr) \in \mathcal{B}_1 \times \mathcal{B}_2 \subseteq \mathbb{R}^2$.
\end{enumerate}
\end{definition}

\begin{definition}[Nash Equilibrium]\label{def:ne}
    In a game $\Gamma$ where $\mathcal{B} = \mathcal{B}_1 \times \mathcal{B}_2 \subseteq \mathbb{R}^2$ is the domain  of $g(\ql,\qr)$, 
    \begin{enumerate}[label=(\roman*)]
    \item A point $\qvec'$ is a \textit{Nash Equilibrium (NE)} if $$g(\ql',\qr') \geq g(\ql,\qr'), \forall \ql \in \mathcal{B}_1 \text{ and } g(\ql',\qr') \geq g(\ql',\qr), \forall \qr \in \mathcal{B}_2.$$
    \item A point $\mathbf{b'}$ is a \textit{local Nash Equilibrium} if there exists some $\delta > 0$ such that
    \begin{align*}
        & g(\ql',\qr') \geq g(\ql,\qr'), \forall \ql \in \mathcal{B}_1 \cap \parenthesis{\ql'-\delta,\ql'+\delta}, \text{ and}\\
        & g(\ql',\qr') \geq g(\ql',\qr), \forall \qr \in \mathcal{B}_2 \cap \parenthesis{\qr'-\delta,\qr'+\delta}.
    \end{align*}
    \item A point $\mathbf{b'}$ is a \textit{suboptimal (local) Nash Equilibrium} if $\mathbf{b'}$ is an (local) NE but there exists another point $\mathbf{b^\star}$ for which $g(\ql^\star, \qr^\star) > g(\ql',\qr').$
    \end{enumerate}
\end{definition}

\begin{definition}[Saddle Point]\label{def:sp}
    For a function $g(\ql,\qr): \mathcal{B} = \mathcal{B}_1 \times \mathcal{B}_2 \to \mathbb{R}$, assume that its first and second directional derivatives exist in all directions at $\mathbf{b'} \in \mathcal{B}.$ Then, $\mathbf{b'}$ is said to be a saddle point of $g$ if the following conditions are satisfied:
    \begin{enumerate}[label=(\roman*)]
    \item The gradient $\nabla g(\qvec') = \mathbf{0};$
    \item The second directional derivatives $\nabla_{\mathbf{v}}^2 g(\qvec') < 0$ and $\nabla_{\mathbf{u}}^2 g(\qvec') > 0$ in some directions $\mathbf{v}$ and $\mathbf{u} \in \mathbb{R}^2.$
    \end{enumerate}
\end{definition}

To identify suboptimal (local) NE in the models we leverage the results that $g(\ql,\qr)$ is concave in the directions $(0,1)$ and $(1,0)$, and convex in the directions $(-1,1)$ and $(1,-1)$. Because the directional derivative at an NE is zero along all directions, having both concavity and convexity effectively means that the NE is also a saddle point. We formalize these conditions in \cref{lem:ne} and \cref{lem:local_ne}. 

\begin{definition}[Interior of a Set]\label{def:interior}
Let $(\mathcal{B},\tau)$ be a topological space. Then
\begin{enumerate}[label=(\roman*)]
    \item A point $\qvec \in \mathcal{B}$ is an \textit{interior point} of the set $\mathcal{B}$ if there exists an open set $U$ with $\qvec \in U \subseteq \mathcal{B}$.
    \item The \textit{interior} of $\mathcal{B}$, denoted by $int(\mathcal{B})$, consists of all its interior points.
\end{enumerate}
\end{definition}

\begin{lemma}[Suboptimal NE and Saddle Point]\label{lem:ne}
In a game $\Gamma$ where the function $g(\ql,\qr)$ has its domain $\mathcal{B} \subseteq \mathbb{R}^2$, assume that its first and second directional derivatives exist in all directions for any $\qvec \in int(\mathcal{B})$. Suppose some NE $\mathbf{b'} \in int(\mathcal{B})$ satisfies the following conditions:
    \begin{enumerate}[label=(\roman*)]
        \item $g(\ql,\qr)$ is strictly concave in the direction $(0,1)$ at any $\qvec \in int(\mathcal{B})$ such that $\ql = \ql'$;
        \item $g(\ql,\qr)$ is strictly concave in the direction $(1,0)$ at any $\qvec \in int(\mathcal{B})$ such that $\qr = \qr'$;
        \item $g(\ql,\qr)$ is strictly convex in the direction $(1,-1)$ for any $\qvec \in int(\mathcal{B})$ such that $\ql+\qr = \ql'+\qr'$.
    \end{enumerate}
Then $\mathbf{b'}$ is a suboptimal NE and a saddle point.
\end{lemma}

\begin{lemma}[Suboptimal Local NE and Saddle Point]\label{lem:local_ne}
    In a game $\Gamma$ where the function $g(\ql,\qr)$ has its domain $\mathcal{B} \subseteq \mathbb{R}^2$, assume that its first and second directional derivatives exist in all directions at $\mathbf{b'} \in int(\mathcal{B})$. Suppose
    \begin{enumerate}[label=(\roman*)]
        \item $g(\ql,\qr)$ is strictly concave in the directions $(0,1)$ and $(1,0)$ at $\mathbf{b'}$;
        \item $g(\ql,\qr)$ is strictly convex in the direction $(1,-1)$ at $\mathbf{b'}$;
        \item The gradient of $g(\ql,\qr)$ is the zero vector at $\mathbf{b'}$.
    \end{enumerate}
    Then $\mathbf{b'}$ is a suboptimal local NE and a saddle point.
\end{lemma}

We now demonstrate that the balanced allocation is indeed a suboptimal NE and a saddle point for graph models examined in this paper. 

% \begin{proposition}\label{prop:2x2_model}
%     In the $2 \times 2$ model, for any $c \in \parenthesis{\frac{\partial \mu(\budget/2,\budget/2)}{\partial \budget}\Big|_{\budget=2},\frac{\partial \mu(\budget/2,\budget/2)}{\partial \budget}\Big|_{\budget=0}}$, there exists a $\budget' \in (0,2)$ such that $\mathbf{b'} = \parenthesis{\budget'/2,\budget'/2}$ is a suboptimal NE and a saddle point for $g(\ql,\qr)$.
% \end{proposition}

\begin{proposition}\label{prop:local_model}
With $k = 2$ in the local model, there exists $c > 0$ such that $\qvec' = \parenthesis{1/2,1/2}$ is a suboptimal NE and a saddle point for $g(\ql,\qr) = \mu(\ql,\qr) - c \cdot (\ql+\qr)$.
\end{proposition}

\begin{proposition}\label{prop:global_model_ne}
Assume that \(10^{-4} < \anf < 0.64 \af - 0.03\) and \( 0.62 \af + \anf < 1.68 \). Then, there exists $c > 0$ such that $\qvec' = (1/2,1/2)$ is a suboptimal local NE and a saddle point for $\bar{g}(\ql,\qr) := \mubar(\ql,\qr) - c \cdot (\ql+\qr)$.
\end{proposition}

\subsection{Proofs of the Results in Appendix \texorpdfstring{\ref{app:ne}}{Lg} }

\begin{proof}{Proof of \cref{lem:ne}}
Let $\qvec'$ be a Nash Equilibrium (NE) in $\text{int}(\mathcal{B})$ for the function $g(\ql,\qr)$ with domain $\mathcal{B} \subseteq \mathcal{R}^2$. We start by showing that $g(\ql,\qr)$ has a gradient of $\mathbf{0}$ at $\qvec'$. If $\nabla_{(0,1)} g(\qvec') > 0,$ then there exists $\epsilon > 0$ such that $g(\ql',\qr'+\epsilon) > g(\ql',\qr')$, contradicting the definition of NE. Similarly, if $\nabla_{(0,1)} g(\qvec') < 0,$ then there exists $\epsilon > 0$ such that $g(\ql',\qr'-\epsilon) > g(\ql',\qr')$, again contradicting the definition of NE. Thus, $\nabla_{(0,1)}g(\qvec') = 0.$ Applying a similar argument to the direction of $(1,0)$ we find that $\nabla_{(1,0)}g(\qvec') = 0.$ Since $g(\ql,\qr)$ has a directional derivative of $0$ in two orthogonal directions at $\qvec'$, it has a gradient of $\mathbf{0}$ at $\qvec'$. Since we also know that $\nabla_{(0,1)}^2 g(\qvec') <0$ and $\nabla_{(1,-1)}^2 g(\qvec') >0$, $\qvec'$ is a saddle point.

We then show that $\qvec'$ is a suboptimal NE. Since $g(\ql,\qr)$ is strictly convex in the direction of $(1,-1)$ at any $\qvec$ such that $\ql+\qr = \ql'+\qr'$ and $g(\ql,\qr)$ has a gradient of $\mathbf{0}$ at $\qvec'$, $\qvec'$ is a global minimum in the direction of $(1,-1).$ That is, there exists $\epsilon > 0$ such that $g(\ql' + \epsilon, \qr'-\epsilon) > g(\ql',\qr').$ Thus, $g(\ql,\qr)$ is a suboptimal NE.
\end{proof}

\begin{proof}{Proof of \cref{lem:local_ne}}

Since $g(\ql,\qr)$ is strictly concave in the direction of $(0,1)$ and $(1,0)$ but also strictly convex in the direction of $(1,-1)$ at $\qvec'$, $\qvec'$ is neither a local maximum nor a local minimum. Combined with the condition that $g(\ql,\qr)$ has a gradient of $\mathbf{0}$ at $\qvec'$, we conclude that $\qvec'$ is a saddle point. 

% \kzedit{
Now, since $\nabla_{(0,1)} g(\qvec') = 0$ and $\nabla_{(0,1)}^2 g(\ql,\qr) <0$, we know that $\qvec'$ is a local maximum in the direction of $(0,1)$ and we can find $\epsilon_1 > 0$ such that $$g(\ql',\qr') \geq g(\ql',\qr), \forall \qr \in \mathcal{B}_2 \cap \parenthesis{\qr'-\epsilon_1,\qr'+\epsilon_1}.$$ Similarly, from $\nabla_{(1,0)} g(\qvec') = 0$ and $\nabla_{(1,0)}^2 g(\ql,\qr) <0$ we find $\epsilon_2 > 0$ such that $$g(\ql',\qr') \geq g(\ql,\qr'), \forall \ql \in \mathcal{B}_1 \cap \parenthesis{\ql'-\epsilon_2,\ql'+\epsilon_2}.$$ Taking $\delta = \min\parenthesis{\epsilon_1, \epsilon_2},$ we find that $\qvec'$ is a local NE.

We then show that $\qvec'$ is a suboptimal local NE. Since $\nabla_{\parenthesis{1,-1}} g(\qvec') = 0$ and $\nabla_{\parenthesis{1,-1}}^2 g(\ql,\qr) >0$, $\qvec'$ is a local minimum in the direction of $(1,-1).$ That is, there exists $\epsilon > 0$ such that $g(\ql' + \epsilon, \qr'-\epsilon) > g(\ql',\qr').$ Thus, $\qvec'$ is a suboptimal local NE.
\end{proof}

\begin{proof}{Proof of \cref{prop:local_model}}

    As established in \cref{thm:curvature_local_model}, we have verified the local convexity of \( \mu(\ql,\qr) \) in the direction \( (1,-1) \) and local concavity in the directions \( (0,1) \) and \( (1,0) \) along diagonals that intersect at \( \qvec' = (1/2,1/2) \). Consequently, the function \( g(\ql,\qr) \) exhibits the same local convexity and concavity properties. According to \cref{lem:ne}, to demonstrate the existence of a constant \( c > 0 \) such that \( \qvec' = (1/2,1/2) \) is a suboptimal Nash Equilibrium (NE) for \( \bar{g}(\ql,\qr) \), it is sufficient to show that \( (1/2,1/2) \) is indeed an NE.

    We construct $c = \nabla_{(1,0)} \mu(1/2,1/2) = \frac{\partial \mu(\ql,1/2)}{\partial \ql}\Big|_{\ql=1/2}$. Then, we have $$\nabla_{(1,0)} g(1/2,1/2) = \nabla_{(1,0)} \mu(1/2,1/2) - c = 0.$$ Since $g(\ql,\qr)$ is strictly concave in the direction of $(1,0)$ when $\qr = 1/2,$ $\qvec' = (1/2,1/2)$ is a global maximum in the direction $(1,0).$ That is, $g(1/2,1/2) \geq g(\ql,1/2), \forall \ql \in (0,1).$
    By symmetry, the same result holds in the direction \( (0,1) \). Thus, according to \cref{def:ne} (i), we conclude that \( \qvec' = (1/2,1/2) \) is indeed an NE for the selected constant \( c \). This completes the proof.
\end{proof}

\begin{proof}{Proof of \cref{prop:global_model_ne}}

    Given that \cref{thm:convex_concave} has established the local convexity of \(\mubar(\ql,\qr)\) in the direction \((1,-1)\) and its local concavity in the directions \((0,1)\) and \((1,0)\) within the specified region of $\af$ and $\anf$, the same local convexity and concavity properties hold for \(\bar{g}(\ql,\qr)\). By \cref{lem:local_ne}, to show that there exists $c > 0$ such that $\qvec' = (1/2,1/2)$ is a suboptimal local NE for $\bar{g}(\ql,\qr)$, it suffices to show that there exists $c>0$ such that the gradient of $\bar{g}(\ql,\qr)$ is the zero vector at $\qvec = (1/2,1/2).$ 
    
    Since $\nabla_{(1,-1)}^2 \bar{g}(\ql,\qr) > 0$ at $(1/2,1/2)$ and by symmetry we have $\bar{g}(\ql,\qr) = \bar{g}(\qr,\ql), \forall (\ql,\qr) \in (0,1)^2,$ $\nabla_{(1,-1)} \bar{g}(\ql,\qr)$ is well-defined at $(1/2,1/2)$. Now, if the directional derivative is strictly positive, we know that there exists $\epsilon > 0$ such that $\bar{g}(1/2+\epsilon,1/2-\epsilon) > \bar{g}(1/2-\epsilon,1/2+\epsilon)$, which contradicts the symmetry condition. By the same argument, the directional derivative cannot be strictly negative, so we find that $\nabla_{(1,-1)} \bar{g}(\ql,\qr) = 0$ at $(1/2,1/2).$ 
    
    We then construct $c = \frac{\partial \mubar(\ql,1/2)}{\partial \ql}\Big|_{\ql=1/2}$. Since $\nabla_{(1,0)}^2 \mubar(\ql,\qr) < 0$ at $(1/2,1/2), \nabla_{(1,0)} \mubar(\ql,\qr)$ is well-defined and equal to $\frac{\partial \mubar(\ql,1/2)}{\partial \ql}\Big|_{\ql=1/2}.$ Now, since $c = \frac{\partial \mubar(\ql,1/2)}{\partial \ql}\Big|_{\ql=1/2}$, we have $$\nabla_{(1,0)} \bar{g}(1/2,1/2) = \nabla_{(1,0)} \mubar(1/2,1/2) - c = 0.$$

    Since the derivative of \(\bar{g}(\ql,\qr)\) at \(\qvec' = (1/2,1/2)\) is zero in two independent directions, it implies that the gradient of \(\bar{g}(\ql,\qr)\) is a zero vector at \(\qvec' = (1/2,1/2)\), thereby completing the proof.
\end{proof}

\section{Proofs of the Main Results}

\subsection{Proofs of the Results in \texorpdfstring{\cref{sec:cannibalization}}{Lg}}\label{sec:coupling_proof}

In this section, we first provide the proof of \cref{thm:compare_1}. To keep it readable, abstract away the technical details into two lemmas, and first present a more readable proof sketch for both of them, before proving them formally.

\noindent\textbf{Proof of \cref{thm:compare_1}.} 
We prove that the one-sided allocation dominates the balanced one when $B=1$ and $\anf=0$. Below, we state two lemmas and prove they imply the theorem, deferring the proofs and constructions for the lemmas to \cref{sec:step1}-\ref{sec:step2} and the corresponding appendices.

We first introduce a new bipartite random graph distribution, denoted $G_n^b$. This distribution is easier to analyze than the balanced allocation random graph (denoted $G_n(1/2,1/2)$) but has the same asymptotic matching probability. In $G_n^b$, exactly $n/2$ nodes are flexible on each side (see \cref{fig:coupling0}). Each flexible node generates directed edges to the nodes on the other side (flexible or not), independently and with a probability~$\pf_n$ for each edge. This means that an edge between two flexible nodes can be generated in both directions {(as shown in the right plot of \cref{fig:coupling0})}. When computing a maximum matching in $G_n^b$ we ignore the directionality of the edges and treat such double edges between nodes as just a single edge. We introduce $G_n^b$ as its realizations can be more easily coupled with the random graph of the one-sided allocation. Denoting the size of a realized maximum matching in $G_n^b$ by the random variable $\mathcal{M}_n^b$ and that of $G_n(1/2,1/2)$ by $\mathcal{M}_n(1/2,1/2)$, we show nodes in $G_n^b$ and $G_n(1/2,1/2)$ have the same asymptotic matching probability.
\begin{lemma}\label{lem:graph_close}
   With the above construction, $\lim_{n\to\infty}\EE{\mathcal{M}_n(1/2,1/2) - \mathcal{M}_n^b}/n=0.$ 
\end{lemma}

Now, we compare $G_n^b$ to the random graph with one-sided allocation. We denote the latter by~$G_n^o$ and its maximum matching size by $\mathcal{M}_n(1,0)$. \dfedit{The next lemma first compares $\mathcal{M}_n^b$ and $\mathcal{M}_n(1,0)$ in a non-asymptotic way and then bounds their expected difference as $n$ goes to infinity. This is the key step of the proof, and relies on an intricate coupling of independent interest.}
\dfedit{
\begin{lemma}\label{lem:bundled_coupling}
     With the above construction, $\liminf_{n\to\infty}\EE{\mathcal{M}_n(1,0)-\mathcal{M}_n^b}/n \geq \parenthesis{\af}^3/2^5 \cdot e^{-7\af}.$
\end{lemma}

%For $\anf = 0,$ 
In the following derivation, \cref{lem:graph_close} gives us the second equality and \cref{lem:bundled_coupling} the inequality: 
\begin{align*}
    \mu(1,0) - \mu(1/2,1/2)\;&\geq 
\liminf_{n \to \infty}\EE{\frac{\mathcal{M}_n(1,0) - \mathcal{M}_n(1/2,1/2)}{n}} \\
&= \liminf_{n \to \infty}\EE{\frac{\mathcal{M}_n(1,0) - \mathcal{M}_n^b}{n}} \geq \parenthesis{\af}^3/2^5 \cdot e^{-7\af},
\end{align*}
which completes the proof of \cref{thm:compare_1}.} \hfill\Halmos
%\end{proof}

\subsection{Proof sketch of \texorpdfstring{\cref{lem:graph_close}}{Lg}}\label{sec:step1}

As illustrated in \cref{fig:coupling0}, the graph $G^b_n$ is a directed random graph that contains edges generated from left to right (denoted~$R^l_{ij}$) and edges generated from right to left ($R^r_{ij}$). The edge probabilities are given by:
\begin{align}\label{eq:gbn}
    \PP{R^l_{ij} = 1} = \pf_n, \forall i \in [n/2], j \in [n] \text{ and } \PP{R^r_{ij} = 1} = \pf_n, \forall j \in [n/2], i \in [n].
\end{align}

$G^b_n$ differs from $G_n(1/2,1/2)$ in two ways: (i) $G^b_n$ contains $n/2$ flexible nodes on each side of the bipartite graph, whereas every node in $G_n(1/2,1/2)$ is flexible with probability $1/2$; (ii) in $G^b_n$ an edge between $\vl_i$ and $\vr_j, i,j \in [n/2],$ is generated from each side with probability $\pf_n,$ instead of being generated only once with probability $2 \pf_n$. It is intuitive that neither (i) or (ii) significantly change the asymptotic matching size: standard concentration bounds guarantee that (i) affects $o(n)$ nodes, and (ii) affects $\sum_{i,j \in [n/2]} \parenthesis{\pf_n}^2 = \sum_{i,j \in [n/2]} \parenthesis{\af/n}^2 \in \mathcal{O}(1)$ possible edges in expectation. In Appendix \ref{app:graph_close}, we formalize this intuition.

\subsection{Proof sketch of \texorpdfstring{\cref{lem:bundled_coupling}}{Lg}}\label{sec:step2}

In our proof, we construct a coupling between \emph{pairs} of realizations of $G_n^b$ and of $G_n^o$ to compare the maximum matching sizes therein. 
First, we show that this coupling is valid in the sense that the coupled realizations occur with the same probability in their respective graphs. Second, we show that the sum of the maximum matching size in the pair of realizations in $G_n^b$ is smaller-equal to that in $G_n^o$ for \emph{any} realization. \dfedit{Finally, we lower bound their expected difference as $n$ scales large.} We present the key steps of our proof here and defer the complete proof to Appendix \ref{app:bundled_coupling}.

\textbf{Coupling the Realizations of Graphs.} We partition the directed edges in a realization of $G_n^b$ into sets $X_1,X_2,X_3$ and $X_4$, depending on whether they are from left/right to top/bottom (see \cref{fig:coupling1}~(A)). 

We couple each realization of edges, i.e., of sets $X_1,X_2,X_3$ and $X_4$, with a second realization (B), also from $G_n^b$, that occurs with the same probability (\cref{fig:coupling1}~(B)). Essentially, we ``flip'' the edges in $X_3$ and $X_4$ across the vertical axis to obtain the sets $\widetilde{X}_3$ and $\widetilde{X}_4$. Then, we couple (A) and (B) with two realizations, (C) and (D) (see \cref{fig:coupling1}~(C) and~(D)), of $G_n^o$. There, we ``flip" the edges in $X_1$ from the upper subgraph in (A) and (B) to the lower subgraph in (C) and (D). Intuitively, as flexibility cannibalization can happen in the upper subgraph of $G_n^b$, where the edges in $X_1$ and $X_2$ are concentrated, we want our coupling to ``flip'' the edges in $X_1$ to the less dense lower subgraph and thereby hopefully increase the number of matches.  However, denoting by $M_A,M_B,M_C,M_D$ the maximum matching sizes in the respective graphs, it is not always true that $M_A \le M_C$ or $M_B \le M_D$. \dfedit{Instead, we first show that $M_A+M_B \leq M_C+M_D$ holds for all $X_1, X_2, X_3$, and~$X_4$; this guarantees that $\EE{\mathcal{M}_n^b} \leq \EE{\mathcal{M}_n(1,0)}$. Finally, we prove the lemma by explicitly identifiying additional edges that can be added to the matching in (C)  with constant probability (as $n$ scales large).} We include the formal coupling in Appendix~\ref{app:bundled_coupling}.% and show that the realizations of (A) and (B) occur with the same probability in $G_n^b$ as (C) and~(D) do in $G_n^o$.

\textbf{Proving the Dominance of One-sided Allocation.} 
Our proof \dfedit{next shows} that the required property $M_A+M_B \leq M_C+M_D$ holds for arbitrary $X_1, X_2, X_3, X_4$, which shows that the flexibility cannibalization in the upper graphs of (A) and (B) indeed induces a lower number of matches. As illustrated in \cref{fig:coupling2}, in (A) we denote by sets $Y_i\subset X_i$ the edges that are part of a given maximum matching; similarly, in (B), we denote by sets $Y_1' \subset X_1, Y_2' \subset X_2, \tilde{Y}_3 \subset \tilde{X}_3$ and $\tilde{Y}_4 \subset \tilde{X}_4$ the edges that are part of a maximum matching. We then injectively map all edges of $M_A$ and $M_B$ (i.e., those in $Y_1, Y_2, Y_3, Y_4, Y_1', Y_2', \tilde{Y}_3$ and $\tilde{Y}_4$) into existing edges of $(C)$ and $(D)$ that also form a matching; this immediately proves $M_A+M_B \leq M_C+M_D$. \dfedit{Finally, we will provide a sufficient condition under which one can construct additional matches in $(C)$ and $(D)$ and lower bound the number of such matches as $n$ scales large.} We construct this mapping in two steps.

\emph{Step 1: mapping $Y_1, Y_2, Y_1'$ and $Y_2'$.} We start by directly copying the matched edges from $Y_1, Y_2, Y_1'$ and $Y_2'$ into (C) and (D), following the coupling rules \dfedit{(see the red, blue, pink, and navy edges in \cref{fig:coupling2})}.

\begin{figure}[ht]
    \centering
    \includegraphics[width=0.8\textwidth]{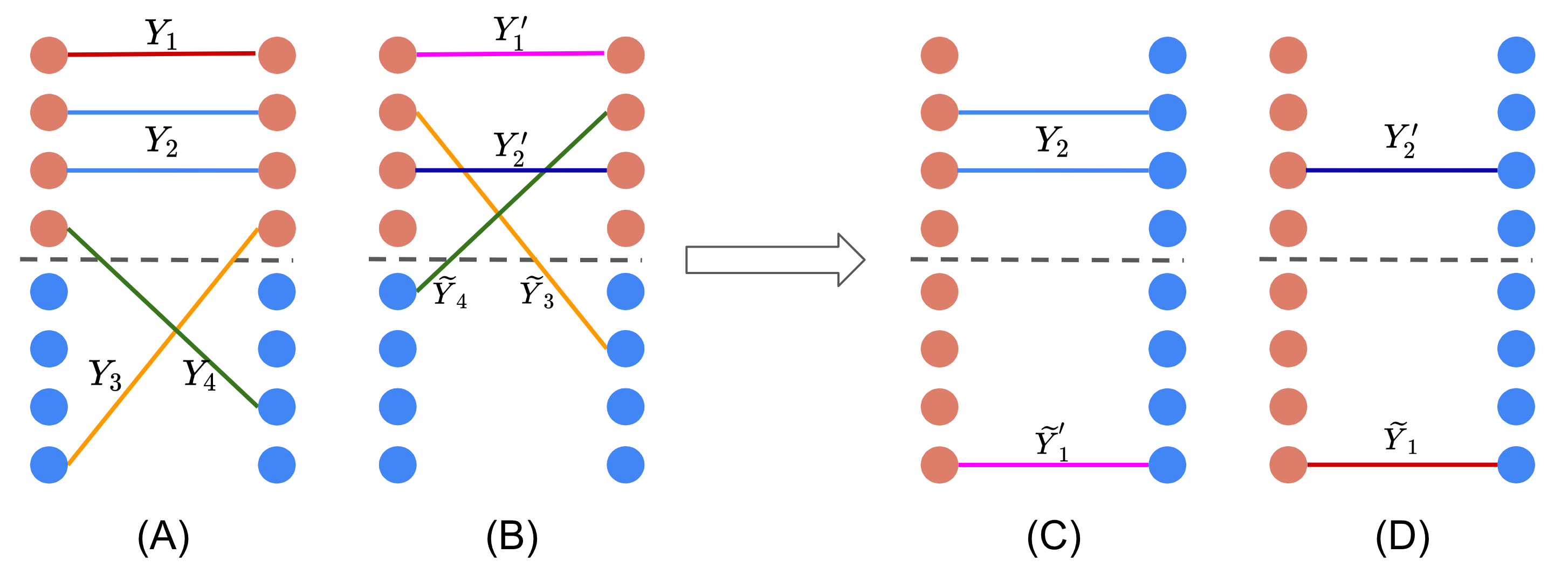}
    \caption{Illustration of the matches in (A) and (B), and the position they are copied into in (C) and (D). }
    \label{fig:coupling2}
    \vspace{-0.2in}
\end{figure}

\begin{figure}[ht]
    \centering
    \includegraphics[width=1\textwidth]{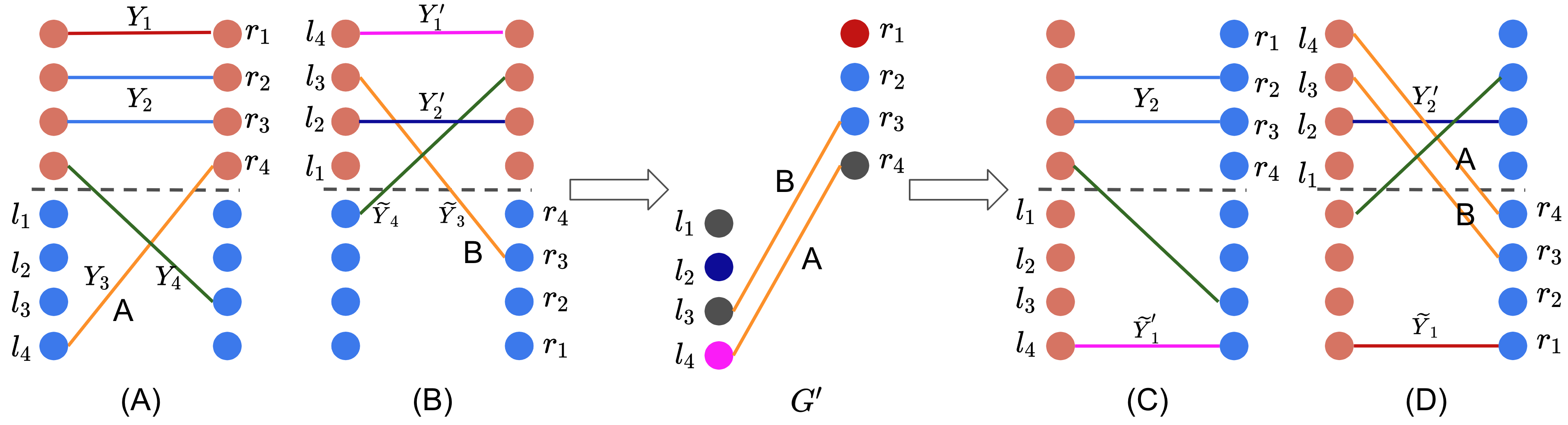}
    \caption{The plot illustrates the mapping of edges in $Y_3$ and $\tilde{Y}_3$ (the yellow edges) to (C) and (D) through the constructed graph $G'$. The labels indicate the correspondence between nodes/edges in $G'$ and those in graphs (A)-(D). A second graph $G''$ can be constructed to map the (green) edges in $Y_4$ and $\tilde{Y}_4$ into the indicated positions in (C) and~(D).}
    \label{fig:coupling3}
\vspace{-.2in}
\end{figure}

\emph{Step 2: mapping $Y_3, Y_4, \tilde{Y}_3$ and $\tilde{Y}_4$.} The \dfedit{remaining} matched edges (the yellow and green edges) can also be mapped into (C) and (D), but this mapping is not static and depends on the matches that are already copied into the graphs. As the nodes in (C) and (D) that are matched through these copied edges can no longer be matched to any other node in the graphs, we denote the remaining nodes in~(C) and~(D) by~$\bar{C}$ and~$\bar{D}$ and the set of edges among these nodes by~$E(\bar{C})$ and~$E(\bar{D})$. Then, \dfedit{to show that $\EE{\mathcal{M}_n^b} \leq \EE{\mathcal{M}_n(1,0)}$,} it suffices to show that we can injectively map all other matches (that we have not copied already) in (A) and (B) to $M(\bar{C}) \cup M(\bar{D})$, where $M(\bar{C})$ and $M(\bar{D})$ are respectively matchings that we construct in  $E(\bar{C})$ and $E(\bar{D})$. We construct such a mapping for edges in $Y_3$ and $\tilde{Y}_3$ based on a~$\frac{n}{2} \times \frac{n}{2}$ colored bipartite multigraph $G'$ (see \cref{fig:coupling3}). $G'$ includes all edges from $Y_3$ and $\tilde{Y}_3$ that occur in graph (A) and (B); we label edges in $G'$ that come from $Y_3$ as type A edges and edges from~$\tilde{Y}_3$ as type~B edges (there can be two edges, one of type A and one of type B, between a pair of nodes in $G'$). We color the nodes in $G'$ based on whether the corresponding nodes in (A) and (B) are incident to $Y_1, Y_2, Y_1'$ and $Y_2'.$ Analogous to $G'$, we create a second graph $G''$ that contains all the edges from $X_4$ that are part of maximum matchings in $(A)$ and $(B)$. We show that edges in $G', G''$ can be mapped into graphs (C) and/or (D) based on their types and the colors of their incident nodes so that, together with the already copied edges, they produce feasible matchings in (C) and (D). 
As a result, each edge from $M_A$ and $M_B$ can be found in a matching in either (C) or (D), implying that $M_A+M_B \leq M_C+M_D$. Thus, $\EE{\mathcal{M}_n^b} \leq \EE{\mathcal{M}_n(1,0)}, \forall n$ when $\anf = 0.$ We formalize these constructions in Appendix~\ref{app:bundled_coupling}.

\emph{Step 3: constructing additional matches.} In the last step of our proof of \cref{lem:bundled_coupling}, we provide an explicit construction under which the combined matching size in graphs (C) and (D) strictly exceeds that in (A) and (B). In particular, starting from an arbitrary flexible node on the right-hand side, we construct a sufficient condition under which all matches from (A) and (B) are mapped to (C) and (D), and yet one additional edge can be matched in (C). This condition holds with probability $(\af/2)^3 e^{-7\af}$; as a result, summing over all flexible nodes on the right-hand side, we obtain an expected difference that scales on the order of $n$. The full construction is provided in Appendix~\ref{app:bundled_coupling}.

\subsubsection{Proof of \texorpdfstring{\cref{lem:graph_close}}{Lg}}\label{app:graph_close}
    Recall that we have constructed a random graph $G^b_n$ in \eqref{eq:gbn} that decomposes the $2 \pf_n$ edges as two groups of directed edges, in each of which an edge exists with probability $\pf_n.$ Notice that when constructing a maximum matching in $G^b_n$, we do not differentiate between edges of different directions. However, we maintain the requirement that no two edges (of either direction) can share a node in the matching. In the rest of the proof we show that neither difference (i) nor (ii) identified in \cref{sec:step1} changes the asymptotic matching size.
    
    We assume without loss of generality that $n$ is an even number (else, we can ignore nodes $\vl_n$ and $\vr_n$ without changing the asymptotic matching probability). We start by applying standard concentration bounds to show that assuming $n/2$ flexible nodes on each side leads to $o(n)$ error in the asymptotic matching size. Specifically, we define the event that in $G_n(1/2,1/2)$ $$E_1 := \bracket{|\sum_i F_i^l - n/2| \leq n^{5/8} \text{ and } |\sum_j F_j^r - n/2| \leq n^{5/8}}.$$ Specifically, letting $E_1^c$ be the complement of event $E_1,$ we have
    \begin{align*}
        \EE{\mathcal{M}_n(1/2,1/2) - \mathcal{M}_n^b} =&\; \EE{\mathcal{M}_n(1/2,1/2) \Bigl| E_1^c} \PP{E_1^c} +\EE{\mathcal{M}_n(1/2,1/2) \Bigl| E_1} \PP{E_1} -\EE{\mathcal{M}_n^b}\\
        \leq&\; n \cdot e^{-\Omega(n^{1/4})} + \EE{\mathcal{M}_n(1/2,1/2) \Bigl| E_1} - \EE{\mathcal{M}_n^b}\\
        \leq&\; \EE{\mathcal{M}_n(1/2,1/2) \Bigl| E_1} - \EE{\mathcal{M}_n^b} + \mathcal{O}(1)\\
        \leq&\; \EE{\mathcal{M}_n(1/2,1/2) \Bigl| \sum_i F_i^l = \sum_j F_j^r = n/2} + o(n) - \EE{\mathcal{M}_n^b} + \mathcal{O}(1)\\
        =&\; \EE{\mathcal{M}_n(1/2,1/2)-\mathcal{M}_n^b \Bigl| \sum_i F_i^l = \sum_j F_j^r = n/2} + o(n).
    \end{align*}
    Notice that the first inequality above is a concentration result that follows from the Chernoff bound, and the third inequality above follows from the fact that having $n^{5/8}$ additional flexible nodes on each side of $G_n(1/2,1/2)$ creates at most $o(n)$ additional matches.

    We next show that drawing two edges, each with probability $\pf_n$, is close to drawing a single edge with probability $2 \pf_n$ in the asymptotic regime we study. With $\sum_i F_i^l = \sum_j F_j^r = n/2,$ we reorder the nodes such that the first $n/2$ nodes on each side of $G_n(1/2,1/2)$ are flexible.
    We then couple the edges in the two graphs ($G^b_n$ and a balanced graph with $\sum_i F_i^l = \sum_j F_j^r = n/2$) by drawing a random variable $\omega_{ij}\sim U(0,1)$ for every $i,j \in [n]$ and using $\omega_{ij}$ to generate the edge between $i$ and $j$. Specifically, we set $R_{ij} = 1$ if and only if $\omega_{ij} \leq \PP{R_{ij} = 1}$ in the respective graph.\footnote{In $G^b_n$, $R_{ij} = 1$ if and only if $R^l_{ij} + R^r_{ij} \geq 1.$}
    In the balanced graph, in which the first $n/2$ nodes are flexible on each side, we obtain $\PP{R_{ij} = 1} = 2 \pf_n, \forall i,j \in [n/2]$ and in $G^b_n$ we get $$\PP{R_{ij} = 1} = \PP{R^l_{ij} + R^r_{ij} \geq 1} = 2\pf_n - \parenthesis{\pf_n}^2, \forall i,j \in [n/2].$$ The probabilities for all other edges are the same in both graphs. This characterization of the probability of $R_{ij}=1$ in each graph implies for a given realization of $\boldsymbol{\omega}$ that 
    \begin{align*}
        \EE{\mathcal{M}_n(1/2,1/2)-\mathcal{M}_n^b \Bigl| \boldsymbol{\omega}, \sum_i F_i^l = \sum_j F_j^r = n/2} \leq \sum_{i,j \in [n/2]} \mathbbm{1}_{\omega_{ij} \in \squarebracket{2\pf_n - \parenthesis{\pf_n}^2, 2 \pf_n}}
    \end{align*}
    because the number of additional matches in $\mathcal{M}_n(1/2,1/2)$ is upper bounded by the number of additional edges in the graph. Taking expectation over $\boldsymbol{\omega}$, we find that 
    \begin{align*}
        \EE{\mathcal{M}_n(1/2,1/2)-\mathcal{M}_n^b \Bigl|\sum_i F_i^l = \sum_j F_j^r = n/2}\leq\sum_{i,j \in [n/2]} \PP{\omega_{ij} \in \squarebracket{2\pf_n - \parenthesis{\pf_n}^2, 2 \pf_n}}= \;\sum_{i,j \in [n/2]} \parenthesis{\pf_n}^2 = O(1),
    \end{align*}
    \begin{align*}
    \text{which implies }\quad    \EE{\mathcal{M}_n(1/2,1/2) - \mathcal{M}_n^b} \leq \EE{\mathcal{M}_n(1/2,1/2)-\mathcal{M}_n^b \Bigl| \sum_i F_i^l = \sum_j F_j^r = n/2} + o(n) = o(n).
    \end{align*}
    Thus, when $\anf = 0$ we have $\lim_{n\to\infty}\EE{\mathcal{M}_n(1/2,1/2) - \mathcal{M}_n^b}/n=0.$ \hfill\Halmos
% \end{proof}

\subsubsection{Proof of \texorpdfstring{\cref{lem:bundled_coupling}}{Lg}}\label{app:bundled_coupling}    Recall from \cref{sec:step2} that we start by constructing a valid coupling of realizations of $G_n^b$ and $G_n^o$, and then compare the matching sizes among the coupled graphs. In $G_n^b,$ we denote an edge from $\vl_i$ to $\vr_j$ by $(\vl_i, \vr_j)$ and an edge from $\vr_j$ to $\vl_i$ by $(\vr_j, \vl_i).$ Then, we partition the realized edges in $G_n^b$ into four groups:
    \begin{align*}
        &X_1 := \bracket{(\vl_i, \vr_j) | i,j \in [n/2], R^l_{ij} = 1}, \quad X_2 := \bracket{(\vr_j, \vl_i) | i,j \in [n/2], R^r_{ij} = 1},\\
        &X_3 := \bracket{(\vr_j, \vl_i) | j \in [n/2], i \in \bracket{n/2+1,...,n}, R^r_{ij} = 1}, X_4 := \bracket{(\vl_i, \vr_j) | i \in [n/2], j \in \bracket{n/2+1,...,n}, R^l_{ij} = 1}.
    \end{align*}
    In \cref{fig:coupling1} (A) we illustrate the  edges in $X_1, X_2, X_3, X_4$ as red, blue, yellow, and green, respectively. 

    Fix a realization of $X_1, X_2, X_3$ and $X_4$. We start by flipping $X_1, X_3$ and $X_4$ vertically around the middle of the bipartite graph and swapping the directions accordingly, defining
    \begin{align*}
        \widetilde{X}_1 := & \bracket{(\vl_{n+1-i}, \vr_{n+1-j}) \text{ for each } (\vl_i, \vr_j) \in X_1},\quad
        \widetilde{X}_3 := & \bracket{(\vl_{n+1-i}, \vr_{n+1-j}) \text{ for each } (\vr_j, \vl_i) \in X_3},\\
        \text{and } \widetilde{X}_4 := & \bracket{(\vr_{n+1-j}, \vl_{n+1-i}) \text{ for each } (\vl_i, \vr_j) \in X_4}.
    \end{align*}
    Then, we construct the following graphs: graph (A) contains edges in $X_1, X_2, X_3$ and $ X_4;$ graph (B) contains edges in $X_1, X_2, \widetilde{X}_3$ and $\widetilde{X}_4;$ graph (C) contains edges in $\widetilde{X}_1, X_2, X_3, X_4,$ (and drop their directions); and graph (D) contains edges in $\widetilde{X}_1, X_2, \widetilde{X}_3, \widetilde{X}_4,$ (and drop their directions). \cref{fig:coupling1} provides an illustration of the different graphs. As before, the edges in $X_1, X_2, X_3, X_4$ are colored in red, blue, yellow, and green, respectively, and the coloring is maintained for their (flipped) copies in graphs (B)-(D). %We use the same color for the flipped edges to highlight that they mirror the original edges. 
    Essentially, we flip $X_3$ and $X_4$ to construct graph (B), and then we flip $X_1$ in (A) and (B) to construct (C) and (D). Our goal is to couple the realizations of $G_n^b$ as (A) or (B) with the realizations of $G_n^o$ as (C) or (D), so that it suffices to compare the combined size of the matchings in graph (A) and (B) with that of (C) and (D). This means that our proof builds on a coupling between two pairs of graphs rather than just one pair; we explain below how this allows us to find for each edge in a matching of either (A) or (B) a corresponding edge that can be part of a matching in (C) or (D). 

    We denote the sizes of a maximum matching in the four graphs, (A)-(D), by $M_A, M_B, M_C$, and $M_D$. Notice that this is a slight abuse of notation because we omitted the dependency of these quantities on $X_1, X_2, X_3$ and $X_4$ for notational convenience. We now argue that graph (A) and (B) are possible realizations of $G_n^b$, while graph (C) and (D) are possible realizations of $G_n^o$, all of which occur with the same probability in the respective random graphs. With $\pnf_n = \anf/n = 0,$ in $G^o_n$ (as $\anf = 0$), we have $\PP{R_{ij} = 1} =  \pf_n, \forall i,j \in [n].$ Combined with \eqref{eq:gbn}, we thus know that, given $X_1, X_2, X_3$ and $X_4$, 
    \begin{align*}
        &\PP{G_n^b \text{ realizes as (A)}} = \PP{G_n^b \text{ realizes as (B)}} = \PP{G_n^o \text{ realizes as (C)}} = \PP{G_n^o \text{ realizes as (D)}} \\
        =&\;\parenthesis{\pf_n}^{|X_1|+|X_2|+|X_3|+|X_4|} \parenthesis{1-\pf_n}^{n^2 - \parenthesis{|X_1|+|X_2|+|X_3|+|X_4|}}.
    \end{align*}

    We now prove the coupling based on $X_1, X_2, X_3$ and $X_4$ is valid. When $X_3 = \widetilde{X}_3$ and $X_4 = \widetilde{X}_4$,\footnote{This may occur when all edges in $X_3$ and $X_4$ are symmetric around the middle of the bipartite graph, i.e., they ``flip" to themselves.} i.e., (A) and (B) are identical, we trivially have $M_A = M_B,$ and the maximum matching size in (A) can be written as $(M_A+M_B)/2.$ On the other hand, when $X_3 \neq \widetilde{X}_3$ or $X_4 \neq \widetilde{X}_4$, since $G_n^b$ realize as (A) and (B) with the same probability the weighted average maximum matching size in (A) and (B) is also $(M_A+M_B)/2.$ Thus, for any $n,$
    \begin{align*}
        \EE{\mathcal{M}_n^b} = \sum_{\substack{\text{all realizations of} \\ X_1, X_2, X_3, X_4}} \left(\pf_n\right)^{|X_1|+|X_2|+|X_3|+|X_4|} \left(1-\pf_n\right)^{n^2 - \left(|X_1|+|X_2|+|X_3|+|X_4|\right)} \cdot \frac{M_A+M_B}{2}.
    \end{align*}
    Similarly, we find that
    \begin{align*}
        \EE{\mathcal{M}_n(1,0)} = \sum_{\substack{\text{all realizations of} \\ X_1, X_2, X_3, X_4}} \left(\pf_n\right)^{|X_1|+|X_2|+|X_3|+|X_4|} \left(1-\pf_n\right)^{n^2 - \left(|X_1|+|X_2|+|X_3|+|X_4|\right)} \cdot \frac{M_C+M_D}{2}.
    \end{align*}
    \dfedit{Thus, for the remainder of the proof of \cref{lem:bundled_coupling}, it suffices to compare $M_A+M_B$ with $M_C+M_D$. Specifically, we first show that
    \begin{equation}\label{eq:inequality_to_prove}
        \forall X_1, X_2, X_3 \text{ and } X_4: M_A+M_B \leq M_C+M_D.
    \end{equation}
    We then explicitly characterize when the inequality is strict and show that 
    \begin{align}\label{eq:lemma2ii}
        \liminf_{n \to \infty}\EE{\squarebracket{\parenthesis{M_C + M_D} - \parenthesis{M_A + M_B}}}/n 
        \;\geq\; \parenthesis{\af/2}^3/2 \cdot e^{-7\af},
    \end{align}
    where the expectation is taken over all realizations of $X_1,\dots,X_4$. This would immediately yield
    \[
        \liminf_{n\to\infty}\EE{\mathcal{M}_n(1,0)-\mathcal{M}_n^b}/n 
        \;\geq\; \parenthesis{\af}^3/2^5 \cdot e^{-7\af}.
    \]

    We start by verifying \eqref{eq:inequality_to_prove}.} We pick an arbitrary matching in (A) and denote the edges in $X_1, X_2, X_3$ and $X_4$ that are involved in the maximum matching by $Y_1, Y_2, Y_3$ and $Y_4.$ Similarly, we pick any matching in (B) and denote the edges in $X_1, X_2, \widetilde{X}_3$ and $\widetilde{X}_4$ that are involved in the maximum matching by $Y'_1, Y'_2, \widetilde{Y}_3$ and $\widetilde{Y}_4$.\footnote{Notice that the distinction between $Y_1, Y_2$ and $Y'_1, Y'_2$ arises from the fact that the edges in $X_1$ and $X_2$ that are involved in a maximum matching for (A) may be different from those for (B).} Our proof proceeds by constructing feasible matchings in (C) and (D) that have a combined size that is greater-equal to the combined size of the matchings in (A) and (B). We drop the direction of the edges as $(\vl_i, \vr_j)$ and $(\vr_j, \vl_i)$ cannot appear in the same matching, and with a slight abuse of notation, we denote an undirected edge between $\vl_i$ and $\vr_j$ by $(\vl_i, \vr_j)$. \cref{fig:coupling2} (A) and (B) illustrate $Y_1, Y_1', Y_2, Y_2'$ as red, pink, blue, and navy; moreover, the plots illustrate $Y_3$ and $\widetilde{Y}_3$ as yellow, and $Y_4$ and $\widetilde{Y}_4$ as green.

    To construct matchings in (C) and (D), we flip $Y_1, Y_1'$ vertically and define
    \begin{align*}
        \widetilde{Y}_1 := & \bracket{(\vl_{n+1-i}, \vr_{n+1-j}) \text{ for each } (\vl_i, \vr_j) \in Y_1}\quad \text{and}\quad
        \widetilde{Y}_1' := & \bracket{(\vl_{n+1-i}, \vr_{n+1-j}) \text{ for each } (\vl_i, \vr_j) \in Y_1'}.
    \end{align*}
    Since graph (C) contains all edges in $\widetilde{X}_1$ and $X_2$, $\widetilde{Y}'_1$ and $Y_2$ are part of a feasible matching in (C). Similarly, since graph (D) also contains all edges in $\widetilde{X}_1$ and $X_2,$ $Y_2'$ and $\widetilde{Y}_1$ are part of a feasible matching in (D). As illustrated in \cref{fig:coupling2} (C) and (D), we copy $\widetilde{Y}'_1$ and $Y_2$ into the construction of a matching in (C), and copy $Y_2'$ and $\widetilde{Y}_1$ into a matching in (D) (as before, the figure maintains consistent coloring for the flipped edges in different subgraphs). 
    
    Then, it suffices to show that all edges in $Y_3, \widetilde{Y}_3, Y_4$ and $\widetilde{Y}_4$ can also be mapped into (C) and (D). We denote by $\bar{C}$ and $\bar{D}$ the remaining nodes in (C) and (D) that are not incident to the already copied matches, and by $E(\bar{C})$ and $E(\bar{D})$ the available edges among $\bar{C}$ and $\bar{D}$. Below we construct a mapping that injectively maps all edges in $Y_3, \widetilde{Y}_3, Y_4$ and $\widetilde{Y}_4$ to
    two matchings $M(\bar{C})$ in (C) and $M(\bar{D})$ in (D), where $M(\bar{C})\subseteq E(\bar{C})$ and $M(\bar{D}) \subseteq E(\bar{D})$. This then immediately implies that $M_A+M_B \leq M_C+M_D$. 

    % Then, we will show that, excluding all edges in/are incident to $\widetilde{Y}'_1, Y_2, \widetilde{Y}_1$ and $Y_2'$ because these edges are already used in the matching of (C) and (D), we can still map the remaining matches in (A) and (B) (those in $Y_3, \widetilde{Y}_3, Y_4$ and $\widetilde{Y}_4$) to a maximum matching. This then immediately implies that $M_A+M_B \leq M_C+M_D$. 

    Since the edges in $X_3$ ($\widetilde{X}_3$) are not incident to those in $X_4$ ($\widetilde{X}_4$), the resulting matches $Y_3, \widetilde{Y}_3$ and $Y_4, \widetilde{Y}_4$ can be analyzed separately. We next show that the matches in $Y_3$ and $\widetilde{Y}_3$ can be injectively mapped to $\parenthesis{M(\bar{C}) \cap X_3} \cup \parenthesis{M(\bar{D}) \cap \widetilde{X}_3}$ in graph (C) and (D) by constructing a multigraph $G'$. The injective mapping from $Y_4$ and $\widetilde{Y}_4$ to $\parenthesis{M(\bar{C}) \cap X_4} \cup \parenthesis{M(\bar{D}) \cap \widetilde{X}_4}$ follows from symmetry through a similarly constructed graph $G''$, and thus our focus is on $G'$ for the rest of this proof. Specifically, we construct $G'$ as a bipartite graph with $n/2$ nodes on each side, indexing nodes on the left as $v_1^l$ through $v_{n/2}^l$ and on the right as $v_1^r$ through $v_{n/2}^r$. The edge set of $G'$ consists of\footnote{As $G'$ is a multigraph, this union may contain two copies of the same edge.} $$\bracket{(\vl_{i-n/2},\vr_j) \text{ for each } (\vl_i,\vr_j) \in Y_3}\cup\bracket{(\vl_{n/2+1-i},\vr_{n+1-j}) \text{ for each } (\vl_i,\vr_j) \in \widetilde{Y}_3}.$$ We refer to the former set as type $A$ edges (as they come from $Y_3$ in graph (A)) and the later set as type $B$ edges. Finally, we color the nodes in $G'$:\footnote{Notice that the coloring is based on the edges, rather than the colors of the flexible/regular nodes that we used for illustrations in (A)-(D).} %\dfcomment{add an explanation of why nodes cannot be colored in two colors}
    \begin{itemize}
        \item We color $\vr_j$ in $G'$ by red (blue) if in graph (A) $\vr_j$ is matched by an edge in $Y_1$ ($Y_2$);
        \item We color $\vl_i$ in $G'$ by pink (navy) if in graph (B) $\vl_{n/2+1-i}$ is matched by an edge in $Y'_1$ ($Y'_2$).
    \end{itemize}
    \cref{fig:coupling3} provides an illustration of $G'$ that is constructed based on \cref{fig:coupling2} (A) and (B). Notice that a node on the left of $G'$ cannot be colored twice because it is matched by at most one edge in $Y_1' \cup Y_2',$ and similarly those on the right cannot be colored twice because they are matched by at most one edge in $Y_1 \cup Y_2$.

    We begin by analyzing the degree of nodes in $G'$. Recall that every edge in $G'$ comes from either $Y_3$ or $\widetilde{Y}_3$. Thus, each node in $G'$ has degree at most $2$, as otherwise at least two incident edges would come from either  $Y_3$ or $\widetilde{Y}_3$, contradicting that $Y_3$ and $\widetilde{Y}_3$ are subsets of matchings in (A) and (B), respectively. Indeed, colored nodes in $G'$ have a degree of at most $1$ because, if the colored node is already matched by an edge in graph (A), then it cannot connect to any type A edges; if it is matched by an edge in graph (B), then it cannot connect to any type B edges. Thus, in $G'$ each colored node can connect to at most one edge from either type A or type B.

    \begin{figure}[ht]
        \vspace{-0.2in}
        \centering
        \includegraphics[width=0.9\textwidth]{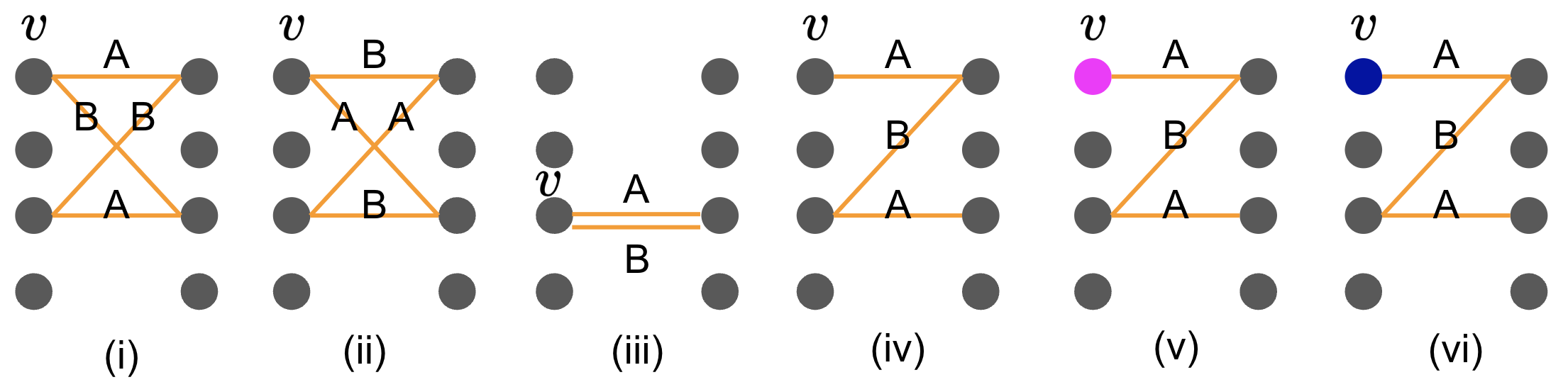}
        \caption{Illustrations of possible connected components in $G'$.}
        \label{fig:coupling4}
        \vspace{-0.2in}
    \end{figure}
    
    Since nodes in $G'$ have a degree of at most $2,$ any connected component in $G'$ is either a path or a cycle (page 109 of \citet{west2001introduction}). \cref{fig:coupling4} illustrates examples of connected components in $G'$.\footnote{The result in \cite{west2001introduction} applies only to simple graphs; as nodes in $G'$ have degree at most 2, whenever there are multiple edges between 2 nodes, this means that these 2 nodes have no other edges incident to them (see \Cref{fig:coupling4} (iii)). Therefore, such a pair of nodes also forms a cycle.} We next construct the matchings $M(\bar{C})$ and $M(\bar{D})$ based on the structure of paths and cycles in $G'$ through two mappings that respectively map the edges in $G'$ to $M(\bar{C})$ and $M(\bar{D})$. The mappings ensure that all edges in $G'$ of either type A or B are mapped to either $M(\bar{C}) \cap X_3$ or $M(\bar{D}) \cap \widetilde{X}_3$.%, and (ii) all edges of both type A and B are mapped to both $M(\bar{C}) \cap X_3$ and $M(\bar{D}) \cap \widetilde{X}_3$. 
    This then immediately completes the proof. Specifically, we use the following bijective mappings from edges in $G'$ to $M(\bar{C}) \cap X_3$ and $M(\bar{D}) \cap \widetilde{X}_3$, which we denote by $f_{C}$ and $f_{D},$ respectively:
    \begin{align}\label{eq:mapping1}
        f_{C}: (\vl_i, \vr_j) \to (\vl_{i+n/2}, \vr_j), f_{D}: (\vl_i, \vr_j) \to (\vl_{n/2+1-i}, \vr_{n+1-j}), \forall i,j \in [n/2].
    \end{align}
    %Similarly, we define the following bijective mappings from type B edges to $X_3$ in (C) and $\widetilde{X}_3$ in (D):
    % \begin{align}\label{eq:mapping2}
    %     f_{B \to C}: (\vl_i, \vr_j) \to (\vl_{i+n/2}, \vr_j), f_{B \to D}: (\vl_i, \vr_j) \to (\vl_{n/2+1-i}, \vr_{n+1-j}), \forall i,j \in [n/2].
    % \end{align}
    In the rest of the proof, we show that every edge in $G'$ is mapped by either $f_C$ or $f_D$ to its respective image in either $M(\bar{C}) \cap X_3$ or $M(\bar{D}) \cap \widetilde{X}_3$, i.e., the edge is mapped to part of a feasible matching solution.

    %To find appropriate mappings for edges in $G',$ 
    We begin by considering the case of cycles in $G'$. Since all nodes in a cycle have a degree of $2,$ no node in the cycle can be colored. Moreover, in a bipartite graph, all cycles are of even length. Since no two type A edges or two type B edges may share the same node, the edges in the cycle must be alternating in type A and B. As illustrated in \cref{fig:coupling4} (i) and (ii), as one traverses through the cycle starting from node $v$ on the top left and moves to the top right, the edges must either be of (1) type A, B, A, B, ..., or (2) type B, A, B, A, ... To create feasible matchings in (C) and (D), we need to ensure that the edges that are incident to the same node in $G'$ are not both mapped by $f_{C}$ (resp. $f_{D}$), since they would otherwise share a node in graph (C) (resp. (D)). Thus, for structure (1), we choose for each type A edge $e$ the edge $f_{C}(e)$ to become part of the matching in (C) and for each type B edge $e$ the edge $f_{D}(e)$ to become part of the matching in graph (D).\footnote{An alternative option is to map all type A edges using $f_{D}$ and type B edges using $f_{C}$.} These choices always lead to a feasible matching because the allocated edges are not incident to any colored nodes, and thus not incident to any matches already copied from $Y_1, Y_2, Y_1'$, and $Y_2'$. The construction based on structure (2) is symmetric. We remark that the structure in \cref{fig:coupling4} (iii), with two edges between a pair of nodes, is a special case of a cycle in $G'$.
    
    Now, we consider the case of paths. Since all but the endpoints of a path have a degree of $2,$ only the two endpoints of a path may be colored. Thus, it suffices to consider the following three subcases: (1) the two endpoints of the path are both uncolored, (2) one of the endpoints is colored, and (3) both of the endpoints of the path are colored.
    \begin{itemize}
        \item In subcase (1), % if the path has a length of one, as illustrated in \cref{fig:coupling4} (iii), there could be two edges between the same pair of nodes. In this case, the edges are included in both $M(\bar{C})$ and $M(\bar{D})$ using both $f_{C}$ and $f_{D}$. In all other situations, 
        as illustrated in \cref{fig:coupling4} (iv), the path must alternate between edges of type A and B. Thus, we can iteratively include all edges in the path in $M(\bar{C})$ or $M(\bar{D})$, as %we did 
        in the case of cycles. 
        \item In subcase (2), if an end-point $v$ is colored pink, as illustrated in \cref{fig:coupling4} (v), then the edge that connects to the endpoint must be of type A. Any subsequent edge to $v$ must then alternate between type B, A, B, .... To avoid the first edge sharing a node with $\widetilde{Y}_1'$ in graph (C), we include all type A edges by using $f_{D}$ to map them into $E(\bar{D})$, and all type B edges by using $f_{C}$ to map them into $E(\bar{C})$. In contrast, if an end-point $v$ is navy, as illustrated in \cref{fig:coupling4} (vi), then the edge that connects to the endpoint must be of type A. Thus, the subsequent edges follow type B,A,B, .... To avoid the first edge sharing a node with $Y_2'$ in graph (D), we then include for each type A edge its image in $E(\bar{C})$ under $f_{C}$ and for each type B edge its image under $f_{D}$ in $E(\bar{D})$.
        All subsequent edges in the path are incident to uncolored nodes and thus their images under $f_C$ and $f_D$ are not incident to any edges already copied from $Y_1, Y_2, Y_1'$, and $Y_2'$.
        The cases with one of the endpoints being red or blue are symmetric. \cref{fig:coupling3} shows how type A and type B edges are mapped to graph (C) and (D) based on colors in $G'$.
        \item Finally, we observe that it is not possible for both endpoints of a path to be colored. If the path is of odd length, the two endpoints must be on different sides of the bipartite graph. Thus, one of the endpoints is colored pink/navy and the other is colored red/blue. Since the edges, starting from the endpoint colored pink/navy, must alternate between type A and B, with an odd number of edges the last edge must be of type A. This contradicts the feasibility of the matching in graph (A) because the colored node is already occupied by $Y_1$ or $Y_2$ in graph (A). On the other hand, if the path is of even length, both of the endpoints must be on the same side of the bipartite graph. Assume without loss of generality that both endpoints are colored pink/navy (the other case is exactly symmetric). Then, starting from one of the endpoints, the path must alternate between edges of type A and B and end with an edge of type B. This contradicts the feasibility of the matching in graph (B) because the colored node is already occupied by $Y_1'$ or $Y_2'$ in graph (B). Thus, subcase (3) is not possible.
    \end{itemize}

    Therefore, in all possible subcases the matches in $Y_3$ and $\widetilde{Y}_3$ can be injectively mapped to $\parenthesis{M(\bar{C}) \cap X_3} \cup \parenthesis{M(\bar{D}) \cap \widetilde{X}_3}$ in graph (C) and (D). This shows that $M_A+M_B \leq M_C+M_D$ for any $X_1, X_2, X_3$ and $X_4$. %and thus $\EE{\mathcal{M}_n^b} \leq \EE{\mathcal{M}_n(1,0)} \; \forall n$ when $\anf = 0.$

    \dfedit{Next, we build upon \eqref{eq:inequality_to_prove} and establish \eqref{eq:lemma2ii} by providing an explicit construction in which $M_C + M_D$ is strictly larger than $M_A + M_B$: We start from an arbitrary flexible node $\vr_i$ on the right-hand side of the bipartite graph (see \cref{fig:coupling_ext1}), and show that after applying the coupling argument above, we always obtain one additional feasible match in (C). We then sum over all possible choices of $\vr_i$ and take the expectation. 

    To formalize the argument, we pick a maximum matching in (A) that includes every $X_2$ edge incident to a degree-1 node on the left-hand side. Such a maximum matching always exists because any edge incident to a degree-1 vertex must belong to some maximum matching \citep{bohman2011karp}. Analogously, we pick a maximum matching in (B) that includes every $\widetilde{X}_3$ edge incident to a degree-1 on the left-hand side. This careful selection, while not necessary for \eqref{eq:inequality_to_prove}, simplifies the proof of \cref{eq:lemma2ii}.
    
    Now, we specify sufficient conditions for $\vr_i$ to ensure that the combined maximum matching size in (A) and (B) is strictly smaller than that in (C) and (D):
    \begin{enumerate}[label=(\arabic*)]
        \item $\vr_i$ generates exactly two edges: one to some $\vl_j$ with $j \leq n/2$ (an $X_2$ edge) and one to some $\vl_{j'}$ with $j'>n/2$ and $j'\neq n+1-j$ (an $X_3$ edge that is not incident to the flip of $\vl_{j}$). Moreover, no node in $V_l$ generates an edge to $\vr_i$. This event occurs with probability $$\parenthesis{1-\af/n}^{n-2}\cdot \parenthesis{\af/n}^2 \cdot (n/2) \cdot (n/2-1) \cdot \parenthesis{1-\af/n}^{n/2}.$$
        \item $\vl_j$ generates no edge and receives no edge other than from $\vr_i$. Conditioned on the event above, this condition holds with probability $$\parenthesis{1-\af/n}^{n} \cdot \parenthesis{1-\af/n}^{n/2-1}.$$
        \item $\vl_{j'}$ receive no other edge. Moreover, its flipped node $\vl_{n+1-j'}$ receives no edge and generates exactly one edge to some $\vr_{i'}$ with $i'\neq i$. Conditioned on the events above, this condition holds with probability $$\parenthesis{1-\af/n}^{n/2-1} \cdot \parenthesis{1-\af/n}^{n/2} \cdot (n/2-1) \cdot \parenthesis{1-\af/n}^{n-1} \cdot \af/n.$$
        \item $\vr_{i'}$ receives no other edge from $V_l$ and generates no edge. Conditioned on the events above, this holds with probability $\parenthesis{1-\af/n}^{n/2-1} \cdot \parenthesis{1-\af/n}^{n}$.
        \item $\vr_{n+1-i'}$ receives no edge from $V_l$. Conditioned on the events above, this holds with probability $\parenthesis{1-\af/n}^{n/2}$.
    \end{enumerate}
    
    \begin{figure}[ht]
    \centering
    \includegraphics[width=0.8\textwidth]{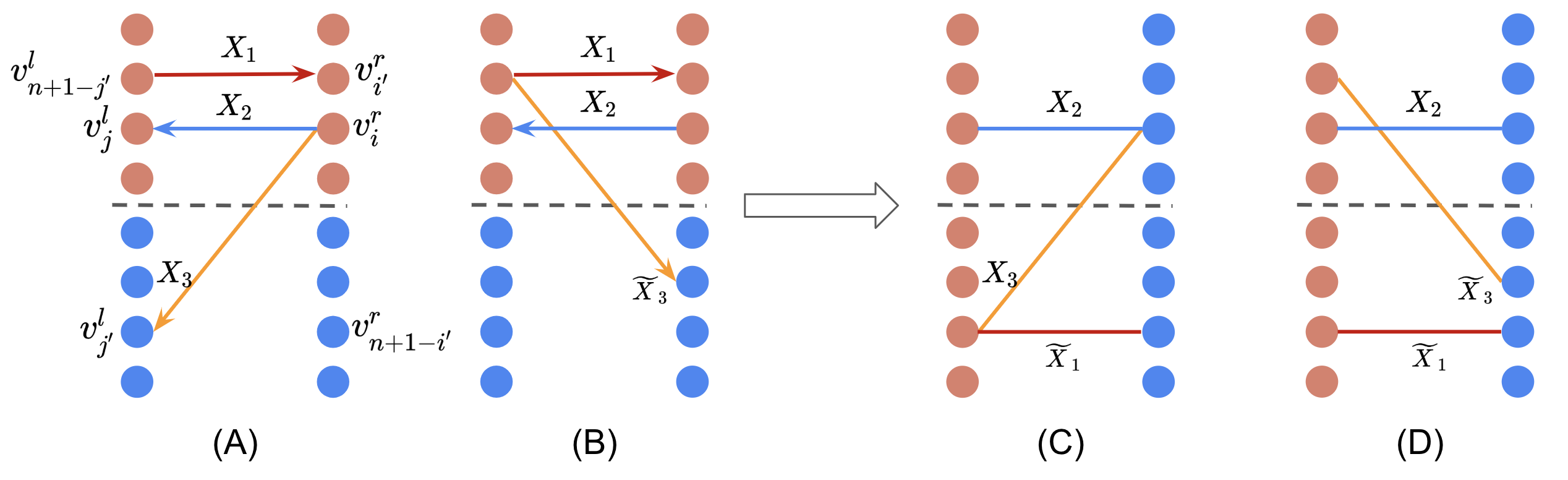}
    \caption{Illustration of the edges in (A) and (B), and the position they are copied into in (C) and (D). }
    \label{fig:coupling_ext1}
    \vspace{-0.35in}
\end{figure}

\begin{figure}[ht]
    \centering
    \includegraphics[width=0.8\textwidth]{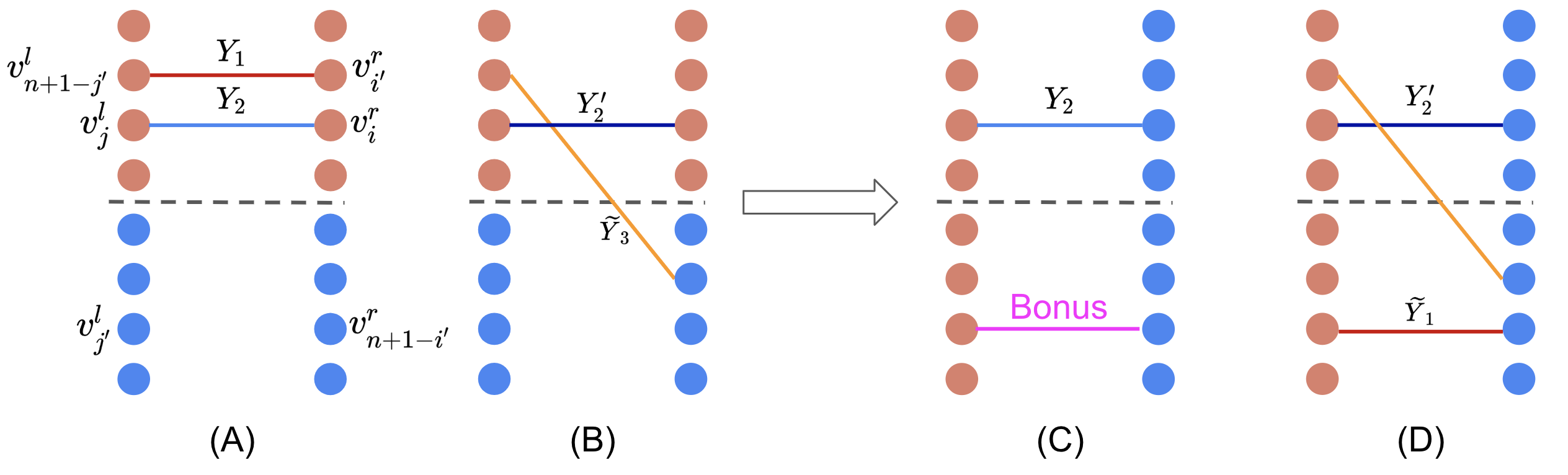}
    \caption{Illustration of the matches in (A), (B), (C), and (D). The additional match in (C) is highlighted in pink.}
    \label{fig:coupling_ext2}
    \vspace{-.25in}

\end{figure}

The combined probability of conditions (1)–(5) is
\[
    \parenthesis{1-\af/n}^{7n-6}\cdot \parenthesis{\af/n}^3 \cdot (n/2)\cdot (n/2-1)^2.
\]

We now show that conditions (1)--(5) guarantee that the combined matching size in (C) and (D) is strictly larger than in (A) and (B). We begin by applying the coupling argument above, which maps all matches in (A) and (B) to (C) and (D). By condition (2), $\vl_j$ has degree~1. Together with condition (1) and the construction of the maximum matching in (A), this implies that $\parenthesis{\vl_{j'}, \vr_{i}}$ (the yellow edge) must belong to the maximum matching in (A). Moreover, since $\vr_i$ generates no additional $X_3$ edges (condition (1)), its flipped copy $\vr_{n+1-i}$ has degree~1 in (B). By condition (3) and the construction of the maximum matching in (B), it follows that $\parenthesis{\vl_{n+1-j'}, \vr_{n+1-i}}$ belongs to the maximum matching in (B) (see \cref{fig:coupling_ext2}), whereas $\parenthesis{\vl_{n+1-j'}, \vr_{i'}}$ does not. Consequently, in the constructed graph $G'$, the edge $(\vl_{n+1-j'}, \vr_{n+1-i})$ is mapped into (D), not into (C). Finally, conditions (3)--(5) ensure that both $\vl_{j'}$ and $\vr_{n+1-i'}$ remain unmatched after all matches from (A) and (B) have been copied into (C) and (D). Hence, we can add the extra edge $(\vl_{j'}, \vr_{n+1-i'})$ into the maximum matching in (C), yielding one additional match. This construction guarantees that 
\[
    M_C + M_D > M_A + M_B.
\]

Taking expectation over all possible positions of $\vr_j$ (of which there are $n/2$), and invoking linearity of expectation, we conclude the proof by arguing that
\begin{align*}
    \liminf_{n \to \infty}\EE{\squarebracket{(M_C + M_D) - (M_A + M_B)}}/n 
    &\;\geq\; \liminf_{n \to \infty} \frac{1}{2} \cdot \parenthesis{1-\af/n}^{7n-6}\cdot \parenthesis{\af/n}^3 \cdot (n/2)\cdot (n/2-1)^2 \\
    &\;\geq\;\parenthesis{\af}^3/2^5 \cdot e^{-7\af}.\qquad\Halmos
\end{align*}
}
% \end{proof}
% \vspace{-.5in}

\subsection{Proofs of the Results in \texorpdfstring{\cref{sec:asymmetry}}{Lg}}\label{sec:proof_global_bounds}

\subsubsection{Proof of \texorpdfstring{\cref{lem:n1_lb}}{Lg}}\label{app:n1_lb}
    Denote the number of nodes in $V_l$ and $V_r$ that have a degree of $d$ by random variables $q_d^l$ and $q_d^r,$ respectively. We start by showing that, in any realization of the $(1-B/2)n \times (1-B/2)n$ bipartite graph, 
    \begin{equation}\label{eq:edge_degree_lb}
        m_1 \geq \sum_{d} q_d^l \cdot d - \sum_{d = 2}^{(1-B/2)n} q_d^l \cdot (d-1) - \sum_{d = 2}^{(1-B/2)n} q_d^r \cdot (d-1).
    \end{equation}
    Notice that $\sum_{d} q_d^l \cdot d = \sum_{d} q_d^r \cdot d$ is the number of edges in the graph, and the second and third term in \eqref{eq:edge_degree_lb} respectively capture the extra edges, i.e., those incident to nodes with degree $> 1$, on the left and right-hand side of the graph. This lower bound holds because, after deleting $d-1$ edges from all nodes with degree $d > 1$ in $V_l$ and $V_r$, all remaining edges in the graph would have degree $1$ on both ends. Hence, the remaining edges are not incident to each other and thus trivially form a (not necessarily maximum) matching. 

    Notice that the probability for a node $v$ on the left or right-hand side of the graph to have degree $d$ is the same in this symmetric bipartite graph. Taking expectations over the lower bound in \eqref{eq:edge_degree_lb}, we find that
    \begin{equation}\label{eq:edge_degree_lb2}
        \begin{split}
            \EE{m_1} \geq& (1-B/2)n \sum_{d = 1}^{(1-B/2)n} \PP{deg(v) = d} \cdot d - 2 (1-B/2)n \sum_{d = 2}^{(1-B/2)n} \PP{deg(v) = d} \cdot (d-1)\\
            =&(1-B/2)n \EE{deg(v)} - 2 (1-B/2)n \sum_{d = 2}^{(1-B/2)n} \PP{deg(v) = d} \cdot (d-1)\\
            =&\parenthesis{1-B/2}^2 n 2 \anf - 2 (1-B/2)n \sum_{d = 2}^{(1-B/2)n} \PP{deg(v) = d} \cdot (d-1).
        \end{split}
    \end{equation}
    We simplify the second term by substituting $t=(1-B/2)n$ and observing that
    \begin{align*}
        \lim_{n \to \infty}\sum_{d = 2}^{(1-B/2)n} \PP{deg(v) = d} \cdot (d-1) = & \lim_{n \to \infty}\sum_{d = 2}^{(1-B/2)n} \binom{(1-B/2)n}{d} \parenthesis{2 \anf/n}^d \cdot \parenthesis{1-2 \anf/n}^{(1-B/2)n -d} \cdot (d-1)\\
        =&\lim_{t \to \infty}\sum_{d = 2}^{t} \binom{t}{d} \parenthesis{2 \anf (1-B/2)/t}^d \cdot \parenthesis{1-2 \anf(1-B/2)/t}^{t -d} \cdot (d-1)\\
        =&\lim_{t \to \infty}\sum_{d = 1}^{t} \binom{t}{d} \parenthesis{2 \anf (1-B/2)/t}^d \cdot \parenthesis{1-2 \anf(1-B/2)/t}^{t -d} \cdot (d-1)\\
        =&2 \anf (1-B/2) - \lim_{t \to \infty}\sum_{d = 1}^{t} \binom{t}{d} \parenthesis{2 \anf (1-B/2)/t}^d \cdot \parenthesis{1-2 \anf(1-B/2)/t}^{t -d}\\
        =&2 \anf (1-B/2) + e^{-2 \anf (1-B/2)} -1.
    \end{align*}
    \begin{align*}
    \text{ Plugging this into \eqref{eq:edge_degree_lb2}, we bound}\quad
        \EE{m_1} \geq& \parenthesis{1-B/2}^2 n 2 \anf - 2(1-B/2)n \parenthesis{2 \anf (1-B/2) + e^{-2 \anf (1-B/2)} -1}\\
        =&2 \cdot \parenthesis{1-B/2} n \squarebracket{1-(1-B/2) \anf - e^{-2 \anf (1-B/2)}} \text{ as } n \to \infty. \Halmos
    \end{align*}

\subsubsection{Proof of \texorpdfstring{\cref{thm:balanced_better}}{Lg}}\label{app:balanced_better}

    We first provide an upper bound on $\mu(B,0)$ and then derive a lower bound on $\mu(B/2,B/2).$ When $\qvec = (B,0),$ a node $v \in V_l$ is regular with probability $1-B$ and all nodes in $V_r$ are regular nodes. Thus, a regular node $v$ forms an edge with a node $u \in V_r$ with probability $2 \anf/n.$ Then, 
    \begin{align}
        \mu(B,0) \leq&\; 1 - \lim_{n \to \infty} \PP{v \in V_l \text{ is regular and has degree $0$}} \notag\\
        =&\; 1-\lim_{n \to \infty} (1-B) \parenthesis{1-2 \anf/n}^n = 1-(1-B) \cdot e^{-2 \anf}.\label{eq:balanced_bound1}
    \end{align}

    Then, to lower bound $\mu(B/2,B/2),$ we adopt a greedy matching scheme: in the first stage we only match the regular nodes in $V_l$ with the regular nodes in $V_r,$ and in the second stage we greedily match the rest of the flexible nodes. Denote the number of matches formed in stage $1$ and $2$ by $n_1$ and $n_2$, respectively. Since each node $v \in V_l$ is regular with probability $1-B/2,$ we can use a Chernoff bound to find that the event $$E_1 := \bracket{|\sum_i F_i^l - B/2 \cdot n| \leq n^{5/8} \text{ and } |\sum_j F_j^r - B/2 \cdot n| \leq n^{5/8}}$$ occurs with a probability of at least $1-e^{-\Omega\parenthesis{n^{1/4}}}.$ 
    By \cref{lem:n1_lb}, a graph of $(1-B/2)n$ regular nodes on each side has a maximum matching among its nodes of size at least $2 \cdot \parenthesis{1-B/2} n \squarebracket{1-(1-B/2) \anf - e^{-2 \anf (1-B/2)}}-o(n)$. Correspondingly, a subgraph of $(1-B/2)n-n^{5/8}$ regular nodes on each side has the same asymptotic maximum matching size (up to $o(n)$ nodes that are removed). 
    Conditioning on $E_1$ \cref{lem:n1_lb} thus implies that as $n \to \infty$ 
    \begin{align*}
        \EE{n_1} = \EE{n_1|E_1} \PP{E_1} + \EE{n_1|E_1^c} \PP{E_1^c} \geq&\; \parenthesis{\EE{m_1} - o(n)} \cdot \parenthesis{1-e^{-\Omega\parenthesis{n^{1/4}}}} - n^{5/8} \cdot e^{-\Omega\parenthesis{n^{1/4}}}\\
        \geq&\;2 \cdot \parenthesis{1-B/2} n \squarebracket{1-(1-B/2) \anf - e^{-2 \anf (1-B/2)}} - o(n).
    \end{align*}

    \dfedit{Through a similar conditioning, we find that: 
    \begin{align*}
        &\lim_{n \to \infty} \PP{\text{regular node } v \in V_l \text{ not connected to any regular node in } V_r}\\
        \geq&\;\lim_{n \to \infty}\parenthesis{1-2 \anf/n}^{(1-B/2) n + n^{5/8}}\cdot\parenthesis{1-e^{-\Omega\parenthesis{n^{1/4}}}} + 0 \cdot e^{-\Omega\parenthesis{n^{1/4}}} \\
        \geq&\;\lim_{n \to \infty} \parenthesis{1-2\anf/n}^{(1-B/2) \cdot n + n^{5/8}} = e^{-2\anf(1-B/2)}.
    \end{align*}
    Thus, as $n \to \infty$ we can also upper bound 
    \begin{align*}
        \EE{n_1} =&\; \EE{n_1|E_1} \PP{E_1} + \EE{n_1|E_1^c} \PP{E_1^c} \\\leq&\;\parenthesis{1-e^{-2\anf(1-B/2)}} \parenthesis{(1-B/2) n + n^{5/8}} \cdot\parenthesis{1-e^{-\Omega\parenthesis{n^{1/4}}}} + n \cdot e^{-\Omega\parenthesis{n^{1/4}}}\\
        <&\; \parenthesis{1-e^{-2\anf(1-B/2)}} \cdot (1-B/2) \cdot n + o(n).
    \end{align*}}
    % $$\EE{n_1} \leq \parenthesis{1-e^{-0.1(1-B/2)}} \parenthesis{(1-B/2) n + o(n)} \leq \parenthesis{1-e^{-0.1(1-B/2)}} (1-B/2) \cdot n + o(n).$$

    Now we examine the flexible nodes and argue that, even if we greedily match the flexible nodes to any unmatched regular nodes on the opposite side, almost all of the flexible nodes will be matched in the second stage. Given $n'$ unmatched regular nodes in $V_r,$ the number of edges between a flexible node $v \in V_l$ and these $n'$ nodes is governed by $\bin\parenthesis{n',\frac{\af + \anf}{n}}$. For any $n' \in \Theta(n),$ by the Poisson Limit Theorem we find that $\bin\parenthesis{n',\frac{\af + \anf}{n}}$ converges in distribution to $\poisson\parenthesis{\frac{n'}{n} \parenthesis{\af + \anf}}$ as $n \to \infty.$ Thus, as $n \to \infty$ $$\PP{v \text{ not connected to any unmatched regular node in } V_r|n'} = e^{-\frac{n'}{n} \parenthesis{\af + \anf}}.$$ 
    
    \dfedit{We now apply this bound to the flexible nodes in $V_l.$ Observe that the event $$E_2 := E_1 \cap \bracket{n_1 \leq \parenthesis{1-e^{-2\anf(1-B/2)}} (1-B/2) n + 2n^{5/8}}$$ occurs with a probability of at least $1-e^{-\Omega\parenthesis{n^{1/4}}}$ by Chernoff bound.\footnote{With high probability there are at most $(1-B/2) n + n^{5/8}$ regular nodes on both sides; each such node is isolated with probability $e^{-2\anf(1-B/2)}$, so we expect to match at most $\parenthesis{(1-B/2) n + n^{5/8}}(1-e^{-2\anf(1-B/2)})$ of them in the first stage.} Under $E_2$, we greedily match each flexible node to any unmatched regular node in $V_r.$ In particular, for the $i$th flexible node under consideration, even if all previous $i-1$ flexible nodes are already matched to regular nodes in $V_r,$ there will still be at least $$(1-B/2) \cdot n - \parenthesis{1-e^{-2\anf(1-B/2)}} (1-B/2) n - i - 3 \cdot n^{5/8}$$} unmatched regular nodes in $V_r$. This allows us to bound:\dfedit{
    \begin{align*}
        &\PP{\text{$i$th flexible node not connected to any unmatched regular node in } V_r|E_2} \\
        \leq&\; e^{-\frac{(1-B/2) \cdot n - \parenthesis{1-e^{-2\anf(1-B/2)}} (1-B/2) n - i - 3 \cdot n^{5/8}}{n} \parenthesis{\af + \anf}} \leq e^{-\parenthesis{(1-B/2)  - \parenthesis{1-e^{-2\anf(1-B/2)}} (1-B/2) - i/n} \af} \text{ as } n \to \infty.
    \end{align*}}
    Thus, as $n \to \infty,$ through the greedy algorithm that iteratively matches each flexible \dfedit{node} to any unmatched regular node, the $i$th flexible node ends up matched with probability at least \dfedit{$$1-e^{-\parenthesis{(1-B/2) - \parenthesis{1-e^{-2\anf(1-B/2)}} (1-B/2)- i/n} \af} = 1-e^{-\parenthesis{- i/n +e^{-2\anf(1-B/2)} (1-B/2)} \af}.$$} This is a lower bound on the matching probability for any $i \in \bracket{1,2,...,B/2 \cdot n - n^{5/8}}$. %In particular, with $B \in [0.4,0.8]$ the expression $- i/n +e^{-0.1(1-B/2)} (1-B/2)$ is strictly positive, and thus the above probability is monotonically increasing in $\af$. 
    The argument for matching flexible nodes in $V_r$ with regular nodes in $V_l$ is symmetric, and \dfedit{we find that:
    \begin{align}
        \lim_{n \to \infty} \EE{\frac{n_2}{n}|E_2} \geq& 2 \cdot \lim_{n \to \infty} \squarebracket{\sum_{i = 1}^{B/2 \cdot n- n^{5/8}} \parenthesis{1-e^{-\parenthesis{- i/n +e^{-2\anf(1-B/2)} (1-B/2)} \af}}/n} \notag\\
        =& 2 \cdot B/2 - 2 \cdot \lim_{n \to \infty} \squarebracket{\sum_{i = 1}^{B/2 \cdot n} e^{-\parenthesis{- i/n +e^{-2\anf(1-B/2)} (1-B/2)} \af}}/n\notag\\
        =& B - 2 \cdot \lim_{n \to \infty} \frac{e^{\af (1/n - (1-B/2)e^{-2\anf(1-B/2)} )} \parenthesis{e^{\af B/2} - 1}}{n \parenthesis{e^{\af/n} - 1}}\notag\\
        =& B - 2 \cdot  \frac{\lim_{n \to \infty}e^{\af (1/n - (1-B/2)e^{-2\anf(1-B/2)} )} \parenthesis{e^{\af B/2} - 1}}{\lim_{n \to \infty}n \parenthesis{e^{\af/n} - 1}}\notag\\
        =& B - 2 \cdot  \frac{e^{- \af (1-B/2)e^{-2\anf(1-B/2)}} \parenthesis{e^{\af B/2} - 1}}{\lim_{n \to \infty}n \parenthesis{e^{\af/n} - 1}}\notag\\
        =&B - \frac{2}{\af}\squarebracket{\parenthesis{e^{\af \cdot B/2} - 1} e^{-\af (1-B/2) e^{-2\anf(1-B/2)}}}, \label{eq:monotonicity}
    \end{align}}
    where the second equality comes from the sum of a geometric sequence, the third from the quotient rule, and the fifth from an application of the L'Hôpital's rule on $\lim_{n \to \infty} \parenthesis{e^{\af/n} -1}/\parenthesis{1/n}$. \dfedit{Thus,%Moreover, from the monotonicity result for $1-e^{-\parenthesis{- i/n +e^{-0.1(1-B/2)} (1-B/2)} \af}$ with respect to $\af$ we know that \eqref{eq:monotonicity} is also monotonically increasing in $\af.$

    % $$\EE{n_2|E_1} \geq 2 \cdot B/2 \cdot n \parenthesis{1-e^{-0.12 \parenthesis{\af + \anf}}} \geq \parenthesis{B - 0.8 \cdot e^{-0.12 \parenthesis{\af + \anf}}} \cdot n > \parenthesis{B - 0.0001} \cdot n$$ when $\anf \in [0.01, 0.05]$ and $\af \geq 100.$ 
    
    %Denote the event that $$n_2 \geq 2 \cdot \parenthesis{0.4 n - 0.401 n e^{-0.16 \parenthesis{\af + \anf}}}$$ by $E_2$ and its complement by $E_2^c.$ Then, by Chernoff bound we find that $$\PP{E_2^c} \leq 2 e^{-2 \cdot 0.001 n \cdot e^{-2 \cdot 0.16 \parenthesis{\af + \anf}}/(400 \cdot 3)} = e^{-\Omega(n)}.$$ 

     %when $\anf \in [0.01, 0.05]$,
    \begin{align}
        \mu(B/2,B/2) \geq& \lim_{n \to \infty} \EE{\frac{n_1 + n_2}{n}} \notag\\
        \geq & \lim_{n \to \infty} \EE{\frac{n_1}{n}} + \lim_{n \to \infty} \parenthesis{\EE{\frac{n_2}{n}|E_2} \cdot \PP{E_2}} + 0 \cdot \lim_{n \to \infty} \parenthesis{1-\PP{E_2}} \notag\\
        \geq & 2 \cdot \parenthesis{1-B/2} \squarebracket{1-(1-B/2) \anf - e^{-2 \anf (1-B/2)}} \notag\\
        &+B - \frac{2}{\af}\squarebracket{\parenthesis{e^{\af \cdot B/2} - 1} e^{-\af (1-B/2) e^{-2\anf(1-B/2)}}}. \label{eq:balanced_bound2}
    \end{align}
    % for any $\anf \in [0.01, 0.05]$ and $\af \geq 100.$ 

    Combining \eqref{eq:balanced_bound1} and \eqref{eq:balanced_bound2}, the condition for $\mu\parenthesis{B/2,B/2} > \mu(B,0)$ is:
    \begin{equation}\label{eq:main_ineq}
    \begin{split}
        1-(1-B) \cdot e^{-2 \anf} < & 2 \cdot \parenthesis{1-B/2} \squarebracket{1-(1-B/2) \anf - e^{-2 \anf (1-B/2)}}\\
        &+B - \frac{2}{\af}\squarebracket{\parenthesis{e^{\af \cdot B/2} - 1} e^{-\af (1-B/2) e^{-2\anf(1-B/2)}}}
    \end{split}
    \end{equation}

    Now we verify that the conditions stated in \cref{thm:balanced_better} guarantee \eqref{eq:main_ineq}. Observe that the left-hand side of \eqref{eq:main_ineq} equals $(1-B)\cdot\parenthesis{1-e^{-2\anf}} + B$. Hence, \eqref{eq:main_ineq} is equivalent to
    \begin{equation}\label{eq:main_ineq_reduced}
    \begin{split}
        \underbrace{(1-B)\cdot\parenthesis{1-e^{-2\anf}}}_{(I)} 
        + \underbrace{B}_{(II)} 
        < & \underbrace{2 \cdot \parenthesis{1-B/2} \squarebracket{1-(1-B/2)\anf - e^{-2 \anf (1-B/2)}}}_{(III)} \notag\\
        &+\underbrace{B - \frac{2}{\af}\squarebracket{\parenthesis{e^{\af B/2} - 1} e^{-\af (1-B/2) e^{-2\anf(1-B/2)}}}}_{(IV)} .
    \end{split}
    \end{equation}
    Thus, it suffices to compare $(I)$ with $(III)$ and $(II)$ with $(IV)$ separately.

    For $(I)$ and $(III)$, by Taylor expansion we use the elementary bounds
    \[
    1-e^{-u} \leq u 
    \quad\text{and}\quad 
    e^{-u} \leq 1-u+u^2/2, 
    \quad \forall u \geq 0.
    \]
    Setting $x := 1-B/2 \in (1/2,1)$, we obtain
    \[
    (I) \leq (1-B)\cdot 2\anf = 2\anf(2x-1),
    \quad\text{and}\quad
    (III) \geq 2x\parenthesis{x\anf-2\anf^2 x^2} 
    = 2\anf x^2 (1-2\anf x).
    \]
    Solving $(I)=(III)$ yields
    \[
    \anf = 0 \text{ or } \anf = \frac{(x-1)^2}{2x^3} 
    = \frac{B^2}{8(1-B/2)^3}.
    \]
    Thus, a sufficient condition for $(I) < (III)$ is $\anf \in \parenthesis{0, \frac{B^2}{8(1-B/2)^3}}$, i.e., imposing $\anf \in \parenthesis{0, \frac{B^2}{8(1-B/2)^3}}$ ensures
    \[
    (III)-(I) \;\geq\; 2\anf\squarebracket{x^2(1-2\anf x)-(2x-1)} > 0.
    \]

    Thus, to prove \eqref{eq:main_ineq} under $\anf \in \parenthesis{0, \frac{B^2}{8(1-B/2)^3}}$, it remains to show $$(IV)-(II) > -2\anf\squarebracket{x^2(1-2\anf x)-(2x-1)}.$$ 
    \[\text{We first require}\qquad
    \anf < \frac{1}{2(1-B/2)} \ln\parenthesis{\frac{2-B}{B}},
    \]
    so that $x e^{-2\anf x} > B/2$. This ensures that the exponential term $e^{\af B/2} \cdot e^{-\af (1-B/2) e^{-2\anf(1-B/2)}}$ in $(IV)$ decreases with $\af$. Using the inequality $e^u-1 \leq u e^u \;\forall u$, we find
    \[
    (IV) \;\geq\; B - B\cdot e^{-\af \parenthesis{xe^{-2\anf x}-B/2}}.
    \]
    Therefore, a sufficient condition for $(IV)-(II) > -2\anf\squarebracket{x^2(1-2\anf x)-(2x-1)}$ is
    \[
    B \cdot e^{-\af \parenthesis{xe^{-2\anf x}-B/2}} \;<\; 2\anf\squarebracket{x^2(1-2\anf x)-(2x-1)}.
    \]
    \[\text{This condition is equivalent to}\quad
    \af \;>\; \af_\star(B,\anf) 
    := \frac{\ln\parenthesis{B}-\ln\parenthesis{2 \anf \squarebracket{\parenthesis{B/2}^2 - 2 \anf \parenthesis{1-B/2}^3}}}{\parenthesis{1-B/2} e^{-2 \anf \parenthesis{1-B/2}}-B/2}.
    \]

    To conclude the proof we obtain $(I)+(II) < (III)+(IV)$  by requiring
    \[
    0 < \anf < \anf_\star(B) 
    := \min\bracket{\frac{B^2}{8(1-B/2)^3}, \;\frac{1}{2(1-B/2)} \ln\parenthesis{\frac{2-B}{B}}}
    \quad\text{and}\quad
    \af > \af_\star(B,\anf).\Halmos
    \]
    }

\subsection{Analysis Based on the KS Algorithm}\label{sec:proof_global_add}

\dfedit{
We begin with the definition of the KS algorithm in \cref{alg:ks}. Intuitively, the KS algorithm proceeds in two phases: in the first phase, it repeatedly matches and removes nodes of degree~$1$ until no such nodes remain; in the second phase, it randomly selects edges to match until all edges are pruned. This algorithm has two advantages: first, its simple ``greedy'' structure makes it amenable to theoretical study, and prior works have described its behavior on random graphs. Second, it can be optimal in sparse settings. For example, it is always optimal if the graph is a tree and yields asymptotically optimal matchings for some classes of sparse random graphs \citep{karp1981maximum, balister2015controllability}. These two facts combined make it a valuable tool to study maximum matching in sparse settings. Though prior works do not encapsulate the random graphs we study (see Section \ref{sec:rel_work}), we show that the KS algorithm is amenable to our model.

\begin{algorithm}
\caption{Karp-Sipser's (KS) Algorithm}\label{alg:ks}
\begin{algorithmic}[1]
\State \textbf{Input:} Graph $G$
\State Initialize graph $G'$ as an empty graph on the same set of nodes as $G$
\While{$G$ has edges}
    \If{there exists a node of degree 1 in $G$}
        \State Choose  an edge $e$ that is incident to a node of degree 1 uniformly at random
    \Else
        \State Choose an edge $e$ from all remaining edges uniformly at random 
    \EndIf
    \State Add edge $e$ to graph $G'$
    \State Delete edge $e$ and all edges incident to $e$ from graph $G$
\EndWhile
\State \textbf{Output:} The number of edges in $G'$
\end{algorithmic}
\end{algorithm}

The KS-based analysis underlies two key results of our paper. First, it allows us to analytically compare one-sided and balanced flexibility allocations in the parameter regimes identified in \cref{thm:compare_q}. Second, it motivates the study of $\mubar(\ql,\qr)$ as a proxy performance measure and facilitates the characterization of its structural properties at $\qvec=(1/2,1/2)$ in \cref{thm:convex_concave}.

The rest of this section proceeds as follows. In Appendix \ref{app:ks_derivation}, we start by constructing a quantity $\mubar(\ql,\qr)$ based on analyses of the KS algorithm. Under \cref{cond:phase2}, \cref{thm:global_matching} shows that $\mubar(\ql,\qr)$ is equal to $\mu(\ql,\qr)$. We then verify \cref{cond:phase2} for a subset of instances, leading to \cref{thm:equivalence}, which establishes the equivalence of $\mu(\ql,\qr)$ and $\mubar(\ql,\qr)$ over a range of parameters. However, to overcome the additional complexity in comparing different flexibility allocations, we need to leverage computer-aided proofs. We show that $\mubar(\ql,\qr)$ can be approximated to arbitrary precision by analyzing a system of non-linear equations that underpins this quantity. This enables us to compute $\mubar(\ql,\qr)$—and hence $\mu(\ql,\qr)$—at a provable level of accuracy and compare different allocations in \cref{thm:compare_q}. We then provide the proofs of all the results in Appendices \ref{app:phase_1}-\ref{app:auxiliary_proofs}.
}

 %Then, in the supercritical regime we leverage probabilistic results to establish upper bound and lower bounds for $\mu(B,0)$ and $\mu(B/2,B/2),$ proving that the balanced allocation can be optimal for a wide range of $B \in (0,1).$ 

% \cref{fig:global_model} illustrates the main results and proof schema of this section. 

% \begin{figure}[ht]
%     \centering
%     \includegraphics[width=1\textwidth]{Plots/global_model.png}
%     \caption{Gray boxes in the diagram highlight the main results for the global model, and the rest of the diagrams highlight the schema of the proofs \kzcomment{This will need to be updated when we settle with Section 3}}
%     \label{fig:global_model}
% \end{figure}

\subsubsection{KS Derivations.} \label{app:ks_derivation}

Throughout the section we fix an arbitrary $(\ql,\qr)$ and use $G$ as the shorthand notation for $G^{glb}_n(\ql,\qr)$. In $G$, the \textit{degree} of a node $v$ is the number of edges that are incident to $v$, and we denote this number by $deg(v)$. %We now define the KS algorithm in \cref{alg:ks}.

% \begin{definition}
%     In graph $G,$ 
%     \begin{enumerate}[label=(\roman*)]
%         \item An edge $e$ is said to be \textit{incident} to a node $v$ if $v$ is one of the endpoints of $e$;
%         \item The \textit{degree} of a node $v$ is the number of edges that are incident to $v$, and we denote this number by $deg(v)$;
%         \item A node $v$ is said to be \textit{isolated} if $deg(v) = 0.$
%     \end{enumerate}
% \end{definition}

Based on \cref{alg:ks}, the edges in $G'$ form a matching; our goal will be to characterize the size of this matching. When all edges incident to a node \( v \) are deleted from \( G \) and yet $v$ has degree $0$ in \( G' \), we know that $v$ will not be part of the resulting matching and call it an \textit{isolated} node. The key to finding the size of a matching based on the KS algorithm is to count the number of nodes that either become matched or isolated as edges are deleted from the graph. We denote the iterations before the first occurrence where no nodes have a degree of 1 in \( G \) as \textit{Phase 1} of the KS algorithm. The subsequent iterations are referred to as \textit{Phase 2} of the KS algorithm. {A key property of the KS algorithm is that it is optimal in its handling of degree-$1$ vertices: given an edge $e$ that is incident to a degree-$1$ vertex, there is always a maximum matching that contains $e$ \citep{bohman2011karp, balister2015controllability}. This result implies that the KS algorithm is optimal until the end of phase 1.} Let $M^l_1, M^r_1, M^l_2$ and $M^r_2$ respectively denote the set of nodes in $V_l$ and $V_r$ that enter the matching (i.e., incident to edges in $G'$) during phase 1 and 2. By symmetry, we know that $m_1 := |M^l_1| = |M^r_1|$ and $m_2 := |M^l_2| = |M^r_2|$. Similarly, let $\delete^l_1, \delete^r_1, \delete^l_2$ and $\delete^r_2$ respectively represent the set of nodes that become isolated in $V_l$ and $V_r$ during phase 1 and 2, where $\delete^l_1$ and $\delete^r_1$ also include the nodes that are already isolated initially in graph $G$. We define $\psi_1 := \max\{|\delete^l_1|,|\delete^r_1|\}$ and $$\psi_2 := n - m_1 - m_2 - \psi_1 = \min\bracket{n - |M^l_1| - |M^l_2| - |\delete^l_1|, n - |M^r_1| - |M^r_2| - |\delete^r_1|} =  \min\{|\delete^l_2|,|\delete^r_2|\}.$$ Intuitively, \( \psi_2 \) represents the number of nodes that become isolated in Phase 2 of the KS algorithm (excluding those already accounted for in Phase 1). It has been demonstrated, for some types of sparse random graph settings, that the expected number of nodes becoming isolated in Phase 2 of the KS algorithm is \( o(n) \), i.e., \( \EE{\psi_2} \in o(n) \). %The proof typically relies on showing the probability of isolation remains small along the trajectory of an evolving system of differential equations, and verifying such results along the trajectory can be tedious. 
We state this as \cref{cond:phase2}, which we later verify for some instances of our model.

\begin{condition}\label{cond:phase2}
    When the KS algorithm is applied to a random graph \( G \) of our model, \( \EE{\psi_2} \in o(n) \).
\end{condition}

\noindent {As the KS algorithm is optimal in Phase 1, \cref{cond:phase2} guarantees that it is asymptotically optimal. In particular, in Phase 2, the expected fraction of nodes that become isolated, and thus unmatched, is vanishingly small, i.e., it involves~$o(n)$ nodes.} Thus, to identify the number of unmatched nodes in both phases, it suffices to evaluate $\psi_1$, for which we evaluate the probability of a node becoming isolated in Phase 1 of the KS algorithm. For $\qvec = (\ql,\qr),$ we will show that the following set of equations determines this probability:

\begin{equation}\label{eq:w}
        \begin{split}
            \wonef{\qvec} =& e^{-2 \qr \af \parenthesis{1-\wtwohatf{\qvec}}-(1-\qr) \cdot (\af + \anf)\parenthesis{1-\wtwohatnf{\qvec}}},\\
            \wonenf{\qvec} =&e^{-\qr\cdot (\af + \anf) \parenthesis{1-\wtwohatf{\qvec}}-2(1-\qr)\anf \parenthesis{1-\wtwohatnf{\qvec}}},\\
            \wtwof{\qvec} =&1-e^{-2 \qr \af \wonehatf{\qvec}-(1-\qr)(\af+\anf) \wonehatnf{\qvec}},\\
            \wtwonf{\qvec} =&1-e^{-\qr(\af+\anf) \wonehatf{\qvec}-2(1-\qr)\anf \wonehatnf{\qvec}},\\
            \wonehatf{\qvec} =&e^{-2 \ql \af \parenthesis{1-\wtwof{\qvec}}-(1-\ql)(\af + \anf) \parenthesis{1-\wtwonf{\qvec}}},\\
            \wonehatnf{\qvec} =&e^{-\ql(\af+\anf) \parenthesis{1-\wtwof{\qvec}}-2(1-\ql)\anf \parenthesis{1-\wtwonf{\qvec}}},\\
            \wtwohatf{\qvec} =&1-e^{-2\ql\af \wonef{\qvec}-(1-\ql)(\af + \anf) \wonenf{\qvec}},\\
            \wtwohatnf{\qvec} =&1-e^{-\ql(\af+\anf) \wonef{\qvec}-2(1-\ql)\anf \wonenf{\qvec}}.
        \end{split}
\end{equation}

\dfedit{Such equations are common in KS-based analyses: whereas Karp and Sipser's original analysis focused on a homogeneous Erdős–Rényi graph and thus relied on just 2 of these equations, ours requires $8$ equations to characterize the probability for flexible or regular nodes on either side to be isolated or not.} We denote the smallest set of solutions\footnote{In line with terminology in \citet{karp1981maximum}, the smallest set of solutions refers to the least fixed point of the system of equations in \eqref{eq:w}. Note that this is well defined since all variables in $\mathbf{w}$ are increasing functions of each other.} %\dfcomment{footnote needs fixing: ``the least fixed point'' is poor english }
$$\mathbf{w} = \parenthesis{\wonef{\qvec}, \wonenf{\qvec}, \wtwof{\qvec}, \wtwonf{\qvec}, \wonehatf{\qvec}, \wonehatnf{\qvec}, \wtwohatf{\qvec}, \wtwohatnf{\qvec}}$$ to \eqref{eq:w} by $\mathbf{y} = \parenthesis{\yonef{\qvec}, \yonenf{\qvec}, \ytwof{\qvec}, \ytwonf{\qvec}, \yonehatf{\qvec}, \yonehatnf{\qvec}, \ytwohatf{\qvec}, \ytwohatnf{\qvec}}.$

\begin{theorem}\label{thm:global_matching}
    Let
    \begin{equation}\label{eq:xi}
        \begin{split}
            \xi(\ql,\qr) = 2&-\ql \yonef{\qvec} -\qr\parenthesis{1-\ytwohatf{\qvec}}\\
            &-\qr\parenthesis{1-\ytwohatf{\qvec}}\parenthesis{2\ql\af \yonef{\qvec}+(1-\ql)(\af + \anf) \yonenf{\qvec}} \\
            &- (1-\ql) \yonenf{\qvec}-(1-\qr)\parenthesis{1-\ytwohatnf{\qvec}}\\
            &-(1-\qr) \parenthesis{1-\ytwohatnf{\qvec}}\parenthesis{\ql(\af+\anf) \yonef{\qvec}+2(1-\ql)\anf \yonenf{\qvec}},
        \end{split}
    \end{equation}
    \begin{equation}\label{eq:xi_hat}
        \begin{split}
            \hat{\xi}(\ql,\qr) =2&-\qr \yonehatf{\qvec} -\ql\parenthesis{1-\ytwof{\qvec}}\\
            &-\ql\parenthesis{1-\ytwof{\qvec}}\parenthesis{2 \qr \af \yonehatf{\qvec}+(1-\qr)(\af+\anf) \yonehatnf{\qvec}}\\
            &- (1-\qr) \yonehatnf{\qvec}-(1-\ql)\parenthesis{1-\ytwonf{\qvec}}\\
            &- (1-\ql) \parenthesis{1-\ytwonf{\qvec}}\parenthesis{\qr(\af+\anf) \yonehatf{\qvec}+2(1-\qr)\anf \yonehatnf{\qvec}}.
        \end{split}
    \end{equation}
    Define $\mubar(\ql,\qr) = \min\parenthesis{\xi(\ql,\qr),\hat{\xi}(\ql,\qr)}.$ Then, under \cref{cond:phase2}, $\mu(\ql,\qr) = \mubar(\ql,\qr).$
\end{theorem}

We now loosen \cref{cond:phase2} into a sufficient (not necessary) condition that is easier to verify.

\begin{lemma}\label{lem:verify_condition} \cref{cond:phase2} holds when the solution to \eqref{eq:w} is unique.
\end{lemma}

\dfedit{\cref{lem:verify_condition} is instrumental for \cref{thm:equivalence} below. It states that, when $B = 1$, $\mu(\ql,\qr)$ equals $\mubar(\ql,\qr)$ for the one-sided and balanced allocations in almost all of the \textit{subcritical regime}, classically defined \citep{aronson1998maximum} as the setting where the average expected degree of a node, $\af + \anf$ in our case, is smaller than Euler's number $e$. 

\begin{theorem}\label{thm:equivalence}
    When $10^{-4} < \anf < \af, \af + \anf < e,$ and $\qvec = (1,0)$ or $(1/2,1/2)$, $\mu(\ql,\qr) =~\mubar(\ql,\qr).$
\end{theorem}

Though \cref{thm:equivalence} characterizes a region in which $\mu(\cdot,\cdot) = \mubar(\cdot,\cdot)$, this does not suffice to make formal comparisons between $\mu(1,0)$ and $\mu(1/2,1/2)$; since there are no closed-form solutions to these nonlinear equations, we need to solve \eqref{eq:w} to provable numerical precision for the region specified by the theorem ({see \eqref{eq:solution_bounds} in Appendix \ref{app:compare_q2}}). This then allows us to compare $\mu(1,0)$ and $\mu(1/2,1/2)$ for these values of $\af$ and $\anf$. Moreover, we derive a continuity property of $\mubar$ in $\af$ and $\anf$, that lets us construct local lower bounds for $\mu(1,0) - \mu(1/2,1/2)$ ({see \eqref{eq:lb_one_sided} and \eqref{eq:ub_balanced}}). We conclude by verifying in a computer-aided proof that these lower bounds exceed~$0$ across the parameters specified in the following theorem:

\begin{theorem}\label{thm:compare_q}
    $\mu(1,0) > \mu(1/2,1/2)$ when (i) $\af + \anf < e$ and (ii) $10^{-4} < \anf< 0.77 \af - 0.16$.\footnote{The boundary in condition (ii) arises from the ability for a computer-aided proof to verify the inequality within a reasonable runtime: for $\delta > 0$, we construct and compute a lower bound the value of $\mu(1,0) - \mu(1/2,1/2)$ within each set of $[\af,\af + \delta) \times [\anf,\anf + \delta)$ within the subcritical regime. Taking $\delta = 0.001$ yields the boundary in \cref{thm:compare_q} (ii) and runs in about $20$ hours.}
\end{theorem}

\cref{thm:compare_q} allows us to prove the dominance of the one-sided allocation for a wider set of parameters in which $\af + \anf$ remains relatively small (recall that \cref{thm:compare_1} allows for arbitrarily large $\af$ but requires $\anf=0$). Beyond the subcritical regime, KS-style analyses have fundamental limitations for two reasons: first, the asymptotic optimality of KS is not known for bipartite graphs beyond the subcritical regime (see \citet{Bollobas1995, mastin2013greedy}); and secondly, our computer-aided comparison of different flexibility allocations requires the nonlinear equations to have a unique set of solutions, which is known \citep[Lemma 1]{karp1981maximum} to require the subcritical regime.

\begin{remark}
An alternative approach to characterizing the expected maximum matching size $\mu$ is based on the analysis of a Monomer--Dimer model in the zero-temperature limit $z\to0$, which leads to recursive distributional equations (RDEs) \citep{bordenave2010rank, bordenave2013matchings}. In this framework, there is a one-to-one correspondence between solutions to the RDEs and the ``historical records'' of a suitably defined function $F(t)$. The value of $\mu$ can then be expressed explicitly in terms of $\max_{t\in[0,1]}F(t)$ (see Theorems~2 and~3 of \citet{bordenave2013matchings}), thereby extending KS-style results to the supercritical regime. However, explicitly computing $\mu$ via this approach requires analyzing the first-order condition of $F(t)$, which admits multiple solutions in the supercritical regime; only the smallest of these solutions corresponds to the global maximizer of $F(t)$. This procedure recovers the KS-style formulas for $G(n,c/n)$ (see \citet{bordenave2013matchings}), but at the cost of substantial analytical complexity. For the purpose of explicitly computing $\mu$ in our setting, we thus focus on the KS-style equations rather than the RDE-based approach.
\end{remark}

The proofs of the KS-based results are structured as follows:}

\begin{itemize}
    \item In Appendix \ref{app:phase_1} we prove \cref{thm:global_matching}, \cref{lem:verify_condition}, and \cref{thm:equivalence}. This requires us to first state the definitions and auxiliary results commonly associated with KS-style analyses, and we then prove each of these three results.
    \item In Appendix \ref{app:compare_q2} we provide our computer-aided proof of \cref{thm:compare_q}. Our proof partitions the region of interest into small cells and numerically derives lower bounds on $\mu(1,0)$ and upper bounds on $\mu(1/2,1/2)$ across each such cell. For each cell, we identify a particular point $(\anf,\af)$ at which we numerically solve \eqref{eq:w} for both $\qvec=(1,0)$ and $\qvec=(1/2,1/2)$ within a given tolerance $\epsilon$ (see \cref{cl:nlsolve}). Crucially, that tolerance guarantees that the solution to  \eqref{eq:w} is within $\epsilon$ of the true solution; we analytically translate this bound into a bound on the gap between the numerically computed value and $\mubar$. We also show that the smallest solution to \eqref{eq:w} is continuous in $\anf$ and $\af$, which implies that the bound holds, with an additional error term, within a $\delta$-neighborhood of $(\anf,\af)$, which represents the cell that $(\anf,\af)$ is part of (\cref{cl:continuity}). We iterate over cells to verify for the entire region that our lower bounds on $\mu(1,0)$ are greater than our upper bounds on $\mu(1/2,1/2)$. 
    \item {In Appendix \ref{app:convex_concave} we provide the computer-aided proof of \cref{thm:convex_concave}. We take directional second-order derivatives (SOD) of $\mubar(\ql,\qr)$ and evaluate them at $\qvec = (1/2,1/2)$ to prove concavity and convexity results in the respective directions. In particular, the SODs depend only on $\anf, \af$, and solutions to \eqref{eq:w}. Then, similar to the approach in Appendix \ref{app:phase_1}, we lower and upper bound the solution to \eqref{eq:w} within every small cell and iterate over cells to verify the signs of the directional SODs.}
    \item {In Appendix \ref{app:auxiliary_proofs} we provide the proof of the auxiliary results in Appendix \ref{app:phase_1}.}
    % \item In Appendix \ref{sec:num_construction} we explain the setup of the computational results in Section \ref{sec:ks} based on $\munum(\ql,\qr)$, as well as the range of parameters that we experiment with.
    %(i) we use our characterizations of $\mubar(\ql,\qr)$ to provide lower bounds on $\mu(1,0)$ and upper bounds on $\mubar(1/2,1/2)$
\end{itemize}

\subsubsection{Phase 1 of the KS Algorithm}\label{app:phase_1} In this section, we analyze phase 1 of the KS algorithm for $G^{glb}_n(\ql,\qr)$, hereafter referred to as $G$ for notational simplicity. The set of edges in $G$ is denoted by $E$, and the set of nodes is denoted by $V := V_l \cup V_r$. Our analysis extends the results for sparse random graphs presented in \citet{karp1981maximum} to random bipartite graphs. Similar to \citet{balister2015controllability}, we analyze bipartite graphs in which the degree distributions for nodes are heterogeneous; however, the ``configuration model" considered in their paper does not capture our setting with flexible and regular nodes, and we require different probabilistic computations to handle the heterogeneous edge probabilities between nodes of different flexibility types (i.e., $2 \anf$ between two regular nodes, $\af + \anf$ between a flexible and a regular node, and $2 \af$ between two flexible nodes). We next present all auxiliary results needed for \cref{thm:global_matching}, \cref{lem:verify_condition} and \cref{thm:equivalence}, the three main technical results based on KS-style analyses.

We begin by introducing the concept of a \textit{derivation}, which is essential for computing the asymptotic size of a maximum matching. {We shall show that nodes that appear in a derivation become either matched or isolated in Phase 1 of the KS algorithm. Moreover, depending on their positions in the derivation, nodes can be classified as either a \textit{target} or a \textit{loser}, which determines the number of nodes that become matched or isolated in Phase 1.}

\begin{definition}\label{def:derivation}
    A \textit{derivation} is a sequence $a_1,b_1,a_2,b_2,...,$ of distinct nodes such that, for $i = 1,2,...:$
    \begin{enumerate}[label=(\arabic*)]
        \item $\bracket{a_i,b_i} \in E$;
        \item $\bracket{a_i,b} \in E$ implies $b \in \bracket{b_1,b_2,...,b_i}.$
    \end{enumerate}
\end{definition}

For example, the sequence $a_1,b_1,a_2,b_2,a_3$ in \cref{fig:derivation} (a) is a derivation: node $a_1$ fulfills condition (1) because $\bracket{a_1,b_1} \in E$, and fulfills condition (2) because it is connected to no other node. Next, we verify that $a_2$ fulfills condition (1) because $\bracket{a_2,b_2} \in E$, and fulfills condition (2) because it is only connected to $b_1$ and $b_2.$ The sequence ends with node $a_3,$ which is again only connected to $b_1$ and $b_2.$ On the other hand, in \cref{fig:derivation} (b) there is no derivation involving the nodes $\{a_1,a_2,b_1,b_2\}$: \dfedit{condition (2) implies that a node cannot appear as $a_i$ unless all but one of its neighbors have already appeared in the derivation; in \cref{fig:derivation} (b), for $a_1$ and $a_2$ there exists no ordering of $b_1$ and $b_2$ such that this is true (and vice versa for $b_1$ and $b_2$).}

%As illustrated in \cref{fig:derivation}, it can be verified that the sequence $a_1,b_1,a_2,b_2,a_3$ is a derivation in \cref{fig:derivation} (a), whereas there is no derivation in \cref{fig:derivation} (b). 
%In particular, we will demonstrate 
In the upcoming proofs we will demonstrate that by following the KS algorithm one can optimally match nodes that appear in a derivation by starting with nodes of degree $1$ and then iteratively resolving the remaining nodes. 

\begin{figure}%
    \centering
    \subfloat[\label{fig:derivation1}\centering]{{\includegraphics[width=0.2\textwidth]{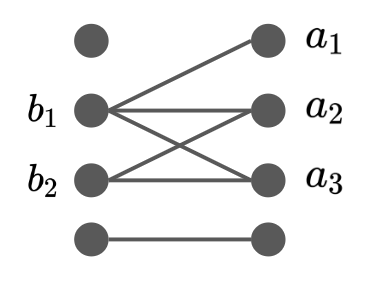} }}%
    \qquad
    \subfloat[\label{fig:derivation2}\centering]{{\includegraphics[width=0.2\textwidth]{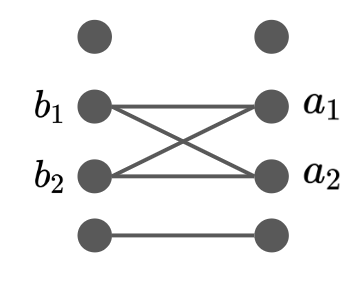} }}%
    \caption{{In \cref{fig:derivation} (a), the sequence $a_1,b_1,a_2,b_2,a_3$ is a derivation. Notice that all nodes in this derivation are either matched or isolated in phase $1$ of the KS algorithm: the edge $(a_1, b_1)$ is added to graph $G'$ as $a_1$ is the only node of degree $1$ within the connected component;  we then delete all edges incident to $a_1$ and $b_1$, leaving $(a_2, b_2)$ and $(b_2, a_3)$ as the only remaining edges in the graph; next, by adding $(a_2, b_2)$ into $G'$ (the case of $(b_2, a_3)$ is symmetric as the sequence $a_1,b_1,a_3,b_2,a_2$ is also a derivation), $a_1,b_1,a_2$ and $b_2$ become matched while $a_3$ becomes isolated. In contrast, there is no derivation in \cref{fig:derivation} (b) that involves the connected component $\{a_1, b_1, a_2, b_2\}$ and none of these nodes become matched or isolated during phase 1 of the KS algorithm.}}
    \label{fig:derivation}
\end{figure}

Within a derivation, we categorize nodes into \textit{target} and \textit{loser} based on the following definition:

\begin{definition}\label{def:target_loser}
    We define the following relation $\bigotimes \subseteq V \times V$: $v \bigotimes u$ if there exists a derivation $a_1,b_1,a_2,b_2,...$ and an index $i$ such that $v = a_i$ and $u = b_i.$ We call $u$ a \textit{target} if for some $v, v \bigotimes u,$ and we call $u$ a \textit{loser} if (1) for some $v, u \bigotimes v$ or (2) $u$ is the last element of an odd length derivation. 
\end{definition}

Based on \cref{def:target_loser}, all members of derivations are targets or losers or both. For instance, in the derivation $a_1,b_1,a_2,b_2,a_3$ in \cref{fig:derivation} (a), $a_1, a_2$ and $a_3$ are losers while $b_1$ and $b_2$ are targets. The next result characterizes the {nodes that are \textit{processed}, i.e., that become either matched or isolated in phase 1} of the KS algorithm. The result is an immediate application of Theorem $8$ in \citet{karp1981maximum} to bipartite graphs. We defer the proofs of \cref{prop:m1} and all auxiliary results in this section to Appendix \ref{app:auxiliary_proofs}.

\begin{proposition}[Theorem $8$ in \citet{karp1981maximum}]\label{prop:m1}
    Consider any execution of the KS algorithm on $G.$ Denote by $M_1$ the set of edges $(v,u)$ that are added to $G'$ in Phase $1.$ Then:
    \begin{enumerate}[label=(\roman*)]
        \item a node $v$ is {processed in phase 1} iff $v$ occurs in some derivation;
        \item if $u$ is a target then $M_1$ contains exactly one edge $(v,u)$ such that $v \bigotimes u;$
        \item if edge $(v,u) \in M_1$ then $v \bigotimes u$ or $u \bigotimes v;$
        \item if $v \bigotimes u$ and $u \bigotimes v$ then edge $(v,u) \in M_1$;
        \item $\psi_1 = \max\parenthesis{\left|\left\{v \in V_l \mid v \text{ is a loser}\right\}\right| - \left|\left\{v \in V_r \mid v \text{ is a target}\right\}\right|,\left|\left\{v \in V_r \mid v \text{ is a loser}\right\}\right| - \left|\left\{v \in V_l \mid v \text{ is a target}\right\}\right|}.$
    \end{enumerate}
\end{proposition}

Thus, the key to finding $\EE{\psi_1}$ and the asymptotic matching probability lies in determining the probability of a node $v$ being a target and/or a loser. We provide asymptotic answers to these questions by (1) conducting a probabilistic analysis of derivations in random trees and (2) demonstrating that a random tree is a good approximation to the structure obtained by selecting a node $v$ in $G$ and conducting a breadth-first search from $v$.

{We now construct a random tree $\bar{G}_n(\ql,\qr)$ to approximate the structure obtained from a breadth-first search from a node $v$ in $G$. As illustrated in \cref{fig:branching}, we construct the layers of the tree sequentially, mimicking a breadth-first search from the root node. The flexibility types of all nodes in a given layer are sampled according to the same distribution, and the distribution for each layer alternates between $\text{Bernoulli}(\ql)$ or $\text{Bernoulli}(\qr)$. This mimics the alternation between nodes in $V_l$ and $V_r$ in the bipartite graph $G$.} Specifically, the construction of $\bar{G}_n(\ql,\qr)$ follows a branching process: assume for simplicity that the flexibility type of the root node $v$ is drawn from the $\text{Bernoulli}(\ql)$ distribution, so that $v$ is a flexible node (i.e., $F_v = 1$) with probability $\ql$ and a regular node with probability $1-\ql$. Then, $v$ has $n$ potential children, each being a flexible node (i.e., $F_u = 1$) with probability $\qr$ and a regular node with probability $1-\qr$. A potential child becomes a realized child of $v$ with probability $2\pnf_n +(F_v + F_u) \cdot (\pf_n - \pnf_n)$. Each realized child $u$ then has $n$ potential children, with flexible and regular probabilities of $\ql$ and $1-\ql$, respectively. This branching process continues until no further child exists for a tree layer, a process that can be either finite or infinite. We omit the dependency on $n$ and $\qvec$ in $\bar{G}_n(\ql,\qr)$ whenever it is clear from the context.

\begin{figure}[ht]
    \centering
    \includegraphics[width=0.5\textwidth]{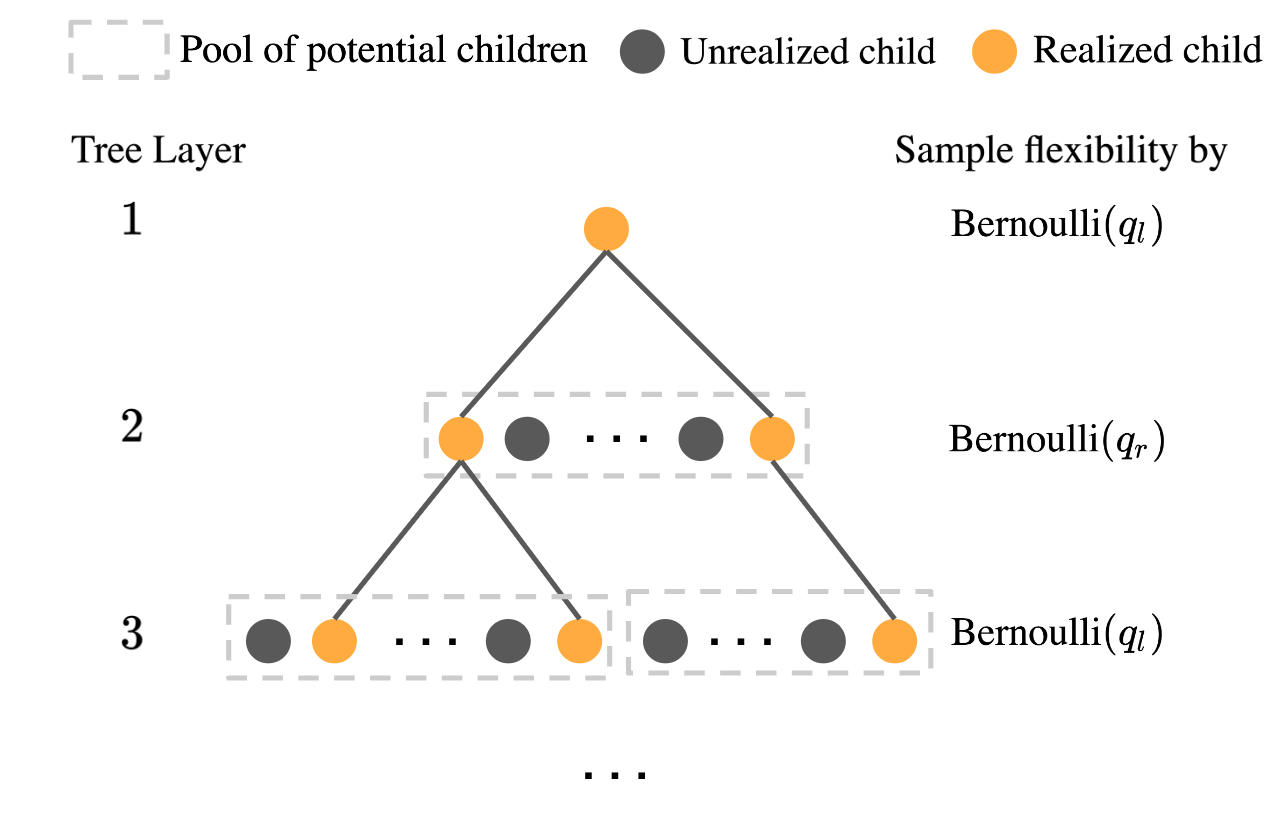}
    \caption{Illustration of the branching process for $\bar{G}$}
    \label{fig:branching}
\end{figure}
\vspace{-.15in}

To analyze the structure of $\bar{G}$ and connect it to $G$, we define two subsets of nodes, $L$ and $H$, through the procedure outlined in \cref{alg:lonely_popular}. The set $L_d$ contains all nodes added into set $L$ in the first $d$ repetitions of line 5 of Algorithm \ref{alg:lonely_popular}; similarly, we denote by $H_d$ the nodes added into set $H$ in the first $d$ repetitions of line 4 of Algorithm \ref{alg:lonely_popular}. Since $L$ contain all leaves of $\bar{G}$ and all other nodes in $\bar{G}$ have at least one child, every node in $\bar{G}$ is added into either $H$ or~$L.$ The classification of nodes into sets $L$ and $H$ is crucial for our study because it determines whether a node $v$ is a target, a loser, or both, as stated in \cref{lem:target_loser}.

\begin{algorithm}
\caption{Classification of Nodes in \tree}\label{alg:lonely_popular}
\begin{algorithmic}[1]
\State \textbf{Input:} A random tree \tree rooted at $v$.
\State \textbf{Initialize:} \lonely = $\{\text{Nodes in \tree with no children in \tree}\}$, \popular = $\emptyset$
\Repeat
    \State Add to \popular\; those nodes that have at least one child in \lonely.
    \State Add to \lonely\; those nodes that have only children in \popular.
\Until{No new nodes are added to either set}
\end{algorithmic}
\end{algorithm}

\vspace{-.15in}

\begin{lemma}[Lemma $3$ in \citet{karp1981maximum}]\label{lem:target_loser}
    Let \tree be a random tree rooted at $v.$
    \begin{enumerate}[label=(\roman*)]
        \item $v$ is a target iff $v$ is in $H$;
        \item $v$ is a loser iff either $v$ is in $L$ or $v$ has exactly $1$ child which is not in $H$.
    \end{enumerate}
\end{lemma}

Given that the nodes across different tree layers exhibit heterogeneity, we define two sets of nodes: the set $\setl$ contains all nodes sampled from layers with $\text{Bernoulli}(\ql)$ and the set $\setr$ contains all nodes from layers with $\text{Bernoulli}(\qr)$. For a flexible node in $\setl$, we denote the probabilities of it being in $L$ and $H$ as $\yonef{\qvec}$ and $\ytwof{\qvec}$, respectively. Similarly, for a regular node in $\setl$, we denote the probabilities as $\yonenf{\qvec}$ and $\ytwonf{\qvec}$. 
 We use an additional hat symbol to denote the counterparts of these probabilities in $\setr$, a notation consistently applied throughout this paper to differentiate quantities associated with $\setr$ from those associated with $\setl$. In \cref{lem:w}, we state that the vector $$\mathbf{y} = \parenthesis{\yonef{\qvec}, \yonenf{\qvec}, \ytwof{\qvec}, \ytwonf{\qvec}, \yonehatf{\qvec}, \yonehatnf{\qvec}, \ytwohatf{\qvec}, \ytwohatnf{\qvec}}$$ can be computed as the smallest set of solutions to the equations in \eqref{eq:w}.

\begin{lemma}\label{lem:w}
    As $n \rightarrow \infty$ the probabilities encoded in $\mathbf{y}$ converge to the smallest solution to \eqref{eq:w}.
\end{lemma}

Now, combining \cref{lem:target_loser} with \cref{lem:w}, we obtain the probability for nodes in \tree to be targets or losers.

{
\begin{lemma}\label{lem:target_loser2}
    Let \tree be a random tree rooted at $v.$ Then, as $n \rightarrow \infty$,
    \begin{enumerate}[label=(\roman*)]
        \item if $v$ is a flexible node in $\setl$, $v$ is a target with probability $\ytwof{\qvec}$ and a loser with probability $$\yonef{\qvec} + \yonef{\qvec}\parenthesis{(1+\qr) \af + (1-\qr) \anf -\qr 2\af \ytwohatf{\qvec} - \parenthesis{1-\qr} \parenthesis{\af + \anf} \ytwohatnf{\qvec}}.$$
        \item if $v$ is a regular node in $\setl$, $v$ is a target with probability $\ytwonf{\qvec}$ and a loser with probability $$\yonenf{\qvec} + \yonenf{\qvec}\parenthesis{\qr \af + (2-\qr) \anf -\qr \parenthesis{\af + \anf} \ytwohatf{\qvec} - \parenthesis{1-\qr} 2 \anf \ytwohatnf{\qvec}}.$$
        \item if $v$ is a flexible node in $\setr$, $v$ is a target with probability $\ytwohatf{\qvec}$ and a loser with probability $$\yonehatf{\qvec} + \yonehatf{\qvec}\parenthesis{(1+\ql) \af + (1-\ql) \anf -\ql 2\af \ytwof{\qvec} - \parenthesis{1-\ql} \parenthesis{\af + \anf} \ytwonf{\qvec}}.$$
        \item if $v$ is a regular node in $\setr$, $v$ is a target with probability $\ytwohatnf{\qvec}$ and a loser with probability $$\yonehatnf{\qvec} + \yonehatnf{\qvec}\parenthesis{\ql \af + (2-\ql) \anf -\ql \parenthesis{\af + \anf} \ytwof{\qvec} - \parenthesis{1-\ql} 2 \anf \ytwonf{\qvec}}.$$
    \end{enumerate}
\end{lemma}}

Equipped with \cref{lem:target_loser2}, we are ready to compute the probability that $v$ appears in a derivation as $n \rightarrow \infty.$

\begin{proposition}[Extension of Theorem $9$ (4) in \citet{karp1981maximum}]\label{prop:derivation}
    Let $v$ be a random node in $G$. Then, as $n \rightarrow \infty$:
    \begin{align*}
        \lim_{n \rightarrow \infty} &\PP{v \text{ in a derivation} \mid v \text{ is a flexible node in } V_l}\\
        = &\ytwof{\qvec} + \yonef{\qvec} + \yonef{\qvec} \squarebracket{(1+\qr) \af + (1-\qr) \anf}\cdot \\
        &\squarebracket{1-\frac{\qr 2 \af}{(1+\qr) \af + (1-\qr) \anf} \parenthesis{\yonehatf{\qvec} + \ytwohatf{\qvec}}-\frac{(1-\qr) \parenthesis{\af + \anf}}{(1+\qr) \af + (1-\qr) \anf} \parenthesis{\yonehatnf{\qvec} + \ytwohatnf{\qvec}}},\\
        \lim_{n \rightarrow \infty} &\PP{v \text{ in a derivation} \mid v \text{ is a regular node in } V_l}\\
        = &\ytwonf{\qvec} + \yonenf{\qvec} + \yonenf{\qvec} \squarebracket{\qr \af + (2-\qr) \anf}\cdot \\
        &\squarebracket{1-\frac{\qr \parenthesis{\af + \anf}}{\qr \af + (2-\qr) \anf} \parenthesis{\yonehatf{\qvec} + \ytwohatf{\qvec}}-\frac{(1-\qr) 2\anf}{\qr \af + (2-\qr) \anf} \parenthesis{\yonehatnf{\qvec} + \ytwohatnf{\qvec}}},\\
        \lim_{n \rightarrow \infty} &\PP{v \text{ in a derivation} \mid v \text{ is a flexible node in } V_r}\\
        = &\ytwohatf{\qvec} + \yonehatf{\qvec} + \yonehatf{\qvec} \squarebracket{(1+\ql) \af + (1-\ql) \anf}\cdot \\
        &\squarebracket{1-\frac{\ql 2 \af}{(1+\ql) \af + (1-\ql) \anf} \parenthesis{\yonef{\qvec} + \ytwof{\qvec}}-\frac{(1-\ql) \parenthesis{\af + \anf}}{(1+\ql) \af + (1-\ql) \anf} \parenthesis{\yonenf{\qvec} + \ytwonf{\qvec}}},\\
        \lim_{n \rightarrow \infty} &\PP{v \text{ in a derivation} \mid v \text{ is a regular node in } V_r}\\
        = &\ytwohatnf{\qvec} + \yonehatnf{\qvec} + \yonehatnf{\qvec} \squarebracket{\ql \af + (2-\ql) \anf}\cdot \\
        &\squarebracket{1-\frac{\ql \parenthesis{\af + \anf}}{\ql \af + (2-\ql) \anf} \parenthesis{\yonef{\qvec} + \ytwof{\qvec}}-\frac{(1-\ql) 2\anf}{\ql \af + (2-\ql) \anf} \parenthesis{\yonenf{\qvec} + \ytwonf{\qvec}}}.\\
    \end{align*}
\end{proposition}

Now, equipped with the auxiliary results, we are ready to prove \cref{thm:global_matching}, \cref{lem:verify_condition} and \cref{thm:equivalence}. 

\emph{Proof of \cref{thm:global_matching}} %Since the KS algorithm trivially provides a lower bound for $\mu(\ql,\qr),$ here we continue to show that under \cref{cond:phase2} the algorithm is provably optimal. By \cref{cond:phase2}, $\EE{\psi_2} \in o(n)$ and the algorithm is optimal (up to an error of $o(n)$) in phase 2. Since we also know that the algorithm is provably optimal until the end of phase 1, the KS algorithm is optimal under \cref{cond:phase2}. Specifically, by \cref{cond:phase2} we find
    Recall that the KS algorithm is asymptotically optimal under \cref{cond:phase2}, i.e., %. Specifically, by \cref{cond:phase2} we find
    \begin{align*}
        \mu(\ql,\qr) &= \lim_{n \to \infty} \frac{\EE{\mathcal{M}_n(\ql,\qr)}}{n} = \lim_{n \to \infty} \frac{\EE{|M_1^l| + |M_2^l|}}{n} = \lim_{n \to \infty} \frac{\EE{n - \psi_1 - \psi_2}}{n}\\
        &=1-\lim_{n \to \infty} \frac{\EE{\psi_1}}{n} - \lim_{n \to \infty} \frac{\EE{\psi_2}}{n} = 1-\lim_{n \to \infty} \frac{\EE{\psi_1}}{n}
    \end{align*}
    provided that these limits exist. From \cref{prop:m1} (v), we know
    \begin{align*}
        \mu(\ql,\qr) = 1-\lim_{n \to \infty} \frac{\EE{\psi_1}}{n} =&\; 1-\lim_{n \to \infty} \max\bracket{\PP{v \in \delete^l_1|v \in V_l}, \PP{v \in \delete^r_1|v \in V_r}}\\
        =&\; 1-\max\Bigl\{\lim_{n \rightarrow \infty} \PP{v \text{ is a loser} \mid v \in V_l} - \lim_{n \rightarrow \infty} \PP{v \text{ is a target} \mid v \in V_r},\\
        &\lim_{n \rightarrow \infty} \PP{v \text{ is a loser} \mid v \in V_r} - \lim_{n \rightarrow \infty} \PP{v \text{ is a target} \mid v \in V_l}\Bigr\}
    \end{align*}
    provided that these limits exist.
    By the law of iterated expectations,  provided that these limits exist, we have 
    \begin{align*}
        &\lim_{n \rightarrow \infty} \PP{v \text{ is a loser} \mid v \in V_l} - \lim_{n \rightarrow \infty} \PP{v \text{ is a target} \mid v \in V_r}\\
        =&\;\lim_{n \rightarrow \infty} \PP{v \text{ is a loser} \mid v \text{ is a flexible node in } V_l} \cdot \PP{v \in v_l \text{ is a flexible node}} \\
        &+ \lim_{n \rightarrow \infty} \PP{v \text{ is a loser} \mid v \text{ is a regular node in } V_l} \cdot \PP{v \in v_l \text{ is a regular node}}\\ 
        &- \lim_{n \rightarrow \infty} \PP{v \text{ is a target} \mid v \text{ is a flexible node in } V_r} \cdot \PP{v \in v_r \text{ is a flexible node}}\\
        &-\lim_{n \rightarrow \infty} \PP{v \text{ is a target} \mid v \text{ is a regular node in } V_r} \cdot \PP{v \in v_r \text{ is a regular node}}.
    \end{align*}

    Now we can plug in probabilities derived in \cref{lem:target_loser2}, which we have shown in \cref{cl:limit_exchange} to be equal to the corresponding probabilities in random graphs:
    \begin{align*}
        &\lim_{n \rightarrow \infty} \PP{v \text{ is a loser} \mid v \in V_l} - \lim_{n \rightarrow \infty} \PP{v \text{ is a target} \mid v \in V_r}\\
        = & \ql \parenthesis{\yonef{\qvec} + \yonef{\qvec}\parenthesis{(1+\qr) \af + (1-\qr) \anf -\qr 2\af \ytwohatf{\qvec} - \parenthesis{1-\qr} \parenthesis{\af + \anf} \ytwohatnf{\qvec}}} - \qr \ytwohatf{\qvec}\\
        & +(1-\ql)\parenthesis{\yonenf{\qvec} + \yonenf{\qvec}\parenthesis{\qr \af + (2-\qr) \anf -\qr \parenthesis{\af + \anf} \ytwohatf{\qvec} - \parenthesis{1-\qr} 2 \anf \ytwohatnf{\qvec}}} - (1-\qr) \ytwohatnf{\qvec}\\
        = & \ql \parenthesis{\yonef{\qvec} + \yonef{\qvec}\parenthesis{\qr 2\af \parenthesis{1-\ytwohatf{\qvec}} + \parenthesis{1-\qr} \parenthesis{\af + \anf} \parenthesis{1-\ytwohatnf{\qvec}}}} -1+ \qr \parenthesis{1-\ytwohatf{\qvec}}\\
        & +(1-\ql)\parenthesis{\yonenf{\qvec} + \yonenf{\qvec}\parenthesis{\qr \parenthesis{\af + \anf} \parenthesis{1-\ytwohatf{\qvec}} + \parenthesis{1-\qr} 2 \anf \parenthesis{1-\ytwohatnf{\qvec}}}} + (1-\qr) \parenthesis{1-\ytwohatnf{\qvec}}\\
        =&\; \ql \yonef{\qvec}+\qr\parenthesis{1-\ytwohatf{\qvec}}\parenthesis{2\ql\af \yonef{\qvec}+(1-\ql)(\af + \anf) \yonenf{\qvec}} -1+ \qr \parenthesis{1-\ytwohatf{\qvec}} \\
        &+(1-\ql) \yonenf{\qvec}+(1-\qr) \parenthesis{1-\ytwohatnf{\qvec}}\parenthesis{\ql(\af+\anf) \yonef{\qvec}+2(1-\ql)\anf \yonenf{\qvec}}+ (1-\qr) \parenthesis{1-\ytwohatnf{\qvec}}\\
        =:& 1-\xi(\ql,\qr).
    \end{align*}

    Similarly, we find that
    \begin{align*}
        &\lim_{n \rightarrow \infty} \PP{v \text{ is a loser} \mid v \in V_r} - \lim_{n \rightarrow \infty} \PP{v \text{ is a target} \mid v \in V_l}\\
        =&\; \qr \yonehatf{\qvec} + \ql\parenthesis{1-\ytwof{\qvec}}\parenthesis{2 \qr \af \yonehatf{\qvec}+(1-\qr)(\af+\anf) \yonehatnf{\qvec}} -1 + \ql\parenthesis{1-\ytwof{\qvec}}\\
        &+(1-\qr) \yonehatnf{\qvec}+(1-\ql) \parenthesis{1-\ytwonf{\qvec}}\parenthesis{\qr(\af+\anf) \yonehatf{\qvec}+2(1-\qr)\anf \yonehatnf{\qvec}}+(1-\ql)\parenthesis{1-\ytwonf{\qvec}}\\
        =:& 1-\hat{\xi}(\ql,\qr).
    \end{align*}

    {Thus, under \cref{cond:phase2}
    \begin{align*}
        \mu(\ql,\qr) =&\; 1-\max\Bigl\{\lim_{n \rightarrow \infty} \PP{v \text{ is a loser} \mid v \in V_l} - \lim_{n \rightarrow \infty} \PP{v \text{ is a target} \mid v \in V_r},\\
        &\lim_{n \rightarrow \infty} \PP{v \text{ is a loser} \mid v \in V_r} - \lim_{n \rightarrow \infty} \PP{v \text{ is a target} \mid v \in V_l}\Bigr\}\\
        =&\; 1-\max\parenthesis{1-\xi(\ql,\qr),1-\hat{\xi}(\ql,\qr)}=\min\parenthesis{\xi(\ql,\qr),\hat{\xi}(\ql,\qr)} := \mubar(\ql,\qr).\hfill\Halmos
    \end{align*}}
% \end{proof}

\emph{Proof of \cref{lem:verify_condition}.}
    From \cref{prop:m1} (i), it is known that $$\lim_{n \rightarrow \infty}\frac{\EE{m_1 + \psi_1}}{n} = \max\bracket{\lim_{n \rightarrow \infty}\PP{v \text{ in a derivation} \mid v \in V_l}, \lim_{n \rightarrow \infty}\PP{v \text{ in a derivation} \mid v \in V_r}}.$$ {We next demonstrate that if the solution to \eqref{eq:w} is unique, then 
    \begin{align}\label{eq:derivation_all}
        \lim_{n \rightarrow \infty}\PP{v \text{ in a derivation} \mid v \in V_l} = \lim_{n \rightarrow \infty}\PP{v \text{ in a derivation} \mid v \in V_r} = 1.
    \end{align}}

    To establish \eqref{eq:derivation_all}, we leverage \cref{prop:derivation} and consider a random node $v$ in $G$, which can be either flexible or regular and either in $V_l$ or in $V_r$.  be a flexible node in $V_l$, a regular node in $V_l$, a flexible node in $V_r$, or a regular node in $V_r$. We shall demonstrate that $\lim_{n \rightarrow \infty}\PP{v \text{ in a derivation}} = 1$ when the solution to \eqref{eq:w} is unique. 

    Note that for any solution vector $\mathbf{w}$ to \eqref{eq:w}, it is always feasible to construct
    \begin{equation}\label{eq:w12sumtoone}
        \begin{split}
            x_L^{f}(\ql,\qr) &= \wonef{\qvec}, x_H^{f}(\ql,\qr) = 1-\wonef{\qvec},\\
            x_L^{nf}(\ql,\qr) &= \wonenf{\qvec}, x_H^{nf}(\ql,\qr) = 1-\wonenf{\qvec},\\
            \hat{x}_L^{f}(\ql,\qr) &= \wonehatf{\qvec}, \hat{x}_H^{f}(\ql,\qr) = 1-\wonehatf{\qvec},\\
            \hat{x}_L^{nf}(\ql,\qr) &= \wonehatnf{\qvec}, \hat{x}_H^{nf}(\ql,\qr) = 1-\wonehatnf{\qvec},\\
        \end{split}
    \end{equation}
    so that $\mathbf{x}$ is provably a solution to \eqref{eq:w}.  Consequently, when \eqref{eq:w} admits a unique solution, the smallest set of solutions $\mathbf{y}$ must satisfy \eqref{eq:w12sumtoone}. We next substitute \eqref{eq:w12sumtoone} into the expressions derived in \cref{prop:derivation} and simplify to find that
    \begin{align*}
        \lim_{n \rightarrow \infty} &\PP{v \text{ in a derivation} \mid v \text{ is a flexible node in } V_l}\\
        = &\ytwof{\qvec} + \yonef{\qvec} + \yonef{\qvec} \squarebracket{(1+\qr) \af + (1-\qr) \anf}\cdot \\
        &\squarebracket{1-\frac{\qr 2 \af}{(1+\qr) \af + (1-\qr) \anf} \parenthesis{\yonehatf{\qvec} + \ytwohatf{\qvec}}-\frac{(1-\qr) \parenthesis{\af + \anf}}{(1+\qr) \af + (1-\qr) \anf} \parenthesis{\yonehatnf{\qvec} + \ytwohatnf{\qvec}}}\\
        =&\; 1+ \yonef{\qvec} \squarebracket{(1+\qr) \af + (1-\qr) \anf}\cdot\squarebracket{1-\frac{\qr 2 \af}{(1+\qr) \af + (1-\qr) \anf} -\frac{(1-\qr) \parenthesis{\af + \anf}}{(1+\qr) \af + (1-\qr) \anf}}\\
        =&\; 1+  \yonef{\qvec} \squarebracket{(1+\qr) \af + (1-\qr) \anf}\cdot 0
        \quad = 1.
    \end{align*}
    One can show analogously that nodes in $V_r$ or regular nodes in $V_l$ have a probability of 1, asymptotically, to be in a derivation. 
    Consequently, we can verify \cref{cond:phase2} via
    \begin{align*}
        &\lim_{n \rightarrow \infty}\frac{\EE{\psi_2}}{n} \leq 1- \lim_{n \rightarrow \infty}\frac{\EE{m_1 + \psi_1}}{n}\\
        =&\; 1- \max\bracket{\lim_{n \rightarrow \infty}\PP{v \text{ in a derivation} \mid v \in V_l}, \lim_{n \rightarrow \infty}\PP{v \text{ in a derivation} \mid v \in V_r}} = 0.\Halmos
    \end{align*}

\begin{proof}{Proof of \cref{thm:equivalence}}
    By \cref{thm:global_matching} and \cref{lem:verify_condition}, to establish \cref{thm:equivalence} it is sufficient to demonstrate that when $10^{-4} < \anf < \af$ and $\af + \anf < e$ the solution to \eqref{eq:w} is unique at  $\qvec\in\{ (1,0),(0,1),(1/2,1/2)\}$. 
    
    When $\qvec = (1,0)$, all nodes in $V_l$ are flexible nodes, while those in $V_r$ are regular nodes. Thus, it suffices to analyze $\wonef{\qvec}, \wtwof{\qvec}, \wonehatnf{\qvec}$ and $\wtwohatnf{\qvec}$. Then, \eqref{eq:w} reduces to 
    \begin{align}
        &\wonef{\qvec} = e^{-\parenthesis{\af + \anf} \parenthesis{1-\wtwohatnf{\qvec}}}, \wtwohatnf{\qvec} = 1-e^{-\parenthesis{\af + \anf} \wonef{\qvec}}, \label{eq:one_sided_reduction1}\\
        &\wonehatnf{\qvec} = e^{-\parenthesis{\af + \anf} \parenthesis{1-\wtwof{\qvec}}}, \wtwof{\qvec} = 1-e^{-\parenthesis{\af + \anf} \wonehatnf{\qvec}}. \label{eq:one_sided_reduction2}
    \end{align}

    Since \eqref{eq:one_sided_reduction1} and \eqref{eq:one_sided_reduction2} are equivalent, it suffices to show that solution to the pair $\parenthesis{\wonef{\qvec}, \wtwohatnf{\qvec}}$ in \eqref{eq:one_sided_reduction1} is unique. This is a direct application of the following result from \citet{karp1981maximum}, by taking $L = \wonef{\qvec}, W = \wtwohatnf{\qvec}$ and $\lambda = \af + \anf:$

    \begin{claim}[Lemma $1$ in \citet{karp1981maximum}]
        Define $L = e^{-\lambda (1-W)}, W = 1-e^{-\lambda L}.$ Then, $L + W \leq 1,$ with equality if and only if $\lambda \leq e.$
    \end{claim}
    The case for $\qvec = (0,1)$ is symmetric.

    For the case of $\qvec = (1/2, 1/2)$, symmetry implies that
    $$\wonef{\qvec} = \wonehatf{\qvec}, \wonenf{\qvec} = \wonehatnf{\qvec}, \wtwof{\qvec} = \wtwohatf{\qvec}, \wtwonf{\qvec} = \wtwohatnf{\qvec}.$$ Substituting $\qvec = (1/2,1/2)$ into \eqref{eq:w}, we obtain the following equations:
    \begin{equation}\label{eq:smallest_solution_converge}
        \begin{split}
            \wonef{\qvec} =&\; e^{-\frac{1}{2} 2\af \parenthesis{1-\wtwohatf{\qvec}}-\frac{1}{2}\parenthesis{\af + \anf} \parenthesis{1-\wtwohatnf{\qvec}}},\\
            \wonenf{\qvec} =&\; e^{-\frac{1}{2} \parenthesis{\af + \anf} \parenthesis{1-\wtwohatf{\qvec}}-\frac{1}{2} 2\anf \parenthesis{1-\wtwohatnf{\qvec}}},\\
            \wtwohatf{\qvec} =&\; 1-e^{-\frac{1}{2} 2\af \wonef{\qvec}-\frac{1}{2}\parenthesis{\af + \anf} \wonenf{\qvec}},\\
            \wtwohatnf{\qvec} =&\; 1-e^{-\frac{1}{2} \parenthesis{\af + \anf} \wonef{\qvec}-\frac{1}{2} 2\anf \wonenf{\qvec}}.\\
        \end{split}
    \end{equation}
    {We observe that values of $\wonef{\qvec}, \wonenf{\qvec}, \wtwohatf{\qvec}$ and $\wtwohatnf{\qvec}$ are trivially bounded between $0$ and $1,$ and all of the variables are increasing in each other. Thus, we initialize the values of $\parenthesis{\wonef{\qvec,0},\wonenf{\qvec,0},\wtwohatf{\qvec,0},\wtwohatnf{\qvec,0}} = (0,0,0,0)$ and define, for any $d \in \mathbb{Z}^+,$
    \begin{align*}
        \wonef{\qvec,d} =&\; e^{-\frac{1}{2} 2\af \parenthesis{1-\wtwohatf{\qvec,d-1}}-\frac{1}{2}\parenthesis{\af + \anf} \parenthesis{1-\wtwohatnf{\qvec,d-1}}},\\
        \wonenf{\qvec,d} =&\; e^{-\frac{1}{2} \parenthesis{\af + \anf} \parenthesis{1-\wtwohatf{\qvec,d-1}}-\frac{1}{2} 2\anf \parenthesis{1-\wtwohatnf{\qvec,d-1}}},\\
        \wtwohatf{\qvec,d} =&\; 1-e^{-\frac{1}{2} 2\af \wonef{\qvec,d-1}-\frac{1}{2}\parenthesis{\af + \anf} \wonenf{\qvec,d-1}},\\
        \wtwohatnf{\qvec,d} =&\; 1-e^{-\frac{1}{2} \parenthesis{\af + \anf} \wonef{\qvec,d-1}-\frac{1}{2} 2\anf \wonenf{\qvec,d-1}}.
    \end{align*}
    Then, the smallest set of solutions to \eqref{eq:smallest_solution_converge} is given by  $\lim_{d\to\infty}\parenthesis{\wonef{\qvec,d},\wonenf{\qvec,d},\wtwohatf{\qvec,d},\wtwohatnf{\qvec,d}}$.}
    
    For any $\vec{\mathbf{x}} \in \mathbb{R}^2,$ define
    \begin{align*}
        F(\vec{\mathbf{x}}) = 
        \begin{pmatrix}
        e^{-\frac{1}{2} 2 \af x_1 - \frac{1}{2} \parenthesis{\af +\anf} x_2} \\
        e^{-\frac{1}{2} \parenthesis{\af +\anf} x_1 - \frac{1}{2} 2 \anf x_2}
        \end{pmatrix}
    \end{align*}
    {and define $F^t(\vec{\mathbf{x}})$ the $t$th application of the function $F$ on $\vec{\mathbf{x}}.$ That is, $F^0(\vec{\mathbf{x}}) = \vec{\mathbf{x}}, F^1(\vec{\mathbf{x}}) = F(\vec{\mathbf{x}})$ and $F^2(\vec{\mathbf{x}}) = F\parenthesis{F(\vec{\mathbf{x}})}.$ Then, the smallest solutions to $\left( \begin{smallmatrix} \wonef{\qvec} \\ \wonenf{\qvec} \end{smallmatrix} \right)$ and $\vec{1} - \left( \begin{smallmatrix} \wtwohatf{\qvec} \\ \wtwohatnf{\qvec} \end{smallmatrix} \right)$ are respectively given by $\lim_{t \to \infty} F^{2t+1}(\vec{1})$ and $\lim_{t \to \infty} F^{2t}(\vec{1})$.} In particular, $\left( \begin{smallmatrix} \wonef{\qvec} \\ \wonenf{\qvec} \end{smallmatrix} \right) = \vec{1} - \left( \begin{smallmatrix} \wtwohatf{\qvec} \\ \wtwohatnf{\qvec} \end{smallmatrix} \right)$ and the solution is unique if $F(\vec{\mathbf{x}})$ has a unique fixed point, i.e., there exists a unique $\vec{\mathbf{x}^\star}$ such that $F\parenthesis{\vec{\mathbf{x}^\star}} = \vec{\mathbf{x}^\star}.$ Notice that in $F(\vec{\mathbf{x}}) = \vec{\mathbf{x}}$ we have $x_1 = e^{-\af x_1 - \frac{1}{2}\parenthesis{\af +\anf} x_2},$ so $x_2 = -2 \frac{\log(x_1) + \af x_1}{\af + \anf}.$ Plugging this into $x_2 = e^{-\frac{1}{2} \parenthesis{\af +\anf} x_1 - \frac{1}{2} 2 \anf x_2},$ we find that
    \begin{equation}\label{eq:balanced_reduction}
        -2 \frac{\log(x_1) + \af x_1}{\af + \anf} = e^{- \frac{1}{2}\parenthesis{\af +\anf} x_1 +2 \frac{\anf}{\af+\anf}\parenthesis{\log(x_1) + \af x_1}},
    \end{equation}
    so it suffices to show that \eqref{eq:balanced_reduction} has a unique solution when $\af + \anf < e.$  Let 
    \begin{equation}\label{eq:f1_def}
        f_1(x_1) := e^{- \frac{1}{2}\parenthesis{\af +\anf} x_1 +2 \frac{\anf}{\af+\anf}\parenthesis{\log(x_1) + \af x_1}} + 2 \frac{\log(x_1) + \af x_1}{\af + \anf}.
    \end{equation}
    The next result establishes a monotonicity property of this function:
    \begin{claim}\label{cl:uniqueness}
        When $10^{-4} < \anf < \af$ and $\af + \anf < e$, $f_1'(x_1) > 1$ for any $x_1 \in (0,1]$.
    \end{claim}
    
    Since $f_1(0) = -\infty$ and $f_1(1) \geq 0,$ by continuity of $f_1(x_1)$ with respect to $x_1$ we know that $f_1(x_1) = 0$ has at least one solution in $(0,1].$ Since we also know from \cref{cl:uniqueness} that $f_1(x_1)$ is strictly monotonically increasing with respect to $x_1$ in $(0,1]$ when $10^{-4} < \anf < \af$ and $\af + \anf < e$, such solution for $x_1$ is unique. This completes the proof of \cref{thm:equivalence}. 
\end{proof}

\subsubsection{Proof of \texorpdfstring{\cref{thm:compare_q}}{Lg}}\label{app:compare_q2}
    Recall from \cref{thm:equivalence} that, in the stated parameter regimes, $\mu(\ql,\qr) = \mubar(\ql,\qr)$ at $\qvec = (1,0)$ and $(1/2,1/2)$. 
    Since the solution to \eqref{eq:w} is unique at these points, \eqref{eq:w12sumtoone} is satisfied by the smallest set of solutions $\mathbf{y}$. Plugging $$\yonef{\qvec} = 1-\ytwof{\qvec}, \yonenf{\qvec} = 1-\ytwonf{\qvec}$$ into \eqref{eq:xi} and \eqref{eq:xi_hat}, we find that at these values of $\qvec$
    \begin{equation*}
        \begin{split}
            \mu(\ql,\qr) =&\; \xi(\ql,\qr) = \hat{\xi}(\ql,\qr)\\
            =&\; 2- \ql \yonef{\qvec} - \qr e^{-b_1(\ql,\qr)} \parenthesis{1+b_1(\ql,\qr)} - (1-\ql) \yonenf{\qvec} -(1-\qr) e^{-b_2(\ql,\qr)} \parenthesis{1+b_2(\ql,\qr)},
        \end{split}
    \end{equation*}
    \begin{align*}
    \text{where}\qquad    b_1(\ql,\qr) =&\; \ql 2 \af \yonef{\qvec} + (1-\ql) \parenthesis{\af + \anf} \yonenf{\qvec},\\
        b_2(\ql,\qr) =&\; \ql \parenthesis{\af + \anf} \yonef{\qvec} + (1-\ql) 2 \anf \yonenf{\qvec}.
    \end{align*}

    When $\qvec = (1,0),$ $\mu(\ql,\qr)$ depends only on $\yonef{\qvec}$, which can be solved as the unique solution $x^\star$ to
    \begin{align}\label{eq:sol_x}
        x = e^{-\parenthesis{\af + \anf} e^{-\parenthesis{\af + \anf}x}}.
    \end{align}
    
    When $\qvec = (1/2,1/2),$ $\mu(\ql,\qr)$ depends only on $\parenthesis{\yonef{\qvec},\yonenf{\qvec}}$, which can be solved as the unique set of solution $\parenthesis{x^\star_1,x^\star_2}$ to
    \begin{equation}\label{eq:sol_x12}
        \begin{split}
            x_1 =&\; e^{-\frac{1}{2} 2 \af x_1 - \frac{1}{2} \parenthesis{\af +\anf} x_2}, \\
            x_2 = & e^{-\frac{1}{2} \parenthesis{\af +\anf} x_1 - \frac{1}{2} 2 \anf x_2}.
        \end{split}
    \end{equation}

    Our objective is to show that $\mu(1,0) > \mu(1/2,1/2)$ for any $\af$ and $\anf$ satisfying $10^{-4} < \anf < \af$ and $\af + \anf < e$. 
    {To do so, we divide the parameter regions into small cells and prove the inequality by deriving bounds within each cell. We fix a constant $\delta > 0$ and derive for any $\anf, \af,$  a lower bound for the expression $\mu(1,0) - \mu(1/2,1/2)$ for any $(\bar{\alpha}^f, \bar{\anf}) \in \mathcal{I}(\anf, \af, \delta) := \bracket{(\bar{\anf},\bar{\af}): \bar{\anf} \in [\anf, \anf + \delta), \bar{\af} \in [\af, \af + \delta)}$.} We then employ a computer-aided proof to iterate over all cells in the claimed region and verify if this lower bound is positive for the respective cell. To do so, we need to bound $\bar{\alpha}^f, \bar{\anf}$ and the resulting $\bar{x}, \bar{x}_1$ and $\bar{x}_2$ in each cell. In $\mathcal{I}(\anf, \af, \delta)$, if we know the respective lower and upper bounds of $\bar{x}, \bar{x}_1$ and $\bar{x}_2$, which we denote by $\xlb, \xub, \xlb_1, \xub_1, \xlb_2$ and $\xub_2$, then we can lower bound 
    \begin{equation}\label{eq:lb_one_sided}
    \begin{split}
        \mu(1,0) =&\; 2 - \bar{x} - e^{-\parenthesis{\bar{\alpha}^f+\bar{\anf}} \bar{x}} \squarebracket{1+\parenthesis{\bar{\alpha}^f + \bar{\anf}} \bar{x}}\\
        \geq&\; 2 - \xub - e^{-\parenthesis{\af + \anf} \xlb} \squarebracket{1+\parenthesis{\af + \anf + 2 \delta} \xub}, \forall \parenthesis{\bar{\alpha}^f, \bar{\anf}} \in \mathcal{I}(\anf, \af, \delta).
    \end{split}
    \end{equation}
    Similarly, we can upper bound
    \begin{equation}\label{eq:ub_balanced}
        \begin{split}
            \mu(1/2,1/2) =&\; 2 - \frac{1}{2}\bar{x}_1 - \frac{1}{2} \bar{x}_2 - \frac{1}{2} e^{-\parenthesis{\bar{\alpha}^f}\bar{x}_1-\frac{1}{2}\parenthesis{\bar{\alpha}^f + \bar{\anf}} \bar{x}_2} \squarebracket{1+\bar{\alpha}^f \bar{x}_1 + \frac{1}{2}\parenthesis{\bar{\alpha}^f + \bar{\anf}} \bar{x}_2}\\
            &-\frac{1}{2} e^{-\frac{1}{2}\parenthesis{\bar{\alpha}^f + \bar{\anf}} \bar{x}_1-\parenthesis{\bar{\anf}}\bar{x}_2} \squarebracket{1+\frac{1}{2}\parenthesis{\bar{\alpha}^f + \bar{\anf}} \bar{x}_1+\bar{\anf} \bar{x}_2 }\\
            \leq&\; 2 - \frac{1}{2}\xlb_1 - \frac{1}{2} \xlb_2\\
            &- \frac{1}{2} e^{-\parenthesis{\af+\delta}\xub_1-\frac{1}{2}\parenthesis{\af + \anf + 2\delta} \xub_2} \squarebracket{1+\af \xlb_1 + \frac{1}{2}\parenthesis{\af + \anf} \xlb_2}\\
            &-\frac{1}{2} e^{-\frac{1}{2}\parenthesis{\af + \anf + 2\delta} \xub_1-\parenthesis{\anf+\delta}\xub_2} \squarebracket{1+\frac{1}{2}\parenthesis{\af + \anf} \xlb_1+\anf \xlb_2 }, \forall \parenthesis{\bar{\alpha}^f, \bar{\anf}} \in \mathcal{I}(\anf, \af, \delta).
        \end{split}
    \end{equation}
    Thus, it suffices to find $\xlb, \xub, \xlb_1, \xub_1, \xlb_2$ and $\xub_2$ in the corresponding cell. To do this, we start by showing that, for given $\af$ and $\anf,$ the solution returned by \texttt{nlsolve} package in Julia programming language is provably close to the true solution $x^\star, x^\star_1$ and $x^\star_2$, and then provide a continuity argument to bound the solution $\bar{x}, \bar{x}_1$ and $\bar{x}_2$ for any $\parenthesis{\bar{\alpha}^f, \bar{\anf}} \in \mathcal{I}(\anf, \af, \delta)$.

    We start by bounding the value of $x^\star, x^\star_1$, and $x^\star_2$ based on the numerical solutions returned by Julia \texttt{nlsolve}. {We parameterize the tolerance level $ftol$ in \texttt{nlsolve} by $\epsilon$, which guarantees that, when solving the equation $g(x) = 0$ for any function $g: \mathbb{R} \mapsto \mathbb{R}$, \texttt{nlsolve} returns a solution $x$ such that $|g(x)| < \epsilon$. We let $f(x):= x-e^{-\parenthesis{\af + \anf} x}$ and recall from \eqref{eq:f1_def} that $f_1(x_1) = e^{- \frac{1}{2}\parenthesis{\af +\anf} x_1 +2 \frac{\anf}{\af+\anf}\parenthesis{\log(x_1) + \af x_1}} + 2 \frac{\log(x_1) + \af x_1}{\af + \anf}.$ Then, we leverage monotonicity properties of $f(x)$ and $f_1(x_1)$ to derive the following bounds on $x^\star, x^\star_1$ and $x^\star_2$:
    {
    \begin{claim}\label{cl:nlsolve}
    Let $x^{sol}$ and $ x^{sol}_1$ respectively denote the solutions returned by \texttt{nlsolve} for solving $f(x) = 0$ and $f_1(x_1) = 0$ with $ftol = \epsilon.$ When $10^{-4} < \anf < \af$ and $\af + \anf < e$, we have:
        \begin{align*}
        &x^\star \in \squarebracket{x^{sol}-\epsilon, x^{sol} + \epsilon}, x^\star_1 \in \squarebracket{x^{sol}_1-\epsilon, x^{sol}_1 + \epsilon}, \text{ and}\\
        &x^\star_2 \in \squarebracket{-2 \frac{\log(x^{sol}_1+\epsilon) + \af \parenthesis{x^{sol}_1+\epsilon}}{\af + \anf}, -2 \frac{\log(x^{sol}_1-\epsilon) + \af \parenthesis{x^{sol}_1-\epsilon}}{\af + \anf}}.
    \end{align*}
    \end{claim}
    } }

    Having established bounds on $x^\star, x^\star_1$ and $x^\star_2$ for fixed $\anf$ and $\af$, we continue to bound $\bar{x}, \bar{x}_1$ and $\bar{x}_2$ for $\parenthesis{\bar{\alpha}^f, \bar{\anf}} \in \mathcal{I}(\anf, \af, \delta)$. Specifically, we leverage the following result on the continuity of $x^\star, x^\star_1$ and $x^\star_2$ with respect to $\af$ and $\anf.$

    \begin{claim}\label{cl:continuity}
        Let $x^\star, x^\star_1$ and $x^\star_2$ be the solution to \eqref{eq:sol_x} and \eqref{eq:sol_x12} given $\af$ and $\anf.$ Moreover, let $\bar{x}, \bar{x}_1$ and $\bar{x}_2$ be the solution to \eqref{eq:sol_x} and \eqref{eq:sol_x12} given $\bar{\alpha}^f$ and $\bar{\anf}.$ Then, given any $\delta \in (0,1/2)$, we know that for any $\parenthesis{\bar{\alpha}^f, \bar{\anf}} \in \mathcal{I}(\anf, \af, \delta)$ such that $10^{-4} < \bar{\anf} < \bar{\af}$ and $\bar{\alpha}^f + \bar{\anf} < e$:
        \begin{enumerate}[label=(\roman*)]
            \item $\bar{x} \in \squarebracket{x^\star(1-2\delta), x^\star};$
            \item $\bar{x}_1 \in \squarebracket{x^\star_1(1-2 \delta), x^\star_1}$ and $\bar{x}_2 \in \squarebracket{x^\star_2(1-2 \delta), x^\star_2}.$
        \end{enumerate}
    \end{claim}

    Combining \cref{cl:nlsolve} and \cref{cl:continuity}, we can lower and upper bound the solution of $\bar{x}, \bar{x}_1$ and $\bar{x}_2$ for $\parenthesis{\bar{\alpha}^f, \bar{\anf}} \in \mathcal{I}(\anf, \af, \delta)$ where $\bar{\af} + \bar{\anf} < e$. In particular, for given $\delta > 0$, we take 
    {\begin{equation}\label{eq:solution_bounds}
        \begin{split}
            \xlb &= \parenthesis{x^{sol} - \epsilon}\parenthesis{1-2\delta}, \xub = \parenthesis{x^{sol} + \epsilon},\\
            \xlb_1 &= \parenthesis{x^{sol}_1 - \epsilon}\parenthesis{1-2 \delta}, \xub_1 = \parenthesis{x^{sol}_1 + \epsilon},\\
            \xlb_2 &= -2 \frac{\log(x^{sol}_1+\epsilon) + \af \parenthesis{x^{sol}_1+\epsilon}}{\af + \anf}\parenthesis{1 -2 \delta}, \\
            \xub_2 &= -2 \frac{\log(x^{sol}_1-\epsilon) + \af \parenthesis{x^{sol}_1-\epsilon}}{\af + \anf}.
        \end{split}
    \end{equation}}
    Plugging these values into \eqref{eq:lb_one_sided} - \eqref{eq:ub_balanced}, we obtain a lower bound of $\mu(1,0) - \mu(1/2,1/2)$ in each cell $\mathcal{I}(\anf, \af, \delta)$. Whenever this lower bound exceeds $0,$ our reasoning implies that $\mu(1,0) > \mu(1/2,1/2)$ for any $\parenthesis{\bar{\alpha}^f, \bar{\anf}}$ within this cell. In \texttt{Theorem3.ipynb},\footnote{The computer-aided proof can be found at \url{https://bit.ly/3UUVFRX}.} we fix $\epsilon = 10^{-8}$ and compute the value of this lower bound for $\af, \anf = \delta, 2 \delta,..., e$ where $\af + \anf < e$. We find that by taking $\delta = 0.01, 0.005, 0.0025$ and $0.001$ we are able to verify $\mu(1,0) > \mu(1/2,1/2)$ in the respective red regions in \cref{fig:compare}. \Halmos

    \begin{figure}[ht]
    \centering
    \includegraphics[width=0.5\textwidth]{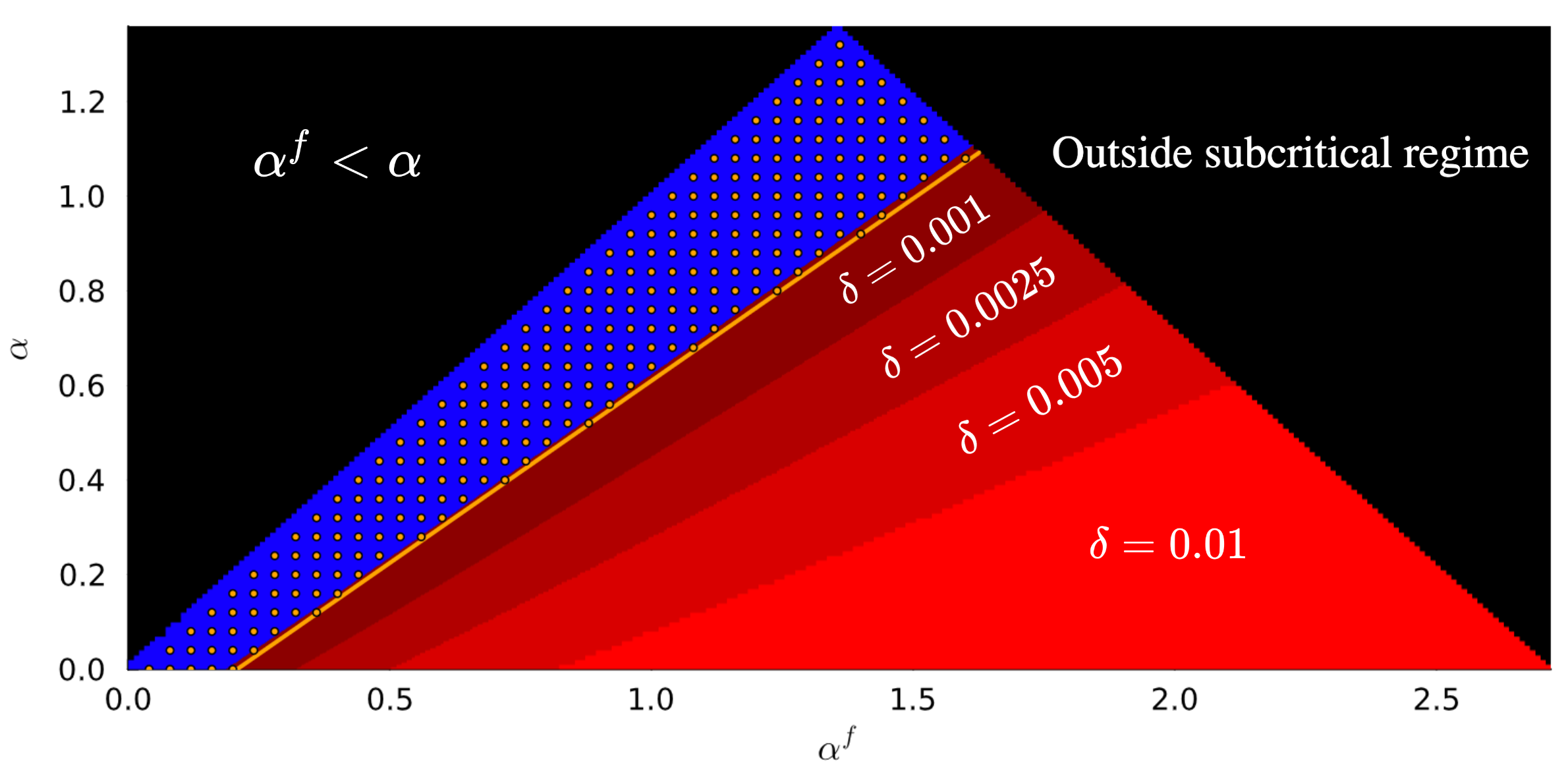}
    \caption{We denote the boundary where \( \anf = 0.77 \af - 0.16 \) by the orange line, and the area of the subcritical regime where one-sided allocation dominates the balanced allocation is displayed in varying shades of red to the right of this boundary. The shades depict growth in the validation area with respect to \( \delta \). The black zone denotes parameters outside the feasible or subcritical regime. Although the inequality cannot be confirmed in the blue region with \( \delta = 0.001 \), we take $\delta = 0$ to verify the inequality for a wide range of $(\af, \anf)$ values highlighted as orange dots.}
    \label{fig:compare}
    \end{figure}
% \end{proof}

\begin{proof}{Proof of \cref{cl:nlsolve}}
For $f(x)= x-e^{-\parenthesis{\af + \anf} x},$ we find that $f'(x) >1 \;\forall x$. When solving $f(x) = 0,$ \texttt{nlsolve} is guaranteed to return a solution \( x^{sol} \) with \( |f(x^{sol})| < \epsilon \). Since \( f'(x) > 1 \;\forall x\), if $x^\star > x^{sol}+ \epsilon$, $$\text{then}\qquad f(x^\star) > f(x^{sol}+ \epsilon) > f(x^{sol}) + \epsilon > 0,$$ contradicting the fact that $f(x^\star) = 0.$ Similarly, we must have $x^\star > x^{sol}- \epsilon$ and thus $x^\star \in [x^{sol}-\epsilon, x^{sol} + \epsilon].$ 

{Since we know from \cref{cl:uniqueness} that $f_1'(x_1) >1$ in the specified parameter regime of $\anf$ and $\af,$ through the same argument on $f_1(x_1)$ we find that $x^\star_1 \in [x^{sol}_1-\epsilon, x^{sol}_1 + \epsilon].$ From \eqref{eq:sol_x12} we know that $x^\star_2 = -2 \frac{\log(x^\star_1) + \af x^\star_1}{\af + \anf},$ which allows us to lower bound $x^\star_2$ by $-2 \frac{\log(x^{sol}_1+\epsilon) + \af \parenthesis{x^{sol}_1+\epsilon}}{\af + \anf}$ and upper bound $x_2$ by $-2 \frac{\log(x^{sol}_1-\epsilon) + \af \parenthesis{x^{sol}_1-\epsilon}}{\af + \anf}.$}
\end{proof}

\begin{proof}{Proof of \cref{cl:continuity}}
    {We start by proving \cref{cl:continuity} (i). We have shown in the proof of \cref{thm:equivalence} that the solution to \eqref{eq:sol_x} is unique in the stated parameter regimes. Thus, we know that $x^\star = e^{-\parenthesis{\af + \anf} x^\star}$ because this construction trivially satisfies \eqref{eq:sol_x}. Similarly, $\bar{x} = e^{-\parenthesis{\bar{\alpha}^f + \bar{\anf}} \bar{x}}$. Now, given that $e^{-\parenthesis{\bar{\alpha}^f + \bar{\anf}}x^\star} \leq e^{-\parenthesis{\af + \anf} x^\star} = x^\star$ for any $\parenthesis{\bar{\alpha}^f, \bar{\anf}} \in \mathcal{I}(\anf, \af, \delta)$, we know that $\bar{x} \leq x^\star.$ Moreover, from the uniqueness of solution we know that it suffices to show that
$$
x^\star (1-2\delta) \leq e^{-\parenthesis{\af + \anf + 2 \delta} x^\star(1-2\delta)} \leq e^{-\parenthesis{\bar{\alpha}^f + \bar{\anf}}x^\star(1-2\delta)},
$$
so then the unique solution $\bar{x} \geq x^\star \parenthesis{1-2\delta}.$ To do so, we observe that
$$
e^{-\parenthesis{\af + \anf + 2 \delta} x^\star(1-2\delta)} \geq e^{-\parenthesis{\af + \anf + 2 \delta} x^\star}
= e^{-\parenthesis{\af + \anf} x^\star - 2\delta x^\star}
= x^\star e^{-2\delta x^\star},
$$
\begin{align}\label{eq:continuity1}\text{so it suffices to show that}\qquad
    x^\star (1-2\delta) \leq x^\star e^{-2\delta x^\star}.
\end{align}
As we trivially have $x^\star \in (0,1]$, we can cancel out $x^\star$ and take logarithm on both sides. 
$$\text{Thus, \eqref{eq:continuity1} is equivalent to}\quad
-\log(1-2\delta) \geq 2\delta x^\star,
$$
$$\text{which is implied by}\quad
-\log(1-2\delta) \geq 2\delta,\quad\text{which holds for all $\delta \in (0,1/2)$.}
$$}

    We next prove \cref{cl:continuity} (ii). Given that in the claimed region
    \begin{align*}
        x^\star_1 =&\; e^{-\frac{1}{2} 2 \af x^\star_1 - \frac{1}{2} \parenthesis{\af +\anf} x^\star_2} \geq e^{-\frac{1}{2} 2 \bar{\alpha}^f x^\star_1 - \frac{1}{2} \parenthesis{\bar{\alpha}^f +\bar{\anf}} x^\star_2}, \\
        x^\star_2 = & e^{-\frac{1}{2} \parenthesis{\af +\anf} x^\star_1 - \frac{1}{2} 2 \anf x^\star_2} \geq e^{-\frac{1}{2} \parenthesis{\bar{\alpha}^f +\bar{\anf}} x^\star_1 - \frac{1}{2} 2 \bar{\anf} x^\star_2}
    \end{align*}
    for any $\parenthesis{\bar{\alpha}^f, \bar{\anf}} \in \mathcal{I}(\anf, \af, \delta)$, we know that $\bar{x}_1 \leq x^\star_1$ and $\bar{x}_2 \leq x^\star_2.$ 

    To prove the claimed lower bound on $\bar{x}_1$ and $\bar{x}_2,$ we show that
    \begin{align}
        x^\star_1 (1-2 \delta) \leq&\; e^{-\frac{1}{2} 2 \bar{\alpha}^f x^\star_1 (1-2 \delta) - \frac{1}{2} \parenthesis{\bar{\alpha}^f +\bar{\anf}} x^\star_2 (1-2 \delta)},\label{eq:continuity2}\\
        x^\star_2 (1-2 \delta) \leq&\; e^{-\frac{1}{2} \parenthesis{\bar{\alpha}^f +\bar{\anf}} x^\star_1(1-2 \delta) - \frac{1}{2} 2 \bar{\anf} x^\star_2(1-2 \delta)}. \notag
    \end{align}
    In particular, \eqref{eq:continuity2} is implied by $$x^\star_1 (1-2 \delta) \leq e^{-\frac{1}{2} 2 \af x^\star_1 - \delta x^\star_1 (1-2\delta) - \frac{1}{2} \parenthesis{\af +\anf} x^\star_2 - \delta x^\star_2 (1-2\delta)} = x^\star_1 \cdot e^{- \delta x^\star_1 (1-2\delta) - \delta x^\star_2 (1-2\delta)}.$$
    
    Since we trivially have $x_1^\star \in (0,1]$, we can cancel out $x_1^\star$ and take logarithm on both sides. We find that \eqref{eq:continuity2} is equivalent to $-\log(1-2 \delta) \geq (x^\star_1 + x^\star_2) \delta (1-2\delta).$ Since $x^\star_1 + x^\star_2 \in [0,2],$ it is then sufficient to show that $\frac{-\log(1-2\delta)}{2\delta(1-2\delta)} \geq 1,$ which holds for all $\delta \in (0,1/2).$ The proof of $\bar{x}_2 \geq x^\star_2(1-2 \delta)$ is symmetric.
\end{proof}

\subsubsection{Proof of \texorpdfstring{\cref{thm:convex_concave}}{Lg}}\label{app:convex_concave}

%\begin{proof}{Proof. }
    We start with the convexity result in \cref{thm:convex_concave} (ii).
    Recall from \eqref{eq:xi} that 
    \begin{equation*}
        \begin{split}
            \xi(\ql,\qr) = 2&-\ql \yonef{\qvec} -\qr\parenthesis{1-\ytwohatf{\qvec}}\\
            &-\qr\parenthesis{1-\ytwohatf{\qvec}}\parenthesis{2\ql\af \yonef{\qvec}+(1-\ql)(\af + \anf) \yonenf{\qvec}} \\
            &- (1-\ql) \yonenf{\qvec}-(1-\qr)\parenthesis{1-\ytwohatnf{\qvec}}\\
            &-(1-\qr) \parenthesis{1-\ytwohatnf{\qvec}}\parenthesis{\ql(\af+\anf) \yonef{\qvec}+2(1-\ql)\anf \yonenf{\qvec}},
        \end{split}
    \end{equation*}
    
    Since we are interested in the direction $(1,-1),$ for ease of notation we denote the sum of flexibility by $\budget$ and replace $\qr$ with $\budget - \ql$. Then, we can re-write $\xi(\ql,\qr) = \xi(\ql,\budget - \ql)$ as
    \begin{equation}\label{eq:xi_convex}
        \begin{split}
            &2-\ql \yonef{\qvec} -(\budget-\ql)\parenthesis{1-\ytwohatf{\qvec}}\\
            &-(\budget-\ql) \parenthesis{1-\ytwohatf{\qvec}}\parenthesis{2\ql\af \yonef{\qvec}+(1-\ql)(\af + \anf) \yonenf{\qvec}} \\
            &- (1-\ql) \yonenf{\qvec}-(1-\budget+\ql)\parenthesis{1-\ytwohatnf{\qvec}}\\
            &-(1-\budget+\ql) \parenthesis{1-\ytwohatnf{\qvec}}\parenthesis{\ql(\af+\anf) \yonef{\qvec}+2(1-\ql)\anf \yonenf{\qvec}}.
        \end{split}
    \end{equation}
    
\noindent    Then the second-order derivative of $\xi(\ql,\qr)$ in the direction $(1,-1)$ equals $\frac{\partial^2 \xi(\ql,\budget - \ql)}{\partial \ql^2}.$ We observe that
    \begin{equation*}
        \xi(\ql,\budget - \ql) = 2-\ql \yonef{\qvec} -(\budget-\ql) e^{-b_1(\ql)}(1+b_1(\ql)) - (1-\ql) \yonenf{\qvec}-(1-\budget+\ql)e^{-b_2(\ql)}(1+b_2(\ql)),
    \end{equation*}
    \begin{align*}\text{where}\qquad
        b_1(\ql) &= \ql \cdot 2\af \cdot \yonef{\qvec}+(1-\ql)(\af + \anf) \yonenf{\qvec},\\
        b_2(\ql) &= \ql(\af+\anf) \yonef{\qvec}+(1-\ql)\cdot 2\anf \cdot \yonenf{\qvec}.
    \end{align*}
\begin{equation}\label{eq:x_def}
    \text{Let}\qquad    x_1(\ql,\qr) = e^{-\ql 2 \af \yonef{\qvec} - (1-\ql)\parenthesis{\af + \anf}\yonenf{\qvec}} \text{ and } x_2(\ql,\qr) = e^{-\ql \parenthesis{\af + \anf} \yonef{\qvec} -(1-\ql) 2\anf \yonenf{\qvec}}
    \end{equation}
    For ease of notation we drop the dependency of $x_1$ and $x_2$ on $\qvec$. Then we simplify $\xi(\ql,\budget - \ql)$ into 
    \begin{align*}
        \xi(\ql,\budget - \ql) = 2&-\ql e^{-(\budget - \ql) 2\af x_1 - (1-\budget+\ql) \parenthesis{\af+\anf} x_2} -(\budget - \ql)x_1(1-\log(x_1))\\
        &-(1-\ql) e^{-(\budget - \ql) \parenthesis{\af+\anf} x_1 - (1-\budget + \ql) 2 \anf x_2}-(1-\budget + \ql)x_2(1-\log(x_2)).
    \end{align*}
    By construction of $x_1,x_2$ and the definition of $\yonef{q},\yonenf{q},$ we have
    \begin{align*}
        x_1 &= e^{-\ql 2 \af e^{-(\budget - \ql) 2\af x_1 - (1-\budget + \ql) (\af+\anf) x_2} - (1-\ql)\parenthesis{\af + \anf} e^{-(\budget - \ql) (\af+\anf) x_1 - (1-\budget + \ql) 2\anf x_2}}\\
        x_2 &= e^{-\ql \parenthesis{\af + \anf} e^{-(\budget - \ql) 2\af x_1 - (1-\budget + \ql) (\af+\anf) x_2} - (1-\ql) 2\anf e^{-(\budget - \ql) (\af+\anf) x_1 - (1-\budget + \ql) 2\anf x_2}}.
    \end{align*}
    For convenience we write 
    \begin{equation}\label{eq:y_def}
        y_1 := e^{-(\budget - \ql) 2\af x_1 - (1-\budget + \ql) (\af+\anf) x_2} \text{ and } y_2 := e^{-(\budget - \ql) (\af+\anf) x_1 - (1-\budget + \ql) 2\anf x_2},
    \end{equation}
    $$\text{so that }\qquad x_1 = e^{-\ql 2 \af y_1 - (1-\ql)\parenthesis{\af + \anf} y_2}, x_2 = e^{-\ql \parenthesis{\af + \anf} y_1 - (1-\ql) 2\anf y_2}.$$
    
    Now, taking second order derivative of $\xi(\ql,\budget - \ql)$ with respect to $\ql,$ we obtain
     \begin{equation}\label{eq:convex_sod}
        \begin{split}
         \frac{\partial^2 \xi(\ql,\budget - \ql)}{\partial \ql^2}=&\; -\ql\parenthesis{-2\af x_1 + \parenthesis{\af+\anf} x_2 + (\budget - \ql) 2 \af \frac{\partial x_1}{\partial \ql} + (1-\budget + \ql)  \parenthesis{\af+\anf}\frac{\partial x_2}{\partial \ql}}^2 y_1\\
         &-(1-\ql)\parenthesis{-\parenthesis{\af+\anf} x_1 + 2\anf x_2 + (\budget - \ql) \parenthesis{\af+\anf} \frac{\partial x_1}{\partial \ql} + (1-\budget + \ql)  2\anf\frac{\partial x_2}{\partial \ql}}^2 y_2\\
         &+2\parenthesis{-2\af y_1+\parenthesis{\af+\anf} y_2} x_1+2\parenthesis{\parenthesis{\af+\anf} y_1-2\anf y_2} x_2\\
         &+(\budget - \ql)2\parenthesis{2\af y_1-\parenthesis{\af+\anf} y_2} \parenthesis{\frac{\partial x_1}{\partial \ql}}+(1-\budget + \ql)2\parenthesis{\parenthesis{\af+\anf} y_1-2\anf y_2} \parenthesis{\frac{\partial x_2}{\partial \ql}}\\
         &+(\budget - \ql)\parenthesis{\frac{\partial x_1}{\partial \ql}}^2/x_1+(1-\budget + \ql)\parenthesis{\frac{\partial x_2}{\partial \ql}}^2/x_2.
        \end{split}
    \end{equation}

    Moreover, by taking derivative of $x_1$ and $x_2$ with respect to $\ql,$ we find that
    \begin{equation}\label{eq:x1_derivative}
        \begin{split}
            \frac{\frac{\partial x_1}{\partial \ql}}{x_1} =&\; y_1 2 \af \ql 2\af (\budget - \ql)\frac{\partial x_1}{\partial \ql}+y_2 (1-\ql)\parenthesis{\af+\anf} (\budget - \ql) \parenthesis{\af+\anf} \frac{\partial x_1}{\partial \ql}+\\
            & y_1 2 \af \ql \parenthesis{\af+\anf} (1-\budget + \ql)\frac{\partial x_2}{\partial \ql}+y_2 (1-\ql)\parenthesis{\af+\anf} (1-\budget + \ql) 2 \anf \frac{\partial x_2}{\partial \ql}+\\
            &y_1 2 \af \ql \squarebracket{\parenthesis{\af+\anf} x_2 - 2 \af x_1}+y_2 \parenthesis{\af+\anf} (1-\ql) \squarebracket{2 \anf x_2 - \parenthesis{\af+\anf} x_1}-\\
            &y_1 2 \af+y_2 \parenthesis{\af+\anf},\qquad \text{and}
        \end{split} 
    \end{equation}
    
    \begin{equation}\label{eq:x2_derivative}
        \begin{split}
            \frac{\frac{\partial x_2}{\partial \ql}}{x_2} =&\; y_1 \parenthesis{\af+\anf} \ql 2\af (\budget - \ql)\frac{\partial x_1}{\partial \ql}+y_2 (1-\ql) 2\anf (\budget - \ql) \parenthesis{\af+\anf} \frac{\partial x_1}{\partial \ql}+\\
            & y_1 \parenthesis{\af+\anf} \ql \parenthesis{\af+\anf} (1-\budget + \ql)\frac{\partial x_2}{\partial \ql}+y_2 (1-\ql)2 \anf (1-\budget + \ql) 2 \anf \frac{\partial x_2}{\partial \ql}+\\
            &y_1 \parenthesis{\af+\anf} \ql \squarebracket{\parenthesis{\af+\anf} x_2 - 2 \af x_1}+y_2 2\anf (1-\ql) \squarebracket{2 \anf x_2 - \parenthesis{\af+\anf} x_1}-\\
            &y_1 \parenthesis{\af+\anf}+y_2 2\anf.
        \end{split}
    \end{equation}

    When $\qvec = (1/2,1/2),$ $\budget = 1$ and $\ql = 1/2$, we find $x_1 = y_1, x_2 = y_2$, so we can simplify \eqref{eq:x1_derivative} and \eqref{eq:x2_derivative} as:
    \begin{align}
        \frac{\partial x_1}{\partial \ql} =&\; \frac{2 x_1 \parenthesis{\anf^2 x_1 x_2 - 2 \af \anf x_1 x_2 - 2 \anf x_2 + \parenthesis{\af}^2 x_1 x_2 + 4\af x_1 - 2 \af x_2}}{\anf^2 x_1 x_2 - 2 \anf \af x_1 x_2 + 4 \anf x_2 + \parenthesis{\af}^2 x_1 x_2 + 4 \af x_1 - 4}, \label{eq:x1_derivative_reduction}\\
        \frac{\partial x_2}{\partial \ql} =&\; -\frac{2 x_2 \parenthesis{\anf^2 x_1 x_2 - 2 \af \anf x_1 x_2 - 2 \anf x_1 + \parenthesis{\af}^2 x_1 x_2 + 4\anf x_2- 2 \af x_1}}{\anf^2 x_1 x_2 - 2 \anf \af x_1 x_2 + 4 \anf x_2 + \parenthesis{\af}^2 x_1 x_2 + 4 \af x_1 - 4}.\label{eq:x2_derivative_reduction}
    \end{align}

    Then, plugging $x_1 = y_1, x_2 = y_2, \budget = 1$ and $\ql = 1/2$ into \eqref{eq:convex_sod}, we obtain
    \begin{align*}
     \frac{\partial^2 \xi(\ql,\budget - \ql)}{\partial \ql^2}\Bigg|_{\budget=1,\ql=\frac{1}{2}}=&\; -\frac{1}{2}\parenthesis{-2\af x_1 + \parenthesis{\af+\anf} x_2 + \frac{1}{2} 2 \af \frac{\partial x_1}{\partial \ql} + \frac{1}{2}  \parenthesis{\af+\anf}\frac{\partial x_2}{\partial \ql}}^2 x_1\\
     &-\frac{1}{2}\parenthesis{-\parenthesis{\af+\anf} x_1 + 2\anf x_2 + \frac{1}{2} \parenthesis{\af+\anf} \frac{\partial x_1}{\partial \ql} + \frac{1}{2} 2\anf\frac{\partial x_2}{\partial \ql}}^2 x_2\\
     &+2\parenthesis{-2\af x_1+\parenthesis{\af+\anf} x_2} x_1+2\parenthesis{\parenthesis{\af+\anf} x_1-2\anf x_2} x_2\\
     &+\parenthesis{2\af x_1-\parenthesis{\af+\anf} x_2} \parenthesis{\frac{\partial x_1}{\partial \ql}}+\parenthesis{\parenthesis{\af+\anf} x_1-2\anf x_2} \parenthesis{\frac{\partial x_2}{\partial \ql}}\\
     &+\frac{1}{2}\parenthesis{\frac{\partial x_1}{\partial \ql}}^2/x_1+\frac{1}{2}\parenthesis{\frac{\partial x_2}{\partial \ql}}^2/x_2.
    \end{align*}
    
    By plugging the values of $\frac{\partial x_1}{\partial \ql}$ and $\frac{\partial x_2}{\partial \ql}$ from \eqref{eq:x1_derivative_reduction} and \eqref{eq:x2_derivative_reduction} into the above, we find that
    
    \begin{equation}\label{eq:sod_convex}
        \begin{split}
             \frac{\partial^2 \xi(\ql,\budget - \ql)}{\partial \ql^2}\Bigg|_{\budget=1,\ql=\frac{1}{2}}=&\; \frac{1}{\anf^2 x_1 x_2 - 2 \anf \af x_1 x_2 + 4\anf x_2 + \parenthesis{\af}^2 x_1 x_2 + 4 \af x_1 - 4} \cdot\\
             &(4 \anf^2 x_1^2 x_2 + 4 \anf^2 x_1 x_2^2 - 8 \anf \af x_1^2 x_2 - 8 \anf \af x_1 x_2^2 - 16 \anf x_1 x_2 \\
             &+16 \anf x_2^2 + 4 \parenthesis{\af}^2 x_1^2 x_2 + 4 \parenthesis{\af}^2 x_1 x_2^2 + 16 \af x_1^2- 16 \af x_1 x_2)\\
             =&\; \frac{\parenthesis{\af-\anf}^2 4 x_1 x_2 \parenthesis{x_1+x_2} - 16 \parenthesis{x_2-x_1}\parenthesis{\af x_1 -\anf x_2}}{\parenthesis{\af-\anf}^2 x_1 x_2 +4\parenthesis{\anf x_2 +\af x_1 -1}}.
        \end{split}
    \end{equation}

    \begin{equation}\label{eq:x_sol}
    \text{Recall that we have}\quad
        x_1 = e^{- \af x_1 - 1/2\parenthesis{\af + \anf} x_2}, x_2 = e^{-1/2 \parenthesis{\af + \anf} x_1 - \anf x_2}
    \end{equation}
    when $\budget = 1, \ql = 1/2.$ This allows us to solve $x_1$ and $x_2$ and determine the size of convexity numerically. In particular, we know from \eqref{eq:solution_bounds} that we can provide bounds $\xlb_1, \xub_1, \xlb_2, \xub_2$ for any $\parenthesis{\bar{\alpha}^f, \bar{\anf}} \in \mathcal{I}(\anf, \af, \delta)$ where $10^{-4} < \bar{\anf} < \bar{\af}$ and $\bar{\af} + \bar{\anf} < e$. Thus, we can lower bound \eqref{eq:sod_convex} for any $\parenthesis{\bar{\alpha}^f, \bar{\anf}} \in \mathcal{I}(\anf, \af, \delta)$ as:

    \begin{align}
        &\frac{\parenthesis{\bar{\alpha}^f-\bar{\anf}}^2 4 \bar{x}_1 \bar{x}_2 \parenthesis{\bar{x}_1+\bar{x}_2} - 16 \parenthesis{\bar{x}_2-\bar{x}_1}\parenthesis{\bar{\alpha}^f \bar{x}_1 -\bar{\anf} \bar{x}_2}}{\parenthesis{\bar{\alpha}^f-\bar{\anf}}^2 \bar{x}_1 \bar{x}_2 +4\parenthesis{\bar{\anf} \bar{x}_2 +\bar{\alpha}^f \bar{x}_1 -1}} \notag\\
        \geq & \frac{-\parenthesis{\af - \anf + \delta}^2 4 \xub_1 \xub_2 \parenthesis{\xub_1+\xub_2} + 16 \max\parenthesis{0, \xlb_2 - \xub_1} \af \xlb_1 - 16 \parenthesis{\xub_2 - \xlb_1} (\anf + \delta) \xub_2}{-\parenthesis{\max\parenthesis{\af - \anf - \delta,0}}^2 \xlb_1 \xlb_2 + 4 \squarebracket{1-\anf \xlb_2 - \af \xlb_1}}.\label{eq:convex_lb}
    \end{align}

    If this lower bound exceeds $0,$ we know that $\xi(\ql,\qr)$ is strictly convex in the direction $(1,-1)$ at $\qvec = (1/2,1/2)$ for any $\parenthesis{\bar{\alpha}^f, \bar{\anf}} \in \mathcal{I}(\anf, \af, \delta)$. At $\qvec = (1/2,1/2)$, we find by symmetry that $\xi(\ql,\qr) = \hat{\xi}(\ql,\qr),$ $$\frac{\partial \xi(\ql,\budget - \ql)}{\partial \ql}\Bigg|_{\budget=1,\ql=\frac{1}{2}} = \frac{\partial \hat{\xi}(\ql,\budget - \ql)}{\partial \ql} \Bigg|_{\budget=1,\ql=\frac{1}{2}} = 0 \text{ and } \frac{\partial^2 \xi(\ql,\budget - \ql)}{\partial \ql^2}\Bigg|_{\budget=1,\ql=\frac{1}{2}} = \frac{\partial^2 \hat{\xi}(\ql,\budget - \ql)}{\partial \ql^2}\Bigg|_{\budget=1,\ql=\frac{1}{2}}.$$ 
    
    Thus, at $\qvec = (1/2,1/2)$, by \cref{cl:taylor} below we know that to show $\nabla_{(1,-1)}^2 \mubar(\ql,\qr) > 0$ it suffices to verify strict local convexity for $\xi(\ql,\qr)$ in the direction $(1,-1)$. That is, we verify $\frac{\partial^2 \xi(\ql,\budget - \ql)}{\partial \ql^2}\Bigg|_{\budget=1,\ql=\frac{1}{2}} > 0.$

    \begin{claim}\label{cl:taylor}
        If for some $\qvec' \in (0,1)^2$ and direction $v \in \mathbb{R}^2, \hat{\xi}(\qvec') = \hat{\xi}(\qvec'), \nabla_{v} \hat{\xi}(\qvec') = \nabla_{v} \hat{\xi}(\qvec')$ and $\nabla_{v}^2 \hat{\xi}(\qvec') = \nabla_{v}^2 \hat{\xi}(\qvec')$, then $\nabla_{v}^2 \mubar(\qvec') = \nabla_{v}^2 \hat{\xi}(\qvec') = \nabla_{v}^2 \hat{\xi}(\qvec').$
    \end{claim}
    
    In \texttt{Theorem7.ipynb},\footnote{The computer-aided proof can be found at \url{https://bit.ly/3P1f6oi}.} we compute the lower bound of $\frac{\partial^2 \xi(\ql,\budget - \ql)}{\partial \ql^2}\Bigg|_{\budget=1,\ql=\frac{1}{2}}$ in \eqref{eq:convex_lb} for $\af, \anf = \delta, 2 \delta,..., e$ where $\af + \anf < e$. We find that by taking $\delta = 0.01, 0.005, 0.0025$ and $0.001$ we are able to verify strict local convexity of $\mubar(1/2,1/2)$ in the direction $(1,-1)$ in the respective red regions in \cref{fig:verify_convex_concave}.

    \begin{figure}[ht]
    \centering
    \includegraphics[width=0.8\textwidth]{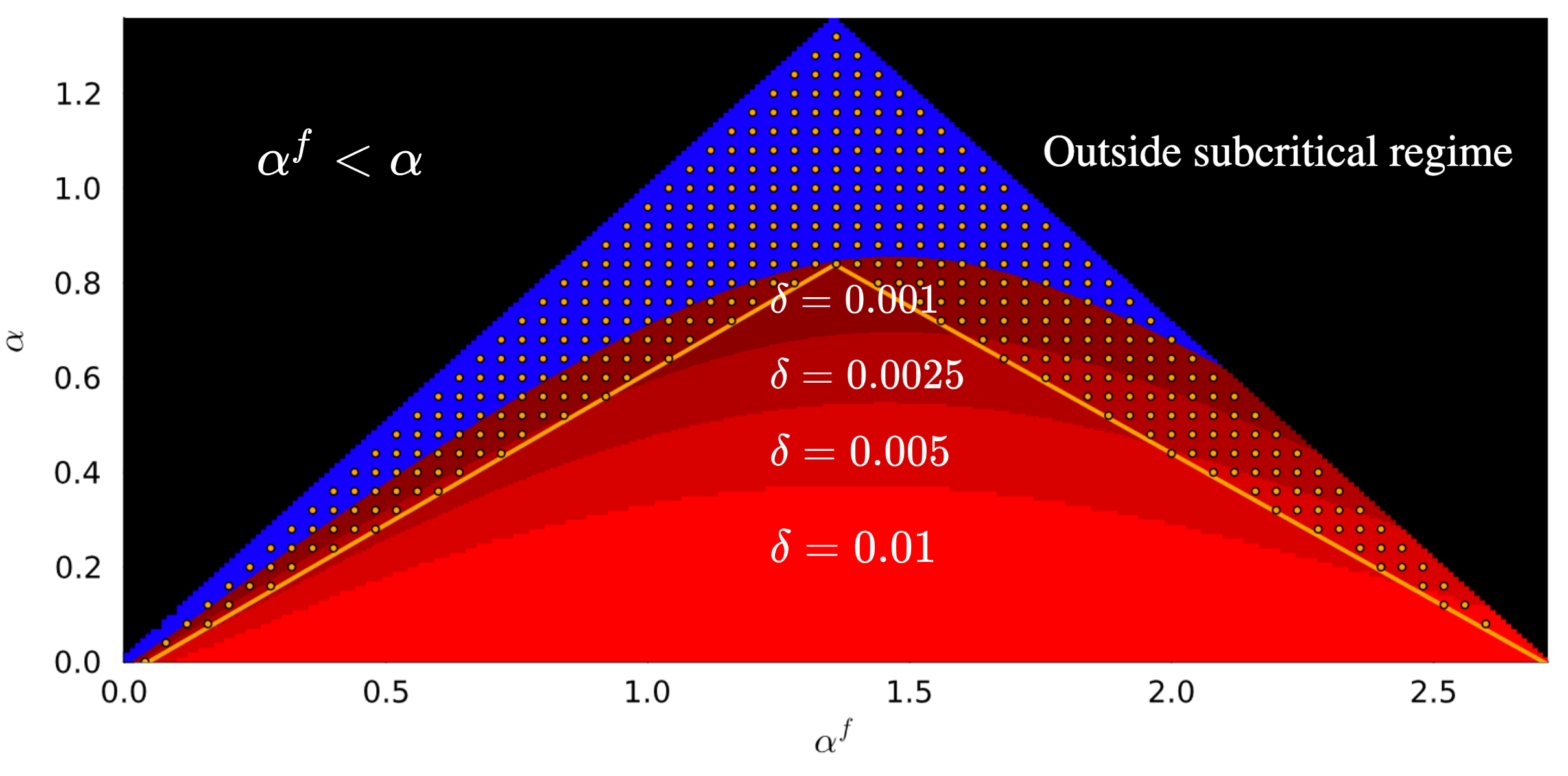}
    \caption{In the figure we denote the boundary where \(10^{-4} < \anf < 0.64 \af - 0.03\) and \( 0.62 \af + \anf < 1.68 \) by the orange lines, and the area of subcritical regime where convexity and concavity properties are verified is displayed in varying shades of red below this boundary. The growth in the validation area with respect to \( \delta \) is depicted through different red gradients. The black zone denotes parameters outside the feasible or subcritical regime. Although the convexity and concavity properties cannot be confirmed in the blue region when \( \delta = 0.001 \), by taking $\delta = 0$ we verify the properties for a wide range of $(\af, \anf)$ values highlighted as orange dots.}
    \vspace{-.15in}\label{fig:verify_convex_concave}
    \end{figure}

    We next prove the concavity result in \cref{thm:convex_concave} (i). We again focus on $\xi(\ql,\qr)$, and the second-order derivative of $\xi(\ql,\qr)$ in the direction $(0,1)$ is equal to $\frac{\partial^2 \xi(\ql,\budget - \ql)}{\partial \budget^2}.$

    Based on $x_1, x_2, y_1, y_2$ constructed in \eqref{eq:x_def} and \eqref{eq:y_def}, we obtain
    \begin{equation}\label{eq:concave_sod}
        \begin{split}
         \frac{\partial^2 \xi(\ql,\budget - \ql)}{\partial \budget^2}=&\; -\ql\parenthesis{2\af x_1 - \parenthesis{\af+\anf} x_2 + (\budget - \ql) 2 \af \frac{\partial x_1}{\partial \budget} + (1-\budget + \ql)  \parenthesis{\af+\anf}\frac{\partial x_2}{\partial \budget}}^2 y_1\\
         &-(1-\ql)\parenthesis{\parenthesis{\af+\anf} x_1 - 2\anf x_2 + (\budget - \ql) \parenthesis{\af+\anf} \frac{\partial x_1}{\partial \budget} + (1-\budget + \ql)  2\anf\frac{\partial x_2}{\partial \budget}}^2 y_2\\
         &+(\budget - \ql)\parenthesis{\frac{\partial x_1}{\partial \budget}}^2/x_1+(1-\budget + \ql)\parenthesis{\frac{\partial x_2}{\partial \budget}}^2/x_2.
        \end{split}
    \end{equation}

    Moreover, by taking derivative of $x_1$ and $x_2$ with respect to $\budget,$ we find that
    \begin{align*}
        \frac{\frac{\partial x_1}{\partial \budget}}{x_1} =&\; y_1 2 \af \ql 2\af (\budget - \ql)\frac{\partial x_1}{\partial \budget}+y_2 (1-\ql)\parenthesis{\af+\anf} (\budget - \ql) \parenthesis{\af+\anf} \frac{\partial x_1}{\partial \budget}+\\
        & y_1 2 \af \ql \parenthesis{\af+\anf} (1-\budget + \ql)\frac{\partial x_2}{\partial \budget}+y_2 (1-\ql)\parenthesis{\af+\anf} (1-\budget + \ql) 2 \anf \frac{\partial x_2}{\partial \budget}-\\
        &y_1 2 \af \ql \squarebracket{\parenthesis{\af+\anf} x_2 - 2 \af x_1}-y_2 \parenthesis{\af+\anf} (1-\ql) \squarebracket{2 \anf x_2 - \parenthesis{\af+\anf} x_1},
    \end{align*}
    \begin{align*}
        \text{and}\qquad \frac{\frac{\partial x_2}{\partial \budget}}{x_2} =&\; y_1 \parenthesis{\af+\anf} \ql 2\af (\budget - \ql)\frac{\partial x_1}{\partial \budget}+y_2 (1-\ql) 2\anf (\budget - \ql) \parenthesis{\af+\anf} \frac{\partial x_1}{\partial \budget}+\\
        & y_1 \parenthesis{\af+\anf} \ql \parenthesis{\af+\anf} (1-\budget + \ql)\frac{\partial x_2}{\partial \budget}+y_2 (1-\ql)2 \anf (1-\budget + \ql) 2 \anf \frac{\partial x_2}{\partial \budget}-\\
        &y_1 \parenthesis{\af+\anf} \ql \squarebracket{\parenthesis{\af+\anf} x_2 - 2 \af x_1}-y_2 2\anf (1-\ql) \squarebracket{2 \anf x_2 - \parenthesis{\af+\anf} x_1}.
    \end{align*}
    When $\budget = 1$ and $\ql = 1/2$, we have $x_1 = y_1, x_2 = y_2.$ Applying these observations we simplify \eqref{eq:concave_sod} as

    \begin{equation}\label{eq:sod_concave}
        \begin{split}
             \frac{\partial^2 \xi(\ql,\budget - \ql)}{\partial \budget^2}\Bigg|_{B=1,q=\frac{1}{2}}=&\; \frac{1}{- \parenthesis{x_1 x_2 \parenthesis{\af-\anf}^2}^2 + 8 x_1 x_2 \parenthesis{\af+\anf}^2 + 16 \parenthesis{\anf^2 x_2^2 + \parenthesis{\af}^2 x_1^2 -1}} \cdot\\
             &(-2\parenthesis{x_1+x_2} \parenthesis{x_1 x_2 \parenthesis{\af-\anf}^2}^2 - 16\parenthesis{x_1+x_2} x_1 x_2 \af \anf \\
             &+8 \parenthesis{\af}^2 x_1 \parenthesis{x_2^2 - 3x_1 x_2+ 4 x_1^2} +8 \anf^2 x_2 \parenthesis{x_1^2 - 3x_1 x_2+ 4 x_2^2}).
        \end{split}
    \end{equation}

    Again, we know from \eqref{eq:solution_bounds} that we can provide bounds $\xlb_1, \xub_1, \xlb_2, \xub_2$ for any $\parenthesis{\bar{\alpha}^f, \bar{\anf}} \in \mathcal{I}(\anf, \af, \delta)$ where $\bar{\af} + \bar{\anf} < e$. Thus, we can upper bound \eqref{eq:sod_concave} for any $\parenthesis{\bar{\alpha}^f, \bar{\anf}} \in \mathcal{I}(\anf, \af, \delta)$ as:
    \begin{equation}\label{eq:concave_ub}
        \begin{split}
            & \frac{1}{-\squarebracket{\xub_1 \xub_2 (\af - \anf + \delta)^2}^2 + 8 (\af + \anf)^2 \xlb_1 \xlb_2 +16\anf^2 \parenthesis{\xlb_2}^2 + 16 \parenthesis{\af}^2 \parenthesis{\xlb_1}^2 - 16}\\
            \cdot& \Biggl(-2\parenthesis{\xlb_1 + \xlb_2} \squarebracket{\xlb_1 \xlb_2 \parenthesis{\max(0,\af-\anf-\delta)}^2}^2 -16 (\xlb_1+\xlb_2) \xlb_1 \xlb_2 \af \anf\\
            +& 8 (\af + \delta)^2 \parenthesis{\xub_1 \parenthesis{\xub_2}^2 + 4\parenthesis{\xub_1}^3} + 8 (\anf + \delta)^2 \parenthesis{\xub_2 \parenthesis{\xub_1}^2 + 4\parenthesis{\xub_2}^3}\\
            -&24 \parenthesis{\af}^2 \parenthesis{\xlb_1}^2 \xlb_2 - 24 \anf^2 \parenthesis{\xlb_2}^2 \xlb_1\Biggr).
        \end{split}
    \end{equation}

    If this upper bound is strictly below $0,$ %we know that 
    $\xi(\ql,\qr)$ is strictly concave in the direction $(0,1)$ at $\qvec = (1/2,1/2)$ for any $\parenthesis{\bar{\alpha}^f, \bar{\anf}} \in \mathcal{I}(\anf, \af, \delta)$. By symmetry at $\qvec = (1/2,1/2)$ we find that $\xi(\ql,\qr) = \hat{\xi}(\ql,\qr),$ $$\frac{\partial \xi(\ql,\budget - \ql)}{\partial \budget}\Bigg|_{\budget=1,\ql=\frac{1}{2}} = \frac{\partial \hat{\xi}(\ql,\budget - \ql)}{\partial \budget} \Bigg|_{\budget=1,\ql=\frac{1}{2}} \text{ and } \frac{\partial^2 \xi(\ql,\budget - \ql)}{\partial \budget^2}\Bigg|_{\budget=1,\ql=\frac{1}{2}} = \frac{\partial^2 \hat{\xi}(\ql,\budget - \ql)}{\partial \budget^2}\Bigg|_{\budget=1,\ql=\frac{1}{2}}.$$ Thus, at $\qvec = (1/2,1/2)$, by \cref{cl:taylor} we know it suffices to verify strict local concavity for $\xi(\ql,\qr)$ in the direction $(0,1)$. That is, we verify $\frac{\partial^2 \xi(\ql,\budget - \ql)}{\partial \budget^2}\Bigg|_{\budget=1,\ql=\frac{1}{2}} < 0.$

    In \texttt{Theorem7.ipynb},\footnote{The computer-aided proof can be found at \url{https://bit.ly/3P1f6oi}.} we compute the upper bound of $\frac{\partial^2 \xi(\ql,\budget - \ql)}{\partial \budget^2}\Bigg|_{B=1,\ql=\frac{1}{2}}$ in \eqref{eq:concave_ub} for $\af, \anf = 10^{-4}, \delta, 2 \delta,..., e$ where $\af + \anf < e$. We find that taking $\delta = 0.01$ is sufficient for verifying \eqref{eq:concave_ub} $< 0$ for all $\af$ anf $\anf$ such that $10^{-4} < \anf < \af$ and $\af + \anf < e$. The concavity in the direction $(1,0)$ is  symmetric. 
%\end{proof}

\emph{Proof of \cref{cl:taylor}}
    By definition, we need to show that $$\nabla_{\mathbf{v}}^2 \mubar(\qvec') = \lim_{h \to 0} \frac{\mubar(\qvec'+ \mathbf{v} h) - 2 \mubar(\qvec') + \mubar(\qvec'- \mathbf{v} h)}{h^2}$$ exists and is equal to the claimed value. By Taylor series expansion, we know that for $h \in \mathbb{R},${
    \begin{align*}
        \xi(\qvec' + \mathbf{v} h) =&\; \xi(\qvec') + h \cdot \nabla_{\mathbf{v}} \xi(\qvec') + h^2/2 \cdot \nabla_{\mathbf{v}}^2 \xi(\qvec') + o(h^2),\\
        \hat{\xi}(\qvec' + \mathbf{v} h) =&\; \hat{\xi}(\qvec') + h \cdot \nabla_{\mathbf{v}} \hat{\xi}(\qvec') + h^2/2 \cdot \nabla_{\mathbf{v}}^2 \hat{\xi}(\qvec') + o(h^2),\\
        \xi(\qvec' - \mathbf{v} h) =&\; \xi(\qvec') - h \cdot \nabla_{\mathbf{v}} \xi(\qvec') + h^2/2 \cdot \nabla_{\mathbf{v}}^2 \xi(\qvec') - o(h^2),\\
        \hat{\xi}(\qvec' - \mathbf{v} h) =&\; \hat{\xi}(\qvec') - h \cdot \nabla_{\mathbf{v}} \hat{\xi}(\qvec') + h^2/2 \cdot \nabla_{\mathbf{v}}^2 \hat{\xi}(\qvec') - o(h^2).
    \end{align*}
    In particular, as $\xi(\qvec') = \hat{\xi}(\qvec'), \nabla_{v} \xi(\qvec') = \nabla_{v} \hat{\xi}(\qvec')$ and $\nabla_{v}^2 \xi(\qvec') = \nabla_{v}^2 \hat{\xi}(\qvec'),$} from the above we know that $$\xi(\qvec' + \mathbf{v} h) -\hat{\xi}(\qvec' + \mathbf{v} h) = o(h^2) \text{ and } \xi(\qvec' - \mathbf{v} h) -\hat{\xi}(\qvec' - \mathbf{v} h) = o(h^2).$$
    \begin{align*}
    \text{ Thus,}\quad\nabla_{\mathbf{v}}^2 \mubar(\qvec') =&\; \lim_{h \to 0} \frac{\mubar(\qvec'+ \mathbf{v} h) - 2 \mubar(\qvec') + \mubar(\qvec'- \mathbf{v} h)}{h^2}\\
        =&\; \lim_{h \to 0}\frac{\min\parenthesis{\xi(\qvec'+ \mathbf{v} h), \hat{\xi}(\qvec'+ \mathbf{v} h)} - 2 \min\parenthesis{\xi(\qvec'), \hat{\xi}(\qvec')} + \min\parenthesis{\xi(\qvec'- \mathbf{v} h), \hat{\xi}(\qvec'- \mathbf{v} h)}}{h^2}\\
        =&\; \lim_{h \to 0}\frac{\xi(\qvec'+ \mathbf{v} h) - 2 \xi(\qvec') + \xi(\qvec'- \mathbf{v} h) +o(h^2)}{h^2}\\
        =&\;\nabla_{\mathbf{v}}^2 \xi(\qvec') = \nabla_{\mathbf{v}}^2 \hat{\xi}(\qvec').\hfill\Halmos
    \end{align*}
% \end{proof}

\subsubsection{Proofs of the Auxiliary Results in \texorpdfstring{Appendix \ref{app:phase_1}}{Lg}}\label{app:auxiliary_proofs}

\begin{proof}{Proof of \cref{prop:m1}}

    Theorem $8$ in \citet{karp1981maximum} (1) - (4) establishes the validity of the statements in \cref{prop:m1} (i) - (iv) for a general graph $G = (V, E)$. Since bipartite graphs are a subset of such general graphs, these results immediately extend.

    For \cref{prop:m1} (v), to determine $|\delete^l_1|$ and $|\delete^r_1|$, we start by finding $m_1$. By (iii), every edge in $M_1$ is connected to at least one target. By (ii), if an edge in $M_1$ is connected to two targets $u$ and $v$, then $v \bigotimes u$ and $u \bigotimes v$. Hence, $$m_1 = |M_1| \leq \left|\left\{v \in V_l \mid v \text{ is a target}\right\}\right| + \left|\left\{v \in V_r \mid v \text{ is a target}\right\}\right| - \left|\left\{(v,u) \mid v \bigotimes u \text{ and } u \bigotimes v \right\}\right|.$$ The equality follows from (ii) because every target is connected to an edge in $M_1$.

    By Theorem 9 (4) in \citet{karp1981maximum}, a node $v$ appears in a derivation if and only if it is a target or a loser or both. Furthermore, $v$ is both a target and a loser if and only if there exists a unique $u$ such that $v \bigotimes u$ and $u \bigotimes v$. Hence, the number of nodes in $V_l$ that appear in a derivation is given by $$\left|\left\{v \in V_l \mid v \text{ is a target}\right\}\right| + \left|\left\{v \in V_l \mid v \text{ is a loser}\right\}\right| - \left|\left\{(v,u) \mid v \bigotimes u \text{ and } u \bigotimes v \right\}\right|.$$

    Then, we find $\delete^l_1$ as the set of nodes in $V_l$ that appear in a derivation but do not belong to $M^l_1$, i.e.,
    \begin{align*}
        |\delete^l_1| =&\; \left|\left\{v \in V_l \mid v \text{ is a target}\right\}\right| + \left|\left\{v \in V_l \mid v \text{ is a loser}\right\}\right| - \left|\left\{(v,u) \mid v \bigotimes u \text{ and } u \bigotimes v \right\}\right| \\
        & - \parenthesis{\left|\left\{v \in V_l \mid v \text{ is a target}\right\}\right| + \left|\left\{v \in V_r \mid v \text{ is a target}\right\}\right|- \left|\left\{(v,u) \mid v \bigotimes u \text{ and } u \bigotimes v \right\}\right|}\\
        =&\; \left|\left\{v \in V_l \mid v \text{ is a loser}\right\}\right| - \left|\left\{v \in V_r \mid v \text{ is a target}\right\}\right|.
    \end{align*}

\noindent    Computing $|\delete^r_1|$ is symmetric. Since $\psi_1 = \max\bracket{|\delete^l_1|, |\delete^r_1|}$ by definition,  \cref{prop:m1} (v) follows.
\end{proof}

\begin{proof}{Proof of \cref{lem:w}}

    We start from the leaf of a random tree, i.e., $d = 1,$ and iteratively trace back to the root of the tree as $d$ scales large. For a flexible node $u \in \setl,$ the number of its children follows a Binomial distribution $\bin\parenthesis{n,(1+\qr) \pf_n + (1-\qr) \pnf_n}$. Thus, the probability for it to have $k$ children is given by $$\degreef{\qvec}{k} := \binom{n}{k} \parenthesis{(1+\qr) \pf_n + (1-\qr) \pnf_n}^k\parenthesis{1-(1+\qr) \pf_n - (1-\qr) \pnf_n}^{n-k}\; \forall k.$$ 
    
    Moreover, since the probability that $u$ connects with a flexible node is $2 \pf$ and the probability that it connects with a regular node is $\pf + \pnf,$ by Bayes' Theorem we have
    \begin{align*}
        \PP{u' \text{ is flexible}|u \text{ is flexible}, u' \text{ is a child of } u} =&\; \frac{\qr \cdot 2 \pf_n}{\qr \cdot 2 \pf_n + (1-\qr) \cdot (\pf_n + \pnf_n)}\\
        =&\; \frac{\qr \cdot 2 \pf_n}{(1+\qr)\pf_n + (1-\qr)\pnf_n}
    \end{align*}
     $$\text{and similarly}\qquad\PP{u' \text{ is regular}|u \text{ is flexible}, u' \text{ is a child of } u} = \frac{(1-\qr) \cdot (\pf_n + \pnf_n)}{(1+\qr)\pf_n + (1-\qr)\pnf_n}.$$

    By definition, $u$ is in $L$ if all of its children are in $H$, including when it has no children. Thus,

    \begin{align*}
    \yonef{\qvec,d} =&\; \sum_{k = 0}^{n} \degreef{\qvec}{k} \cdot \parenthesis{\frac{2 \qr \pf_n}{(1+\qr)\pf_n + (1-\qr)\pnf_n} \ytwohatf{\qvec,d-1} + \frac{(1-\qr) \cdot (\pf_n + \pnf_n)}{(1+\qr)\pf_n + (1-\qr)\pnf_n} \ytwohatnf{\qvec,d-1}}^k\\
    =&\;\sum_{k = 0}^{n} \binom{n}{k} \parenthesis{(1+\qr) \pf_n + (1-\qr) \pnf_n}^k\parenthesis{1-(1+\qr) \pf_n - (1-\qr) \pnf_n}^{n-k}\\
    &\cdot \parenthesis{\frac{2 \qr \pf_n}{(1+\qr)\pf_n + (1-\qr)\pnf_n} \ytwohatf{\qvec,d-1} + \frac{(1-\qr) \cdot (\pf_n + \pnf_n)}{(1+\qr)\pf_n + (1-\qr)\pnf_n} \ytwohatnf{\qvec,d-1}}^k\\
    =&\;\sum_{k = 0}^{n} \binom{n}{k} \parenthesis{1-(1+\qr) \pf_n - (1-\qr) \pnf_n}^{n-k} \cdot \parenthesis{2 \qr \pf_n \ytwohatf{\qvec,d-1} + (1-\qr) \cdot (\pf_n + \pnf_n)\ytwohatnf{\qvec,d-1}}^k\\
    =&\;\squarebracket{2 \qr \pf_n \ytwohatf{\qvec,d-1} + (1-\qr) \cdot (\pf_n + \pnf_n)\ytwohatnf{\qvec,d-1} + 1-(1+\qr) \pf_n - (1-\qr) \pnf_n }^{n}\\
    =&\;\squarebracket{1-2 \qr \pf_n \parenthesis{1-\ytwohatf{\qvec,d-1}} - (1-\qr) \cdot (\pf_n + \pnf_n)\parenthesis{1-\ytwohatnf{\qvec,d-1}}}^{n}\\
    =&\; \squarebracket{1-\frac{2 \qr \af \parenthesis{1-\ytwohatf{\qvec,d-1}} + (1-\qr) \cdot (\af + \anf)\parenthesis{1-\ytwohatnf{\qvec,d-1}}}{n}}^{n}\\
    =&\; e^{-2 \qr \af \parenthesis{1-\ytwohatf{\qvec,d-1}}-(1-\qr) \cdot (\af + \anf)\parenthesis{1-\ytwohatnf{\qvec,d-1}}} \text{ as } n \rightarrow \infty.
    \end{align*}

    Notice that the fourth equality is an application of the Binomial Theorem, and the last one follows from $$\lim_{n \rightarrow \infty} \parenthesis{1-\frac{x}{n}}^n = e^{-x}, \forall x.$$ The expressions for $\yonenf{\qvec,d}, \yonehatf{\qvec,d}$ and $\yonehatnf{\qvec,d}$ can be derived in a similar fashion.

    Next, by definition, $u$ is in $H$ if it has at least one child in $L$. Thus,
    \begin{align*}
    \ytwof{\qvec,d} =&\; 1-\sum_{k = 0}^{n} \degreef{\qvec}{k} \cdot \parenthesis{\frac{2 \qr \pf_n}{(1+\qr)\pf_n + (1-\qr)\pnf_n} \parenthesis{1-\yonehatf{\qvec,d-1}} + \frac{(1-\qr) \cdot (\pf_n + \pnf_n)}{(1+\qr)\pf_n + (1-\qr)\pnf_n} \parenthesis{1-\yonehatnf{\qvec,d-1}}}^k\\
    =&\;1-\sum_{k = 0}^{n} \binom{n}{k} \parenthesis{(1+\qr) \pf_n + (1-\qr) \pnf_n}^k\parenthesis{1-(1+\qr) \pf_n - (1-\qr) \pnf_n}^{n-k}\\
    &\cdot \parenthesis{\frac{2 \qr \pf_n}{(1+\qr)\pf_n + (1-\qr)\pnf_n} \parenthesis{1-\yonehatf{\qvec,d-1}} + \frac{(1-\qr) \cdot (\pf_n + \pnf_n)}{(1+\qr)\pf_n + (1-\qr)\pnf_n} \parenthesis{1-\yonehatnf{\qvec,d-1}}}^k\\
    =&\;1-\sum_{k = 0}^{n} \binom{n}{k} \parenthesis{1-(1+\qr) \pf_n - (1-\qr) \pnf_n}^{n-k} \\
    &\cdot \parenthesis{2 \qr \pf_n \parenthesis{1-\yonehatf{\qvec,d-1}} + (1-\qr) \cdot (\pf_n + \pnf_n)\parenthesis{1-\yonehatnf{\qvec,d-1}}}^k\\
    =&\;1-\squarebracket{2 \qr \pf_n \parenthesis{1-\yonehatf{\qvec,d-1}} + (1-\qr) \cdot (\pf_n + \pnf_n)\parenthesis{1-\yonehatnf{\qvec,d-1}} + 1-(1+\qr) \pf_n - (1-\qr) \pnf_n }^{n}\\
    =&\;1-\squarebracket{1-2 \qr \pf_n \yonehatf{\qvec,d-1} - (1-\qr) \cdot (\pf_n + \pnf_n)\yonehatnf{\qvec,d-1}}^{n}\\
    =&\; 1-\squarebracket{1-\frac{2 \qr \af \yonehatf{\qvec,d-1} + (1-\qr) \cdot (\af + \anf)\yonehatnf{\qvec,d-1}}{n}}^{n}\\
    =&\; 1-e^{-2 \qr \af \yonehatf{\qvec,d-1}-(1-\qr) \cdot (\af + \anf)\yonehatnf{\qvec,d-1}} \text{ as } n \rightarrow \infty.
    \end{align*}

    The expressions for $\ytwonf{\qvec,d}, \ytwohatf{\qvec,d}$ and $\ytwohatnf{\qvec,d}$ can be derived in a similar fashion. Since all leaf nodes are in $L$, we have $$\ytwof{\qvec,1} = \ytwonf{\qvec,1} = \ytwohatf{\qvec,1} = \ytwohatnf{\qvec,1} = 0.$$ Moreover, since $\yonef{\qvec,d}, \yonenf{\qvec,d}, \ytwof{\qvec,d}, \ytwonf{\qvec,d}, \yonehatf{\qvec,d}, \yonehatnf{\qvec,d}, \ytwohatf{\qvec,d}, \ytwohatnf{\qvec,d}$ are all bounded increasing sequences with respect to $d,$ these sequences converge as $d \rightarrow \infty$ and $\mathbf{y}$ is given by the smallest solution to \eqref{eq:w}.
\end{proof}

\begin{proof}{Proof of \cref{lem:target_loser}}
    Lemma 3.1 and 3.2 in \citet{karp1981maximum} respectively establish the statements in \cref{lem:target_loser}(i) and (ii) for a general tree rooted at vertex $v$. As \tree under consideration is also rooted at vertex $v$, these results immediately extend.
\end{proof}

\emph{Proof of \cref{lem:target_loser2}}
    By \cref{lem:target_loser} (i), we have $$\PP{v \text{ is a target}|v \text{ is a flexible node in }\setl} = \PP{v \in H|v \text{ is a flexible node in }\setl} = \ytwof{\qvec}.$$

    To find the probability for a flexible node $v$ in $\setl$ to be a loser, we need to sum the probability that $v$ is in $L$ and that $v$ has exactly $1$ child which is not in $H$. The former is simply given by $\yonef{\qvec}$, while the latter can be computed as 
    \begin{align*}
        &\PP{v \text{ has exactly 1 child that is not in } H|v \text{ is a flexible node in }\setl} \\
        =&\; \sum_{k = 0}^{n} \degreef{\qvec}{k} \cdot k \cdot \parenthesis{\frac{2 \qr \pf_n}{(1+\qr)\pf_n + (1-\qr)\pnf_n} \ytwohatf{\qvec} + \frac{(1-\qr) \cdot (\pf_n + \pnf_n)}{(1+\qr)\pf_n + (1-\qr)\pnf_n}  \ytwohatnf{\qvec}}^{k-1}\\
        &\cdot \parenthesis{1-\frac{2 \qr \pf_n}{(1+\qr)\pf_n + (1-\qr)\pnf_n} \ytwohatf{\qvec} - \frac{(1-\qr) \cdot (\pf_n + \pnf_n)}{(1+\qr)\pf_n + (1-\qr)\pnf_n} \ytwohatnf{\qvec}}\\
        =&\;\sum_{k = 0}^{n} \binom{n}{k} \parenthesis{(1+\qr) \pf_n + (1-\qr) \pnf_n}^k\parenthesis{1-(1+\qr) \pf_n - (1-\qr) \pnf_n}^{n-k}\\
        & \cdot k \cdot \parenthesis{\frac{2 \qr \pf_n}{(1+\qr)\pf_n + (1-\qr)\pnf_n} \ytwohatf{\qvec} + \frac{(1-\qr) \cdot (\pf_n + \pnf_n)}{(1+\qr)\pf_n + (1-\qr)\pnf_n}  \ytwohatnf{\qvec}}^{k-1}\\
        & \cdot \parenthesis{1-\frac{2 \qr \pf_n}{(1+\qr)\pf_n + (1-\qr)\pnf_n} \ytwohatf{\qvec} - \frac{(1-\qr) \cdot (\pf_n + \pnf_n)}{(1+\qr)\pf_n + (1-\qr)\pnf_n} \ytwohatnf{\qvec}}\\
        =&\;\sum_{k = 1}^{n} n \binom{n-1}{k-1} \parenthesis{(1+\qr) \pf_n + (1-\qr) \pnf_n}^{k-1}\parenthesis{1-(1+\qr) \pf_n - (1-\qr) \pnf_n}^{n-k}\\
        & \cdot \parenthesis{\frac{2 \qr \pf_n}{(1+\qr)\pf_n + (1-\qr)\pnf_n} \ytwohatf{\qvec} + \frac{(1-\qr) \cdot (\pf_n + \pnf_n)}{(1+\qr)\pf_n + (1-\qr)\pnf_n} \cdot \ytwohatnf{\qvec}}^{k-1}\\
        & \cdot \parenthesis{(1+\qr) \pf_n + (1-\qr) \pnf_n-2 \qr \pf_n \ytwohatf{\qvec} - (1-\qr) \cdot (\pf_n + \pnf_n)\ytwohatnf{\qvec}}\\
        =&\;\sum_{k = 0}^{n} \binom{n-1}{k} \parenthesis{(1+\qr) \pf_n + (1-\qr) \pnf_n}^{k}\parenthesis{1-(1+\qr) \pf_n - (1-\qr) \pnf_n}^{n-1-k}\\
        & \cdot \parenthesis{\frac{2 \qr \pf_n}{(1+\qr)\pf_n + (1-\qr)\pnf_n} \ytwohatf{\qvec} + \frac{(1-\qr) \cdot (\pf_n + \pnf_n)}{(1+\qr)\pf_n + (1-\qr)\pnf_n} \cdot \ytwohatnf{\qvec}}^{k} \vspace{1in}\\
        & \cdot n \cdot \parenthesis{(1+\qr) \pf_n + (1-\qr) \pnf_n-2 \qr \pf_n \ytwohatf{\qvec} - (1-\qr) \cdot (\pf_n + \pnf_n)\ytwohatnf{\qvec}}\\
        =&\;\sum_{k = 0}^{n} \binom{n-1}{k} \parenthesis{1-(1+\qr) \pf_n - (1-\qr) \pnf_n}^{n-1-k} \cdot \parenthesis{2 \qr \pf_n \ytwohatf{\qvec} + (1-\qr) \cdot (\pf_n + \pnf_n) \ytwohatnf{\qvec}}^{k}\\
        & \cdot \parenthesis{(1+\qr) \af + (1-\qr) \anf-2 \qr \af \ytwohatf{\qvec} - (1-\qr) \cdot (\af + \anf)\ytwohatnf{\qvec}}\\
        =&\; \parenthesis{1-(1+\qr) \pf_n - (1-\qr) \pnf_n + 2 \qr \pf_n \ytwohatf{\qvec} + (1-\qr) \cdot (\pf_n + \pnf_n) \ytwohatnf{\qvec}}^{n-1}\\
        & \cdot \parenthesis{(1+\qr) \af + (1-\qr) \anf-2 \qr \af \ytwohatf{\qvec} - (1-\qr) \cdot (\af + \anf)\ytwohatnf{\qvec}}\\
        =&\; \squarebracket{1-\frac{2 \qr \af \parenthesis{1-\ytwohatf{\qvec}} + (1-\qr) \cdot (\af + \anf)\parenthesis{1-\ytwohatnf{\qvec}}}{n}}^{n-1}\\
        & \cdot \parenthesis{(1+\qr) \af + (1-\qr) \anf-2 \qr \af \ytwohatf{\qvec} - (1-\qr) \cdot (\af + \anf)\ytwohatnf{\qvec}}\\
        =&\; e^{-2 \qr \af \parenthesis{1-\ytwohatf{\qvec}}-(1-\qr) \cdot (\af + \anf)\parenthesis{1-\ytwohatnf{\qvec}}} \\
        & \cdot \parenthesis{(1+\qr) \af + (1-\qr) \anf-2 \qr \af \ytwohatf{\qvec} - (1-\qr) \cdot (\af + \anf)\ytwohatnf{\qvec}} \text{ as } n \rightarrow \infty\\
        =&\; \yonef{\qvec}\parenthesis{(1+\qr) \af + (1-\qr) \anf -\qr 2\af \ytwohatf{\qvec} - \parenthesis{1-\qr} \parenthesis{\af + \anf} \ytwohatnf{\qvec}} \text{ as } n \rightarrow \infty.
    \end{align*}

    The third equality above follows from $k \binom{n}{k} = n \binom{n-1}{k-1}$, the fourth equality substitutes $k$ with $k-1$ everywhere and starts summation from $k = 0$, and the sixth equality is an application of the Binomial Theorem.

    Thus, $\PP{v \text{ is a loser}|v \text{ is a flexible node in }\setl}$ is given by
    \begin{align*}
        \yonef{\qvec} + \yonef{\qvec}\parenthesis{(1+\qr) \af + (1-\qr) \anf -\qr 2\af \ytwohatf{\qvec} - \parenthesis{1-\qr} \parenthesis{\af + \anf} \ytwohatnf{\qvec}}.
    \end{align*}

    The probabilities conditional on $v$ being a regular node or being from $\setr$ can be derived analogously.\Halmos

\emph{Proof of \cref{prop:derivation}}
    Following closely the proof of Theorem $9$ in \citet{karp1981maximum}, we start by showing that a random tree is a good approximation to the structure obtained by conducting a breadth-first search from $v.$ Denote the subgraph of $G$ induced by vertices at most distance~$d$ from~$v$ as the $d$-neighborhood of~$v$. A vertex~$v$ is referred to as a $d$-target if there exists a derivation proving $v$ to be a target within the $d$-neighborhood of~$v$. Note that if $d-$neighborhood proves that $v$ is a target then $v$ is a target in any other graph that yields the same $d-$neighborhood.
Let \begin{align*}
        Y_n :=&\; \PP{v \text{ is a target in } G_n \mid v \text{ is a flexible node in } V_l},\\
        Y^d_n :=&\; \PP{v \text{ is a $d-$target in } G_n \mid v \text{ is a flexible node in } V_l},\\
        Y^d :=&\; \PP{v \text{ is a $d-$target root in } \bar{G} \mid v \text{ is a flexible node in } \setl}.
    \end{align*}

    % We next leverage a result from \citet{karp1981maximum} that shows, for large $n$, the probability that a $d-$neighborhood occurs in a random graph approaches the probability of that $d-$neighborhood occurring in a random tree.
    
    \cref{cl:limit_exchange} shows that, for large $n$, the probability that a $d-$neighborhood occurs in a random graph approaches the probability of that $d-$neighborhood occurring in a random tree.

    \begin{claim}\label{cl:limit_exchange}
        $\lim_{n \rightarrow \infty} Y_n = \lim_{n \rightarrow \infty} \lim_{d \rightarrow \infty} Y^d_n = \lim_{d \rightarrow \infty} \lim_{n \rightarrow \infty} Y^d_n = \lim_{d \rightarrow \infty} Y^d = \ytwof{\qvec}.$
    \end{claim}
    
    That is, $\PP{v \text{ is a target} \mid v \text{ is a flexible node in } V_l} = \ytwof{\qvec}$ as $n \rightarrow \infty$. Similarly, the probabilities for $v$ to be a target or a loser, when $ v \in V_l$ or $V_r,$ and when $v$ is a flexible or regular node, follow those derived for random trees in \cref{lem:target_loser2}.

    Since all members of derivations are targets or losers or both, we next find the probability for a random edge $(v,u)$ in $G$ to satisfy both $v \bigotimes u$ and $u \bigotimes v$ i.e., $v$ is both a target and a loser. By Theorem 9 (3) in \citet{karp1981maximum}, in \tree this occurs if and only if both $u$ and $v$ are in $L$. We compute this probability conditional on the types of root nodes $v$, and the extension from \tree to $G$ follows from \cref{cl:limit_exchange}. 
    \begin{align*}
        \lim_{n \rightarrow \infty} &\PP{v \bigotimes u \text{ and } u \bigotimes v \mid v \text{ is a flexible node in } V_l} \\
        =&\; \PP{v, u \text{ are both in } L \mid v \text{ is a flexible node in } V_l}\\
        =&\; \yonef{\qvec} \PP{u \text{ is a flexible node} \mid (v,u) \in E} \yonehatf{\qvec} + \yonef{\qvec} \PP{u \text{ is a regular node} \mid (v,u) \in E} \yonehatnf{\qvec} \\
        =&\; \yonef{\qvec} \frac{\qr 2 \af}{(1+\qr) \af + (1-\qr) \anf}\yonehatf{\qvec} + \yonef{\qvec} \frac{(1-\qr) (\af + \anf)}{(1+\qr) \af + (1-\qr) \anf}\yonehatnf{\qvec}.
    \end{align*}

    Similarly, we find 
    \begin{align*}
        \lim_{n \rightarrow \infty} &\PP{v \bigotimes u \text{ and } u \bigotimes v \mid v \text{ is a regular node in } V_l} \\
        =&\; \yonenf{\qvec} \frac{\qr (\af + \anf)}{\qr \af + (2-\qr) \anf}\yonehatf{\qvec} + \yonenf{\qvec} \frac{(1-\qr)2 \anf}{\qr \af + (2-\qr) \anf}\yonehatnf{\qvec},\\
        \lim_{n \rightarrow \infty} &\PP{v \bigotimes u \text{ and } u \bigotimes v \mid v \text{ is a flexible node in } V_r} \\
        =&\; \yonehatf{\qvec} \frac{\ql 2 \af}{(1+\ql) \af + (1-\ql) \anf}\yonef{\qvec} + \yonehatf{\qvec} \frac{(1-\ql) (\af + \anf)}{(1+\ql) \af + (1-\ql) \anf}\yonenf{\qvec},\\
        \lim_{n \rightarrow \infty} &\PP{v \bigotimes u \text{ and } u \bigotimes v \mid v \text{ is a regular node in } V_r} \\
        =&\; \yonehatnf{\qvec} \frac{\ql (\af + \anf)}{\ql \af + (2-\ql) \anf}\yonef{\qvec} + \yonehatnf{\qvec} \frac{(1-\ql)2 \anf}{\ql \af + (2-\ql) \anf}\yonenf{\qvec}.
    \end{align*}
    \begin{align*}
        \text{Thus,}\quad&\lim_{n \rightarrow \infty} \PP{v \text{ in a derivation} \mid v \text{ is a flexible node in } V_l}\\
        = & \lim_{n \rightarrow \infty} \PP{v \text{ is a target} \mid v \text{ is a flexible node in } V_l} + \lim_{n \rightarrow \infty} \PP{v \text{ is a loser} \mid v \text{ is a flexible node in } V_l} \\
        &- \lim_{n \rightarrow \infty} \PP{v \bigotimes u \text{ and } u \bigotimes v \mid v \text{ is a flexible node in } V_l} \cdot \EE{\left|u \text{ s.t. } (v,u) \in E\right| \biggr\lvert v \text{ is a flexible node in } V_l}\\
        =&\; \ytwof{\qvec} + \yonef{\qvec} + \yonef{\qvec}\parenthesis{(1+\qr) \af + (1-\qr) \anf -\qr 2\af \ytwohatf{\qvec} - \parenthesis{1-\qr} \parenthesis{\af + \anf} \ytwohatnf{\qvec}}\\
        &-\squarebracket{\yonef{\qvec} \frac{\qr 2 \af}{(1+\qr) \af + (1-\qr) \anf}\yonehatf{\qvec} + \yonef{\qvec} \frac{(1-\qr) (\af + \anf)}{(1+\qr) \af + (1-\qr) \anf}\yonehatnf{\qvec}} \cdot \parenthesis{(1+\qr) \af + (1-\qr) \anf}\\
        = &\ytwof{\qvec} + \yonef{\qvec} + \yonef{\qvec} \squarebracket{(1+\qr) \af + (1-\qr) \anf}\cdot \\
        &\squarebracket{1-\frac{\qr 2 \af}{(1+\qr) \af + (1-\qr) \anf} \parenthesis{\yonehatf{\qvec} + \ytwohatf{\qvec}}-\frac{(1-\qr) \parenthesis{\af + \anf}}{(1+\qr) \af + (1-\qr) \anf} \parenthesis{\yonehatnf{\qvec} + \ytwohatnf{\qvec}}}.
    \end{align*}

    The probabilities conditional on other types of nodes are computed similarly. \Halmos

%\end{proof}

\begin{proof}{Proof of \cref{cl:limit_exchange}}
    % The proof of this claim follows the proof of Theorem 9 (1) in \citet{karp1981maximum}. 
    The first and last equalities in the claim arise directly from our definitions and \cref{lem:target_loser2}. We now prove the third and the second equality. By the Poisson Limit Theorem, for any flexible node $v \in V_l,$ the distribution of its degrees follows $\poisson\parenthesis{\qr 2 \af + (1-\qr) (\af +\anf)}$ as $n \to \infty.$ Thus, for any $\epsilon > 0$ there is a constant $k$ such that $\lim_{n \to \infty} \PP{\text{there is a node of degree $>k$ in the $d-$neighborhood of $v$}} < \epsilon.$ The rest of the analyses exactly follow the proof of Theorem 9 (1) in \citet{karp1981maximum}, which shows $\lim_{n \rightarrow \infty} Y^d_n = Y^d$ because (i) there are finitely many $d-$neighborhoods that lack a vertex of degree $>k$, and (ii) the probability of encountering each such $d-$neighborhood around a node $v$ in $G_n$ is close to that around a root $v$ in $\bar{G}$ as $n \to \infty$. This proves the third equality.
    
    Finally, we justify the limit exchange in the second equality. We leverage the result from Theorem 9 (1) in \citet{karp1981maximum} that, for every positive $\epsilon$, there exists a $d$ such that for all sufficiently large $n$, \(\PP{v \text{ is a target but not a \(d\)-target}} < \epsilon\).\footnote{This result relies on probabilistic bounds on the length of a shortest derivation that proves $v$ to be a target.} This implies that for all sufficiently large $n, n' d, d', Y^d_n$ is close to $Y^{d'}_{n'}$ and thus we may exchange limit. 
\end{proof}

\emph{Proof of \cref{cl:uniqueness}}
    First, we calculate the derivative of \( f_1(x_1) \):
    \begin{align*}
        f_1'(x_1) =&\; \frac{\partial e^{- \frac{1}{2}\parenthesis{\af +\anf} x_1 +2 \frac{\anf}{\af+\anf}\parenthesis{\log(x_1) + \af x_1}} + 2 \frac{\log(x_1) + \af x_1}{\af + \anf}}{\partial x_1}\\
        =&\; e^{- \frac{1}{2}\parenthesis{\af +\anf} x_1 +2 \frac{\anf}{\af+\anf}\parenthesis{\log(x_1) + \af x_1}} \parenthesis{- \frac{1}{2}\parenthesis{\af +\anf} +2 \frac{\anf}{\af+\anf}\parenthesis{\af + 1/x_1}} + \frac{2}{\af+\anf}\parenthesis{\af + 1/x_1}.
    \end{align*}
    We next employ a computer-aided proof to verify that $f_1'(x_1) > 1$ for any $x_1 \in (0,1]$ when $10^{-4} < \anf < \af$ and $\af + \anf < e$. Fixing $\delta_1, \delta_2 > 0$, we establish a lower bound on the value of \( f_1'(x_1) \) for any $\parenthesis{\bar{\alpha}^f, \bar{\anf}, \bar{x}_1}$ in the set \( [\af,\af + \delta_1) \times [\anf,\anf + \delta_1) \times [x_1,x_1 + \delta_2) \), where $10^{-4} < \anf < \af$ and \( \af + \anf < e \). We find that:
    \begin{equation}\label{eq:f_derivative_lb}
        \begin{split}
            f_1'(x_1) \geq & \ e^{- \frac{1}{2}\parenthesis{\af +\anf} x_1 +\frac{2 \parenthesis{\af+\delta_1} \parenthesis{\anf+\delta_1} \parenthesis{x_1+\delta_2}}{\af+\anf}+\frac{2 \anf \log(x_1 + \delta_2)}{\af+\anf+2 \delta_1}} \parenthesis{- \frac{1}{2}\parenthesis{\af +\anf + 2 \delta_1}}\\
            &+e^{- \frac{1}{2}\parenthesis{\af +\anf + 2 \delta_1} \parenthesis{x_1+\delta_2} + \frac{2 \af \anf x_1}{\af+\anf + 2\delta_1}+\frac{2 \parenthesis{\anf + \delta_1} \log(x_1 )}{\af+\anf}} \parenthesis{\frac{2\anf \parenthesis{\af + \frac{1}{x_1 + \delta_2}}}{\af+\anf + 2 \delta_1}}\\
            &+\frac{2 \parenthesis{\af + \frac{1}{x_1 + \delta_2}}}{\af+\anf + 2 \delta_1} \text{ when } x_1 > 0,\\
            f_1'(x_1) \geq & \ e^{- \frac{1}{2}\parenthesis{\af +\anf} x_1 +\frac{2 \parenthesis{\af+\delta_1} \parenthesis{\anf+\delta_1} \parenthesis{x_1+\delta_2}}{\af+\anf}+\frac{2 \anf \log(x_1 + \delta_2)}{\af+\anf+2 \delta_1}} \parenthesis{- \frac{1}{2}\parenthesis{\af +\anf + 2 \delta_1}}\\
            &+\frac{2 \parenthesis{\af + \frac{1}{x_1 + \delta_2}}}{\af+\anf + 2 \delta_1} \text{ when } x_1 = 0.
        \end{split}
    \end{equation}

    For given \( \delta_1, \delta_2 >0 \), if the lower bound in \eqref{eq:f_derivative_lb} is greater than $1$, then \( f_1'(x_1) > 1 \) for any \( (\bar{\alpha}^f, \bar{\anf}, \bar{x}_1) \) in the corresponding set \( [\af, \af + \delta_1) \times [\anf,\anf + \delta_1) \times [x_1,x_1 + \delta_2) \). In the computational notebook titled \texttt{Claim3.ipynb},\footnote{The computer-aided proof can be found at \url{https://bit.ly/3VhfumL}.} we compute the value of \eqref{eq:f_derivative_lb} for \( x_1 = 0, \delta_2, 2 \delta_2, \ldots, 1 \) and \( \af, \anf = 10^{-4}, \delta_1, 2 \delta_1, \ldots, e \), under the constraint \( \af + \anf < e \). Through this computation, we find that taking $\delta_1 = \delta_2 = 0.01$ is sufficient for verifying \eqref{eq:f_derivative_lb} $> 1$ for all $x_1 \in (0,1], 10^{-4} < \anf < \af$ and $\af + \anf < e$.\Halmos
%\end{proof}

\subsubsection{Computational results based on \texorpdfstring{$\munum(\ql,\qr)$}{Lg}.}\label{sec:num_construction} To compute \(\munum(\ql,\qr)\), we resort to the \texttt{NLsolve} package available in the Julia Programming Language to solve \eqref{eq:w} at a tolerance level \(ftol = 10^{-8}\). Specifically, \texttt{NLsolve} iteratively refines candidate solutions using the Trust Region Method until the infinite norm of the residuals of the current solution falls below the threshold \(ftol = 10^{-8}\) \citep{NLsolve.jl2023}. When $10^{-4} < \anf < \af$ and $\af + \anf < e$, at $\qvec = (1/2,1/2), (1,0)$ and $(0,1)$ we find that \eqref{eq:w} reduces to a single non-linear equation that exhibits strict monotonicity on both sides and can be solved to provable precision using this method (see the proof of \cref{thm:compare_q} for details). In other cases where there are no theoretical guarantees for the closeness of \(\munum(\ql,\qr)\) to $\mu(\ql,\qr)$, numerical studies indicate that \(\muemp_n(\ql,\qr)\) still tends to converge to \(\munum(\ql,\qr)\) as \(n\) scales large. We thus employ \(\munum(\ql,\qr)\) to compare the one-sided and the balanced allocations for a wide range of parameters. Specifically, the findings presented in \cref{sec:ks} are based on values of $\parenthesis{\budget, \ql, \af, \anf}$ in a set $S$ that contains all $B \in \bracket{0.1,0.2,\cdots,1}, \ql \in \bracket{0,0.01,\cdots,1}, \af \in \bracket{0.05,0.10,\cdots,7.45,7.50}$ and $\anf \in \bracket{0, 0.05,0.10,\cdots,2.95,3.00}$ such that $B \geq \ql$ and $\af > \anf$.

\section{Degree-0 Metric (Proof of \Cref{thm:deg0_metric})}\label{app:deg0}

% \dfcomment{no need to have a subsection...}

\dfedit{

%\subsection{Proof of \texorpdfstring{\cref{thm:deg0_metric}}{Lg}}\label{app:deg0_metric_proof}
% \emph{. }
We first derive the closed-form expression for $\phi(\ql,\qr)$, then prove that it always takes a maximum at either $(B,0)$ or $\parenthesis{B/2, B/2}$, and finally %followed by the necessary and sufficient condition for 
compare the two solutions.

\noindent\textbf{Step 1: Deriving $\phi(\ql,\qr)$.} 
For a node $\vl_i \in V_l$, the probability of being isolated is the probability that no edge realizes to any of the $n$ right nodes. Conditioning on its flexibility type, this equals
\begin{align*}
\PP{\deg(v_i^l)=0}
&=\;\underbrace{\ql}_{\text{$\vl_i$ is flexible}}
\Big(
\underbrace{\qr}_{\substack{\vr_j\text{ is flexible}}}
\Bigl(1-\tfrac{2\af}{n}\Bigr)
+
\underbrace{(1-\qr)}_{\substack{\vr_j\text{ is regular}}}
\Bigl(1-\tfrac{\af+\anf}{n}\Bigr)
\Big)^{n} \\
&\quad+
\underbrace{(1-\ql)}_{\text{$\vl_i$ is regular}}
\Big(
\underbrace{\qr}_{\substack{\vr_j\text{ is flexible}}}
\Bigl(1-\tfrac{\af+\anf}{n}\Bigr)
+
\underbrace{(1-\qr)}_{\substack{\vr_j\text{ is regular}}}
\Bigl(1-\tfrac{2\anf}{n}\Bigr)
\Big)^{n}\\
&=\;\ql \Bigl(
1-\tfrac{2\af\qr+(\af+\anf)(1-\qr)}{n}
\Bigr)^{n} +
(1-\ql)
\Bigl(
1-\tfrac{(\af+\anf)\qr+2\anf(1-\qr)}{n}
\Bigr)^{n}.
\end{align*}
% \noindent\textbf{Step 1: Deriving $\phi(\ql,\qr)$.} 
% For a node $\vl_i \in V_l$, the probability of being isolated is the probability that no edge to any of the $n$ right nodes forms. Conditioning on its flexibility type, this equals
% \begin{align*}
% \PP{\deg(v_i^l)=0}
% &=\;\underbrace{\ql}_{\text{$\vl_i$ is flexible}}
% \underbrace{\Bigl(1-\tfrac{2\af}{n}\Bigr)^{\,\qr n}}_{\substack{\text{no edge to flexible }\vr_j}}
% \underbrace{\Bigl(1-\tfrac{\af+\anf}{n}\Bigr)^{\,(1-\qr)n}}_{\substack{\text{no edge to regular }\vr_j}}
% +\underbrace{(1-\ql)}_{\text{$\vl_i$ is regular}}
% \underbrace{\Bigl(1-\tfrac{\af+\anf}{n}\Bigr)^{\,\qr n}}_{\substack{\text{no edge to flexible }\vr_j}}
% \underbrace{\Bigl(1-\tfrac{2\anf}{n}\Bigr)^{\,(1-\qr)n}}_{\substack{\text{no edge to regular }\vr_j}}.
% \end{align*}
By symmetry, the expected isolated fraction on the right is obtained by swapping $\ql$ and $\qr$.  

Taking $n\to\infty$ and using $(1-x/n)^n\to e^{-x}$ gives
\begin{align*}
\phi_1(\ql,\qr)
&=\ql\,e^{-\bigl(2\af\qr+(\anf+\af)(1-\qr)\bigr)}+(1-\ql)\,e^{-\bigl((\anf+\af)\qr+2\anf(1-\qr)\bigr)},\\
\phi_2(\ql,\qr)
&=\qr\,e^{-\bigl(2\af\ql+(\anf+\af)(1-\ql)\bigr)}+(1-\qr)\,e^{-\bigl((\anf+\af)\ql+2\anf(1-\ql)\bigr)}.
\end{align*}
Thus, the limiting maximum fraction of isolated nodes is $\max\{\phi_1(\ql,\qr),\phi_2(\ql,\qr)\}$, and therefore
\[
\phi(\ql,\qr)=1-\max\{\phi_1(\ql,\qr),\ \phi_2(\ql,\qr)\}.
\]
% For $\ql,\qr\in[0,1]$ define
% \begin{align*}
% \phi_1(\ql,\qr)
% &:=\ql\,e^{-\parenthesis{2\af\,\qr+(\anf+\af)(1-\qr)}}
% +(1-\ql)\,e^{-\parenthesis{(\anf+\af)\qr+2\anf(1-\qr)}},\\
% \phi_2(\ql,\qr)
% &:=\qr\,e^{-\parenthesis{2\af\,\ql+(\anf+\af)(1-\ql)}}
% +(1-\qr)\,e^{-\parenthesis{(\anf+\af)\ql+2\anf(1-\ql)}}.
% \end{align*}
% We take
% \[
% \phi(\ql,\qr)\;:=\;1-\max\{\phi_1(\ql,\qr),\ \phi_2(\ql,\qr)\}.
% \]
\noindent\textbf{Step 2: Optimality occurs at either the balanced or the one-sided allocation.} Let $d:=\af-\anf$ and $x:=e^{-d}>0$. If $d = 0,$ we trivially have $\phi(\ql, B- \ql) = \phi(\ql', B- \ql') \; \forall \ql, \ql' \in [0,B],$ so both the balanced and the one-sided allocations are optimal. Thus, we focus on the case where $d > 0$, i.e., $\af > \anf.$ 

We start by showing that $\phi_1(\ql, B-\ql)$ is monotonically increasing in $\ql \in [0,\ql^\star]$ for some $\ql^\star \geq B/2$ and monotonically decreasing thereafter. This property will later allow us to prove the optimality of the two special allocations. First, a direct algebraic rearrangement yields that
the two exponentials appearing in $\phi_1(\ql,B-\ql)$ have a constant ratio (independent of $\ql$):
\[
\frac{e^{-\,[2\af(\budget-\ql)+(\anf+\af)(1-\budget+\ql)]}}
     {e^{-\,[\,(\anf+\af)(\budget-\ql)+2\anf(1-\budget+\ql)\,]}}
= e^{-(\af-\anf)} \;=\; x.
\]

Hence we can factor one exponential and write
\begin{equation}\label{eq:phi1_factor}
\phi_1(\ql, B - \ql)
= C_\budget\,e^{d\ql}\,\Big(1-(1-x)\,\ql\Big), 
\qquad\text{where}\quad
C_\budget:=\exp\!\big(-[(\anf+\af)\budget+2\anf(1-\budget)]\big)>0.
\end{equation}
Differentiating \eqref{eq:phi1_factor} with respect to $\ql$ gives
\begin{align}
\frac{\partial \phi_1(\ql, B-\ql)}{\partial \ql}
&= C_\budget\,e^{d\ql}\Big(d\big(1-(1-x)\ql\big)-(1-x)\Big).\label{eq:phi1_derivative}
\end{align}
Since $C_\budget\,e^{d\ql}>0$, the sign of $\frac{\partial \phi_1(\ql, B-\ql)}{\partial \ql}$ is determined by the \emph{affine}
function
\[
\;d\big(1-(1-x)\ql\big)-(1-x)\;=\;-d(1-x)\,\ql\;+\;
% \underbrace{
\big(d-(1-x)\big)
%}_{\text{constant}}.
\]
As the second summand is independent of $\ql$, this function is linear in $\ql$, and its first-order condition yields:
\begin{equation}\label{eq:qstar_def}
\ql^\star
\;=\;\min\bracket{\frac{d-(1-x)}{d(1-x)}, B},
\qquad\text{whenever }d\neq 0.
\end{equation}
This proves that $\phi_1(\ql, B-\ql)$ is monotone increasing in $\ql \in [0,\ql^\star]$ and monotone decreasing thereafter.

We next show that $\frac{d-(1-x)}{d(1-x)} \geq \tfrac{1}{2} \geq \tfrac{B}{2}$. Since $B \geq B/2,$ this would immediately imply that $\ql^\star \ge \tfrac{B}{2}$. Using $x=e^{-d}$ and simple algebra,
\[
\frac{d-(1-x)}{d(1-x)} \;\ge\;\tfrac{1}{2}
\iff
2\big(d-(1-x)\big)\;\ge\;d(1-x)
\iff
2d-(2+d)(1-x)\;\ge\;0.
\]
{Since $x=e^{-d} \ge \frac{2-d}{2+d}$ for any $d \geq 0$, we have
\[
(2+d)(1-x)\;\le\;(2+d)\Big(1-\frac{2-d}{2+d}\Big)\;=\;2d,
\]}
which proves $2d-(2+d)(1-x)\ge 0$ and thus $\ql^\star\ge \tfrac{B}{2}$.
%Because $\budget\le 1$, this implies $\ql^\star\ge \tfrac{1}{2}\ge \tfrac{\budget}{2}$.

We next leverage these properties of $\phi_1(\ql,B-\ql)$ to prove that optimality of $\phi(\ql,B-\ql)$ occurs at either the balanced or the one-sided allocation. %Notice that when $\ql^\star \geq B$, $\phi_1(\ql,B-\ql)$ is monotonically increasing for all $\ql \in [0,B]$ and $\phi_2(\ql,B-\ql) = \phi_1(B-\ql,\ql)$ is monotonically decreasing for all $\ql \in [0,B]$. Thus, $\max\bracket{\phi_1(\ql, B-\ql), \phi_2(\ql, B-\ql)}$ is minimized at the balanced allocation $(B/2,B/2)$, and $\phi(\ql,B-\ql)$ is maximized at $(B/2, B/2)$.
As $\ql^\star \leq B,$ from $\phi_2(\ql, B-\ql) = \phi_1(B-\ql, \ql)$ we obtain $$\max\bracket{\phi_1(\ql, B-\ql), \phi_2(\ql, B-\ql)} = \phi_1(\ql, B-\ql) \; \forall \ql \in [B/2, \ql^\star].$$ By the monotonicity of $\phi_1(\ql, B-\ql)$ in this same region, we conclude that $$\max\bracket{\phi_1(B/2,B/2), \phi_2(B/2,B/2)} \leq \max\bracket{\phi_1(\ql, B-\ql), \phi_2(\ql, B-\ql)} \; \forall \ql \in [B/2, \ql^\star]$$ and thus $\phi(B/2,B/2) \geq \phi(\ql,B-\ql) \; \forall \ql \in [B/2, \ql^\star]$. By symmetry of the function $\phi(\ql,B-\ql)$ around $(B/2,B/2)$, we obtain:
\begin{align}\label{eq:deg0_analysis1}
    \phi(B/2,B/2) \geq \phi(\ql,B-\ql) \; \forall \ql \in [B-\ql^\star, \ql^\star].
\end{align}

Also, since both $\phi_1(\ql, B-\ql)$ and $\phi_2(\ql,B-\ql)$ are monotonically decreasing for $\ql \in [\ql^\star, B]$, we have $$\max\bracket{\phi_1(0,B), \phi_2(0,B)} \leq \max\bracket{\phi_1(\ql, B-\ql), \phi_2(\ql, B-\ql)} \; \forall \ql \in [\ql^\star,B].$$ Combined with the symmetry of function $\phi(\ql,B-\ql)$, we obtain 
\begin{align}\label{eq:deg0_analysis2}
\phi(B,0) = \phi(0,B) \geq \phi(\ql,B-\ql) \; \forall \ql \in [0,B-\ql^\star] \cup [\ql^\star,B].
\end{align}
Combining \eqref{eq:deg0_analysis1} and \eqref{eq:deg0_analysis2}, we find that one of $(B/2,B/2)$ or $(B,0)$ and $(0,B)$ optimizes $\phi(\ql,B-~\ql)$.

\noindent\textbf{Step 3: Comparing one-sided and balanced allocations.} It suffices to then compare the one-sided and the balanced allocations. By the definition of $\phi$,
\begin{equation}
\phi(\budget,0)\ \ge\ \phi\!\left(\tfrac{\budget}{2},\tfrac{\budget}{2}\right)
\iff
\max\big\{\phi_1(\budget,0),\ \phi_2(\budget,0)\big\}\ \le\
\phi_1\!\left(\tfrac{\budget}{2},\tfrac{\budget}{2}\right).
\label{eq:max-to-Phi-macro}
\end{equation}
A direct substitution gives
\begin{align*}
\phi_1(\budget,0)&=\budget\,e^{-(\anf+\af)}+(1-\budget)\,e^{-2\anf},\\
\phi_2(\budget,0)&=e^{-\parenthesis{2\anf+\budget(\af-\anf)}} = e^{-2 \anf} x^B,\\
\phi_1\!\Big(\tfrac{\budget}{2},\tfrac{\budget}{2}\Big)
&=\tfrac{\budget}{2}\,e^{-\parenthesis{\af \budget+(\anf+\af)(1-\tfrac{\budget}{2})}}
+\Big(1-\tfrac{\budget}{2}\Big)\,e^{-\parenthesis{(\anf+\af)\tfrac{\budget}{2}+2\anf(1-\tfrac{\budget}{2})}}
= e^{-2\anf} x^{B/2} \squarebracket{\parenthesis{1-B/2} + B/2 \cdot x}.
\end{align*}

Thus, the inequality
$\phi_2(\budget,0)\le \phi_1(\tfrac{\budget}{2},\tfrac{\budget}{2})$
is equivalent to
$x^{\budget/2}\ \le\ \Big(1-\tfrac{\budget}{2}\Big)+\tfrac{\budget}{2}\,x,$
which holds for any $\af \geq \anf$ and $B$. Hence from \eqref{eq:max-to-Phi-macro} we obtain the simplification
\begin{equation}
\phi(\budget,0)\ \ge\ \phi\!\left(\tfrac{\budget}{2},\tfrac{\budget}{2}\right)
\iff
\phi_1(\budget,0)\ \le\ \phi_1\!\left(\tfrac{\budget}{2},\tfrac{\budget}{2}\right).
\label{eq:iff-core-macro}
\end{equation}

With $x=e^{-(\af-\anf)}$,
\[
\phi_1(\budget,0)=e^{-2\anf}\big[(1-\budget)+\budget x\big],\qquad
\phi_1\!\Big(\tfrac{\budget}{2},\tfrac{\budget}{2}\Big)
= e^{-2\anf}\,x^{\budget/2}\!\Big[\Big(1-\tfrac{\budget}{2}\Big)+\tfrac{\budget}{2}x\Big].
\]
As $e^{-2\anf}>0$, \eqref{eq:iff-core-macro} is equivalent to
$
(1-\budget)+\budget x\ \le\ x^{\budget/2}\Big(1-\tfrac{\budget}{2}+\tfrac{\budget}{2}x\Big),$ so for $\budget\in(0,1]$ and any $(\anf,\af)$,
\[
\phi(\budget,0)\ \ge\ \phi\!\left(\tfrac{\budget}{2},\tfrac{\budget}{2}\right)
\iff
(1-\budget)+\budget\,e^{-(\af-\anf)}
\ \le\
e^{-\frac{\budget}{2}(\af-\anf)}\!\left(1-\tfrac{\budget}{2}+\tfrac{\budget}{2}\,e^{-(\af-\anf)}\right) \iff \phi^\star\parenthesis{B, \af, \anf} \geq 0.
\]
where $\phi^\star\parenthesis{B, \af, \anf} := e^{-\frac{\budget}{2}(\af-\anf)}\!\left(1-\tfrac{\budget}{2}+\tfrac{\budget}{2}\,e^{-(\af-\anf)}\right) - (1-\budget)-\budget\,e^{-(\af-\anf)}.$

\noindent\textbf{Step 4: The balanced allocation always dominates when $B \leq 2/3$.} Consider the auxiliary univariate function on $(0,1]$,
\[
h_\budget(u)
:= u^{\budget/2}\Bigl(1-\tfrac{\budget}{2}+\tfrac{\budget}{2}\,u\Bigr)
-(1-\budget)-\budget\,u,\qquad u\in(0,1],
\]
\[\text{so that}\qquad
\phi^\star\parenthesis{\budget,\af,\anf}=h_\budget\!\parenthesis{e^{-(\af-\anf)}}.
\]
Note that $h_\budget(1)=0$ and $\lim_{u\downarrow 0}h_\budget(u)=\budget-1<0$ for all $\budget\in(0,1]$.

Taking the first- and second-order derivatives of $h_\budget(u)$ yields:
\[
h_\budget'(u)
=\tfrac{\budget}{2}u^{\budget/2-1}\Bigl(1-\tfrac{\budget}{2}+\tfrac{\budget}{2}u\Bigr)
+u^{\budget/2}\cdot\tfrac{\budget}{2}-\budget, \text{ and}
\]
\begin{align*}
h_\budget''(u)
&=\tfrac{\budget}{2}\!\left(\tfrac{\budget}{2}-1\right)u^{\budget/2-2}\Bigl(1-\tfrac{\budget}{2}+\tfrac{\budget}{2}u\Bigr)
+\tfrac{\budget}{2}u^{\budget/2-1}\cdot\tfrac{\budget}{2}
+\tfrac{\budget}{2}\!\left(\tfrac{\budget}{2}\right)u^{\budget/2-1} \\[2mm]
&=\frac{\budget}{8}\,u^{\budget/2-2}\,\Bigl((\budget^2+2\budget)\,u+(-\budget^2+4\budget-4)\Bigr).
\end{align*}
For each fixed $\budget>0$, the bracket is affine and increasing in $u$; thus for $u\in(0,1]$ it is upper bounded by its value at $u=1$:
\[
(\budget^2+2\budget)\,u+(-\budget^2+4\budget-4)\ \le\ 6\budget-4.
\]
\[\text{Hence,
}\qquad
\budget\le\tfrac{2}{3}\ \Longrightarrow\ h_\budget''(u)\le 0\quad\text{for all }u\in(0,1],
\]
i.e., $h_\budget$ is concave on $(0,1]$ whenever $\budget\le\tfrac{2}{3}$. Because $h_\budget$ is concave and continuous on $[0,1]$ with $h_\budget(1)=0$ and $h_\budget(0^+)=\budget-1<0$, its maximum over $[0,1]$ equals $0$ and is attained at $u=1$. Therefore,
\[
\budget\le\tfrac{2}{3}\ \Longrightarrow\ h_\budget(u)\le 0\quad\text{for all }u\in(0,1].
\]
Evaluating at $u=e^{-(\af-\anf)}\in(0,1]$ yields
\[
\budget\le\tfrac{2}{3}\ \Longrightarrow\ \phi^\star\parenthesis{\budget,\af,\anf}\le 0\ \Longrightarrow\  \phi(\budget,0)\ \leq\ \phi\!\left(\tfrac{\budget}{2},\tfrac{\budget}{2}\right) \quad\text{for all }(\af,\anf).
\]

In contrast, when $\budget>\tfrac{2}{3}$, we have:
\[
h_\budget(1)=0,\qquad h_\budget'(1)=0,\qquad
h_\budget''(1)=\frac{\budget(6\budget-4)}{8}>0,
\]
so $u=1$ is a strict local \emph{minimum} of $h_\budget$. Therefore there exists $u\in(0,1)$ sufficiently close to $1$ with $h_\budget(u)>0$. Since every such $u$ can be written as $u=e^{-(\af-\anf)}$ for some $(\af,\anf)$, we obtain
\[
\phi^\star\parenthesis{\budget,\af,\anf}=h_\budget\!\parenthesis{e^{-(\af-\anf)}}>0
\]
for a suitable choice of $(\af,\anf)$. This shows that $\phi(\budget,0)\ \leq\ \phi\!\left(\tfrac{\budget}{2},\tfrac{\budget}{2}\right)$ for all $(\af,\anf)$ if and only if $B \leq 2/3.$

\noindent\textbf{Step 5: The one-sided allocation always dominates when $B = 1$.}
At $B = 1$, the optimality condition for the one-sided allocation reduces to 
$$e^{-\parenthesis{\af - \anf}} \leq \frac{1}{2} e^{-\frac{1}{2}\parenthesis{\af - \anf}} \parenthesis{1+e^{-\parenthesis{\af - \anf}}}.$$ 
We divide both sides by \(e^{-\parenthesis{\af - \anf}/2}>0\) to obtain
$e^{-\parenthesis{\af - \anf}/2}\ \le\ \tfrac{1}{2}\Big(1+e^{-\parenthesis{\af - \anf}}\Big).$
Defining the auxiliary variable \(y:=e^{-\parenthesis{\af - \anf}/2}\in(0,1]\), this is equivalent to $
y\ \le\ \tfrac{1}{2}\big(1+y^2\big)$,
which always holds because $\tfrac{1}{2}\big(y^2-2y+1\big)\;=\;\tfrac{1}{2}\,(y-1)^2 \geq 0.$ Thus, at $B = 1$ we have $$\phi(B,0) = \max_{\ql \in [0,B]} \phi(\ql,B - \ql), \forall \af \geq \anf, B \in [0,1].\qquad\Halmos$$
%\end{proof}
}

\section{Proofs of the Local Model}\label{sec:proof_local}

In this section, we prove the results for the local model. We derive a closed-form expression for \(\mu(\ql,\qr)\) as a rational function, in which both the denominator and numerator are eighth-order polynomials in terms of \(\ql\), \(\qr\), \(\pf\), and \(\pnf\). This expression facilitates the comparison between one-sided and balanced allocation, as well as the analysis of the convexity and concavity properties of \(\mu(\ql,\qr)\) along specific diagonals where analyses of directional second-order derivatives are tractable.

\subsection{Proof of \texorpdfstring{\cref{thm:compare_local_model}}{Lg}}\label{app:compare_local_model}

% \begin{proof}{Proof. }
    We begin by characterizing the asymptotic fraction of nodes matched in the local model. Let $E$ denote the set of all edges. We propose \cref{alg:local} to construct a matching \( M \), and argue that it is at most $1$ below the size of a maximum matching (thus, the matching probability under \cref{alg:local} is the same, asymptotically, as that under a maximum matching).

    \begin{algorithm}\caption{Maximum Matching Construction in Local Model}\label{alg:local}
    \begin{algorithmic}[1]
    \State Initialize the matching set $M \gets \emptyset$
    \If{$\parenthesis{\vl_1, \vr_1} \in E$}
        \State Add $\parenthesis{\vl_1, \vr_1}$ to $M$
    \EndIf
    \For{each subsequent node $\vr_i$ with $i > 1$}
        \If{$\parenthesis{\vl_{i-1}, \vr_i} \in E$ and $\vl_{i-1}$ is not already matched in $M$}
            \State Add $\parenthesis{\vl_{i-1}, \vr_i}$ to $M$
        \ElsIf{$\parenthesis{\vl_i, \vr_i} \in E$}
            \State Add $\parenthesis{\vl_i, \vr_i}$ to $M$
        \EndIf
    \EndFor
    \State \textbf{return} $M$
    \end{algorithmic}
    \end{algorithm}
    \vspace{-.15in}

    Our analysis focuses on the nodes in $V_r$ that are being matched through \cref{alg:local}, as opposed to the ones that could be matched in a maximum matching. Recall that each node $\vl_i$ in $V_l$ can only connect to its two neighbors in $V_r$. If $(\vl_i, \vr_i) \notin E$ and $(\vl_i, \vr_{i +1}) \in E$, then there exists a maximum matching that contains $(\vl_i, \vr_{i+1})$. This is because node $\vl_i$ cannot be matched otherwise, and not using $(\vl_i, \vr_{i + 1})$ in the matching would at most save $\vr_{i+1}$ for one additional match. Hence, this algorithm is provably optimal for all nodes in $V_r$ except $\vr_{1}$, which is myopically matched to $\vl_1$ if an edge exists. As we are interested in computing \(\mu(\ql,\qr) = \lim_{{n \to \infty}} \EE{\frac{\mathcal{M}_n(\ql,\qr)}{n}}\), the resulting asymptotic error in $\mu(\ql,\qr)$ approaches $0$. % in the asymptotic setting where $n \to \infty$.

    Therefore, to compute $\mu(\ql,\qr)$ it is sufficient to calculate the asymptotic fraction of nodes matched through this algorithm. Observe that whether a node $\vr_{i} \in V_r$ is matched to $\vl_{i-1} \in V_l$ depends only on edges incident to $\vl_{i-1}$ and is independent of $\vl_{i}$. For each $\vr_{i}$ with $i \in \{2,...,n\},$ we define
    \begin{equation*}
        \begin{split}
            x^f_i :=\; \PP{\parenthesis{\vl_{i-1}, \vr_{i}} \in M \mid \vr_{i} \text{ is flexible}} \text{ and } x^n_i :=\; \PP{\parenthesis{\vl_{i-1}, \vr_{i}} \in M \mid \vr_{i} \text{ is regular}}.
        \end{split}
    \end{equation*}
    % Notice that as $n \to \infty$ for a random $\vr_{i+1}$ with $i \in [n-1]$ we also have $x^f = \PP{\parenthesis{\vl_{i}, \vr_{i+1}} \in M \mid \vr_{i+1} \text{ is flexible}}$ and $x^n = \PP{\parenthesis{\vl_{i}, \vr_{i+1}} \in M \mid \vr_{i+1} \text{ is regular}}$ because the inclusion of $\vr_1$ does not change $x^f$ or $x^n$ asymptotically.
    % \begin{equation*}
    %     \begin{split}
    %         x^f :=&\; \PP{\parenthesis{\vl_{i-1}, \vr_{i}} \in M \mid \vr_{i} \text{ is flexible}} = \PP{\parenthesis{\vl_{i}, \vr_{i+1}} \in M \mid \vr_{i+1} \text{ is flexible}},\\
    %         x^n :=&\; \PP{\parenthesis{\vl_{i-1}, \vr_{i}} \in M \mid \vr_{i} \text{ is regular}} = \PP{\parenthesis{\vl_{i}, \vr_{i+1}} \in M \mid \vr_{i+1} \text{ is regular}}.
    %     \end{split}
    % \end{equation*}
    We now establish a system of equations to compute $(x^f_{i+1}, x^n_{i+1})$ based on $(x^f_{i}, x^n_{i})$, which hold for any $i \in \{2,...,n\}.$ 
    To compute $x^f_{i+1}$, we consider all possible scenarios in which $\vr_{i+1}$ gets matched to $\vl_i$: (1) $\parenthesis{\vl_{i-1},\vr_{i}} \in M$ and~$\parenthesis{\vl_i, \vr_{i+1}} \in E$, or (2) $\parenthesis{\vl_{i-1},\vr_{i}} \notin M$, $\parenthesis{\vl_i, \vr_{i}} \notin E$ and~$\parenthesis{\vl_i, \vr_{i+1}} \in E$. Suppose $\vl_i$ and $\vr_{i}$ are both flexible nodes; if~$\vr_{i+1}$ is also flexible then (1) occurs \emph{w.p.} $x^f_i 2 \pf$ and (2) occurs \emph{w.p.} $(1-x^f_i) (1-2\pf) 2\pf$.
    Thus, conditioned on $\vl_i$, $\vr_{i}$, and $\vr_{i+1}$ all being flexible nodes, $\parenthesis{\vl_{i},\vr_{i+1}} \in M$ with probability $x^f_i 2 \pf + (1-x^f_i) (1-2\pf) 2 \pf$.
    
%    To compute $x^f_{i+1}$, we consider the cases where $\vl_i$ and $\vr_{i}$ are flexible/regular nodes: when $\vl_i$ and $\vr_{i}$ are both flexible nodes, a flexible node $\vr_{i+1}$ will only be matched with $\vl_i$ if (1) $\parenthesis{\vl_{i-1},\vr_{i}} \in M$ and $\parenthesis{\vl_i, \vr_{i+1}} \in E$, which occurs \emph{w.p.} $x^f_i 2 \pf$, or (2) $\parenthesis{\vl_{i-1},\vr_{i}} \notin M$, $\parenthesis{\vl_i, \vr_{i}} \notin E$ and $\parenthesis{\vl_i, \vr_{i+1}} \in E$, which occurs \emph{w.p.} $(1-x^f_i) (1-2\pf) 2\pf$. Thus, conditional on $\vl_i$, $\vr_{i}$, and $\vr_{i+1}$ all being flexible nodes, $\parenthesis{\vl_{i},\vr_{i+1}} \in M$ with probability $x^f_i 2 \pf + (1-x^f_i) (1-2\pf) 2 \pf.$ 
For the cases where $\vl_i$ or $\vr_{i}$ are of other node types, we compute (conditioned on node types) the respective probabilities of the events $$\bracket{\bracket{\parenthesis{\vl_{i-1},\vr_{i}} \in M} \cap \bracket{\parenthesis{\vl_i, \vr_{i+1}} \in E}} \text{ and } \bracket{\bracket{\parenthesis{\vl_{i-1},\vr_{i}} \notin M} \cap \bracket{\parenthesis{\vl_i, \vr_{i}} \notin E} \cap \bracket{\parenthesis{\vl_i, \vr_{i+1}} \in E}}$$ accordingly and find that:
    \begin{equation}\label{eq:sol_xfn}
        \begin{split}
            x^f_{i+1} =&\; f_1(x^f_{i}, x^n_{i}):= \ql \qr \squarebracket{x^f_i 2 \pf + (1-x^f_i) (1-2\pf) 2 \pf} + (1-\ql) \qr \squarebracket{x^f_i (\pf+\pnf) + (1-x^f_i) (1-\pf-\pnf) (\pf+\pnf)}\\
            &+\ql (1-\qr) \squarebracket{x^n_i 2 \pf + (1-x^n_i) (1-\pf-\pnf) 2 \pf} + (1-\ql)(1-\qr) \squarebracket{x^n_i (\pf+\pnf) + (1-x^n_i) (1-2\pnf) (\pf+\pnf)}.
        \end{split}
    \end{equation}
    Similarly, for $x^n_{i+1}$ we find that:
    \begin{equation}\label{eq:sol_xfn2}
        \begin{split}
            x^{n}_{i+1} =&\;f_2(x^f_{i}, x^n_{i}) := \ql \qr \squarebracket{x^f_i (\pf+\pnf) + (1-x^f_i) (1-2\pf) (\pf+\pnf)} + (1-\ql) \qr \squarebracket{x^f_i 2\pnf + (1-x^f_i) (1-\pf-\pnf) 2\pnf}\\
            &+\ql (1-\qr) \squarebracket{x^n_i (\pf+\pnf)+ (1-x^n_i) (1-\pf-\pnf) (\pf+\pnf)} + (1-\ql)(1-\qr) \squarebracket{x^n_i 2\pnf + (1-x^n_i) (1-2\pnf) 2\pnf}.
        \end{split}
    \end{equation}

    Since $f_1(x^f_{i}, x^n_{i})$ and $f_2(x^f_{i}, x^n_{i})$ are linear with respect to $x^f_{i}$ and $x^n_{i}$, they are trivially continuous. Since $\ql, \qr \in [0,1]$ and $0\leq 2 \pnf < \pnf + \pf < 2 \pf \leq 1$, we also find that $\frac{\partial f_1(x^f_{i}, x^n_{i})}{\partial x^f_{i}}, \frac{\partial f_1(x^f_{i}, x^n_{i})}{\partial x^n_{i}},\frac{\partial f_2(x^f_{i}, x^n_{i})}{\partial x^f_{i}}, \frac{\partial f_2(x^f_{i}, x^n_{i})}{\partial x^n_{i}} \in [0,1).$ Thus, by applying the Banach fixed-point theorem, we know that as $i \to \infty$ the fixed-point iteration $(x^f_{i+1}, x^n_{i+1}) = \parenthesis{f_1(x^f_{i}, x^n_{i}),f_2(x^f_{i}, x^n_{i})}$ converges to the unique fixed point $(x^f, x^n)$ such that $(x^f, x^n) = \parenthesis{f_1(x^f, x^n),f_2(x^f, x^n)}$. This is a linear system of equations, the solutions of which provide the limiting values for $x^f$ and $x^n$ (see below in \eqref{eq:local_sol}). 

    We next compute $\mu(\ql,\qr)$ using $x^f$ and $x^n$. The asymptotic fraction of matched nodes is equal to the asymptotic probability that a random node $\vr_{i} \in V_r$ is matched. If $\vr_{i}$ is a flexible node, it is matched with $\vl_{i-1}$ with probability $x^f_i$ and matched with $\vl_i$ with probability $(1-x^f_i) \squarebracket{(1+\ql) \pf + (1-\ql) \pnf}$. If it is a regular node, it is matched with $\vl_{i-1}$ with probability $x^n_i$ and matched with $\vl_i$ with probability $(1-x^n_i) \squarebracket{\ql \pf + (2-\ql) \pnf}$. Thus,
    \begin{equation}\label{eq:local_sol2}
        \begin{split}
            \mu(\ql,\qr) =&\; \lim_{i \to \infty} \qr x^f_i + \qr (1-x^f_i) \squarebracket{(1+\ql) \pf + (1-\ql) \pnf}+(1-\qr) x^n_i + (1-\qr) (1-x^n_i) \squarebracket{\ql \pf + (2-\ql) \pnf}\\
            =&\;\qr x^f + \qr (1-x^f) \squarebracket{(1+\ql) \pf + (1-\ql) \pnf}+(1-\qr) x^n + (1-\qr) (1-x^n) \squarebracket{\ql \pf + (2-\ql) \pnf}.
        \end{split}
    \end{equation}

    Solving $x^f$ and $x^n$ from \eqref{eq:sol_xfn}-\eqref{eq:sol_xfn2} and plugging into \eqref{eq:local_sol2}, we obtain
    \begin{align}
        \mu(\ql,\qr) =\Biggl(&2 (\ql + \qr)^2 \pnf^4 \ql^2 - 2 (\ql + \qr)^2 \pnf^4 \ql - 8 (\ql + \qr)^2 \pnf^3 \pf \ql^2 + 8 (\ql + \qr)^2 \pnf^3 \pf \ql + 12 (\ql + \qr)^2 \pnf^2 \parenthesis{\pf}^2 \ql^2 \notag\\
        &- 12 (\ql + \qr)^2 \pnf^2 \parenthesis{\pf}^2 \ql - (\ql + \qr)^2 \pnf^2 - 8 (\ql + \qr)^2 \pnf \parenthesis{\pf}^3 \ql^2 + 8 (\ql + \qr)^2 \pnf \parenthesis{\pf}^3 \ql + 2 (\ql + \qr)^2 \pnf \pf \notag\\
        &+ 2 (\ql + \qr)^2 \parenthesis{\pf}^4 \ql^2 - 2 (\ql + \qr)^2 \parenthesis{\pf}^4 \ql - (\ql + \qr)^2 \parenthesis{\pf}^2 - 4 (\ql + \qr) \pnf^4 \ql^3 + 2 (\ql + \qr) \pnf^4 \ql^2 \notag\\
        &+ 2 (\ql + \qr) \pnf^4 \ql + 16 (\ql + \qr) \pnf^3 \pf \ql^3 - 8 (\ql + \qr) \pnf^3 \pf \ql^2 - 8 (\ql + \qr) \pnf^3 \pf \ql - 24 (\ql + \qr) \pnf^2 \parenthesis{\pf}^2 \ql^3 \notag\\
        &+ 12 (\ql + \qr) \pnf^2 \parenthesis{\pf}^2 \ql^2 + 12 (\ql + \qr) \pnf^2 \parenthesis{\pf}^2 \ql - 2 (\ql + \qr) \pnf^2 \ql + 7 (\ql + \qr) \pnf^2 + 16 (\ql + \qr) \pnf \parenthesis{\pf}^3 \ql^3 \notag\\
        &- 8 (\ql + \qr) \pnf \parenthesis{\pf}^3 \ql^2 - 8 (\ql + \qr) \pnf \parenthesis{\pf}^3 \ql + 4 (\ql + \qr) \pnf \pf \ql - 6 (\ql + \qr) \pnf \pf - 2 (\ql + \qr) \pnf \notag\\
        &- 4 (\ql + \qr) \parenthesis{\pf}^4 \ql^3 + 2 (\ql + \qr) \parenthesis{\pf}^4 \ql^2 + 2 (\ql + \qr) \parenthesis{\pf}^4 \ql - 2 (\ql + \qr) \parenthesis{\pf}^2 \ql - (\ql + \qr) \parenthesis{\pf}^2 \notag\\
        &+ 2 (\ql + \qr) \pf + 2 \pnf^4 \ql^4 - 2 \pnf^4 \ql^2 - 8 \pnf^3 \pf \ql^4 + 8 \pnf^3 \pf \ql^2 + 12 \pnf^2 \parenthesis{\pf}^2 \ql^4 - 12 \pnf^2 \parenthesis{\pf}^2 \ql^2 \notag\\
        &+ 2 \pnf^2 \ql^2 - 8 \pnf^2 - 8 \pnf \parenthesis{\pf}^3 \ql^4 + 8 \pnf \parenthesis{\pf}^3 \ql^2 - 4 \pnf \pf \ql^2 + 4 \pnf + 2 \parenthesis{\pf}^4 \ql^4 - 2 \parenthesis{\pf}^4 \ql^2 + 2 \parenthesis{\pf}^2 \ql^2\Biggr)/\notag\\
        \Biggl(&(\ql + \qr)^2 \pnf^4 \ql^2 - (\ql + \qr)^2 \pnf^4 \ql - 4 (\ql + \qr)^2 \pnf^3 \pf \ql^2 + 4 (\ql + \qr)^2 \pnf^3 \pf \ql + 6 (\ql + \qr)^2 \pnf^2 \parenthesis{\pf}^2 \ql^2 \notag\\
        &- 6 (\ql + \qr)^2 \pnf^2 \parenthesis{\pf}^2 \ql - 4 (\ql + \qr)^2 \pnf \parenthesis{\pf}^3 \ql^2 + 4 (\ql + \qr)^2 \pnf \parenthesis{\pf}^3 \ql + (\ql + \qr)^2 \parenthesis{\pf}^4 \ql^2  \notag\\
        &- (\ql + \qr)^2 \parenthesis{\pf}^4 \ql - 2 (\ql + \qr) \pnf^4 \ql^3 + (\ql + \qr) \pnf^4 \ql^2 + (\ql + \qr) \pnf^4 \ql + 8 (\ql + \qr) \pnf^3 \pf \ql^3 - 4 (\ql + \qr) \pnf^3 \pf \ql^2 \notag\\
        &- 4 (\ql + \qr) \pnf^3 \pf \ql - 12 (\ql + \qr) \pnf^2 \parenthesis{\pf}^2 \ql^3 + 6 (\ql + \qr) \pnf^2 \parenthesis{\pf}^2 \ql^2 + 6 (\ql + \qr) \pnf^2 \parenthesis{\pf}^2 \ql - 2 (\ql + \qr) \pnf^2 \ql \notag\\
        &+ 3 (\ql + \qr) \pnf^2 + 8 (\ql + \qr) \pnf \parenthesis{\pf}^3 \ql^3 - 4 (\ql + \qr) \pnf \parenthesis{\pf}^3 \ql^2 - 4 (\ql + \qr) \pnf \parenthesis{\pf}^3 \ql + 4 (\ql + \qr) \pnf \pf \ql \notag\\
        &- 2 (\ql + \qr) \pnf \pf - 2 (\ql + \qr) \parenthesis{\pf}^4 \ql^3 + (\ql + \qr) \parenthesis{\pf}^4 \ql^2 + (\ql + \qr) \parenthesis{\pf}^4 \ql - 2 (\ql + \qr) \parenthesis{\pf}^2 \ql \notag\\
        &- (\ql + \qr) \parenthesis{\pf}^2 + \pnf^4 \ql^4 - \pnf^4 \ql^2 - 4 \pnf^3 \pf \ql^4 + 4 \pnf^3 \pf \ql^2 + 6 \pnf^2 \parenthesis{\pf}^2 \ql^4 - 6 \pnf^2 \parenthesis{\pf}^2 \ql^2 + 2 \pnf^2 \ql^2 - 4 \pnf^2 \notag\\
        &- 4 \pnf \parenthesis{\pf}^3 \ql^4 + 4 \pnf \parenthesis{\pf}^3 \ql^2 - 4 \pnf \pf \ql^2 + \parenthesis{\pf}^4 \ql^4 - \parenthesis{\pf}^4 \ql^2 + 2 \parenthesis{\pf}^2 \ql^2 + 1\Biggr). \label{eq:local_sol}
    \end{align}

    Given the closed-form expression for \( \mu(\ql,\qr) \) in \eqref{eq:local_sol}, to prove \cref{thm:compare_local_model} it suffices to compute $\mu(\budget,0) - \mu(\budget/2,\budget/2)$ and verify that the difference is strictly positive. Using Wolfram Mathematica in \texttt{Theorem2.nb},\footnote{The codes can be found at \url{https://bit.ly/3uS7wW6}.} we verify that \( \mu(\budget,0) - \mu(\budget/2,\budget/2) > 0 \) for all \( \budget \in (0,1] \) and \( 0 \leq \pnf < \pf \leq~1/2\). \Halmos
% \end{proof}

\subsection{Proof of \texorpdfstring{\cref{thm:curvature_local_model}}{Lg}}\label{app:curvature_local_model}

% \begin{proof}{Proof. }
    For the concavity result, we evaluate \( \nabla_{(0,1)}^2 \mu(\ql,\qr) \) with \( \ql = 1/2 \), which is equivalent to \( \frac{\partial^2 \mu(1/2,\qr)}{\partial \qr^2} \). In \texttt{Theorem6.nb},\footnote{The codes can be found at \url{https://bit.ly/3uRh6bQ}.} we again use Wolfram Mathematica to verify that \( \frac{\partial^2 \mu(1/2,\qr)}{\partial \qr^2} < 0 \) for all \( \qr \in (0,1) \) and \( 0 \leq \pnf < \pf \leq 1/2 \). The case for the direction \( (1,0) \) is symmetric. The proof for \( \frac{\partial^2 \mu(0,\qr)}{\partial \qr^2} < 0 \) for all \( \qr \in (0,1) \) and \( 0 \leq \pnf < \pf \leq 1/2 \) follows from the same analyses by taking \( \ql = 0 \).

    The convexity result is easier to analyze by hands. We set \( \qr = 1-\ql \) to simplify \eqref{eq:local_sol} to
    $$\mu(\ql,1-\ql) = 2 \frac{\parenthesis{\pf}^2 \ql^2 - \parenthesis{\pf}^2 \ql - 2 \pf \pnf \ql^2 + 2 \pf \pnf \ql +\pf +\pnf^2 \ql^2 - \pnf^2 \ql + \pnf}{\parenthesis{\pf}^2 \ql^2 - \parenthesis{\pf}^2 \ql - 2 \pf \pnf \ql^2 + 2 \pf \pnf \ql +\pf +\pnf^2 \ql^2 - \pnf^2 \ql + \pnf + 1}.$$
    Taking the second-order derivative with respect to \( \ql \), we obtain
    $$\frac{\partial^2 \mu(\ql,1-\ql)}{\partial \ql^2} = \frac{4 \parenthesis{\pf-\pnf}^2 \underbrace{\squarebracket{-\parenthesis{\pf}^2 - 3 \ql^2 \parenthesis{\pf-\pnf}^2 + 3 \ql \parenthesis{\pf - \pnf}^2 + 2 \pf \pnf + \pf - \pnf^2 +\pnf+1 }}_{(I)}}{\parenthesis{\underbrace{\ql^2 \parenthesis{\pf -\pnf}^2 - \ql \parenthesis{\pf -\pnf}^2 + \pf+\pnf + 1}_{II}}^3}.$$
    We demonstrate that both \( (I) \) and \( (II) \) are strictly positive for all \( \ql \in (0,1) \) and \( 0 \leq \pnf < \pf \leq 1/2 \), ensuring \( \frac{\partial^2 \mu(\ql,1-\ql)}{\partial \ql^2} > 0 \). For $(I)$ we have
    \begin{align*}
        (I) =&\; 3 \ql (1-\ql) \parenthesis{\pf - \pnf}^2 - \parenthesis{\pf - \pnf}^2 + \pf + \pnf + 1\\
        =&\;\parenthesis{3 \ql - 3\ql^2 - 1} \parenthesis{\pf -\pnf}^2 + \pf + \pnf + 1\\
        \geq & - \parenthesis{\pf - \pnf}^2 + \pf + \pnf+ 1 \geq -0.25 + \pf + \pnf+ 1 > 0,
    \end{align*}
    where the first inequality comes from $\ql - \ql^2 \geq 0$ and the second from $\parenthesis{\pf - \pnf}^2 \leq 0.25.$
For $(II)$ we have $$(II) = (\ql^2 -\ql) \parenthesis{\pf - \pnf}^2 + \pf + \pnf + 1 \geq -0.25 + \pf + \pnf + 1 > 0,$$ where the first inequality comes from $\ql^2 - \ql \geq -1$ and $\parenthesis{\pf - \pnf}^2 \leq 0.25.$ (since $\pf\leq 1/2$). Since \( \nabla_{(1,-1)}^2 \mu(\ql,\qr) \) with \( \ql + \qr = 1 \) is equivalent to \( \frac{\partial^2 \mu(\ql,1-\ql)}{\partial \ql^2} > 0 \), we conclude the proof.\Halmos

% \end{proof}

\section{Additional Simulation Results}\label{sec:simu_additional}

\dfedit{In this section, we consider an imbalanced model that extends the spatial model in \cref{sec:spatial}. Assuming that there are $n$ drivers and $\lambda \cdot n$ riders, we again find that such an imbalance strengthens the case for the one-sided allocation because the allocation $(B,0)$ creates more flexible agents (in expectation) than any other allocation. As illustrated in \cref{fig:ratio4_additional} (a) and (b), this makes the better of the one-sided allocations outperform the balanced allocation for any sufficiently large $\anf$ and $\af$, regardless of whether $B < 1$. %Thus, the asymmetry effect disappears, while the cannibalization effect remains. 
At the same time, we still observe the balanced allocation outperforming the one-sided allocation in the regime with very small $\anf,\af$, where the number of edges is much larger for the balanced allocation (see \cref{sec:spatial}).}

% In this section we consider an alternative model of an imbalanced market in which a flexibility allocation $(\ql,\qr)$ means that drivers and riders are flexible with probability $\ql$ and $\qr,$ respectively. This modeling approach guarantees that the expected number of edges for flexibility allocations $(B,0)$ and $(0,B)$ are the same, though they result in a different expected number of flexible nodes. Indeed, with this modeling approach the one-sided allocation $(B,0)$ produces more flexible nodes in expectation than any other flexibility design with the same flexibility budget~$B,$ and this leads to a significant advantage for the one-sided allocation. \cref{fig:ratio4_additional} (a) and (b) show that the additional flexible nodes ensure that the one-sided allocation $(B,0)$ outperforms the balanced allocation for any sufficiently large $\anf$ and $\af$, regardless of whether $B < 1,$ overshadowing the flexibility asymmetry effect. Thus, the asymmetry effect disappears, while the cannibalization effect remains. At the same time, we still observe the balanced allocation outperforming the one-sided allocation in the regime with very small $\anf,\af$, where the number of edges is much larger for the balanced allocation (see \cref{sec:spatial}).

\begin{figure}[H]%
    \centering
    \subfloat[\centering $B = 0.6$]{{\includegraphics[width=0.45\textwidth]{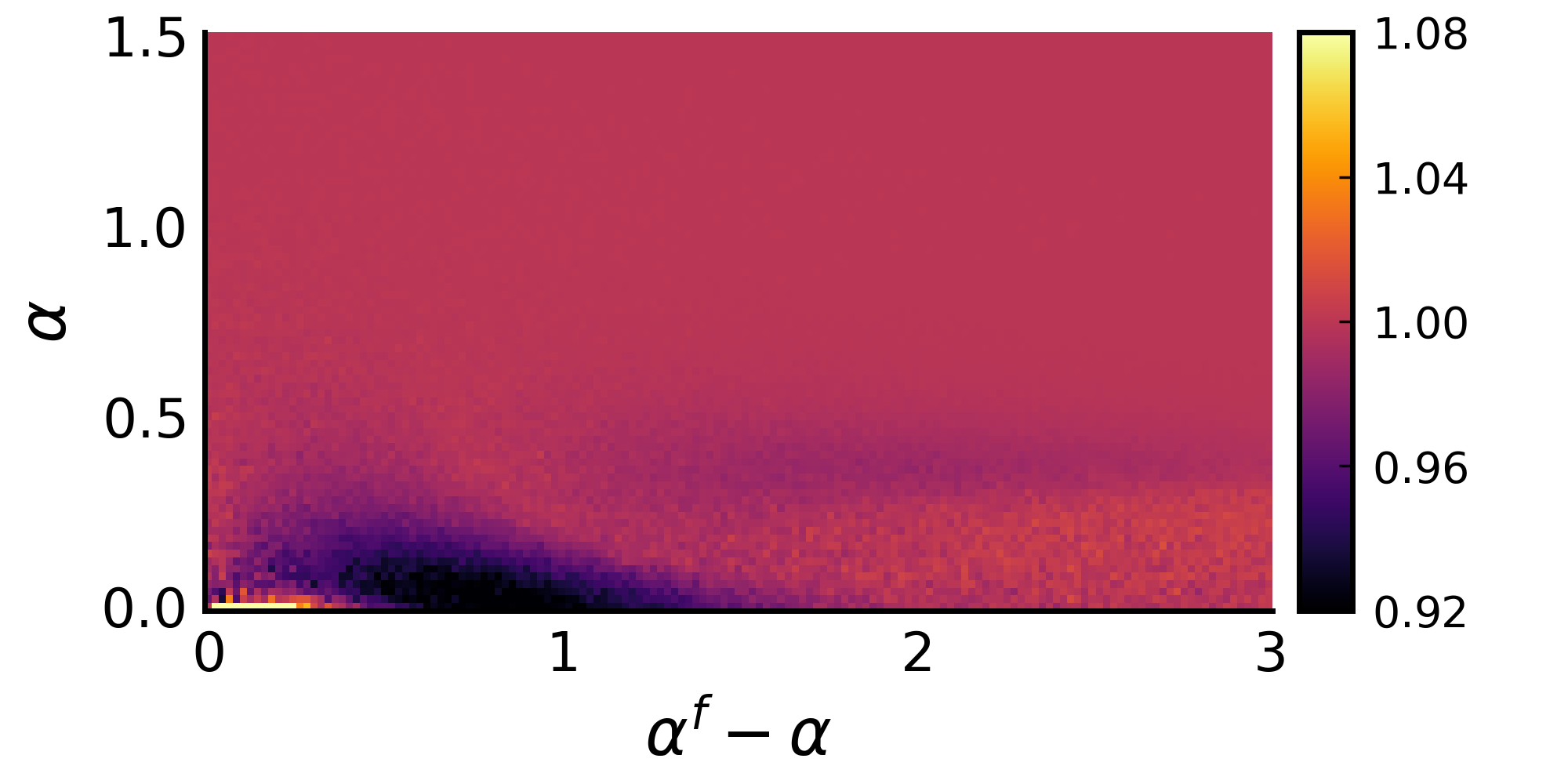} }}%
    \hspace{0cm}
    \subfloat[\centering $B = 1$]{{\includegraphics[width=0.45\textwidth]{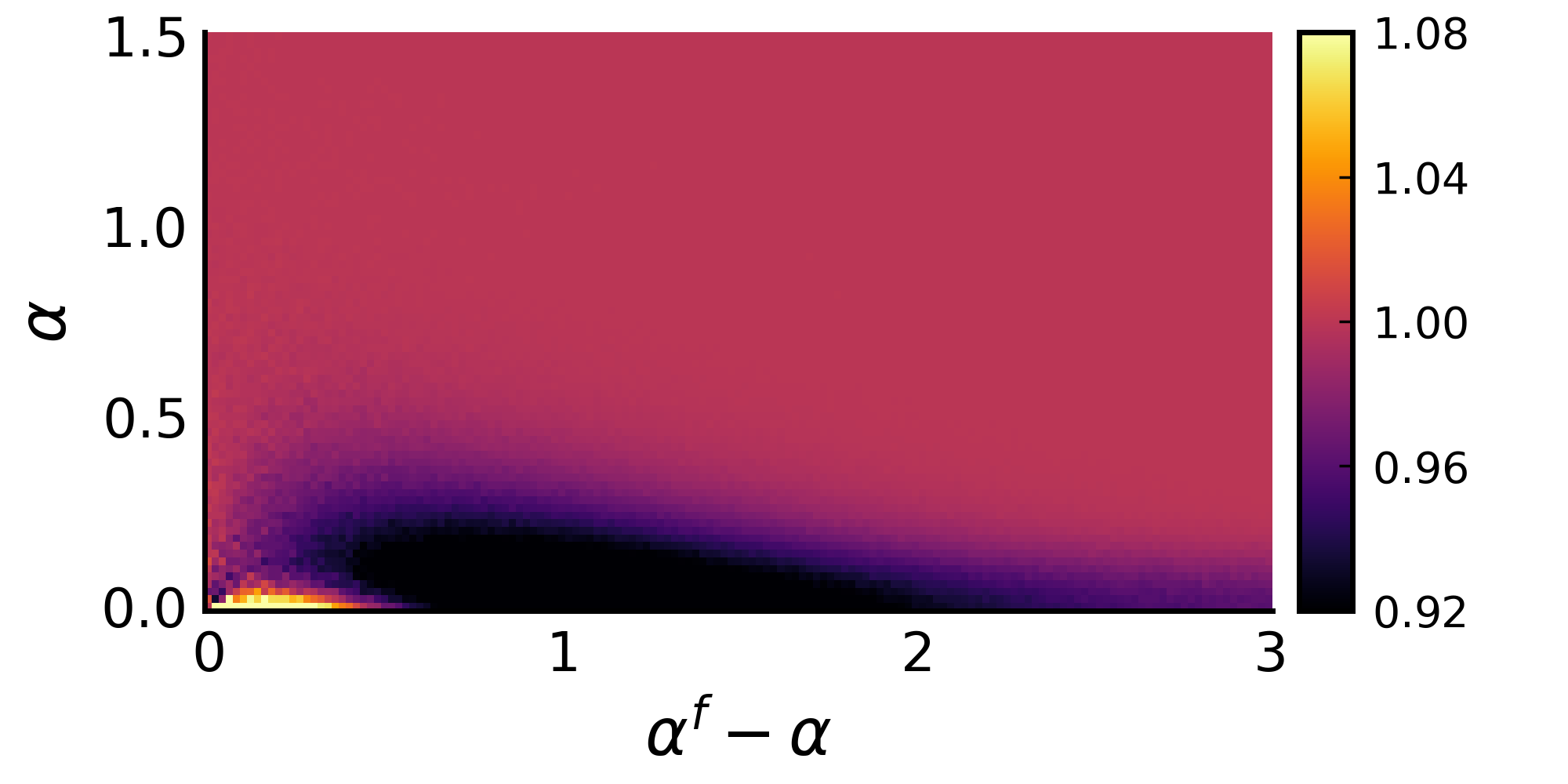} }}%
    \caption{Values of $\frac{\muemp(\budget/2,\budget/2)}{\max\bracket{\muemp(\budget,0),\muemp(0,\budget)}}$ across varying $\anf$ and $\af -\anf$ when $\lambda = 0.8.$}
    \label{fig:ratio4_additional}
\end{figure}

}
\end{APPENDICES}

\end{document}